%% file: Main.tex
\documentclass[
12pt,		
openright,	
twoside,  
a4paper,			
chapter=TITLE,		
brazil,			
french,				
spanish,			
english				
]{USPSC}

\usepackage[section]{placeins} 
\usepackage{listings} 
\usepackage{xcolor}

\definecolor{codegreen}{rgb}{0,0.6,0}
\definecolor{codegray}{rgb}{0.5,0.5,0.5}
\definecolor{codepurple}{rgb}{0.58,0,0.82}
\definecolor{backcolour}{rgb}{0.95,0.95,0.92}

\lstdefinestyle{mystyle}{
  backgroundcolor=\color{backcolour},   commentstyle=\color{codegreen},
  keywordstyle=\color{magenta},
  numberstyle=\tiny\color{codegray},
  stringstyle=\color{codepurple},
  basicstyle=\ttfamily\footnotesize,
  breakatwhitespace=false,         
  breaklines=true,                 
  captionpos=b,                    
  keepspaces=true,                 
  numbers=left,                    
  numbersep=5pt,                  
  showspaces=false,                
  showstringspaces=false,
  showtabs=false,                  
  tabsize=2
}
\usepackage{comment} 
\usepackage{physics}            
\usepackage[T1]{fontenc}		
\usepackage[utf8]{inputenc}		
\usepackage{lmodern}			
\usepackage{lastpage}			
\usepackage{indentfirst}		
\usepackage{color}				
\usepackage{graphicx}			
\usepackage{float} 				
\usepackage{mathtools}
\usepackage{amssymb, amsmath, amsbsy} 
\usepackage[scr=rsfs]{mathalpha} 
\usepackage{xfrac}
\usepackage{bbold}              
\usepackage{empheq}             
\usepackage{cancel}             

\allowdisplaybreaks[1]

\usepackage{microtype} 			
\usepackage{pdfpages}
\usepackage{makeidx}            
\usepackage[hang,small,labelsep=endash]{caption} 

\usepackage{tikz}
\usetikzlibrary{decorations.pathmorphing,patterns}
\usetikzlibrary{patterns,decorations.pathmorphing}
\usetikzlibrary{fadings,arrows,decorations.markings}
\usetikzlibrary{quotes,angles}

\usepackage{graphicx}
\usepackage{subfig}
\usepackage{braket}

\usepackage{lscape}
\usepackage{makecell}

\usepackage{afterpage}

\usepackage{cite} 

\renewcommand{\footnotesize}{\small} 

\usepackage{lipsum}				

\usepackage{multicol}	
\usepackage{multirow}	
\usepackage{longtable}	
\usepackage{threeparttablex}    
\usepackage{array}
{%
	\begin{labeling}
		{\rsnumber{R39/23/24/25}}
	}{%
\end{labeling}%
}%

\siglaunidade{IFSC} 
\programa{MFB}
\definecolor{blue}{RGB}{41,5,195}

\makeatletter
\hypersetup{
	pdftitle={\@title}, 
	pdfauthor={\@author},
	pdfsubject={\imprimirpreambulo},
	pdfcreator={LaTeX with abnTeX2},
	pdfkeywords={abnt}{latex}{abntex}{USPSC}{trabalho acadêmico}, 
	colorlinks=true,       		
	linkcolor=blue,          	
	citecolor=blue,        		
	filecolor=magenta,      		
	urlcolor=blue,
	bookmarksdepth=4
}
\makeatother

\makeindex

\begin{document}

\frenchspacing 

\renewcommand{\ABNTEXchapterfontsize}{\fontsize{12}{12}\bfseries}
\renewcommand{\ABNTEXsectionfontsize}{\fontsize{12}{12}\bfseries}
\renewcommand{\ABNTEXsubsectionfontsize}{\fontsize{12}{12}\normalfont}
\renewcommand{\ABNTEXsubsubsectionfontsize}{\fontsize{12}{12}\normalfont}
\renewcommand{\ABNTEXsubsubsubsectionfontsize}{\fontsize{12}{12}\normalfont}

\imprimircapa
\imprimirfolhaderosto*
\includepdf{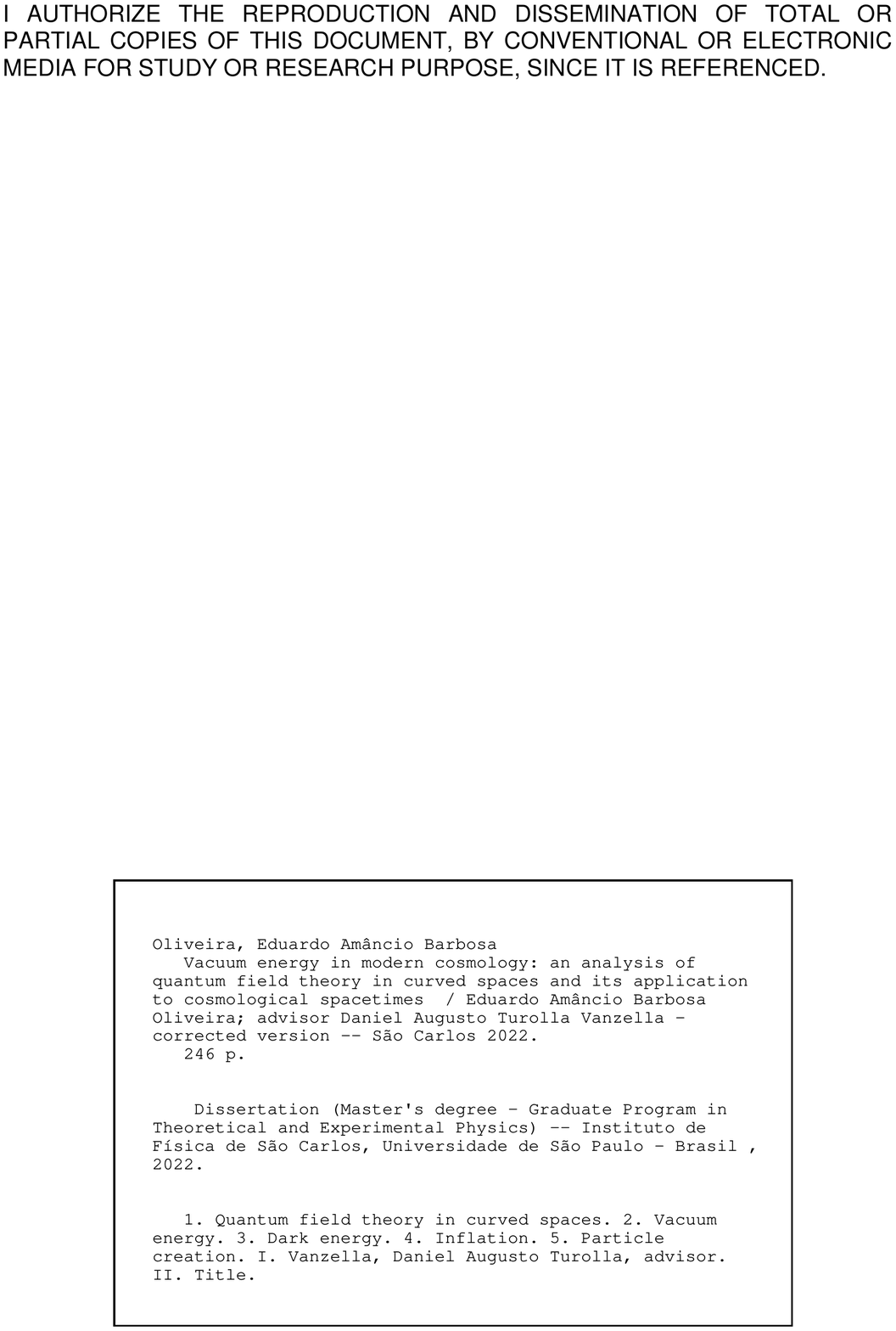}
\include{chap-Agradecimentos}
 
\begin{epigrafe}
    \vspace*{\fill}
	\begin{flushleft}
		\textit{``Era ainda jovem demais para saber \\
		que a memória do coração elimina as más lembranças e enaltece as boas \\
		e que graças a esse artifício conseguimos suportar o passado. \\
		Mas quando voltou a ver do convés do navio \\
		o promontório branco do bairro colonial, \\
		os urubus imóveis nos telhados, \\
		a roupa dos pobres estendida a secar nas sacadas,\\ compreendeu até que ponto tinha sido uma vítima fácil das burlas caritativas da saudade.''}\\ \vspace{-8pt}
    \end{flushleft}
    \begin{flushright}
		Gabriel García Marquez
	\end{flushright}
\end{epigrafe}

%

\include{chap-abstract}
\include{chap-Resumo}





%

\begin{simbolos}
  \item[$\hbar,c,G$] Fundamental physical constants. In this work, we employ natural units, $\hbar=c=1$, in chapters 1-4, and Planck units, $\hbar=c=G=1$, in chapter 5.
  \item[$v^a$, $\omega_a$] Spacetime vectors, dual vectors (covectors) and general tensors are written in the abstract index notation. Abstract indices will be latin letters ranging from \textit{a} to \textit{h}; vectors are denoted with a single upper index, dual vectors with a single lower index, and general tensors may have multiple upper and lower indexes. Exceptionally, we also use a single latin subscript to compactly denote multiple or composite fields: $\phi_a$.
   \item[$v^\mu$, $\omega_\mu$] Greek indexes denote general vector/tensor components, each running from $0$ to $n$, where $n$ denotes the spacetime dimension. Latin indexes from $i$ ownwards denote spatial components, running from $1$ to $n$. Following the Einstein notation, repeated indices denote an implicit sum, unless otherwise stated.
  \item[$ g_{ab} $] Spacetime metric. In this work, we use the metric signature of Birrell \& Davies \cite{birrell}: $(+,-,...-)$. The flat Minkowski metric is denoted by $\eta_{ab}$ and its components in a Global Inertial Frame are $\eta_{\mu\nu} = \operatorname{diag}(1,-1,...-1)$.
  \item[$ \mu $] The greek letter $\mu$ is often used to denote a measure on an arbitrary measure space, such as the spectrum of a linear operator $H$, $\sigma(H)$, or a manifold $\mathcal{M}$. For a metric manifold $(\mathcal{M},g_{ab})$, $d\mu_g(x)$ will denote the natural volume element on $\mathcal{M}$ induced by $g_{ab}$.
  \item[$ T_{(ab)}, T_{[ab]} $] Curly and square brackets are respectively used to denote complete symmetrization and antissymetrization over the encompassed indices. For example, $T_{(ab)}\equiv \frac{1}{2}(T_{ab}+T_{ba})$ and $T_{[ab]}\equiv \frac{1}{2}(T_{ab}-T_{ba})$.
  \item[$\pounds_v$] Lie derivative with respect to a vector field $v^a$.
  \item[$\mathbf{x}$, $\mathbf{y}$] Boldface letters denote vectors in $\mathbb{R}^n$, particularly, ordinary 3-dimensional spatial vectors. From chapter 3 onward, they are often used to compactly denote a set of spatial coordinates $\mathbf{x} \equiv \{x^i\}_{i=1,2,3}$ \textit{even when spatial surfaces $\Sigma \subset \mathcal{M}$ do not possess a linear structure.}
  
\end{simbolos}
\pdfbookmark[0]{\contentsname}{toc}
\tableofcontents*
\cleardoublepage
\textual

\include{chap1}
\include{chap2}
\include{chap3}
\include{chap4}

\include{chap5}
\include{chap6}

\postextual

%
%

\renewcommand\bibname{References}

\include{Apendices}
%
%



\end{document}

%% file: chap-Agradecimentos.tex
\begin{agradecimentos}

Atravessar dois anos de mestrado em meio a uma pandemia que virou todas as nossas vidas do avesso foi uma experiência singular e extremamente desafiadora, e certamente não teria sido possível sem o suporte das muitas pessoas que conviveram e estiveram comigo nesse período tão peculiar. Deixo aqui os meus agradecimentos a algumas delas:

Aos meus pais, Amoacy e Ludmila, pelo apoio incondicional que me permitiu chegar aqui.

Ao meu orientador, Daniel Vanzella, por toda a física que me ensinou desde os anos de graduação, pela oportunidade de mestrado num tema tão interessante, pelas minunciosas discussões e observações, e pela desproporcional confiança depositada em mim ao longo desses últimos 2 anos. E ao meu ex-orientador, Emanuel Henn, pelos ensinamentos edificantes nos primeiros passos da minha vida acadêmica e por todo o apoio posterior.

Ao Japonês, pela grande ajuda com a formatação do trabalho, e pelas das cautelosas e mascaradas aventuras.

Ao Pestana e ao Hot Wheels, pelas longas pedaladas pelas ruas, trilhas e cachoeiras de São Carlos, que tanto contribuíram para o meu bem-estar e sanidade nesse período apocalíptico.

Ao Felipe (Neves), por infernizar a minha vida como ninguém.

À Giórgia, pelo carinho, suporte, e por todo o escopo de experiências e memórias que ajudaram a tornar esses turbulentos anos tão especiais.

E ao Momo (Buendía, Momão), meu onipresente companheiro de pandemia, com quem pude contar em todos os momentos de necessidade, num período em que necessidades não faltaram. Por toda a amizade e companheirismo, pelas irregulares jogatinas e escassas aventuras, e pelo suporte tecnológico e humano para produzir 13 das figuras que ilustram esse texto, meu muito obrigado.

Este trabalho foi financiado pelo Centro Nacional de Desenvolvimento Científico e Tecnológico, CNPq, no processo de número 131012/2020-7. 

\end{agradecimentos}

%% file: chap-abstract.tex
\begin{resumo}[Abstract]
 \begin{otherlanguage*}{english}
	\begin{flushleft} 
		\setlength{\absparsep}{0pt} 
 		\SingleSpacing 
 		\imprimirautorabr~~\textbf{\imprimirtitleabstract}. \imprimirdata.  \pageref{LastPage}p. 
		Dissertation (Master in Science)~-~\imprimirinstituicao, \imprimirlocal, 	\imprimirdata. 
 	\end{flushleft}
	\OnehalfSpacing

The last decades have witnessed an unprecedented advancement in our knowledge of the large scale universe. In particular, increasingly accurate cosmological observations have allowed us to discover a form of ``dark energy'', which presently dominates the expansion of the universe -- making it accelerated. On the other hand, fundamental problems in the standard ($\operatorname{\Lambda CDM}$) cosmological model point towards the possibility of a primordial inflationary period. Both these expansion phases have in common the fact that they should be governed by forms of energy with properties much similar to those of vacuum energy of classical or quantum fields. In the meanwhile, quantum field theory in curved spaces (QFTCS) has proved a rich framework to analyze phenomena of a quantum nature in regimes where spacetime curvature is relevant, but not too extreme, and, particularly, it yields novel insights on the structure and dynamics of quantum vacuum. In this dissertation, we make a thorough exposition of the fundamentals of QFTCS and present some of its applications in cosmological spacetimes. Particular attention is given to the construction of an empirical notion of particles through an idealized model of particle detectors, and to the phenomenon of particle creation in expanding FLRW spacetimes. Further, we develop the procedure of adiabatic renormalization, and use it to compute the renormalized stress tensor in these spacetimes. For a noninteracting scalar field in exponentially expanding (de Sitter) spaces, we find that these results take the form of a cosmological constant, although a quantitatively self-consistent value with the background expansion can only be found at Planckian densities. We also present a construction of a simple inflationary model, driven by a self-interacting classical scalar field, and show how the quantized fluctuations of this field could give rise to a nearly scale-invariant power spectrum, like the one that is currently observed in the Cosmic Microwave Background.

\vspace{\onelineskip}
 \end{otherlanguage*}
\end{resumo}

%% file: chap-Resumo.tex
\begin{resumo}[Resumo]
\begin{flushleft} 
\setlength{\absparsep}{0pt} 
\SingleSpacing 
\imprimirautorabr~ 
~\textbf{Energia de Vácuo na Cosmologia Moderna:} uma análise dos fundamentos de teoria quântica de campos em espaços curvos e suas aplicações a espaçostempos cosmológicos. \imprimirdata. \pageref{LastPage}p. 
\imprimirtipotrabalho~-~\imprimirinstituicao, \imprimirlocal, \imprimirdata. 
\end{flushleft}
\OnehalfSpacing 		

\selectlanguage{portuguese}

As últimas décadas testemunharam um avanço sem precedentes no nosso conhecimento do universo em larga escala. Em particular, medidas cosmológicas cada vez mais precisas nos permitiram descobrir uma forma de ``energia escura'', que atualmente domina a expansão do universo -- tornando-a acelerada. Por outro lado, problemas fundamentais no modelo cosmológico padrão ($\operatorname{\Lambda CDM}$) apontam para a possibilidade de um período inflacionário primordial. Ambas essas fases de expansão têm em comum o fato de que elas deveriam ser governadas por formas de energia com propriedades muito similares àquelas da energia de vácuo de campos clássicos ou quânticos. Enquanto isso, teoria quântica de campos em espaços curvos (TQCEC) se mostrou um rico paradigma para analisar fenômenos de natureza quântica em regimes onde a curvatura do espaçotempo é relevante, mas não demasiado extrema, e, particularmente, ela provê novos \textit{insights} sobre a estrutura e a dinâmica do vácuo quântico. Nesta dissertação, nós fazemos uma exposição detalhada dos fundamentos de TQCEC e apresentamos algumas das suas aplicações a espaçostempos cosmológicos. Particular atenção é dada à construção de uma noção empírica do conceito de partícula através de um modelo idealizado de detectores de partículas, e ao fenômeno de criação de partículas em espaçostempos de FLRW em expansão. Ademais, desenvolvemos aqui o procedimento de renormalização adiabática, e o usamos para computar o tensor energia-momentum renormalizado nesses espaçostempos. Para um campo escalar livre em espaços em expansão exponencial (espaços de de Sitter), encontramos resultados na forma de uma constante cosmológica; esta, todavia, só apresenta um valor quantitativamente autoconsistente com a expansão cósmica de fundo em escalas planckianas. Também apresentamos a construção de um modelo inflacionário simples, governado por um campo escalar clássico autointeragente, e mostramos como as flutuações quantizadas desse campo podem dar origem a um espectro aproximadamente invariante de escala, como o que é atualmente observado na Radiação Cósmica de Fundo.

\textbf{Palavras-chave}: Teoria quântica de campos em espaços curvos. Energia de vácuo. Energia escura. Inflação. Criação de partículas.

\selectlanguage{english}

\end{resumo}

%% file: chap1.tex
\chapter{Introduction}\label{intro}

The last 100 years have witnessed an unprecedented advancement in our understanding of the large scale universe. Einstein's theory of General Relativity -- which radically changed the way we see spacetime, providing it with a dynamical character --, along with Hubble's first observations of departing galaxies \cite{hubble}, paved the way for the realisation (and eventual scientific consensus) that universe itself is in expansion. Already in 1920's and 30's emerged the so-called Friedman-Lemâitre-Robertson-Walker (FLRW) models\footnote{These were found independently and complementarily by many authors; for the seminal works of the 4 mentioned above, see \cite{friedman,lemaitre,robertson,walker}. }; based on the simplifying hypotheses of a spatially homogeneous and isotropic spacetime (which turns out to be quite accurate in very large scales) these models shaped much of the development of cosmology throughout the 20th century, laying the fundamentals for predictions about the universe's large scale structure and evolution based on a few of its average properties and parameters. But it was particularly in the last two or three decades that observation techniques and technologies were sufficiently developed to allow precise measurements of cosmological parameters -- bringing uncertainties that were often of the same order (or grater!) than the measured values themselves down to just a few percentile points -- and put tighter constraints in our models, allowing more rigorous consistency tests, as well as new, more precise and specific predictions.

Up to this date, the standard cosmological model (also known as the $\operatorname{\Lambda CDM}$ model) describes our cosmological observations with astounding precision\footnote{Although, more recently, interesting problems are starting to arise due to increasingly precise measurements, the most prominent of which is the $H_0$ tension. \cite{eleonora}}, relying only in very few fundamental assumptions and parameters. However, in spite of being so observationally successful, this model suffers from many fundamental problems. These precise measurements of cosmological parameters allowed us to infer (assuming that General Relativity holds accurately in very large scales) much information about the matter and energy content of the universe. Astonishingly, the vast majority of the energy in the universe does not seem to be in the form of any known matter (which only seems to amount to about 5\% of it), but rather in forms that we have only been able to detect gravitationally -- and thus, particularly, that do not interact with light --, composing the so-called dark sector. This sector is divided in two major components: on relatively small (astrophysical) scales, the observed behaviour of massive matter structures (particularly, the rotation velocities of galaxies and peculiar velocities of galaxy clusters), as well as the formation of structure itself, requires for the presence of \emph{Cold Dark Matter} (CDM), which composes around 25\% of the total energy content; although exotic and not directly observable, dark matter seems to behave rather regularly gravitationally (in a much similar manner to ordinary baryonic matter). On very large (cosmological) scales, however, the expansion of the universe seems to be presently dominated by a much stranger energy form, the so-called \emph{Dark Energy}, which comprises the remaining 70\% ; it is not only undetectable through any nongravitational means, but it also (i) does not seem to form any types of structures, being distributed in a highly homogeneous way throughout the universe, and (ii) presents extremely negative pressures, with magnitude comparable to that of its energy density, which results in an effectively \textit{repulsive} gravitational behaviour, causing the present cosmic expansion to be \textit{accelerated}, rather than decelerated. Such exotic properties, although notably alien to most known physical systems, do appear naturally elsewhere: the vacuum energy of quantum (and classical) fields. As we shall see later in the present work, it is not unusual for renormalization procedures to yield negative expectation values of energy and/or pressure, even when these are classically positive-definite. Furthermore, it is a form of energy that permeates all space, so that it should be quite natural for it to be homogeneous and devoid of structure\footnote{
 This is indeed found to be the case for simple noninteracting fields \emph{in homogeneous and isotropic spacetimes}, but, as we shall see throughout this dissertation, we should not underestimate how complex vacuum can be.}.
  Finally, although there are numerous difficulties and indeterminacies in the calculation of renormalized energy, momentum and stress observables, the simplest form these quantities can take is precisely that of a cosmological constant $\Lambda$,  which is exactly the form that Dark Energy seems to take.
 
To make its matters worse, the $\operatorname{\Lambda CDM}$ model is also full of `coincidences' that are very difficult to explain from first principles. First and foremost, why is the universe so spatially homogeneous? And notably so in the past, before matter gravitationally collapsed and formed astrophysical structures; particularly, back in the Cosmic Microwave Background (CMB) formation the fluctuations in density and temperature were extremely small, with relative anysotropies of the order of $\mathcal{O}(10^{-5})$. In the standard Hot Big Bang scenario (in which the radiation era extends all the way back to a primordial singularity -- the Big Bang) the patches of the sky that are causally connected should be no larger than about $2^\circ$. If widely separated portions of this early plasma have not had time to thermalize, and come to an equilibrium density, how come do they have such astoundingly similar temperatures and densities?\footnote{Well, one could say they were just extremely homogeneous and uniform to start with. Although not impossible, this extreme level of \textit{fine-tuning} in initial conditions makes for an arguably implausible and artificial explanation. We shall discuss the matter of initial conditions in further detail in section \ref{cinflation}.}
 Furthermore, among all the possible values for spatial curvature in a homogeneous isotropic universe why is it so close (if not exactly equal) to $0$? The matter becomes particularly acute when we note that any nonzero values of curvature (to which we can associate an effective energy density) tend to rapidly become dominant over ordinary matter components with nonnegative pressures\footnote{Of course, this does not apply to dark energy. However, extremely small values of curvature in the past would have made it become dominant on time scales many orders of magnitude below than those for which DE became relevant.}; if we were to adjust initial conditions at the Planck time $t_p \sim 10^{-43}s$, one would have to fine-tune the matter density to its so called critical value $\rho_c$ in about 1 part in $10^{-41}$ so that universe would remain nearly flat up to the present time. One way to address all of these issues is by postulating a very brief primordial inflationary phase, which would have lasted about $\sim \!10^{-33}s$ and during which the universe would have undergone an extremely fast and accelerated expansion, inflating by a factor of at least about $10^{26}$. Such a wild proposal is, of course, extremely hard to probe and highly open to speculation. Nevertheless, it is quite widely accepted, since it provides a unified solution to many issues, as well as an arguably natural framework for studying the CMB fluctuation spectrum as due to primordial vacuum fluctuations of an inflation-driving field (so-called inflaton), stretched to wavelengths greater than the Hubble horizon during inflation. Not surprisingly, in order to bring about this primordial period of accelerated expansion, one needs a form of energy with peculiar properties much similar to those of dark energy, which could also be encompassed by vacuum energy.

In the meanwhile, quantum field theory in curved spacetimes (QFTCS) has proved a very rich and profound paradigm to analyze many intrinsically quantum phenomena in regimes where the spacetime curvature is relevant, but not too extreme (below Planck scales). In this approach, one avoids the (so far overwhelming) difficulties for obtaining a full quantum theory of matter and gravity, by quantizing only matter fields in classical curved background geometries. One of its most notable achievements is the prediction that Black Holes, rather than being perfectly opaque, actually emit thermal radiation -- the well-known Hawking radiation \cite{hawking} --, and can be meaningfully assigned with both temperature and entropy\footnote{ 
 Although these quantities are already computable in the classical framework of GR \cite{wald} (up to a multiplicative constant in each, which is ultimately fixed by $\hbar$, but which disappears in the product $TS$), their physical meaning hardly seems to transcend a mere thermodynamical \emph{analogy} before taking quantum effects in consideration.}.
 Equally noteworthy is the discovery of a flat space analogous to this thermal radiation, the Unruh effect \cite{unruh}. These effects turn out to provide deep insights in the nature of the quantum vacuum, and reveal that the concept of particles is considerably more malleable and observer-dependent than one could intuitively conceive.
 
 In a cosmological context, the theory's novel features regarding vacuum energy provide a variety of theoretical possibilities to investigate dark energy, as well as the very early universe, particularly, a primordial inflationary period and some of its observational consequences today. The estimates that we can make for the necessary duration of inflation indicate that at least a significant portion of it should have occurred in extreme, but yet sub-Planckian regimes, where QFTCS is expected to hold\footnote{
As we shall discuss in further detail in chapter \ref{cosmology}, this restriction is somewhat tautological with our initial assumptions that our models are describable in terms of classical spacetimes, and ultimately reflects our ignorance both regarding the very early universe and what a quantum theory of gravity and matter should be like. Nonetheless, there are reasonable and \textit{self-consistent} arguments that inflation should indeed last up until $\sim\! 10^{-33}s$, quite far from Planckian regimes. If one wishes to go beyond a semiclassical approach, he/she is forced to struggle with the far more intricate problem of finding an adequate and computable theory of quantum gravity; a thorough and up-to-date account of efforts in this sense can be found in \cite{QGreview} (particularly in section 1.3.3, regarding cosmological observables) and references therein.}. 
 In fact, one can find reasonable models for inflation whose average behaviour can be described in terms of the vacuum energy of classical fields. However, considering the quantization of these fields, one is able to find a considerably richer structure for this vacuum energy; among other things, it allows one to draw sensible predictions for the very small fluctuations that we observe in the CMB today.

In the present work, we attempt at providing a thorough and comprehensive exposition of QFTCS, particularly on vacuum energy and its applications in cosmological contexts. The dissertation is divided as follows: in Chapter 2, we lay some fundamentals of Quantum Field Theory (QFT), first taking a section to introduce the subject of Classical Field Theory, upon which it builds, and then providing a brief but fairly comprehensive description of the process of canonical quantization for continuous systems. There, we take the chance to explore some nontrivial effects of vacuum energy that already appear in flat space in the paradigmatic example of the Casimir Effect, and show a first example of renormalization in this simpler context. We also present some formal aspects and apparatus of the theory that will be useful on later discussions. In Chapter 3, after reviewing some general features of curved spacetime and General Relativity, we generalize the procedures presented in the previous chapter to curved, globally hyperbolic spacetimes, laying the basic formulation of QFTCS. Thereupon, we make a thorough discussion of some of its basic features, particularly analyzing the concepts of vacuum and particles, and exploring some novel aspects in the theory's phenomenology, with special attention to the creation of particles in expanding FLRW spacetimes. We then present the notion of adiabatic vacuum, which arises in the analysis of the limits of an infinitely slow expansion, for which particle creation is suppressed and one may obtain a physically meaningful (approximate) notion of vacuum state, analogous to that in Minkowski spacetime. Chapter 4 is concerned with the more technical and convoluted problem of renormalization, which is essential to make physical sense of divergent quantities of the theory, and allow for a number of physically meaningful predictions; particularly it is necessary to obtain finite expectation values for the vacuum energy in curved spacetimes, and thus to analyze its potential effects on the dynamics of the universe. In Chapter 5, we dwell in the subject of cosmology. First, we present some basic features of standard cosmology, its observational successes, and its fundamental issues, both showing the scientific motivation and constructing the necessary framework to introduce and discuss inflation. Next, we show how field theory can account for a finite primordial inflationary period, and comment briefly on symmetry breaking and the roles it could play in the early universe. Finally, we then present basic aspects of inflationary cosmology and how it addresses the problems of standard cosmology; we concretely illustrate some quantitative features of inflation in a simplified model, within the so-called chaotic inflation scenario \cite{andrei}, showing how its average dynamics can give rise to an exponentially expanding phase, and how its quantized fluctuations can give rise to a (nearly) scale-invariant power-spectrum, as we observe in the CMB today. Finally, in chapter 6, we summarize a few conclusions of this work, and make our final remarks regarding the perspectives on this fascinating subject.

Also, for completeness and to keep this text as self-contained as possible, we summarize a few relevant results in the subject of distributions in Appendix \ref{distributions}, and develop some geometrical derivations for curved spaces in Appendix \ref{geometry}.

%% file: chap2.tex
\chapter{Fundamentals of QFT in Minkowski Spacetime} \label{QFTMS}

Throughout this work, we shall be primarily concerned with quantum field theory in curved spacetimes (QFTCS) and some of its cosmological consequences. In many aspects, this theory arises as straightforward generalization of the more well stablished quantum field theory (QFT) in Minkowski spacetime. Thus, we find it constructive and pedagogical to introduce many of the concepts and techniques in this simpler and more familiar framework before diving in QFTCS.

We begin in section \ref{CFT} with a brief outline of classical field theory in Minkowski spacetime, both sketching its similarities with discrete particle mechanics  -- which will later ease the description of canonical quantization, drawing analogies with these simpler systems --, and introducing a few tools required for handling continuous systems, with special emphasis in distributions and functional derivatives.

In section \ref{CQion} we review the procedure of Canonical Quantization for particle systems, and directly generalize it to field theories. The latter is then exemplified in the paradigmatic example of a real scalar field, where we explore the decomposition field modes and deduce the pivotal commutation relations for the creation and ahnilation modes $a_\mathbf{k}$, $a_\mathbf{k}^\dagger$. Further, we write energy-momentum observables in terms of these modes and show how vacuum energy already presents divergences in Minkowski space.

Then, in section \ref{Casimireff}, we explore nontrivial vacuum effects on flat space by means of the paradigmatic Casimir Effect, carrying simpler procedures of regularization and subtraction to obtain a finite, renormalized vacuum energy; we then interpret our results physically.

Finally, in sections \ref{NormalModes} and \ref{2-PFs}, we go over some technical details in the expansions of field solutions in normal modes and, in the light of these results, formally construct and interpret many elementary Green Functions of our theory and draw their connection to vacuum expectation values of two-point functions. Both sections rely heavily on the technical apparatus of Appendix A and, although they are not essential for most direct calculation in chapter 3, they should render the subject of QFT more conceptual clarity, and operationally help with more intricate calculations in chapters 3 and 4.

\section{Classical Field Theory} \label{CFT}

\subsection{From Particle Systems to Relativistic Fields}

In ordinary particle mechanics, one is able to derive the dynamical behaviour of a system with generalized position coordinates $q_i$ ($i\!=\!1,2...N$) and velocities $\dot{q}_i$ through a Lagrangian function $L(q,\dot{q},t)$, by the principle of stationary action. The action functional is defined by:

\begin{align}
S = S[q(t)] \equiv \int_{t_1}^{t_2} L(q,\dot{q},t)dt. \label{1}
\end{align}

By demanding that the physical trajectory $q(t)$ of the system between two arbitrary endpoints $q(t_1) = q_1$ and $q(t_2) = q_2$ is that which lends $S$ stationary, we get the Euler-Lagrange equations of motion:

\begin{align}
\frac{d}{dt} \!\Bigl( \frac{\partial L}{\partial \dot{q}_i} \Bigl) - \frac{\partial L}{\partial q_i} = 0. \label{discreteEL}
\end{align}

And, by solving these equations (a set of $N$ coupled ODE's) with a known set of initial positions and velocities, $q(t_0)$ and $\dot{q}(t_0)$, one is able to predict the complete physical information of the system, given by the trajectory $q(t)$ in configuration space.

Alternatively, in the Hamiltonian formulation, one may eliminate the dependence on the velocities in favor of their canonically conjugated momenta\footnote{
In the scope of the present work, for reasons of clarity and brevity, we shall not develop further on the subtleties and complications involved when there are primary constraints, that is, when one or more of the $p_i$ are identically null and one cannot solve for all $\dot{q}_i$ in function of $p_i$. This leaves out important aspects of the extremely important class of gauge fields; for the reader interested in the suitable extensions to that class, we recommend \cite{nivaldo, mandlshawn, birrell, parker}, ranging through a treatment in Classical Field Theory, Quantum Field Theory, and Quantum Field Theory in Curved Spaces. }
 $p_i \equiv \frac{\partial L}{\partial \dot{q}_i}$, by means of a Legendre transformation. The Hamiltonian is then defined as a function of all positions $q_i$ and momenta $p_i$ -- in the ($2N$-dimensional) domain that is collectively known as phase space -- by:

\begin{align}
H(q,p,t) = \sum_i^N p_i\dot{q}_i - L(q,\dot{q},t).
\end{align}

Making use of this definition and (\ref{discreteEL}), one may easily derive the Hamilton equations of motion:

\begin{align}
\dot{q} = \frac{\partial H}{\partial q}, \qquad \qquad \qquad
\dot{p} = -\frac{\partial H}{\partial p}. \label{discreteHE}
\end{align}

These are generally equivalent to \eqref{discreteEL} (making for a system of $2N$ \emph{first order} ODEs, rather than $N$ \emph{second order} ones), although they may at times be easier to solve than the former (or vice-versa). But, much more than that, the Hamiltonian formulation provides us with a number of geometrical aspects in phase space, which, properly exploited, not only give an entirely new perspective on classical mechanics, but also are at the roots for the procedure of canonical quantization. In particular, one explores the canonical antissimetric bilinear form: Poisson Brackets. These are defined for a pair of functions $f$, $g$ in phase space as:

\begin{align}
\{f,g\} = \sum_i \frac{\partial f}{\partial q_i}\frac{\partial g}{\partial p_i} - \frac{\partial f}{\partial p_i}\frac{\partial g}{\partial q_i}. \label{discretePB}
\end{align}

Particularly, one may build these brackets for any two functions from basic building blocks, the Canonical Poisson brackets of positions $q_i$ and momenta $p_i$:

\begin{align}
\hspace{-0.05\linewidth}\{q_i, q_j \} = 0 = \{p_i, p_j \}, \qquad\qquad \{q_i, p_j \} = \delta_{ij}. \label{papobra}
\end{align}

Among other things, the Poison brackets allow for yet another form to write the time evolution of any dynamical observables in our theory, through the time translation generator, the Hamiltonian\footnote{One may similarly define spatial translations from momenta, rotations from angular momenta, etc.}. Since observables can be written as functions of phase space (plus an eventual explicit time dependance), $f(q,p;t)$, one finds that its total time derivative takes the form:

\begin{align}
\frac{df}{dt} = \{f,H\} + \frac{\partial f}{\partial t},
\end{align}
as a direct consequence of \eqref{discreteHE}. Particularly, one may even reexpress (\ref{discreteHE}) as:

\begin{align}
\dot{q} = \{q,H\}, \qquad \qquad \qquad
\dot{p} = \{p,H\}. \label{poissonHE}
\end{align}

Having such results stablished for discrete particle systems, we are ultimately interested in showing how they can be extended for continuous ones\footnote{Our primary interest, of course, is studying fields, but this analysis is equally applicable to fluid mechanics, or other continuous systems.}. The simplest possible case is that of a single scalar field: we describe its configurations at a given time $t$ by field amplitudes at all space points, rather than by a finite number of coordinates; one may think of this as passing from a discrete label $i$ to a continuous one $\mathbf{x}$: $q_i(t) \rightarrow \phi(\mathbf{x},t)$ (here we have in mind fields defined in ordinary 3-dimensional flat space: $\mathbf{x}\in \mathbb{R}^3 $). Thus, in the absence of any internal constraints, such systems have an infinite number of degrees of freedom, which may be roughly regarded as ``one  degree of freedom per point in space''. Of course, one may also have a theory with multiple or composite fields -- such that several discrete indices might be required to denote different fields or field components -- having instead ``$N$ degrees of freedom per point in space''. We denote discrete indices collectively by the subscript $a$: $\phi_a(x)$.

Now, just as in the case of particle mechanics, we want to derive the dynamical behaviour for these fields through an action principle. Analogously to equation (\ref{1}), we define the action functional of a trajectory\footnote{Although we are no longer speaking of position variables, we still refer to the evolution of configurations of a field in a continous time interval as its 'trajectory' or `path'.}:

\begin{align}
S[\phi_a] = \int_{t_1}^{t_2} L[\phi_a, \dot{\phi}_a;t] dt, \label{LaF1}
\end{align}
whereas the Lagrangian that appears in (\ref{LaF1}) is no longer an ordinary function of a finite number of variables, but rather a functional of the field configurations (and its first time derivatives) at time $t$, which prevents us from obtaining dynamical equations straighforwardly as in (\ref{discreteEL}), through mere partial derivatives in each degree of freedom. 

Apart from that, we have so far said very little about how this Lagrangian functional may depend on field variables. In particle mechanics, we have kinetical terms, usually quadratic in velocities $\dot{q}$, and mutual interaction terms, which usually depend on the separation between particles $V(\mathbf{x_i}-\mathbf{x_j})$. The first can be quite obviously transposed to field time derivatives $\dot{q}$; as of the second, we would like to make an analogue for \textit{local} field interactions, allowing for a dependence in field amplitude \emph{variations}, $\phi((\mathbf{x}+\delta\mathbf{x})-\mathbf{x}) \rightarrow \delta\mathbf{x}\cdot\!\boldsymbol{\nabla}\!\phi \propto \!\boldsymbol{\nabla}\!\phi$. Since both terms involve space-time derivatives of $\phi$, they are often collectively referred to as ``kinetical'' in field theory. Besides these terms, one often uses one-particle potentials $U(q)$, directly dependent on the coordinates\footnote{
 These are often due to the effects of agents considered external to our system, such as external electrostatic potential on charges, or even a spring on a simple mechanical harmonic oscillator, whose only `internal' dynamical variable is the particle's position. Curiously the analogous terms in field theory, most notably, mass terms $\propto \phi^2$ can often be found to emerge from the system field's interactions with `external' fields. We will discuss this point a little further in Chapter \ref{cosmology}.};
 analogously, we allow here for potential terms directly proportional to field amplitudes $\phi$: $V(\phi)$. With that in mind, we introduce the Lagrangian density $\mathscr{L}$ containing any of these contributions; $\mathscr{L}$ will then be a local\footnote{At this point, we mean `local' not (necessarily) in the relativistic sense, but rather in the sense that the values of $\mathscr{L}$ at a given point only depends on the values of $\phi$ in an arbitrarily small vicinity of that point.} function of field amplitudes and its space-time derivatives, in terms of which we write a \emph{spatially global Lagrangian functional} $L$, carrying only a time dependence:

\begin{align}
L_t[\phi_a, \dot{\phi}_a] = \int_t d^3\mathbf{x} \, \mathscr{L}(\phi_a, \partial \phi_a; \mathbf{x}, t),
\end{align}
where $\partial \phi_a$ denotes collectively spatial and temporal first derivatives.
Written in terms of $\mathscr{L}$, the action looks much similar to (\ref{1}):

\begin{align}
S[\phi_a] = \int_{t_1}^{t_2} \!\!\!dt \int_t d^3\mathbf{x} \, \mathscr{L}(\phi_a, \partial \phi_a; \mathbf{x}, t), \label{NRaction}
\end{align}
as there are now only a finite number of variables associated to each space and time points (those being the field amplitudes, as well as a finite number of derivatives, at that point) and an eventual explicit dependency on space and/or time. Thus, one may think of (\ref{NRaction}) as merely a `version with extra integration dimensions' of \eqref{1}. Thereby, it can be extremized in a similar fashion, yielding:

\begin{align}
\frac{\partial}{\partial} \Bigl( \frac{\partial \mathscr{L}}{\partial \dot{\phi_a}} \Bigl) + \,\boldsymbol{\nabla} \!\cdot\! \Bigl( \frac{\partial \mathscr{L}}{\partial (\boldsymbol{\nabla} \!\phi_a)} \Bigl) - \frac{\partial \mathscr{L}}{\partial \phi_a} = 0. \label{notcovEL}
\end{align}

By comparing this equation to (\ref{discreteEL}), we see that, whereas the former is a second-order ODE system, this is a second-order PDE system. Besides being more technically complicated, these equations require not only initial conditions at some time $t_0$, $(\phi(t_0),\dot{\phi}(t_0))$, but ofen also spatial boundary conditions that restrain the physically permitted configurations of the theory. We shall discuss these further in the concrete example of a scalar field.

Of course, we shall be mainly interested in relativistically covariant theories. This imposition is actually simple to implement in an extreme-action (Lagrangian) formulation. All that we must require is that $S$ is invariant under any spacetime transformations (\textit{i.e.}, translations, rotations and boosts), that is, we must require that $S$ is a scalar (in the relativistic sense). Since our theory is also required to be local, this means that $\mathscr{L}$ must be a \emph{scalar field}\footnote{Depending on how it is defined, it could be required to be a \emph{scalar density}; we make a more careful discussion of that point in the next chapter. For now, we are only making use of global inertial coordinates in Minkowski spacetimes, so that the two will coincide.}. By looking at the form of \eqref{NRaction}, we see that we are already integrating in spacetime, so now we express spacetime events by a single variable $x=(\mathbf{x},t)$ and write covariantly:

\begin{align}
S[\phi_a] = \int d^4x \mathscr{L}(\phi_a, \partial \phi_a; x), \label{action}
\end{align}
which is extremized in an identical manner to (\ref{NRaction}), yielding, in Einstein summation convention:

\begin{align}
\partial_\mu \Bigl( \frac{\partial \mathscr{L}}{\partial (\partial_\mu \phi_a)} \Bigl) - \frac{\partial \mathscr{L}}{\partial \phi_a} = 0.
\end{align}

\subsection{Functional Derivatives}

From the beginning of this chapter, we have been working with functionals, but we have only scratched very superficially what they are, and how to operate with them; so far, we have just pointed at a few classical results for extremizing them (on which we have not even elaborated much, relying on the reader's familiarity with those results from analytical mechanics). In order to better exploit them, and allow for a more systematic approach to field theory, it is worth pausing here to lay down a few basic definitions, and develop some tools for operating with functionals. A much more thorough treatment of this topic, in which the present exposition is based, can be found in the final chapter of \cite{nivaldo}.

Generally speaking, a functional $F$ is a function of functions into numbers. That is, it takes as an input a function $f\in \mathcal{F}$, $\mathcal{F}$ being an appropriate function space, and gives a numerical output associated to it (usually a real or complex number):

\begin{align}
F: \mathcal{F} &\rightarrow \mathbb{R}, \mathbb{C}  \nonumber \\[-4pt] f &\rightarrow F[f].
\end{align}

Thus, the domain of a functional is a set of functions $\mathcal{F}$, defined in their own domain and counterdomain; in the case of fields, we are generally interested in (sufficiently smooth) functions of spacetime into a finite-dimensional space (usually numbers, tensors or spinors, depending whether we have scalar, tensor or spinor fields). The examples we have used so far is the action $S[\phi_a]$, which is a scalar function of the field values in a 4-dimensional region of spacetime and the Lagrangian $L_t[\phi_a,\dot{\phi}_a]$ which is a scalar (though not it the relativistic-invariant sense) function of field values in an (equal-time) 3-dimensional surface and their first derivative in the direction orthogonal to that surface.

Operationally, it is very important to be able to evaluate variations of functionals when we vary their arguments (and particularly, to extremize them and find stationary points). For that, as with ordinary functions with a finite-dimensional domain, one must be able to take derivatives. However, there are complications in extending this procedure to an infinite-dimensional domain, acutely so in the continuum, which do not allow for a straightforward application of mere partial derivatives. A proper extension of the concept of derivatives into functional spaces is given by the so-called \emph{functional derivatives}. To define them, we start from the concept of directional derivatives in a finite-dimension domain:

\begin{align}
\lim_{\epsilon \rightarrow 0} \frac{f(\mathbf{x}+ \epsilon \mathbf{u}) - f(\mathbf{x})}{\epsilon} = \mathbf{u} \cdot \nabla f(\mathbf{x}) = \sum_i u^i\frac{\partial f}{\partial x^i}. \label{gradient}
\end{align}

Here, we can see that the derivative $\nabla f$ is a \emph{dual vector} that, acting on a vector $\mathbf{u}$ produces the rate of variation of $f$ along $\mathbf{u}$ (in terms of infinitesimal variations, one may say $\nabla f$ acts on an infinitesimal displacement $\delta \mathbf{u}$ to produce the infinitesimal variation $\delta f = \delta \mathbf{u} \cdot \nabla f$).

Thus, inspired in equation \eqref{gradient}, we define functional derivatives by the equation:

\begin{align}
\lim_{\epsilon \rightarrow 0} \frac{F[f + \epsilon \sigma]-F[f]}{\epsilon} \equiv \int_\Omega dx \, \sigma(x) \frac{\delta F}{\delta f(x)} . \label{funcder}
\end{align}

This produces the variations of $F$ along (the abstract direction of) a function $\sigma$. One can also write the infinitesimal version of a variation:

\begin{align}
\delta F = \int_\Omega dx \frac{\delta F}{\delta f(x)} \delta \sigma(x).
\end{align}

From the definition \eqref{funcder}, one may also easily check that functional differentiation obeys some elementary identities crucial to derivative operators, such as linearity and Leibniz rule. As it happens with ordinary finite-dimensional derivatives, functional derivatives belong to the dual space $\mathcal{F}^*$ -- i.e. they are \emph{linear functionals} acting on $\mathcal{F}$ to produce numbers (namely, rates of variations along them). Thus, in general, they are \emph{distributions}, rather than functions (see appendix \ref{distributions}), although often one can identify them with functions.

Functionals may also depend on one or several parameters. An example above is the Lagrangian $L_t$, which depends on time. These are not usually treated as arguments\footnote{
Although, technically speaking, they are, as the functionals turn out as functions of the form $F: \mathcal{F} \times \mathbb{R}^n \rightarrow \mathbb{R} $ (where, for definiteness, we are representing real parameters and outputs).},
 since their variations can be more straightforwardly analyzed through ordinary calculus techniques. Then, quite naturally, one may think of the function itself (evaluated at a given point) as a functional: $F_x[\phi] = \phi(x)$. This particular functional relates to a very special distribution -- the Dirac delta:

\begin{align}
f(x) = \int_{\Omega} dy f(y)\delta(x-y). \label{deltakernel}
\end{align}

Comparing with our definition (\ref{funcder}), we then immediately obtain that:

\begin{align}
\frac{\delta f(y)}{\delta f(x)} = \delta(x-y). \label{funcderdelta}
\end{align}

It is also worth to lay here a few operational considerations and elementary examples (for more of them, see \cite{nivaldo}):

\textbf{1- Linear functionals with an integration Kernel.} They are extremely straightforward to evaluate, and one can apply \eqref{funcder} directly:

\begin{align}
F_x[f] = \int_{\Omega} dy \,K(x,y)f(y) \quad \Rightarrow \quad \frac{\delta F_x}{\delta f(y)} = K(x,y).
\end{align}

Note also that \eqref{funcderdelta} is just a particular application of it.

\textbf{2- Locally composite functionals.} These can easily be verified to obey a simple chain rule upon functional differentiation:

\begin{align}
F[f] = \int_\Omega dx \,g(f(x)) \quad \Rightarrow \quad \frac{\delta F}{\delta f(x)} = \frac{dg}{df}(x);
\end{align}

\textbf{3- Locally composite functionals involving a finite number derivatives.} Assuming the argument functions to vanish at the boundary of the integration domain $\partial\Omega$ (or constraining the variations to always be null at $\partial\Omega$, as we do with the action), one can compute the functional derivatives through a sequence of derivations by parts, yielding:

\begin{align}
F[f] = &\int_\Omega dx \,g\bigl(f(x), f'(x), f''(x),... f^{(n)}(x)\bigl) \nonumber \\[4pt]
       &\Rightarrow \quad \frac{\delta F}{\delta f(x)} = \frac{dg}{df}(x) - \frac{d}{dx} \!\!\left(\! \frac{dg}{df'}(x)\! \right) + \frac{d^2}{dx^2}\!\! \left(\! \frac{dg}{df''}(x) \!\right) ... + (-1)^n \frac{d^n}{dx^n}\!\! \left( \!\frac{dg}{df^{(n)}}(x)\! \right),
\end{align}
which was written in the form of ordinary derivatives, but the extension to a finite-dimensional domain $\Omega$ is done in the obvious way in terms of partial derivarives.

Particularly, we can immediately apply this last results to write the Euler-Lagrange equations in an elegant and compact manner:

\begin{align}
\frac{\delta S}{\delta \phi_a(x)} = \partial_\mu \Bigl( \frac{\partial \mathscr{L}}{\partial (\partial_\mu \phi_a)} \Bigl) - \frac{\partial \mathscr{L}}{\partial \phi_a} = 0.
\end{align}

\subsection{The Hamiltonian Formulation of Field Theory}

Although the Lagrangian formalism suffices for us to obtain the dynamical equations for the field in a simple and manifestly covariant manner, it is quite more complicated to obtain a quantized theory from it. Although in section \ref{ARenormalization} we will make a brief introduction to the Lagrangian-based path integral approach to quantum mechanics, it proves most convenient in a first approach to introduce a Hamiltonian formalism, which allows for the construction of the more straightforward scheme of canonical quantization, similarly to how it is usually done in ordinary quantum mechanics.

A disadvantage in the Hamiltonian formulation is that, unlike its Lagrangian counterpart, it must `break' manifest spacetime covariance: in order to extract field instantaneous configurations and velocities (or momenta) from its spacetime trajectory, one must single out one time coordinate $t$ and set it apart from space coordinates $\mathbf{x}$. For a given choice for the split of space and time, we define the field's velocities $\dot{\phi}_a(x)\equiv\frac{d}{dt}\phi_a(x)$ and momenta:

\begin{align}
\pi^a(x) \equiv \frac{\partial \mathscr{L}}{\partial \dot{\phi}_a(x)}.
\end{align}

Once again, performing a Legendre transformation, we define the Hamiltonian density in phase space:

\begin{align}
\mathcal{H}(\phi_a,\pi^a,x) = \pi^a(x)\phi_a(x) - \mathscr{L} (\phi_a, \dot{\phi}_a,x),
\end{align}
where the velocities $\dot{\phi}_a$ are implied to be a function of the momenta $\pi^a$.

In direct analogy with \eqref{discretePB}, we also define Poisson Brackets in the continuum:

\begin{align}
\{F,G\} = \int{ d^3 \mathbf{x} \; \frac{\delta F}{\delta \phi_a(\mathbf{x},t)}\frac{\delta G}{\delta \pi^a(\mathbf{x},t)} - \frac{\delta G}{\delta \phi_a(\mathbf{x},t)}\frac{\delta F}{\delta \pi^a(\mathbf{x},t)}} \;. \label{bfipobra}
\end{align}
Particularly, we have the fundamental canonical Poisson Brackets:

\begin{align}
\{\phi_a(\mathbf{x},t),\pi^b(\mathbf{y},t)\} = \delta_a^{\;\,b}\,\delta^{(3)}(\mathbf{x}-\mathbf{y}), \label{fipobra}
\end{align}
which will play a key role in canonical quantization.

\subsection{The free real scalar field}

An example of major importance which we shall explore in extensive detail throughout this work is the free real scalar field (also known as the real Klein-Gordon field). The starting point for defining it is the Lagrangian density:

\begin{align}
\mathscr{L} = \frac{1}{2}\eta^{\mu\nu}(\partial_\mu \phi) (\partial_\nu \phi) - \frac{m^2}{2}\phi^2 .\label{KGlagrangian}
\end{align}
This contains ordinary kinetic terms as well as a simple quadratic (harmonic) potential term. Here, $m^2$ is a positive parameter characterizing the steepness of the potential well (we use this suggestive notation because, as we shall see later, $m$ will be identified with the mass of the quanta -- the particles -- of the quantized field); we also note that, in natural units ($\hbar\!=\!c\!=\!1$), $m$ has units of inverse length\footnote{In regular units, this inverse lenght reads $\mu = \frac{mc}{\hbar}$.}, conferring the field with a characteristic lengh scale, $m^{-1}$.

From this Lagrangian, we easily obtain the dynamical equations:

\begin{align}
[\Box + m^2]\phi = 0, \label{fkge}
\end{align}
where we have defined the D'Alembertian in flat spacetime: $\Box \phi = \eta^{\mu\nu}\partial_\mu \partial_\nu \phi = \partial_t^2\phi -\nabla^2\phi$ (where the last equality is expressed in globally inertial Cartesian coordinates, and $\nabla^2$ represents an ordinary spatial Laplacian: $\nabla^2 = \partial^2_x + \partial^2_y + \partial^2_z$).

Since the field equations are linear, we may expand any solutions in terms of a complete set of modes. One particularly convenient basis for \eqref{fkge} are plane wave modes:

\begin{align}
u_{\mathbf{k}}(\mathbf{x},t) = \frac{1}{\sqrt{2\omega_k}} e^{-i \omega_k t} e^{i \mathbf{k} \cdot \mathbf{x}}. \label{plane waves}
\end{align}

Here, we have defined positive frequencies, $\omega_k = +\sqrt{m^2+\mathbf{k}^2}$, and we have included the normalization factor $({2\omega_k})^{-1/2}$ for later convenience. We may then write the field expansion as:

\begin{align}
\phi(\mathbf{x},t) = \sum_\mathbf{k} a_\mathbf{k} u_\mathbf{k}(\mathbf{x},t) + a^*_\mathbf{k}u^*_\mathbf{k}(\mathbf{x},t). \label{ffe}
\end{align}

Of course, to specify which range of wave vectors $\mathbf{k}$ are allowed (or, more generally, which combinations of $a_\mathbf{k}$ and $a^*_\mathbf{k}$ are permitted), we must also specify (spatial) boundary conditions. These should, of course, depend on the global physical conditions we want to impose to our field. Whereas these are relatively straightforward for spatially bounded systems (a paradigmatic example in field theory is the electromagnetic field confined within a conducting surface, for which one just applies Dirichlet conditions on the boundary), they may raise nontrivial questions for spatially open systems (as will be the case in cosmological contexts with noncompact universes), regarding the behaviour of the field at infinity.
 Nevertheless, such questions are often of little relevance to the local dynamics\footnote{One should bear in mind, however, that quantum theory has important nonlocal features. We shall further analyze the effects of boundary conditions when we discuss the Casimir Effect in section \ref{Casimireff}.},
  so it is a common practice to take artificial boundary conditions that simplify one's calculations. A rather convenient choice is to take periodic boundary conditions in a cube with dimentions $L \times L \times L$, such that the wave vectors $\mathbf{k}$ (and therefore the frequencies $\omega_k$) only take a discrete set of values; in a properly chosen Cartesian grid, their components will be:

\begin{align}
\qquad \qquad \qquad \qquad  k^i = \frac{2\pi}{L}n_i, \qquad i=1,2,3 \,, \qquad n_i \in \mathbb{Z}.
\end{align}
In doing so, we also incorporate a volume factor $V=L^3$ in the normalization of \eqref{plane waves}, defining: 

\begin{align}
u_{\mathbf{k}}(\mathbf{x},t) \equiv \frac{1}{\sqrt{2V\omega_k}} e^{-i \omega_k t} e^{i \mathbf{k} \cdot \mathbf{x}}. \label{V plane waves}
\end{align}

From this construction, a very straightforward way to analyze the continuum limit and drop the periodic conditions is to take $L\!\rightarrow\!\infty$, and make a proper change in normalization (this amounts simply to going from a discrete Fourier series to a continuous Fourier transform; see e.g. \cite{arfken}). The adjustment that we make in the latter aims at bilinear integrals (particularly the orthornormality conditions \eqref{discreteON} \eqref{continuousON} below) and can be motivated as follows: for finite $L$, one has the spectral volume around each mode (the ``volumetric spacing between modes''): $\Delta^3 \mathbf{k} = (2\pi/L)^3$. Thus, we take sums into integrals by making:

\begin{align}
\left(\frac{2\pi}{L} \right)^{\!\!3}\sum_\mathbf{k} \;=\; \sum_\mathbf{k}\Delta^3\mathbf{k} \longrightarrow \int d^3 \mathbf{k}\,. \label{continuumlimit}
\end{align}
We then end up with the following normalized modes in the continuum:

\begin{align}
u_{\mathbf{k}}(\mathbf{x},t) \equiv \frac{1}{\sqrt{2(2\pi)^3\omega_k}} e^{-i \omega_k t} e^{i \mathbf{k} \cdot \mathbf{x}}, \label{C plane waves}
\end{align}
for which one writes the integral field expansion:

\begin{align}
\phi(\mathbf{x},t) = \int \!d^3\mathbf{k} \,a(\mathbf{k}) u_\mathbf{k}(\mathbf{x},t) + a^*\!(\mathbf{k})u^*_\mathbf{k}(\mathbf{x},t). \label{ffce}
\end{align}

Note this particular approach to the continuum also implies a boundary condition at infinity: by restricting the wave vectors to be real, it forces $u_\mathbf{k}$ to remain bounded, not allowing for any exponentially increasing solutions. This will be crucial for mode decomposition, as we want to constrain our modes to have finite projections on integrable field solutions (\textit{e.g.}, wave-packets).

In either case, to obtain the complete physical information about this system, one must solve the field equations, with some given initial conditions\footnote{
 Or boundary and initial conditions, but, as we just mentioned, spatial boundary conditions are usually incorporated in the determination of a complete set of modes}.
 Well, given the expansion (\ref{ffe}) (or \eqref{ffce}), this amounts to finding the coefficients $a_\mathbf{k}$ -- i.e. \emph{the amplitudes for each field mode} (which do not change with time, since the modes are decoupled and evolve independently) -- for which we match the initial conditions:

\begin{align}
\phi(\mathbf{x},t_0) = f(\mathbf{x}), \qquad \dot{\phi}(\mathbf{x},t_0) = g(\mathbf{x}). \label{pic}
\end{align}

Note that the (spectral) mode amplitudes $a_\mathbf{k}$ and the (spatial) field amplitudes $\phi(\mathbf{x},t)$ depend linearly on one another. Since we can already express $\phi(\mathbf{x},t)$ in terms of $a_\mathbf{k}$, the above task amounts to inverting that expression to obtain $a_\mathbf{k}$ in terms of $\phi(\mathbf{x},t)$ (and $\dot{\phi}(\mathbf{x},t)$) at $t=t_0$. More specifically, we want to \emph{project} the initial conditions in all the basis modes. To achieve this, it proves useful to define the following scalar product (sometimes called the Klein-Gordon product):

\begin{align}
(\phi,\psi) \equiv i \int_t d^3\mathbf{x} \; \phi^* \overleftrightarrow{\partial_t} \psi = i \int_t d^3\mathbf{x} \; \phi^* \partial_t \psi - (\partial_t \phi^*)\psi , \label{fsp}
\end{align}
where the integration may be taken in any arbitrary fixed time $t$. One can immediately verify it obeys the following elementary properties:

\begin{subequations} \label{spp}
\begin{align}
(u,\alpha v_1 + v_2) = \alpha(u,v_1) + (u,v_2)&, \qquad \alpha \in \mathbb{C}, \label{linearity}  \\
(v,u) = -(u^*,v^*) = (u,v)^*&. \label{antisimmetry}
\end{align}
\end{subequations}

Note, in particular, that \eqref{antisimmetry} implies:

\begin{align*}
(u,u^*) = 0.
\end{align*}

The definition (\ref{fsp}) suggests that this product is time-dependent (i.e. that it depends on the equal-time surface $\Sigma_t$ chosen to perform the integration). Indeed, that would be the case if we calculated the product of two (completely) arbitrary functions of spacetime. However, we shall show that the product of any two \emph{solutions of the dynamical equations} is actually conserved (presently, we limit our demonstration to equal-time surfaces $\Sigma_t$; we shall give this demonstration in greater generality in chapter \ref{QFTCS}). 

Let us take 2 time instants $t^\prime > t$. Note that the 2 hypersurfaces $\Sigma_{t^\prime}$ and $\Sigma_t$ are the boundary of the spacetime region $\Omega$ between them, and that their outward-pointing normal vectors are $n^a (\frac{\partial}{\partial t})^a$ and $n^a = -(\frac{\partial}{\partial t})^a$, respectively, so that the difference between the product $(\phi,\psi)$ computed at $t'$ and $t$ may be written as a boundary term:

\begin{align}
(\phi,\psi)_{t^\prime} - (\phi,\psi)_t &= \int_{\Sigma_{t'}} d^3\mathbf{x} n^\mu \phi^*(x) \overleftrightarrow{\partial_\mu}  \psi(x) - \int_{\Sigma_t} d^3\mathbf{x} n^\mu \phi^*(x) \overleftrightarrow{\partial_\mu}  \psi(x) \nonumber \\
 &= \int_{\partial \Omega} d^3\mathbf{x} n^\mu \phi^* \overleftrightarrow{\partial_\mu}  \psi . \nonumber \\
\end{align}

Then, applying the Gauss divergence theorem, we obtain:

\begin{align}
(\phi,\psi)_{t^\prime} - (\phi,\psi)_t &= \int_\Omega d^4\!x \partial^\mu \bigl(\phi^*(x) \overleftrightarrow{\partial_\mu}  \psi(x) \bigl) \nonumber \\
 &= \int_\Omega d^4\!x \bigl[\phi^*(x) \Box_x  \psi(x)  - \psi(x) \Box_(x) \phi^*(x) \bigl] \nonumber \\
 &= \int_\Omega m^2 \bigl[\phi^*(x)\psi(x) - \psi(x)\phi^*(x) \bigl] \nonumber \\
 &= 0.
\end{align}

With these basic properties in mind, we may now use this scalar product to solve the initial value problem of the free scalar field. First, note that the field modes \eqref{V plane waves} are orthornormal with respect to it:

\begin{subequations} \label{discreteON}
\begin{align}
(u_\mathbf{k},u_\mathbf{k'}) = \delta_\mathbf{kk'} = -(u_\mathbf{k}^*,u_\mathbf{k'}^*), \\
(u_\mathbf{k},u_\mathbf{k'}^*) = 0,
\end{align}
so that the coefficients of expansion (\ref{ffe}) are simply given by the projections in each field mode:
\end{subequations}

\begin{subequations}
\begin{align}
a_\mathbf{k} = (u_\mathbf{k},\phi), \\ a^*_\mathbf{k} = -(u^*_\mathbf{k},\phi),
\end{align}
\end{subequations}
which can be directly computed at the time $t_0$, using the inicial conditions (\ref{pic}). 

In the continuum case, one can similarly verify that the modes also follow a suitable orthornormality condition:

\begin{subequations}\label{continuousON}
\begin{align}
(u(\mathbf{k}),u(\mathbf{k'}) ) &= \delta(\mathbf{k} - \mathbf{k'})  = -(u^*(\mathbf{k}), u^*(\mathbf{k'})), \\
(u(\mathbf{k}),u^*(\mathbf{k'}) ) &= 0,
\end{align}
\end{subequations}
and the coefficents can be similarly obtained by the projections:

\begin{subequations}
\begin{align}
a(\mathbf{k}) &= (u(\mathbf{k}),\phi), \\ a^*\!(\mathbf{k}) &= -(u^*(\mathbf{k}),\phi).
\end{align}
\end{subequations}

Presently, we limit our presentation of the field modes to just these basic properties. We shall explore them in greater detail for the quantized field, where they will play a central role in our Fock Space representation.

\section{Canonical Quantization} \label{CQion}

Now that we developed some key aspects of the formalism for classical fields, we would like to proceed to their quantization. Here, we shall carry this procedure in the Hamiltonian formalism, drawing close analogy to discrete particle systems.
Recall that in canonical quantization of a particle system with position coordinates $x_i$ and canonically cojugated momenta $p_i$ we promote these classical observables to quantum \emph{operators} -- linear operators acting in a suitable Hilbert space $\mathcal{H}$ (usually taken to be an appropriate subspace of square-integrable functions $\mathcal{F} \subset \mathcal{L}^2(\mathbb{R}^N)$) --, obeying the canonical commutation relations ($\hbar=1$):

\begin{align}
[x_i,x_j] = [p_i,p_j] = 0, \qquad [x_i,p_j] = i \delta_{ij}. \label{cacopa}
\end{align}

These relations are postulated in direct reference to the canonical Poisson brackets (\ref{papobra}), where one substitutes the brackets between two classical observables by $-i$ times the commutator between their corresponding quantum operators. At this point, we stress that these operators do not have direct physical significance on their own; truly observable quantities arise, for example, in the form of expectation values in specified quantum states $\ket{\psi}$, such as $\braket{\psi|x_i|\psi}$, as well as of projections between two states in a given time $\braket{\phi|\psi}$ (which yield transition amplitudes between the states $\ket{\psi}$ and $\ket{\phi}$); more generally, one can consider transition amplitudes of the form $\braket{\phi|A|\psi}$.

Thus, there is an inherent ambiguity in this formalism concerning the definition of state vectors and operators. By performing complementary unitary transformations on both, transforming all state vectors as $\ket{\psi} \rightarrow U \ket{\psi}$ (and dual vectors as $\bra{\psi} \rightarrow \bra{\psi} U^\dagger$) and operators as $A \rightarrow U A U^\dagger$ ($UU^\dagger = \mathbb{1}$), one attains an equivalent physical description of the theory, as all observable quantities remain invariant. Each of these descriptions corresponds to a representation, or \emph{picture} of the theory. 

Two eminent pictures of quantum theory are the Schrödinger and the Heisenberg pictures. In nonrelativistic quantum mechanics, one usually works in the Schrödinger picture, in which the `fundamental' observables $x_i,\, p_i$ are time-independent and state vectors evolve according to the Schrödinger equation $i \frac{d}{dt} \ket{\psi(t)} = H \ket{\psi(t)}$, whose solution for a given initial state $\ket{\psi_0}$ at $t=0$ is given in terms of the unitary evolution operator $U(t)$:

\begin{align}
\ket{\psi(t)} = U(t)\ket{\psi_0}, 
\end{align}
where $U$ obeys the following relations:

\begin{align}
U^\dagger U = UU^\dagger = \mathbb{1}, \qquad  i \frac{d}{dt}U(t) = H U(t), \qquad  U(0) = \mathbb{1}.
\end{align}

Particularly, in the case where $H$ is time-independent, we recover the simple form: $U(t)=e^{-iHt}$.

On the other hand, in the Heisenberg picture, state vectors are kept fixed and the operators evolve in time, in such a manner that all physical observables remain unchanged:

\begin{align}
\ket{\psi^H} = U^\dagger(t) \ket{\psi^S(t)} = \ket{\psi_0}, \qquad x^H(t) = U^\dagger(t) x^S U(t).
\end{align}

Taking the time derivative of the Heisenberg observables, we see that they obey formally identical equations to their classic counterparts (\ref{poissonHE}) (but which must now be interpreted as operator-valued equations):

\begin{align}
i \frac{d}{dt} x(t) = [x(t),H], \qquad \qquad i \frac{d}{dt} p(t) = [p(t),H]. \label{Heinsenberg equation}
\end{align}

As in the classical context, these will be ultimately equivalent to the Euler-Lagrange (operator-valued) equations:

\begin{align}
\frac{d}{dt} \biggl( \frac{d L}{d\dot{x}} \biggl) \;-\; \frac{d L}{d{x}} = 0. \label{QEuler-Lagrange}
\end{align}

Finally, by re-evaluating the commutation relations (\ref{cacopa}), one may easily verify that they still hold for the Heisenberg operators \emph{for equal times}:

\begin{align}
[x_i(t),x_j(t)] = [p_i(t),p_j(t)] = 0, \qquad [x_i(t),p_j(t)] = i \delta_{ij}.
\end{align}

We may then take a similar approach for quantizing field systems. We promote the classical field $\phi$ to a quantum \emph{field operator}, for which -- based on the classical Poisson brackets (\ref{fipobra})-- we postulate the so called \emph{equal-time commutation relations}:

\begin{align}
[\phi_a(\mathbf{x},t),\phi_b(\mathbf{y},t)] = [\pi^a(\mathbf{x},t),\pi^b(\mathbf{y},t)] = 0, \qquad [\phi_a(\mathbf{x},t),\pi^b(\mathbf{y},t)] = i \delta_a^{\;b} \,\delta^{(3)}(\mathbf{x}-\mathbf{y}). \label{fcacore}
\end{align}

These relations -- although not manifestly covariant, as they require singling out a time $t$ -- will actually be Lorentz invariant, provided we have a Lorentz invariant (scalar) Lagrangian density. (For a derivation of covariant commutation relations for various fields in flat spacetime, see for example \cite{mandlshawn}.) We just state the result here that, in this case, eqs. \eqref{fcacore} will imply:

\begin{align}
[\phi(x),\phi(y)] = 0, \label{cocore}
\end{align}
always that $x$ and $y$ have a spacelike separation.

As position observables in particle mechanics, the quantized fields obey, in the Heisenberg picture, analogous (operator-valued) equations as their classical counterparts. In the latter case, however, we often only appeal to a Hamiltonian formalism to outline the bridge in canonical quantization, being more convenient to work directly with the Euler-Lagrange equations to analyze the fields' dynamics.

Finally, we remind that, while in classical mechanics one can in principle completely determine the values of positions and momenta simultaneously at a given time, and thus predict with certainty their values for any other time, this is forbidden in quantum mechanics in virtue of the uncertainty principle. In the latter case, the maximal information that one can ascertain, which can be used to completely determine a \emph{quantum state} is attached to the so-called \emph{Complete Sets of Commuting Observables} (or C.S.C.O.'s), whose eigenstates spawn the entire Hilbert Space $\mathcal{H}$. The most immediate such C.S.C.O's are those given \emph{either} by positions \emph{or} momenta (field configurations or field momenta) at any given time. We write, for instance (in the Heisenberg Picture):

\begin{align}
\ket{\phi'}:& \; \phi(\mathbf{x},t_0) \ket{\phi'} = \phi'(\mathbf{x},t_0) \ket{\phi'}, \quad \forall \mathbf{x} \in \mathbb{R}^3, \\
\ket{\pi'}:& \; \pi(\mathbf{x},t_0) \ket{\pi'} = \pi'(\mathbf{x},t_0) \ket{\pi'}, \quad \forall \mathbf{x} \in \mathbb{R}^3,
\end{align}
where $t_0$ is \emph{fixed} time, and we have used primes to distinguish field/momentum \emph{eigenvalues} from field/momentum operators\footnote{ Rigorously speaking, $\ket{\phi'}$ and $\ket{\pi'}$ are not rigorously states belonging to $\mathcal{H}$, but they can be used to spawn any actual states $\ket{\Psi}\in \mathcal{H}$. See appendix \ref{distributions} for more details. }.

Of course, one can build a myriad of different C.S.C.O's in $\mathcal{H}$, usually by considering combinations and functions of positions and momenta. In the next section, we shall particularly emphasize those associated with field modes.

\subsection{Quantizing the scalar field}

Now that we are armed with a general prescription to quantize field systems we shall illustrate it explicitly in the case of the scalar field $\phi$ \eqref{KGlagrangian}. After obtaining the quantized field, we shall give emphasis to its expansion in normal modes and construct the Fock space based on them, noting how the notion of particles naturally emerges as excitations of these field modes. We also use this expansion to compute energy-momentum observables.

As $\phi$ obeys the operator analogous of equations (\ref{fkge}), it remains useful to expand the field in plane wave modes (\ref{plane waves}). However, we must substitute the classical amplitudes by the field mode operators $a_\mathbf{k}$ and $a^\dagger_\mathbf{k}$:

\begin{align}
\phi(\mathbf{x},t) &= \sum_\mathbf{k} a_\mathbf{k} u_\mathbf{k}(\mathbf{x},t) + a^\dagger_\mathbf{k}u^*_\mathbf{k}(\mathbf{x},t), \\
\pi(\mathbf{x},t) &= \sum_\mathbf{k} -i\omega_k \bigl( a_\mathbf{k} u_\mathbf{k}(\mathbf{x},t) - a^\dagger_\mathbf{k}u^*_\mathbf{k}(\mathbf{x},t) \bigl).
\end{align}

Such expansions express the entire range of canonical observables ($\phi$ and $\pi$, $\forall \mathbf{x}$) in terms of $a$ and $a^\dagger$ ($\forall \mathbf{k}$). Conversely, by using the projections $a_\mathbf{k} = (u_\mathbf{k},\phi)$ -- which involve both $\phi$ and its time derivative --, we may express $a$ and $a^\dagger$ in terms of $\phi$ and $\pi$. Then we can easily derive:

\begin{align}
[a_\mathbf{k}, a_\mathbf{k'}] &= \Bigl[ \int_t\! d^3\mathbf{x} (u^*_\mathbf{k}(x)\pi(x)-i\omega_k u^*_\mathbf{k}(x)\phi(x)) ,\int_t\! d^3 \mathbf{y} (u^*_\mathbf{k'}(y)\pi(y)-i\omega_{k'} u^*_\mathbf{k'}(y)\phi(y)) \Bigl] \nonumber \\
 &= \int_t\! d^3\mathbf{x} \int_t\! d^3\mathbf{y} u^*_\mathbf{k}(x) u^*_\mathbf{k'}(y) \Bigl\{ -i[\pi(\mathbf{x},t),\phi(\mathbf{y},t)] -i[\phi(\mathbf{x},t),\pi(\mathbf{y},t)] \nonumber \\ & \qquad\qquad\qquad\qquad\qquad\quad\;\;  + [\phi(\mathbf{x},t),\phi(\mathbf{y},t)] + i^2[\pi(\mathbf{x},t),\pi(\mathbf{y},t)] \Bigl\} \nonumber \\
 &= \int_t\! d^3\mathbf{x} \int_t\! d^3\mathbf{y} u^*_\mathbf{k}(x) u^*_\mathbf{k'}(y) \Bigl\{ \delta(\mathbf{x}-\mathbf{y}) - \delta(\mathbf{x}-\mathbf{y}) \Bigl\} \nonumber \\
 &= 0,
\end{align}
where we have used the time invariance of \eqref{fsp} to compute the projections for $a_\mathbf{k}$ and $a^\dagger_\mathbf{k}$ at the same time $t$, and be able to use \eqref{fcacore}.

Similarly, one can compute $[a^\dagger_\mathbf{k}, a^\dagger_\mathbf{k'}] = 0$. We then have the nontrivial commutator:

\begin{align}
[a_\mathbf{k}, a^\dagger_\mathbf{k'}] &= \Bigl[ \int_t\! d^3\mathbf{x} (u^*_\mathbf{k}(x)\pi(x)-i\omega_k u^*_\mathbf{k}(x)\phi(x)) ,-\int_t\! d^3 \mathbf{y} (u_\mathbf{k'}(y)\pi(y)-i\omega_{k'} u_\mathbf{k'}(y)\phi(y)) \Bigl] \nonumber \\
 &= \int_t\! d^3\mathbf{x} \int_t\! d^3\mathbf{y}\, u^*_\mathbf{k}(x) u_\mathbf{k'}(y) \Bigl( i\omega_{k'}[\pi(\mathbf{x},t),\phi(\mathbf{y},t)] -i\omega_{k}[\phi(\mathbf{x},t),\pi(\mathbf{y},t)] \Bigl) \nonumber \\
 &= \int d^3\mathbf{x}\, u^*_\mathbf{k}(x) u^*_\mathbf{k'}(x) ( \omega_{k'} + \omega_k ) \nonumber \\
 &= \delta_{\mathbf{kk'}}.
\end{align}

Summarizing:

\begin{align}
[a_\mathbf{k}, a_\mathbf{k'}] = [a^\dagger_\mathbf{k}, a^\dagger_\mathbf{k'}] = 0, \qquad [a_\mathbf{k}, a^\dagger_\mathbf{k'}] = \delta_\mathbf{kk'}. \label{mocorepw}
\end{align}

Classically, a complete set of observables for this field was given by field amplitudes at all spacetime points (which, in their turn, could be obtained by the field amplitudes and their first time derivavatives \emph{for all space points at a given time}, \textit{i.e.}, at surface of simultaneity); equivalently, one could specify the field amplitudes $a_\mathbf{k}$ for all modes. For the quantized field, however, it is impossible to determine such complete information about the field's trajectory (and thus, about its amplitudes and `velocities' at a given time); conversely, in the mode perspective, one cannot attain the full information about the amplitudes $a_\mathbf{k}$ (technically, $a_\mathbf{k}$ is not an observable in the traditional quantum mechanical sense, since it is not a self-adjoint operator, and thus there is no guarantee that it will be possible to build a basis of eigenstates in $\mathcal{H}$ from it\footnote{
 That is not to say that there are no Eingestates of $a_\mathbf{k}$ in $\mathcal{H}$. In fact, the so-called coherent states, taking the form $\ket{\nu}= e^{\nu a^\dagger}\!\ket{0}$, not only form a continuous family of eingenstates of $a$, $a\ket{\nu} = \nu\ket{\nu}$, {but they are also very important in the evaluation of the classical limit of the theory; For more details on them, check chapter 1 of \cite{mandlshawn} (particularly, exercise 1.1) } .}).
 That is, one cannot find simultaneously its real and imaginary parts (note that, for a simple harmonic oscillator, $\operatorname{Re}(a) \propto x$ and $\operatorname{Im}(a) \propto p$). Alternatively, one can decompose $a_\mathbf{k}$ in its magnitude $|a_\mathbf{k}| = (a_\mathbf{k}a^*_\mathbf{k})^\frac{1}{2}$ and phase $\theta_\mathbf{k}$, which, again, cannot be determined simultaneously. In QFT it is their magnitude that takes a proeminent role\footnote{
 For a comprehensive discussion of phase observables, which seldom appear in the QFT literature, see \textit{e.g.} (chapter 5 of) \cite{matheus}.},
 more precisely, their quadratic magnitude, which classically reads $|a_\mathbf{k}|^2 = a_\mathbf{k}a_\mathbf{k}^*$; then, in the quantized theory, we define the occupation observables:
 
\begin{align}
N_\mathbf{k} = a_\mathbf{k} a_\mathbf{k}^\dagger.
\end{align}

It follows from the commutation relations \eqref{mocorepw} that our modes decompose the scalar field as an infinite set of decoupled harmonic oscillators, whence we have immediately that the spectrum for each occupation observable is just natural numbers, \textit{i.e.}, $\sigma(N_\mathbf{k})= \mathbb{N}$; for this reason, these are often called occupation numbers. We can also define the \emph{total} occupation number $N \equiv \sum_\mathbf{k} N_\mathbf{k}$, which will similarly have the spectrum $\sigma(N) = \mathbb{N}$. The description in terms of occupation numbers then gives us a natural framework to define \emph{particles} in our theory: we interpret each quantum in mode $u_\mathbf{k}$ as a particle of momentum $\mathbf{k}$, and energy $E = \sqrt{m^2 + \mathbf{k}^2}$ (this also gives us a natural interpretation of $m$ as the mass of each particle). This representation of our Hilbert Space, based on the occupation numbers for each mode, is called a \emph{Fock Space}. We can construct it starting from the vacuum state, $\ket{0}$, which has no quanta (particles) of any type. More precisely, it is defined by:

\begin{align}
a_\mathbf{k} \ket{0} = 0, \quad \forall \mathbf{k}.
\end{align}

Then, we can obtain the all $n$-particle states by applying the various creation operators $a_\mathbf{k}^\dagger$ to the vaccuum:

\begin{align}
\ket{n_{\mathbf{k_1}},n_{\mathbf{k_2}}...} = \frac{1}{\sqrt{n_{\mathbf{k_1}}!n_{\mathbf{k_2}}!...}}(a_{\mathbf{k_1}}^\dagger)^{n_{\mathbf{k_1}}}(a_{\mathbf{k_2}}^\dagger)^{n_{\mathbf{k_2}}}... \ket{0}.
\end{align}

Furthermore, we can expand any observables of our theory in terms of field modes. Special dynamic interest attaches to energy and momentum of our field. In relativistic theories in the continuum, energy, momentum and stress are all codified in a single tensorial observable, the well-known stress tensor $T_{\mu\nu}$\footnote{
 We shall not construct $T_{\mu\nu}$ here in detail. If the reader is unfamiliar with it, we suggest chapter 10 of \cite{nivaldo} for the construction of a classical $T_{\mu\nu}$ in nonrelativistic and special relativistic theories, as a conserved Noether current associated with the symmetries of space-time translations. See also chapter 4 of \cite{wald} for a discussion of $T_{\mu\nu}$ in Special and General Relativity.}.
 Its time-time and time-space components are interpreted as energy and momentum density, respectively, whereas its space-space components are related to stresses (being its diagonal components related to pressures, and its non-diagonal ones, to shears), all with respect to a stationary observer in the adopted coordinate frame. That is, given an observer $\mathcal{O}$ with a normalized vector $u^a$ tangent to its worldline, and normalized 3-vectors $e_i^a$:
 
\begin{subequations}
 \begin{empheq}[left=\empheqlbrace]{align}
 T_{\mu\nu}u^\mu u^\nu &\equiv T_{00} = \mathcal{H} = \rho ,  \\
 T_{\mu\nu}u^\mu e_i^\nu &\equiv T_{0i} = \pi_i,  \\
 T_{\mu\nu}e_i^\mu e_j^\nu &\equiv T_{ij},
 \end{empheq}
\end{subequations}
being the diagonal spatial components, $T_{ii}$ related to pressures $p_i$ in each direction (for an nonviscous isotropic fluid or field, $T_{ij} = p\delta_{ij}$). We urge the reader not to mistake the \emph{mechanical 3-momentum density} $\pi^i$ with the \emph{canonically conjugated momentum} $\pi$. 

In virtue of Noether's Theorem \cite{nivaldo}, $T_{\mu\nu}$ is a conserved current, that is:

\begin{align}
\partial_\mu T^\mu_{\;\;\nu} = 0.
\end{align}

Using Gauss's Theorem, it is straightforward to verify that (less of any energy-momentum currents at infinity), this will also imply a \emph{global} conservation law in Minkowski spacetime:

\begin{align}
P^\mu(t) = \int_t\! d^3\mathbf{x}\, n^\mu T^\mu_{\;\;\nu}(\mathbf{x},t) = \int_t d^3\mathbf{x} \,T^\mu_{\;\;0}(\mathbf{x},t) \qquad \Rightarrow \qquad \frac{d}{dt}P^\mu(t) = 0,
\end{align}
where $n^\nu$ is the future-directed normal to the equal-time surface $\Sigma_t$. The spatial and temporal components of $P^\mu$ are interpreted as the total energy and momentum:

\begin{subequations}\label{4-momentum}
 \begin{empheq}[left=\empheqlbrace]{align}
 P^0 &= \int\! d^3\mathbf{x} \,\mathcal{H}(\mathbf{x},t) = H ,  \\
 P^i &= \int\! d^3\mathbf{x} \,\pi^i(\mathbf{x},t).
 \end{empheq}
\end{subequations}

For a free scalar field, the canonical stress tensor reads \cite{nivaldo, wald}:

\begin{align}
T_{\mu\nu} = (\partial_\mu \phi)( \partial_{\nu} \phi) - \tfrac{1}{2} \eta_{\mu\nu} \bigl[ (\partial^{\alpha} \phi) (\partial_{\alpha} \phi) - m^2 \phi^2 \bigl].
\end{align}

Particularly, we have the energy density,

\begin{align}
\mathcal{H} =\frac{1}{2}\bigl( \pi^2 + (\mathbf{\nabla}\phi)^2 + m^2 \phi^2 \bigl),
\end{align}
and the 3-momentum density $\boldsymbol{\pi}$ (again, do not misktake it for the conjugated momentum $\pi$):

\begin{align}
\boldsymbol{\pi} = \dot{\phi} \,(\!\boldsymbol{\nabla}\!\phi) = \pi \,(\!\boldsymbol{\nabla}\!\phi).
\end{align}

To evaluate their global (spatially integrated) correspondents \eqref{4-momentum}, we note that the integral for each of their terms can be computed without much difficulty by noting that:

\begin{align}
\frac{1}{V}\int d^3\mathbf{x} e^{i(\mathbf{k}-\mathbf{k}')\cdot \mathbf{x}} = \delta_{\mathbf{k},\mathbf{k'}}, \qquad \frac{1}{(2\pi)^3}\int d^3\mathbf{x} e^{i(\mathbf{k}-\mathbf{k}')\cdot \mathbf{x}} = \delta(\mathbf{k} - \mathbf{k'}),
\end{align}
in the dicrete and in the continuum, respectively. For simplicity, we work in the discrete, for which we find the basic terms of the Hamiltonian:

\begin{align}
  \int d^3\mathbf{x} \,\phi^2(\mathbf{x},t) &= \frac{1}{2} \sum_\mathbf{k} \frac{1}{\omega_k} \Bigl( a_\mathbf{k}a^\dagger_\mathbf{k} + a^\dagger_\mathbf{k}a_\mathbf{k} + a_\mathbf{k}a_{-\mathbf{k}} e^{-2i\omega_k t} + a^\dagger_\mathbf{k}a^\dagger_{-\mathbf{k}} e^{2i\omega_k t} \Bigl), \\[4pt]
  \int d^3\mathbf{x} (\nabla\phi)^2(\mathbf{x},t) &= \frac{1}{2} \sum_\mathbf{k} \frac{\mathbf{k}^2}{\omega_k} \Bigl( a_\mathbf{k}a^\dagger_\mathbf{k} + a^\dagger_\mathbf{k}a_\mathbf{k} + a_\mathbf{k}a_{-\mathbf{k}} e^{-2i\omega_k t} + a^\dagger_\mathbf{k}a^\dagger_{-\mathbf{k}} e^{2i\omega_k t} \Bigl), \\[4pt]
  \int d^3\mathbf{x} \,\dot{\phi}^2(\mathbf{x},t) &= \frac{1}{2} \sum_\mathbf{k} \frac{\omega_k^2}{\omega_k} \Bigl( a_\mathbf{k}a^\dagger_\mathbf{k} + a^\dagger_\mathbf{k}a_\mathbf{k} + a_\mathbf{k}a_{-\mathbf{k}} e^{-2i\omega_k t} + a^\dagger_\mathbf{k}a^\dagger_{-\mathbf{k}} e^{2i\omega_k t} \Bigl).
\end{align}

Combining these, we obtain the total Hamiltonian:

\begin{align}
H = \int d^3 \mathbf{x}\, \mathcal{H}(x) = \sum_\mathbf{k} \omega_k(N_\mathbf{k}+ \tfrac{1}{2}) ,
\end{align}
where we see that the time-dependent terms cancel out, and we have rewritten the time-independent ones in terms of $N_\mathbf{k}$ and the commutator $[a_\mathbf{k},a^\dagger_\mathbf{k}] = 1$.

We can similarly compute the total 3-momentum:

\begin{align}
\mathbf{P} = \int d^3\mathbf{x} \,\dot{\phi}(x)(\!\boldsymbol{\nabla}\!\phi) &= \frac{1}{2} \sum_\mathbf{k} \frac{\omega_k \mathbf{k}}{\omega_k} \Bigl( a_\mathbf{k}a^\dagger_\mathbf{k} + a^\dagger_\mathbf{k}a_\mathbf{k} + a_\mathbf{k}a_{-\mathbf{k}} e^{-2i\omega_k t} + a^\dagger_\mathbf{k}a^\dagger_{-\mathbf{k}} e^{2i\omega_k t} \Bigl) \\
 &= \sum_\mathbf{k} \mathbf{k} \,N_\mathbf{k} ,
\end{align}
where we have exploited the parity symmetry $\mathbf{k}\rightarrow -\mathbf{k}$ to conveniently cancel the last two terms (as well as the one emerging from the commutator) \emph{in this conditionally convergent sum}, and write the result in the last line.

Having constructed these observables, it will be of particular interest to us to evaluate their vacuum expectation value, which should correspond to vacuum energy and momentum, respectively. Not surprisingly, we find the vacuum momentum (in our particular summation convention) to be null:

\begin{align}
\braket{0| \mathbf{P} |0} = \mathbf{0}.
\end{align}

However, if we attempt to evaluate the vacuum energy, we immediately find a divergent result:

\begin{align}
\braket{0|H|0}  =  \frac{1}{2}\sum_\mathbf{k}\omega_k,
\end{align}
as the sum extends to infinitely many modes of arbitrarily high frequencies (therefore this is called an \emph{ultraviolet (UV) divergence}).

It is particularly useful to analyze this divergence in the continuum limit, making the correspondence \eqref{continuumlimit}, for which we obtain:

\begin{align}
\braket{0|H|0}  =  \frac{1}{2}\sum_\mathbf{k}\omega_k \;\longrightarrow\; \lim_{L\rightarrow\infty} \frac{L^3}{2(2\pi)^3}\int\! d^3\mathbf{k}\omega_k =  \biggl( \lim_{V\rightarrow\infty}\frac{V}{4\pi^2}\biggl)\int\limits_0^\infty\! dk \,{k^2}{\omega_k} \,.\label{MinkVE}
\end{align}

Here, we see (i) a divergent \emph{total energy} related to the fact that we are considering a homogeneous energy density in an infinite volume and (ii) a UV-divergent \emph{energy density} for the vacuum; as $\omega_k$ behaves like $\sim k$ at the UV ($k^2 \gg m^2$), we see that the integrand grows cubically, which means the integral diverges quartically.

However, for a free theory in the absence of gravity, only energy differences are observable quantities, not absolute energy values. Thus, one can simply ignore this divergent energy value, by \emph{redefining} the vacuum energy as $0$. This can be sistematically achieved for any observables in the theory through the well-known procedure of \textit{normal ordering}. In a normal-ordered observable, one just sets all annihilation operators to the right, and all creation operators to the left, so as not to have any residual contributions from the commutators; for example, we define the normal-ordered Hamiltonian:

\begin{align}
:\!H\!: \;= \tfrac{1}{2}\sum_\mathbf{k} \omega_k ( :\!a^\dagger_\mathbf{k}a_\mathbf{k}\!: + :\!a_\mathbf{k}a^\dagger_\mathbf{k}\!: ) = \sum_\mathbf{k} \omega_k a^\dagger_\mathbf{k}a_\mathbf{k} = \sum_\mathbf{k} \omega_k N_\mathbf{k}.
\end{align}
Thus:

\begin{align}
\braket{0| \!:\!H\!:\! |0} = 0.
\end{align}

We note, however, that it is only in very special circumstances that this extremely simple procedure works for a physically meaningful cancelation of vacuum energy divergencies. In the next section, we shall see a less trivial example for which it no longer applies.

\section{Vacuum Energy in Flat Space; the Casimir Effect} \label{Casimireff}

Much of the discussion in the present work regards vacuum energy. As we have seen in the last section, the most straightforward and naive approach to calculate it yields a divergent result. For free fields in Minkowski spacetime, this kind of divergency can be eliminated throughout by normal ordering, conventioning vacuum energy to be zero. In general, we will have to find a way to make sense of infinities which appear throughout for many observables through a systematic procedure of renormalization in curved spacetimes, which will be presented in more detail in Chapter \ref{renormalization}. 
Still, even for free fields in flat spacetimes, one may find nontrivial vacuum effects, which cannot be accounted for by mere normal ordering. Thus, in this section, we shall make a preamble of the subject of renormalization, employing a simpler subtraction procedure to account for these nontrivial vacuum effects in flat spaces, and make a connection with one of the few instances where there are experimental results in the subject.

In the original Casimir effect, one explores the physical effect of vacuum energy for the electromagnetic field in the presence of two large parallel conducting plates, separated by small distance $a$, and grounded in a common potential. The situation is depicted as follows:

\begin{figure}[H]
\centering
\includegraphics[width=0.6\linewidth]{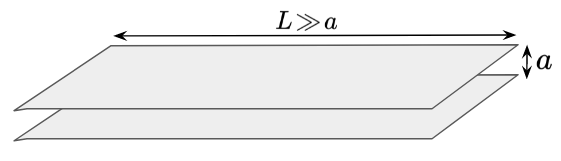}
\caption{Two parallel conducting plates, with characteristic width $L$, separated by a small distance $a \ll L$. Both plates are grounded at a common electric potential $V(z=0) = V(z=a) = cte$, usually conventioned to be $0$. \\  Source: By the author. }
\label{CasimirPlates}
\end{figure}

The presence of these grounded conducting plates creates a nontrivial boundary condition for the electromagnetic field, which is forced to vanish on these surfaces. In practice, we want to analyze the behaviour of the field between the plates and far from their edges, so that we shall just take the continuum limit in both transverse directions $L\rightarrow\infty$. In such a situation, we could use the following field modes:

\begin{align}
A^\mu \propto \epsilon^\mu e^{i \mathbf{k}_\perp \cdot \mathbf{x}_\perp} \sin(n\pi z/a)e^{-i\omega t} \label{EMmodes}
\end{align}
(where $A^\mu$ is the usual electromagnetic 4-potential, and $\epsilon^\mu$ represents a polarization vector).

In this setup, one can predict that there should be an attractive force between the plates due to vacuum fluctuation effects, shedding light on the nontrivial role that vacuum energy plays in Quantum Field Theory. This attractive force due to vacuum fluctuations between conductors is called the Casimir Effect; this effect has actually been measured in laboratory, making this one of the few instances of experimental evidence of vacuum energy (for a few historical details on the discovery and measurement of the Casimir Effect, check the last section of chapter 5 of \cite{fulling}, and references therein).

Inspired by this setup, we shall present here a simplified analogue model for the Casimir effect, using a massless scalar field with periodic boundary conditions. This model already allows us to compute a meaningful form of vacuum energy, and illustrates some of the general features of renormalized energy-momentum-stress observables, such as negative energy densities and pressures. Furthermore, we show that in this model it is more ``energetically favorable'' to have shorter period lengths $a$, mimicking the effects of an attractive force between plates in the actual Casimir setup.

Let us then develop the model in more detail. Since we chose periodic boundary conditions, we should work with modes of the form \eqref{V plane waves}, whose allowed wave vectors are given by:

\begin{align}
k^x, k^y = \frac{2\pi}{L}n^{x,y}, \qquad k^z = \frac{2\pi}{a}n^z, \qquad n^i \in \mathbb{Z}.
\end{align}

They correspond to the allowed frequencies:

\begin{align}
\omega_k = |\mathbf{k}| = 2\pi \sqrt{\frac{i^2+j^2}{L^2}+\frac{l^2}{a^2}},
\end{align}
where we have denoted $n^{x,y,z}$ as $i, j, l$, respectively.

Now, the first step in our analysis of the vacuum energy of this field is to put a divergent expression like \eqref{MinkVE} in a \emph{regularized} form. First, we write the energy density, which we shall regard as a function of the separation $a$, as \emph{a limit of a convergent sum}:

\begin{align}
\rho_0(a) = \frac{1}{V} \braket{0_a|H|0_a} = \frac{1}{2aL^2} \sum_\mathbf{k} \omega_k = -\frac{1}{2aL^2} \lim_{\alpha \rightarrow 0^+} \Bigl[ \frac{d}{d\alpha} \sum_\mathbf{k} e^{-\alpha \omega_k} \Bigl]. \label{fvenergy}
\end{align}

Note that for a finite $\alpha$ the exponential factor acts as a cutoff for arbitrarily high frequencies, taming the ultraviolet (UV), $|\mathbf{k}| \rightarrow \infty$, divergencies in the expression. Of course, we still get a divergent value for $\rho_0(a)$ when we take the limit $\alpha \rightarrow 0^+$. Our procedure then consists in keeping $\alpha$ temporarily finite -- which is called \emph{regularization} --, so that we can more closely identify the structures of the divergencies, and then find a meaningful physical subtraction to cancel them and obtain a finite result -- which is called \emph{renormalization}\footnote{The term \textit{renormalization} is actually associated with a wider procedure in which one absorbs these subtracted infinities in a \textit{redefinition} of basic parameters of the theory, such as masses and elementary charges. We shall discuss renormalization precedures in this more specific sense in chapter 4.3.}.
Once renormalization has been carried out, one may relax the regularization and take the limit $\alpha \rightarrow 0^+$, obtaining what is to be regarded as the physical prediction for that observable, to be compared with experiments\footnote{Sometimes comparisons of this type will involve the adjustment of one or several free renormalization parameters.}.

Now, the regularized expression for $\rho(a)$ reads

\begin{align}
\rho_0(\alpha,a) = -\frac{1}{2aL^2} \Bigl[ \frac{d}{d\alpha} \sum_\mathbf{k} e^{-\alpha \omega_k} \Bigl], \label{RegCasDensity}
\end{align}
for which we recover \eqref{fvenergy} as $\rho_0(a) = \displaystyle{\lim_{\;\;\alpha\rightarrow0^+}} \rho_0(\alpha,a)$.

As our notation suggests, we are particularly interested in determining how the vacuum energy density depends on $a$, and whether we can separate a finite contribution from it out of this divergent expression. To achieve that, we are going to torture our regularized expression \eqref{RegCasDensity}, making quite cumbersome operations on it, so that we may squeeze the $a$ dependence out of the infinities.

First, we define an auxiliary function

\begin{align}
S(\alpha,a) \equiv \frac{1}{L^2} \sum_\mathbf{k} e^{-\alpha \omega_k} \;\longrightarrow\; \frac{1}{(2\pi)^2} \sum_{l=-\infty}^{+\infty} \int d^2\mathbf{k}_\perp \exp[-\alpha( \mathbf{k}_\perp^2 +(\tfrac{2\pi}{a})^2l^2)^{1/2}], \label{Scasimir}
\end{align}
where we have taken the continuum limit ($L\rightarrow\infty$) for the transverse directions. Notice that, like $\rho_0$, this function's dependence on $a$ is given implicitly by how it determines the domain $\Omega$ of allowed wave vectors $\mathbf{k}$ over which we perform the summation. 

We can then write \eqref{Scasimir} as

\begin{align}
 S(\alpha,a) &= \frac{1}{(2\pi)^2} \sum_{l=-\infty}^{+\infty} \int d^2\mathbf{k}_\perp \exp[-\alpha( \mathbf{k}_\perp^2 +(\tfrac{2\pi}{a})^2l^2)^{1/2}] \nonumber \\
&= \frac{1}{2\pi} \int_0^\infty\!\! dk_\perp k_\perp e^{-\alpha k_\perp} + \frac{2}{2\pi}\sum_{l=1}^{\infty} \int_0^\infty\!\! dk_\perp k_\perp e^{-\alpha ( k_\perp^2 +(\tfrac{2\pi}{a})^2l^2)^{1/2} } \nonumber \\
&= \frac{1}{2\pi}\bigl[\,F(0) + 2\textstyle{\sum_l}F(l) \,\bigl], \label{Scasimir2}
\end{align}
where we have defined

\begin{align}
F(l) \equiv \int_0^\infty\!\! dk_\perp k_\perp e^{-\alpha \bigl[ k_\perp^2 +\bigl(\tfrac{2\pi}{a}\bigl)^{\!\!2}l^2\bigl]^{1/2} } = \biggl[ \frac{1}{\alpha^2} + \frac{1}{\alpha} \frac{2\pi l}{a} \biggl] e^{-\frac{2\pi l}{a} \alpha}.
\end{align}

Since we are ultimately interested in taking the limit $\alpha \rightarrow 0^+$, we would like to write some kind of power expansion in $\alpha$, so as to identify divergent, finite and vanishing terms. Here, it is very convenient to work with the Euler-Maclaurin formula for analytic functions:

\begin{align}
\tfrac{1}{2}F(b) + \sum_{l=1}^\infty F(b+l) = \int_b^\infty dl F(l) - \sum_{m=1}^\infty \frac{B_{2m}}{(2m)!}F^{(2m-1)}(b), \label{eulermaclaurin}
\end{align}
being $B_j$ the $j$th Bernoulli number (we have, for instance: $B_2=1/6$, $B_4=-1/30$, $B_6=1/42$, etc.).

We now apply this formula to (\ref{Scasimir2}), setting $b=0$. With a little algebraic effort, one may verify that

\begin{align}
F^{(1)}(0) = 0, \quad F^{(3)}(0) = 2\alpha \Bigl( \frac{2\pi}{a} \Bigl)^3, \quad \text{and} \quad F^{(j)}(0) = \mathcal{O}(\alpha^2), j \geq 5.
\end{align}
Besides, since $F(l)$ only depends on $a$ through the combination $l/a$, we have that:

\begin{align}
\int_0^\infty dl F(l) = a G(\alpha),
\end{align}
where $G(\alpha)$ does not depend on $a$. Thus, we may write

\begin{align}
\pi S(\alpha,a) &= aG(\alpha) - \frac{B_4}{24}F^{(3)}(0) + \mathcal{O}(\alpha^2) = aG(\alpha) + \frac{\pi^3}{45a^3}\alpha + \mathcal{O}(\alpha^2), \nonumber \\ 
  & \qquad \Rightarrow \quad S(\alpha,a) = \frac{a}{\pi}G(\alpha) + \frac{\pi^2}{45a^3}\alpha + \mathcal{O}(\alpha^2). \label{casimirS}
\end{align}

Now, we substitute this result in our original expression for $\rho_0(a)$, \eqref{fvenergy}. Note that when we take a derivative with respect to $\alpha$ and carry the limit $\alpha \rightarrow 0^+$ all of the terms $\mathcal{O}(\alpha^2)$ in \eqref{casimirS} vanish, so that we are left with

\begin{align}
\rho_0(a) = -\frac{1}{2a} \lim_{\alpha \rightarrow 0^+} \Bigl\{ \frac{d}{d\alpha}S(\alpha,a) \Bigl\} = -\frac{1}{2\pi} \lim_{\alpha \rightarrow 0^+} G'(\alpha) - \frac{\pi^2}{90a^4} \label{fvenergyS}
\end{align}
(where the prime $'$ on $G$ denotes its derivative with respect to $\alpha$).

This expression, of course, still presents a divergence when we take the limit (we have just tortured our expression, we have not mutilated it yet). However, we have isolated this divergence in the first term, which does not depend on $a$. Now, since in flat spacetime we are not concerned with absolute values of the energy, but are rather interested in how it \emph{varies} as we change the separation $a$, we have a freedom to redefine our 0-point energy and ignore this constant divergent term. An arguably natural choice for this 0-point is the Minkowski vacuum (corresponding to the limit $a \rightarrow \infty$); thus, we redefine the energy density through our regularized expression as:

\begin{align}
\rho(a) \equiv \lim_{\alpha \rightarrow 0^+} \bigl\{ \rho_0(\alpha,a) - \rho_0(\alpha, \infty) \bigl\}.
\end{align}
Then, we are left just with the second term in \eqref{fvenergyS},

\begin{align}
\rho(a) = - \frac{\pi^2}{90a^4}, \label{ScalarCasimirRho}
\end{align}
which makes for a quite nice expression after those lengthy calculations. Torture is over.

Let us now take a moment to analyze and interpret our results physically. First, note that the energy density we obtained is \emph{negative}. Of course, this is only necessarily so because we have defined the reference Minkowski vaccum energy as 0; we could well add any constant $C$ to that density ($\rho \rightarrow \rho + C$), and obtain a (possibly positive\footnote{
 However, since \eqref{ScalarCasimirRho} is unbounded as we make $a$ arbitrarily small, we can only have a positive $\rho$ for \emph{any} values of $a$ by making $C$ infinite.})
 physically equivalent result in flat spaces. Second, note that $\rho$ \emph{decreases as $a$ decreases}, regardless of the choice of $C$ (for $C=0$, it becomes \emph{more negative}). To look at the consequences of that more closely, it is clarifying to look at the transverse energy surface density, $\sigma = a \times\rho$, or, more conveniently, return to a finite \textit{fixed} transverse size, $L \gg a$, which yields a finite \emph{total} vacuum energy:
 
\begin{align}
E(a) = aL^2 \times \rho(a) = -\frac{\pi^2}{90a^3}L ^2 \qquad \Rightarrow \qquad \sigma(a) = -\frac{\pi^2}{90a^3}.
\end{align}
 
We see that $E$ also \emph{decreases} with $a$, so that it should be more ``energetically favorable'' to have arbitrarily small $a$ values. Analyzing the internal work that would be necessary to vary/expand $a$, we obtain a negative pressure $p$, given by:

\begin{align}
dE = -pdV \qquad \Rightarrow \qquad  p(a) = \frac{1}{L^2} \frac{dE}{da} = - \frac{\pi^2}{30a^4}. \label{ScalarCasimirPressure}
\end{align}

We emphasize that the pressure \eqref{ScalarCasimirPressure} turns out negative regardless of the choice of $C$ for the energy density. Of course, in our periodic condition setup there is no physical boundary to move, and to empirically analyze this pressure; it merely reflects the theoretical exercise of varying an arbitrary field periodicity length. However, we shall find an entirely analogous result in the original Casimir setup, where there is a physical boundary given by the conducting plates. 

In the electromagnetic case, there will be just two relevant differences in the calculation: the modes will be of the form \eqref{EMmodes}, rather than \eqref{V plane waves}, and there are two independent polarizations for the electromagnetic field. The latter will simply give us a $2$ factor, whereas the former has the effects of (i) changing the allowed $k^z$ vectors $2\pi l/a \rightarrow \pi l/a$ -- this will inflict a change in the value of  $F^{(3)}$, $2\alpha (\frac{2\pi}{a})^3 \rightarrow 2\alpha (\frac{\pi}{a})^3 $ --, and (ii) making the modes $+l$ and $-l$ linearly dependent (note that the $l=0$ mode makes no contribution to the renormalized $\rho$). Summarizing these factors, we obtain:
 
\begin{align}
\rho_{EM}(a) &= - \frac{1}{8}\times\frac{1}{2}\times2\times\frac{\pi^2}{90a^4} = - \frac{\pi^2}{720a^4}, \label{EMCasimirRho} \\[8pt]
p_{EM}(a) &= \frac{1}{L^2} \frac{dE_{EM}}{da} = - \frac{\pi^2}{240a^4}. \label{EMCasimirPressure}
\end{align}

In flat space, one often interprets these negative values as being merely relative to the `outside region' -- an embedding Minkowski spacetime --, since there are less modes that ``fit'' in the finite length $a$ (imposing the appropriate boundary conditions). Furthermore, this explanation seems quite natural in both our simplified periodic setup and the original Casimir one, as both can be embedded in a larger, Minkowski space. In the latter, then, one interprets the attractive force between the plates as being due to a higher \emph{positive vacuum pressure outside, pushing the plates together}.
 
However, the situation is radically different in curved spacetimes, where this kind of interpretation is no longer generally attainable, for two basic reasons. First, when we take gravity into account, we must ascribe physical meaning to \emph{absolute values} of energy (more precisely of energy-momentum-stress), which act as the source of curvature in Einstein's Equations, so that one is no longer at liberty of considering only energy differences. Secondly, there is generally no embedding spacetime (or `outside' region) to compare to, so that one is forced to analyze the quantum fluctuation effects of renormalized observables \textit{intrinsically}. In particular, this means that we can end up with physical, renormalized negative energy densities and pressures, even when those are classically positive-definite.

\section{Formal Remarks on Expansions in Normal Modes} \label{NormalModes}

Already from the fact that there are divergencies in the theory, we can anticipate that there are formal issues not fully addressed in the presentation so far (indeed, one can glimpse in Appendix A that these emerge from a forced attempt to make sense of products of distributions). Although the present text does not aim at providing a fully rigorous treatment of QFT, it is the author's personal belief that a more thorough presentation of some of its formal aspects may be very enlightening both operationally and conceptually, especially when we must handle intricate and often physically nebulous topics such as renormalization. A more complete treatment of the topics covered in this and the next section is given in \cite{fulling} (chapters 2-4), from where most of the exposition here is based and supplementary material to this discussion can be found in Appendix \ref{distributions}.

The problem of expansion in normal modes is one of Linear Algebra. Given a Hilbert space that contains all the acceptable physical states in our theory (and excludes all the unphysical ones) -- armed with a complete set of elementary observables which allows us to build any physical observable to probe it --, we typically want a convenient basis in terms of which we may expand any state in it. While this is a relatively trivial task for any finite-dimensional Hilbert space, it involves some subtleties when we go to infinite dimensions.

Already in more elementary instances such as nonrelativistic Quantum Mechanics, one is faced with the problem of nonnormalizable wave functions in the continuum. In this context, one usually breaks the types of eigenvalue problem in two instances:

1 - Discrete spectrum $\{E_j\}$: in this case, one can find a complete set of normalizable wave functions $\psi_j$, which form an ordinary orthornormal basis in the Hilbert space $\mathcal{H} \subset \mathcal{L}^2$ of the theory:

\begin{align}
H \psi_j = E_j \psi_j, \quad \braket{\psi_j,\psi_k} = \delta_{jk}.
\end{align}
In this instance, one can expand any wave function $\phi$ in terms of this basis:

\begin{align}
\phi(j) \equiv \braket{ \psi_j, \phi} \quad \Rightarrow \quad \phi(x) = \sum_j \phi(j) \psi_j(x),
\end{align}
as well as easily evaluate the action of operators:

\begin{align}
H \phi = \sum_j E_j \phi(j) \psi_j.
\end{align}

2 - Continuous spectrum $\{E_\lambda\}$: in this case, the functional solutions $f_\lambda$ to the differential equation,

\begin{align}
H f = E_\lambda f_\lambda \quad \Leftrightarrow \quad (H - E_\lambda) f_\lambda = 0,
\end{align}
\emph{will not generally belong to the Hilbert space $\mathcal{H}$}, and therefore will not be proper wave functions. Nevertheless, the Spectral Theorem (see chapter 2 of \cite{fulling}) assures that one may still project any wave function $\phi\in\mathcal{H}$ in all of these modes and write the expansions

\begin{align}
\phi(\lambda) = \braket{f_\lambda, \phi} \quad \Rightarrow \quad \phi(x) = \int_{\sigma(H)} \!\!\!\! d\lambda \, \phi(\lambda) f_\lambda(x), \label{contvecexpansion}
\end{align}
as well as

\begin{align}
H\phi = \int_{\sigma(H)} \!\!\!\! d\lambda \, \phi(\lambda) E_\lambda f_\lambda.
\end{align}

So far, we have treated both the discrete index $j$ and the continuous one $\lambda$ as nondegenerate. It may well happen that there are degeneracies. For instance, in the continuous case, we could have a free particle with a nonzero spin, which would oblige us to modify an expression like \eqref{contvecexpansion} to

\begin{align}
\phi_i(\lambda) \equiv \braket{f_{i,\lambda}, \phi} \quad \Rightarrow \quad \phi(x) = \int_{\sigma(H)} \! \sum_i \phi_i(\lambda) f_{i,\lambda}(x)d\lambda.
\end{align}

More generally, there may also be continuous degeneracies in each eigenvalue, and/or the dimension of each subspace may depend on the eingenvalue $E$. One may even have a mixture of the two cases considered above (as it happens in hydrogen atom, where one has a point spectrum for bounded states $E<0$ and a continuous spectrum for unbounded states $E>0$). To avoid more cumbersome notations and the need to split between various cases, we condense our notation through a single, nondegenarate index $\lambda$ ($\lambda$ may belong to a multidimensional space, and comprise both continuous and discrete indices) and write our expansions as

\begin{align}
\phi(x) = \int_{\sigma(H)} d\mu(\lambda) \braket{u_\lambda,\phi} u_\lambda(x) \,,
\end{align}
where $\mu(\lambda)$ represents a measure over the spectrum\footnote{If the reader is unfamiliar with the concept of a measure, we recommend section 1.D of \cite{yvone}.}. In continuous portions of $\sigma(H)$, $\mu(\lambda)$ will be a continuous, monotonically increasing function, whereas in the discrete portions, it will be a constant function with discontinuous ``jumps'' at $\lambda \in \sigma(H)$ (so that $d\mu(\lambda)$ will be a countable sum of Dirac deltas).

Having introduced this unified notation, we take the chance to explore slightly more general linear scalar field equations, in the form

\begin{align}
-\frac{\partial}{\partial t^2} \phi(\mathbf{x},t) = H^2_\mathbf{x}\phi(\mathbf{x},t) \qquad \Leftrightarrow \qquad \bigl[ \partial^2_t + H^2_\mathbf{x} \bigl]\phi(\mathbf{x},t) = 0, \label{elliptic}
\end{align}
where $H_\mathbf{x}$ is an elliptic differential operator\footnote{
  For practical purposes, one must not worry with the precise definition of an elliptic operator; here, it will be a technical requirement for the Hamiltonian to be bounded from below, so that there is a stable vacuum state. The reader will find more precise definitions and thorough discussion in \cite{fulling}.}
  acting only in the spatial variables (we recover the familiar Klein-Gordon equation \eqref{fkge} by making $H^2_\mathbf{x} = -\nabla^2_{\!\mathbf{x}} + m^2 $ ). Equation \eqref{elliptic} still allows for a mode decomposition in the form:
 
\begin{align}
u_\lambda(\mathbf{x},t) = \frac{e^{-i\omega_\lambda t}}{\sqrt{2\omega_\lambda}} \psi_\lambda(\mathbf{x}),
\end{align}
where

\begin{align}
H^2_\mathbf{x} \psi_\lambda(\mathbf{x}) = \omega_\lambda^2 \psi_\lambda(\mathbf{x}).
\end{align}
And therefore, we write an expansion for $\phi$ in the form

\begin{align}
\phi(x) = \int_{\sigma(H)} \! d\mu(\lambda) \, a(\lambda) u_\lambda(x) + a^\dagger(\lambda) u^*_\lambda\!(x),
\end{align}
for which the commutation relations read

\begin{subequations} \label{mocore_lambda}
\begin{align}
[a(\lambda),a(\lambda')] = [a^\dagger(\lambda),a^\dagger(\lambda')] &= 0, \\
[a(\lambda),a^\dagger(\lambda')] &= \delta(\lambda,\lambda'),
\end{align}
\end{subequations}
where $\delta(\lambda,\lambda')$ is the delta distribution with respect to the measure $\mu$, that is

\begin{align}
\int_{\sigma(H)}\! d\mu(\lambda') f(\lambda') \delta(\lambda,\lambda') = f(\lambda).
\end{align}

We shall make use of this more general, unified expansion in the next section, where we analyze integral kernels to the wave equation \eqref{elliptic} and two-point functions.

We shall make use of this more general, unified expansion in the next section, where we analyze integral kernels to the wave equation \eqref{elliptic} and two-point functions.

Also, we note that in the continuum case there will be a subtlety in the definition of field modes occupations and Fock space. Just as eigenvectors from positions and momenta, the states determined by application of one creation operator in the continuum, $\ket{1_\lambda} = a^\dagger(\lambda)\ket{0}$, must be understood in a generalized, distributional sense (see Appendix \ref{distributions}). Actual one-particle states will be given by integrals of these generalized states in a continuous interval:

\begin{align}
\ket{ \psi_1 } = \int _{\sigma(H)}\! d\mu(\lambda') \rho(\lambda) a^\dagger(\lambda) \ket{0},
\end{align}
and one can similarly write n-particle states $\ket{\psi_n}$ with integrals of products of $n$ creation operators.

Throughout most of this dissertation, however, we shall simply write our states as discrete sums with the usual notation in Fock spaces, and the appropriate generalization will be implied in them continuum.

\section{Two-point Functions} \label{2-PFs}

To conclude this chapter, we give an overview of a very important class of functions, useful to perform many computations in the theory: two-point functions. These are the (number-valued) expectation values of observables bilinear in field amplitudes at two spacetime events\footnote{As the reader may have noted from appendix \ref{distributions}, these are generally not functions, but distributions. Still, we maintain the terminology throughout the section and the rest of the dissertation.} $x$ and $x'$:

\begin{align}
G(x,x') = \braket{ \Psi| f\bigl( \phi(x),\phi(x') \bigl) | \Psi},
\end{align}
where $f\bigl( \lambda\phi(x), \lambda'\phi(x') \bigl)= \lambda\lambda'f\bigl( \phi(x),\phi(x') \bigl)$.

These will be central in computing very important physical quantities, such as field correlations $f\!=\! \phi(x)\phi(x')$, commutators $f\!=\![\phi(x),\phi(x')]$ and anticommutators $f\!=\!\{\phi(x),\phi(x')\}$. Since many of these bilinear observables are actually proportional to the identity operator (as is the case with the commutator of scalar fields), their corresponding two-point functions will actually be state-indepent. Generally, however, they may bear a state dependence (which is quite natural, for instance, for field correlations), and a state that will have central importance for computing them is the vacuum $\ket{0}$. In fact, it turns out that vacuum expectation values of various of these bilinear observables can be identified with various Green Functions of the field equations.

Turning to the example of the more general scalar field introduced in the previous section, we then begin our analysis by investigating the Green functions of the wave equation \eqref{elliptic}:

\begin{align}
\bigl[ \partial_t^2 + H^2_\mathbf{x} \bigl] G(x,x') = \delta(x,x') = \delta(t-t') \delta(\mathbf{x}-\mathbf{x}'), \label{greenequation}
\end{align}
where we have made explicit use of global inertial coordinates to split $\delta(x,x')$ into spatial and temporal Dirac deltas\footnote{We just need a weaker split between space and time, but we will avoid getting more technical at this point.}.

This integral kernel (Green function) allows us to write the classical field solutions $g(x)$ to our wave equations with a source $J(x)$, $[\partial^2_t + H^2_\mathbf{x}]g(x) = J(x)$, in the form

\begin{align}
g(x) = \int d^4\!x' J(x') G(x,x') + \text{Homogeneous solution}. \label{classgreensol}
\end{align}
If we then take a formal Fourier transform in time $t$ and a spectral transform in space $\mathbf{x}$ \linebreak ( $\int \!dt e^{i\omega t} \!\int \!d^3\mathbf{x} \, \psi^*_\lambda\!(\mathbf{x})$ ) in equation \eqref{greenequation}, we obtain

\begin{align}
(-\omega^2 + \omega_\lambda^2)\tilde{G}(\omega,\lambda; x') = e^{i\omega t'} \psi^*_\lambda (x').
\end{align}
Then, reversing this formal Fourier transform, we obtain the following expansion for $G(x,x')$:

\begin{align}
G(x,x') = -\frac{1}{2\pi} \int d\mu(\lambda) \int d\omega \frac{1}{\omega^2-\omega_\lambda^2}e^{-i\omega(t-t')} \psi^*_\lambda(\mathbf{x'}) \psi_\lambda(\mathbf{x}) \,.\label{GFFourier}
\end{align}

When we attempt to make sense of this formal expression, starting from the $\omega$ integral in the right, we run into trouble due to the poles of the integrand at $\pm \omega_\lambda$. How, then, should we compute (and interpret) this expression? Well, as we are carrying this integral in the real axis, one could propose we displace the poles by a distance $\epsilon$ in the complex plane, carry on the finite (regularized) integral, and then try to take the limit $\epsilon \rightarrow 0$. Equivalently, one could leave the poles fixed and displace the integration contour a little around them\footnote{
 In this case, one does not have to actually take any limits to bring the contour back into the real axis, as the integral will be evaluated by Cauchy's theorem, remaining invariant unless the contour crosses a pole. If the reader is not particularly comfortable with contour integrations, we recommend a brief review, \textit{e.g.}, in chapters 6 and 7 of \cite{arfken}.}.
 As there are numerous ways to displace the contours, we end up with correspondingly numerous Green functions. 

\begin{figure}[H]
\centering
\includegraphics[width=0.7\linewidth]{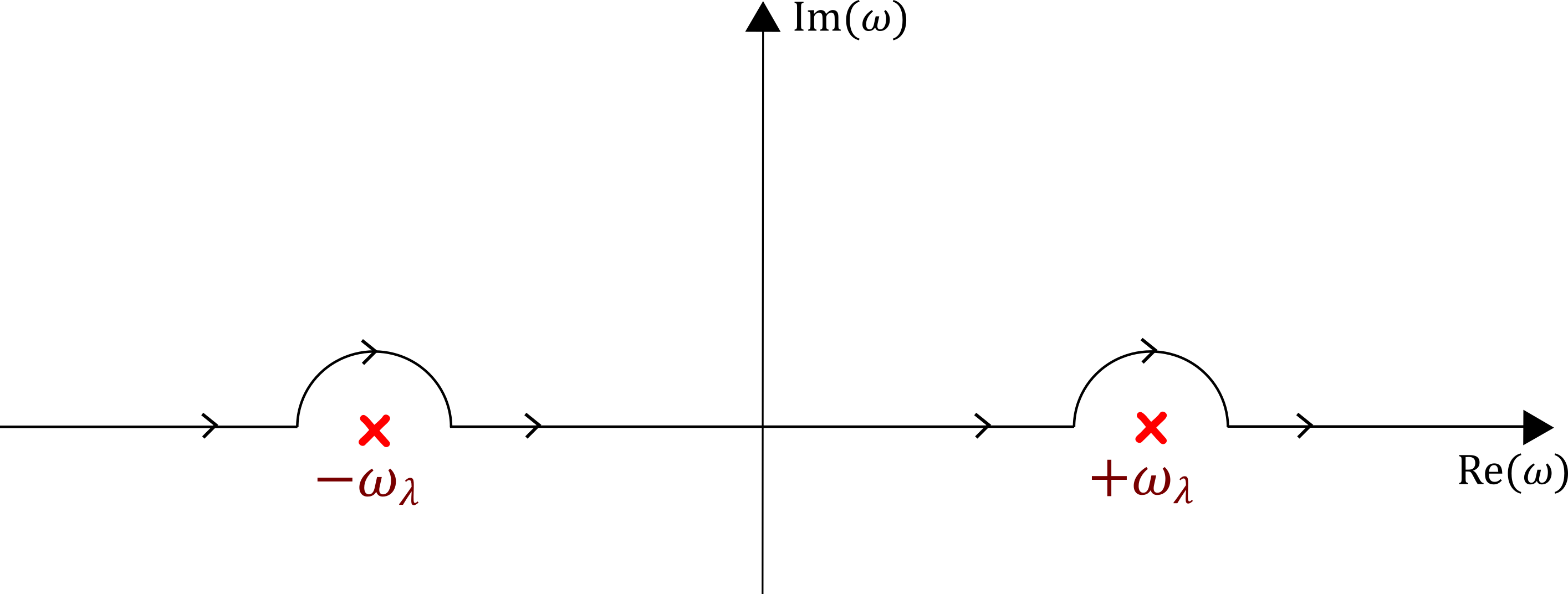}
\caption{Contour for the Retarded Green Function. As the other contours, it must be closed at the upper half of the complex plane for $t<t'$ and at the lower half for $t>t'$. \\ Source: By the author. }
\label{GRETARDED}
\end{figure}

Before we actually carry the various integrations, it is worth noting explicitly that, for $t>t'$, $e^{-i\omega(t-t')}$ will exponentially diverge in the upper half complex plane (``as $\omega \rightarrow +i\infty$'') and exponentially decay in the lower half complex plane (``as $\omega \rightarrow -i\infty$''), and vice-versa for $t<t'$. To assure the contributions outside the real axis will vanish, we \emph{should always close the contour in the decaying region} (\textit{e.g.} by a semicircle at infinity), so that the encompassed poles will depend on the considered times. With those considerations, let us enumerate and compute a few relevant Green function, giving the prescription for their respective contours:

\textbf{1- Retarded Green Function, $G_{ret}$:}  pass the contour \emph{above} both poles (see Figure \ref{GRETARDED}), yielding

\begin{subequations}
\begin{empheq}[left={G_{ret}(x,x') = \empheqlbrace}, right={\;\;.}]{align}
 &  0,  & t<t' \\
 -& 2\pi i \bigl( \operatorname{Res}_I(\omega\!=\!+\omega_\lambda) + \operatorname{Res}_I(\omega\!=\!-\omega_\lambda)  \bigl),  & t>t' 
 \end{empheq}
\end{subequations}

Thus, for $t>t'$:

\begin{align}
G_{ret}(x,x') &= \int d\mu(\lambda) \frac{i}{2\omega_\lambda} \bigl[ e^{-i\omega_\lambda(t-t')} - e^{+i\omega_\lambda(t-t')} \bigl] \psi_\lambda^*(\mathbf{x'})\, \psi_\lambda(\mathbf{x}) \nonumber \\ 
 &= \int \frac{d\mu(\lambda)}{\omega_\lambda} \sin\bigl( \omega_\lambda(t-t') \bigl) \psi_\lambda^*(\mathbf{x'})\, \psi_\lambda(\mathbf{x}). \label{Gretarded}
\end{align}

We can then see that, in eq \eqref{classgreensol}, this kernel would correspond to a solution $g(x)$ of the field equations that incorporates the source $J(x')$ only to the past of $x$ (we shall see ahead that it is actually supported in the past light cone of $x$, which will independ of the arbitrary choice of simultaneity for space separated events).

\begin{figure}[H]
\centering
\includegraphics[width=0.6\linewidth]{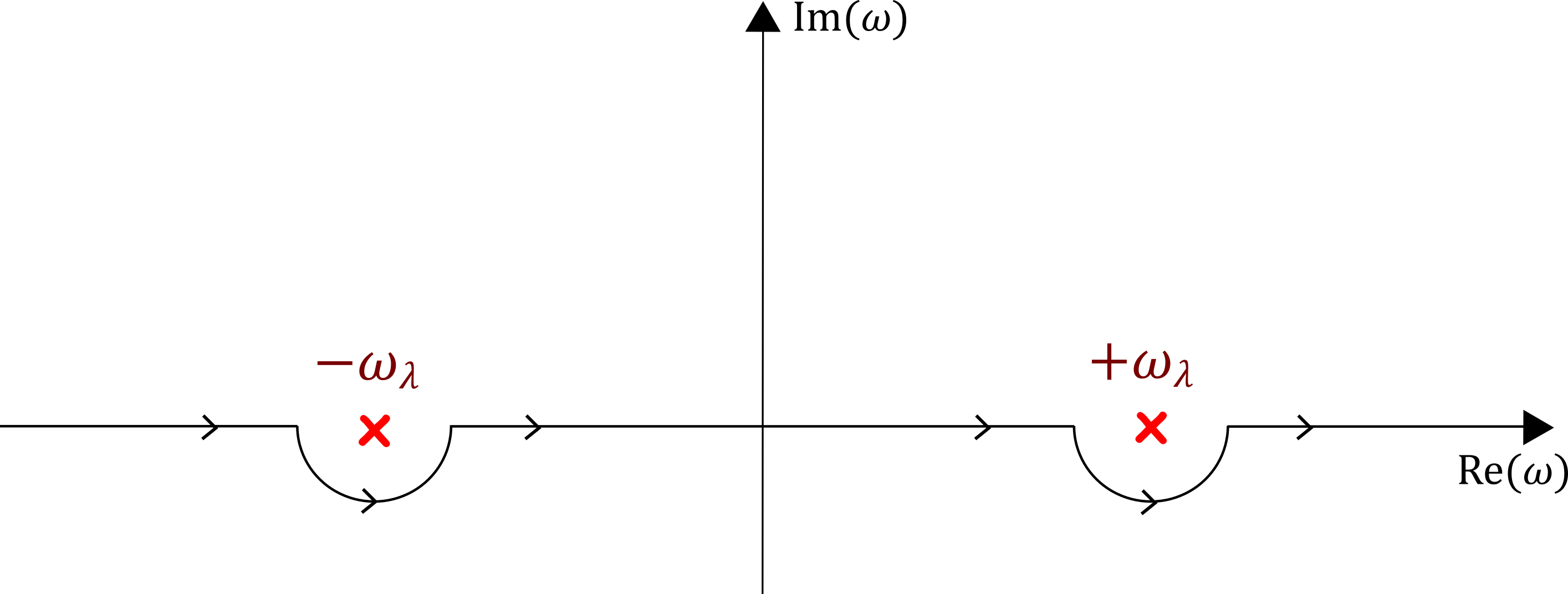}
\caption{Contour for the Advanced Green Function. \\ Source: By the author. }
\label{GADVANCED}
\end{figure}

\textbf{2- Advanced Green Function, $G_{adv}$:} pass the contour \emph{under} both poles (see Figure \ref{GADVANCED}), so that

\begin{subequations}
\begin{empheq}[left={G_{ret}(x,x') = \empheqlbrace}, right={\;\;.}]{align}
 +& 2\pi i \bigl( \operatorname{Res}_I(\omega\!=\!+\omega_\lambda) + \operatorname{Res}_I(\omega\!=\!-\omega_\lambda)  \bigl), & t<t'\\
 & 0,  & t>t'
 \end{empheq}
\end{subequations}

Then, similarly to \eqref{Gretarded}, we have for $t<t'$:

\begin{align}
G_{adv}(x,x') &= -\int \frac{d\mu(\lambda)}{\omega_\lambda} \sin\bigl( \omega_\lambda(t-t') \bigl) \psi^*_\lambda(\mathbf{x'})\, \psi_\lambda(\mathbf{x}). \label{Gadvanced}
\end{align}

Correspondingly, this kernel yields a solution that incorporates the source $J(x')$ only to the future of $x$ (complementary to $G_{ret}$, $G_{adv}$ will only be supported in the future light cone of $x$).

\textbf{3- The Feynman propagator, $G_F$:} go \emph{under the left pole} ($\omega=-\omega_\lambda$), and \emph{over the right one} ($\omega=+\omega_\lambda$), so that one or the other will contribute for $t>t'$ or $t<t'$. 

\begin{subequations}
\begin{empheq}[left={G_{F}(x,x') = \empheqlbrace}, right={\;\;.}]{align}
 +& 2\pi i \operatorname{Res}_I(\omega\!=\!-\omega_\lambda),  & t<t' \\
 -& 2\pi i \operatorname{Res}_I(\omega\!=\!+\omega_\lambda),  & t>t' 
 \end{empheq}
\end{subequations}

\begin{figure}[H]
\centering
\includegraphics[width=0.6\linewidth]{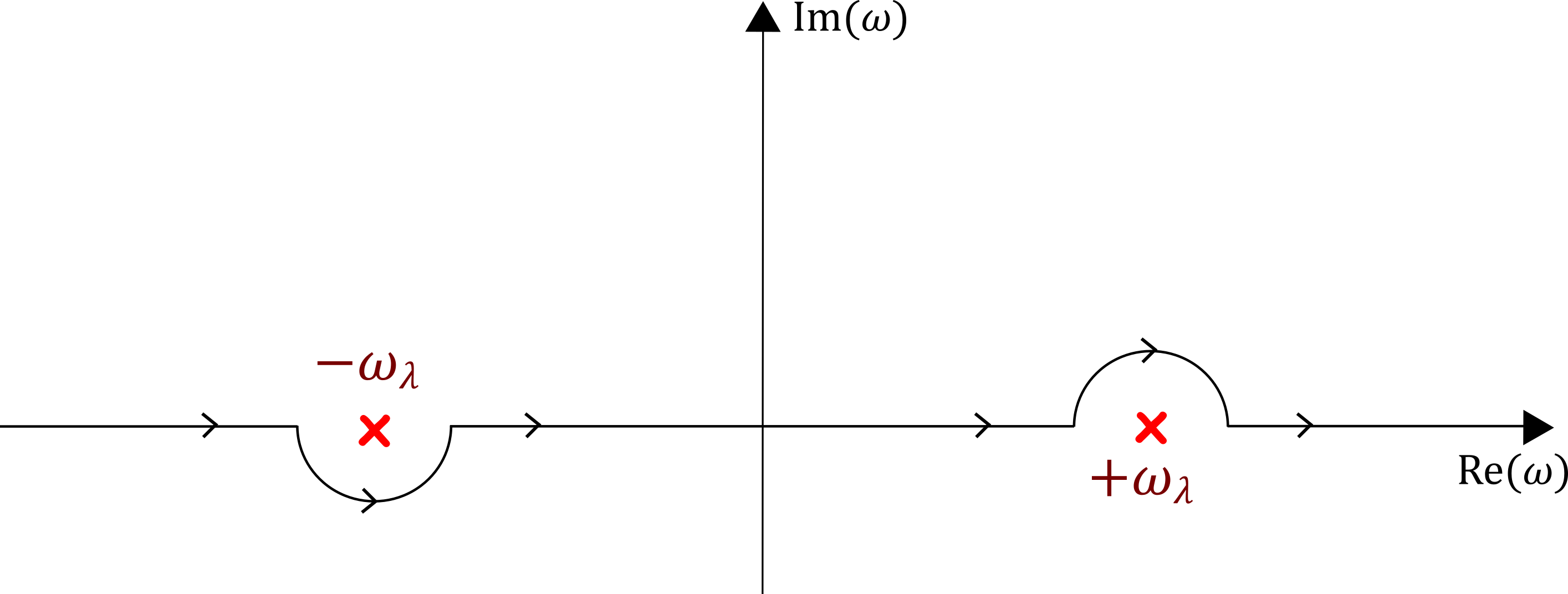}
\caption{Contour for the Feynman propagator. \\ Source: By the author. }
\label{GFEYNMAN}
\end{figure}

Carrying the residue integration, we find:

\begin{align}
G_F(x,x') = i \int \frac{d\mu(\lambda)}{2\omega_\lambda} e^{ -i\omega_\lambda|t-t'|} \psi_\lambda(\mathbf{x'})\, \psi^*_\lambda(\mathbf{x}). \label{GFeynmann}
\end{align}

\begin{figure}[H]
\centering
\includegraphics[width=0.6\linewidth]{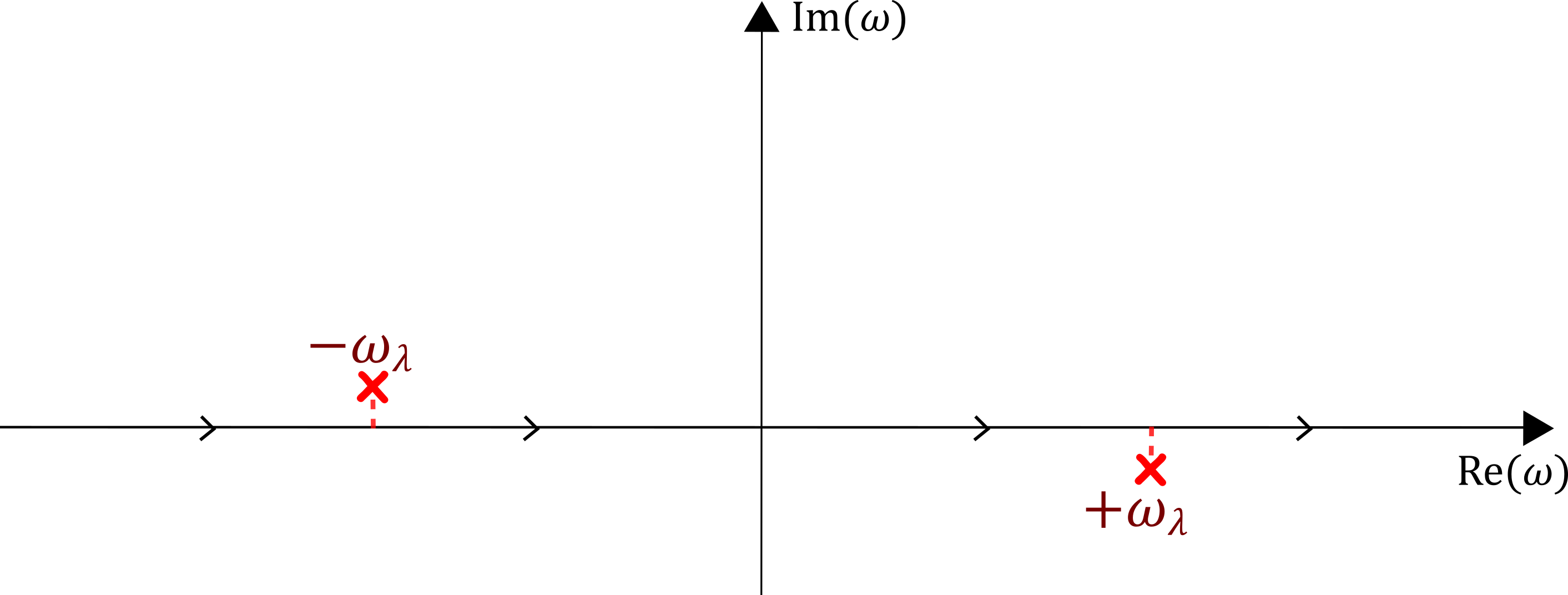}
\caption{Contour for the Feynman propagator in the real axis, with the poles correspondingly displaced in the imaginary plane. \\ Source: By the author. }
\label{GiFEYNMAN}
\end{figure}

This will be a crucial Green Function, as it represents a particularly important integral kernel to the inverse of our differential operator, $[\partial_t^2 + H_\mathbf{x}]^{-1}$, as stated in equation \eqref{greenequation}. To see this more clearly, we cast the residue integral in a different form, displacing both poles an infinitesimal distance from the real axis, as (see Figure \ref{GiFEYNMAN}):

\noindent \strut\vspace{14pt} \hspace{50pt}
\begin{minipage}[t]{0.3\linewidth}
 \begin{empheq}[left=\empheqlbrace]{align*}
  \omega &= -\omega_\lambda \\[4pt]
  \omega &= +\omega_\lambda
 \end{empheq}
\end{minipage}\hspace{-25pt}%
\begin{minipage}[t]{0.1\linewidth}
  \begin{align*}
  \\[-20pt]
  \qquad \scalebox{1.4}{$\longrightarrow$}
  \end{align*} 
\end{minipage}\hspace{35pt}%
\hspace{-20pt}
\begin{minipage}[t]{0.3\linewidth}
 \vspace{0pt}
 \begin{empheq}[left=\empheqlbrace, right = {\;\;,}]{align*}
  \omega &= -\omega_\lambda + i\epsilon \\[4pt]
  \omega &= +\omega_\lambda - i\epsilon
 \end{empheq}
\end{minipage}\\
so that, to first order in $\epsilon$:

\begin{align}
\omega^2 + \omega_\lambda(-\omega_\lambda) \qquad \longrightarrow \qquad \omega^2 + (\omega_\lambda - i\epsilon)(-\omega_\lambda + i\epsilon) = \omega^2 - \omega_\lambda^2 + i\epsilon \quad
\end{align}
(where we have absorbed a positive factor $\omega_\lambda/2$ into $\epsilon$ in the last equality).

Thus, we write $G_F$ as

\begin{align}
G_{F}(x,x') &= \lim_{\epsilon \rightarrow 0^+} \; \Bigl\{ -\frac{1}{2\pi} \int d\mu(\lambda) \int d\omega \frac{e^{-i\omega_\lambda(t-t')}}{\omega^2 - \omega_\lambda^2 + i\epsilon} \psi_\lambda^*(\mathbf{x'})\, \psi_\lambda(\mathbf{x}) \Bigl\} . \label{iGFeynmann}
\end{align}
(Note that, using \eqref{1/iepsilon} -- and paying proper attention to which pole we are encompassing for each $t$ -- one can easily recover \eqref{GFeynmann}.)

From this form, it is easy to see that it will correspond to the following integral kernel (again, working in first order in $\epsilon$):

\begin{align}
G_F = \lim_{\;\epsilon \rightarrow 0^+} \bigl\{ \operatorname{Ker}\bigl( [\partial^2_t + H_\mathbf{x}^2 - i\epsilon]^{-1} \bigl) \bigl\} . \label{imaginary kernel}
\end{align}

Indeed, in chapter \ref{renormalization} we will make use of it in the renormalization of the effective action. Also, this $i\epsilon$ factor will play the role of a regularizer for path integrals, making them well defined when we vary $e^{iS}$ taking field amplitudes up to infinity. 

Finally, we note that this integral kernel will obey a Green equation \textit{with a reversed sign}:

\begin{align}
    [\partial^2_t + H_\mathbf{x}^2]G_F(x,x') = - \delta(x,x') = -\delta(t-t')\delta(\mathbf{x}-\mathbf{x'}) \,. \label{GF minus equation}
\end{align}

\textbf{4- Principal Value Green Function, $\bar{G}$:} pass the integration contour directly through the poles, taking the \emph{principal value} at each one (see Appendix \ref{distributions} for more details on principal-value distributions). 

\begin{figure}[H]
\centering
\includegraphics[width=0.6\linewidth]{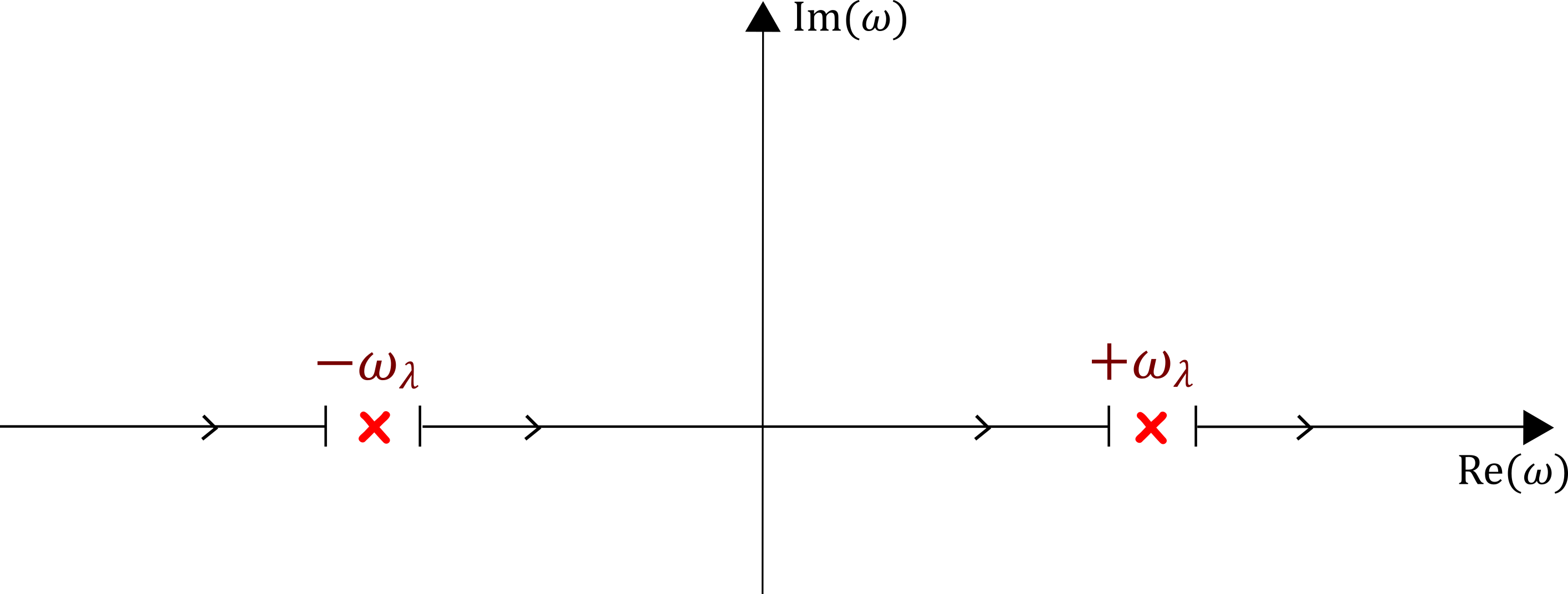}
\caption{Integration representation for the Principal Value Green Function. Here, one must approach each pole symmetrically from both sides, to cancel out the divergent terms. \\ Source: By the author. }
\label{GPV1}
\end{figure}

\begin{figure}[H]
\centering
\includegraphics[width=0.6\linewidth]{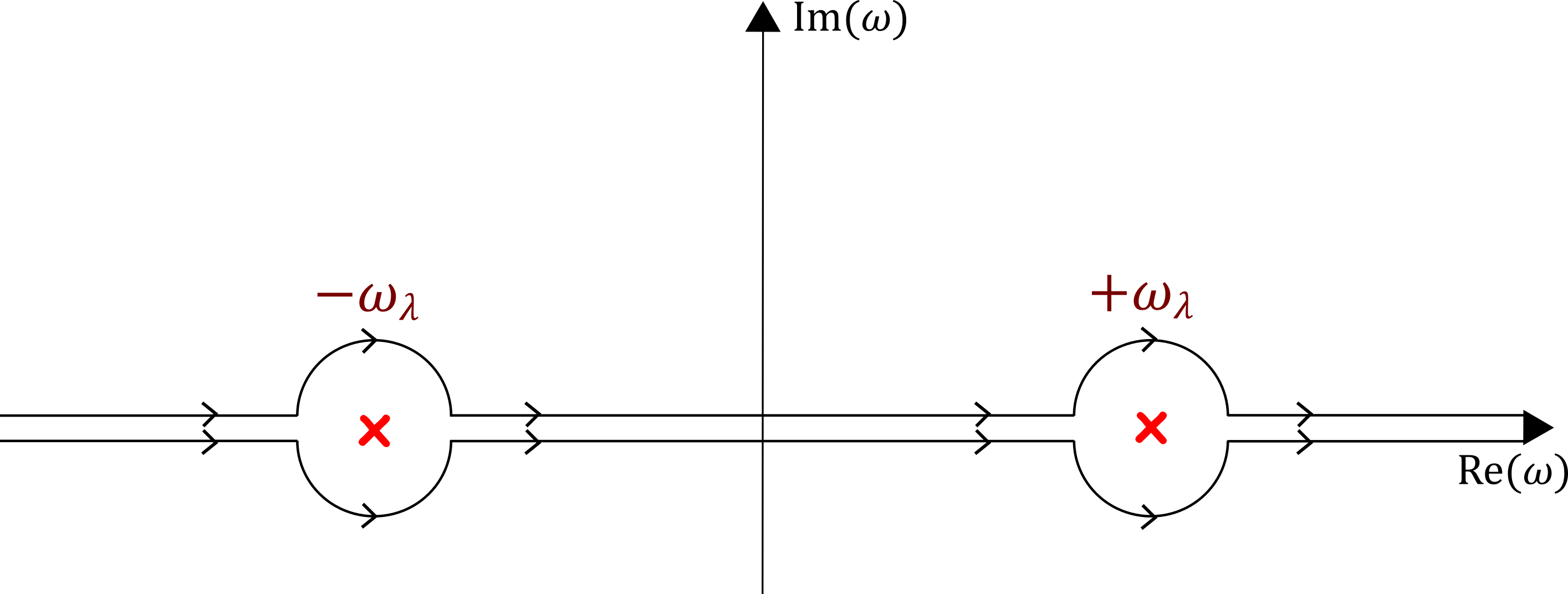}
\caption{The Principal Value contour represented as (half of) the juxtaposition of the advanced and retarded contours. \\ Source: By the author. }
\label{GPV2}
\end{figure}

This contour can be thought of as the juxtaposition of the advanced and retarded contours (more precisely, half of this jusxtaposition, so one does not count the integral twice), as we try to sketch in Figure \ref{GPV2}. Thus, we have:

\begin{align}
\bar{G}(x,x') \equiv \frac{1}{2} \bigl( G_{ret}(x,x') + G_{adv}(x,x') \bigl).
\end{align}

\begin{figure}[H]
\centering
\includegraphics[width=0.6\linewidth]{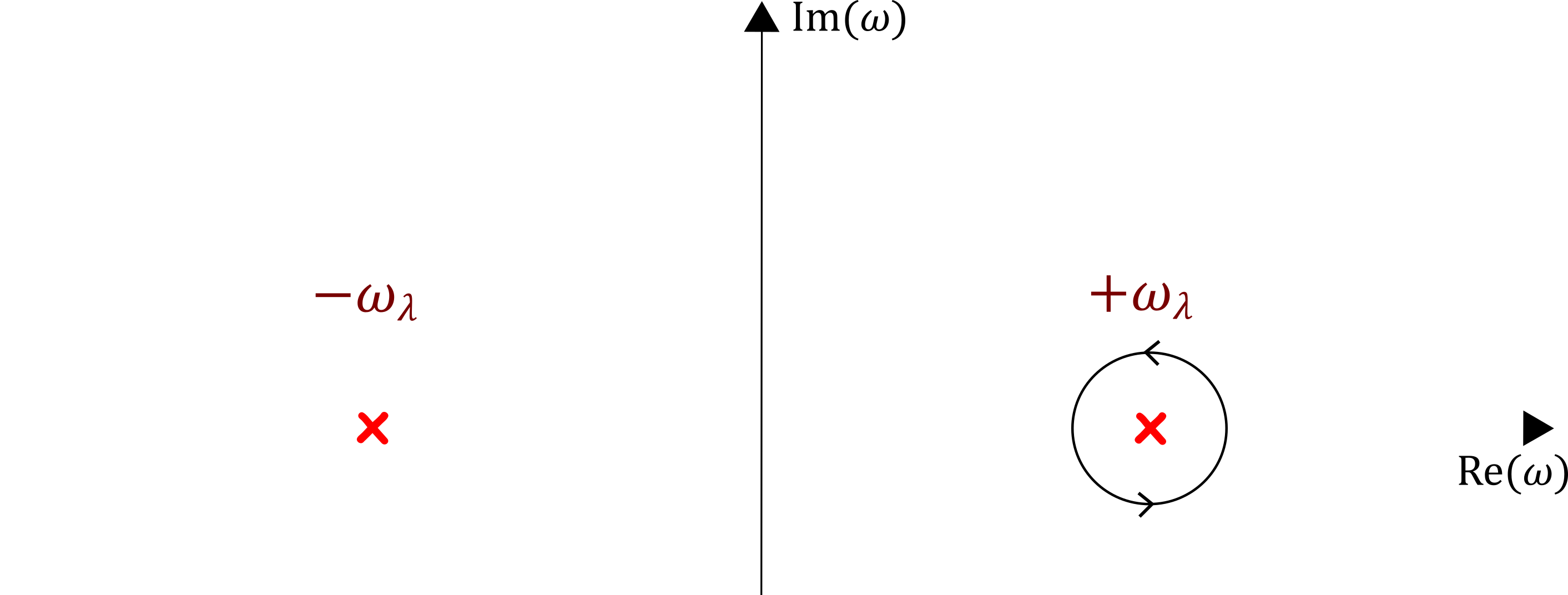}
\caption{Closed contour corresponding $-iG^+$. Here, one goes around the right pole once, counterclockwise. \\  Source: By the author. }
\label{GWIGHTMAN+}
\end{figure}

All of the kernels presented so far are \emph{actual Green functions}, obeying the inhomogeneous equation \eqref{greenequation}. By taking the difference between pairs of them, we arrive at solutions to the \emph{homogeneous} wave equation, corresponding to closed contours in the complex plane. In the literature, these are also referred to as Green functions, so we shall maintain that terminology. We enumerate a few relevant ones:

\textbf{5- Wightman Function $-iG^+$:} take a retarded contour and subtract a (properly adjusted) Feynman one, so that we end up with a closed curve around the right pole (see Figure \ref{GWIGHTMAN+}). This yields

\begin{align}
G^+ = \int \frac{d\mu(\lambda)}{2\omega_\lambda} e^{ -i\omega_\lambda(t-t')} \psi^*_\lambda(\mathbf{x'})\, \psi_\lambda(\mathbf{x}). \label{Gwhightmann+}
\end{align}

\textbf{6- Wightman Function $+iG^-$:} similarly by taking the Feynman contour and subtracting the advanced one, we encompass the left pole, yielding

\begin{align}
G^- = \int \frac{d\mu(\lambda)}{2\omega_\lambda} e^{ +i\omega_\lambda(t-t')} \psi^*_\lambda(\mathbf{x'})\, \psi_\lambda(\mathbf{x}). \label{Gwhightmann-}
\end{align}

We also note here that $G^- = (G^+)^*$.

\begin{figure}[H]
\centering
\includegraphics[width=0.6\linewidth]{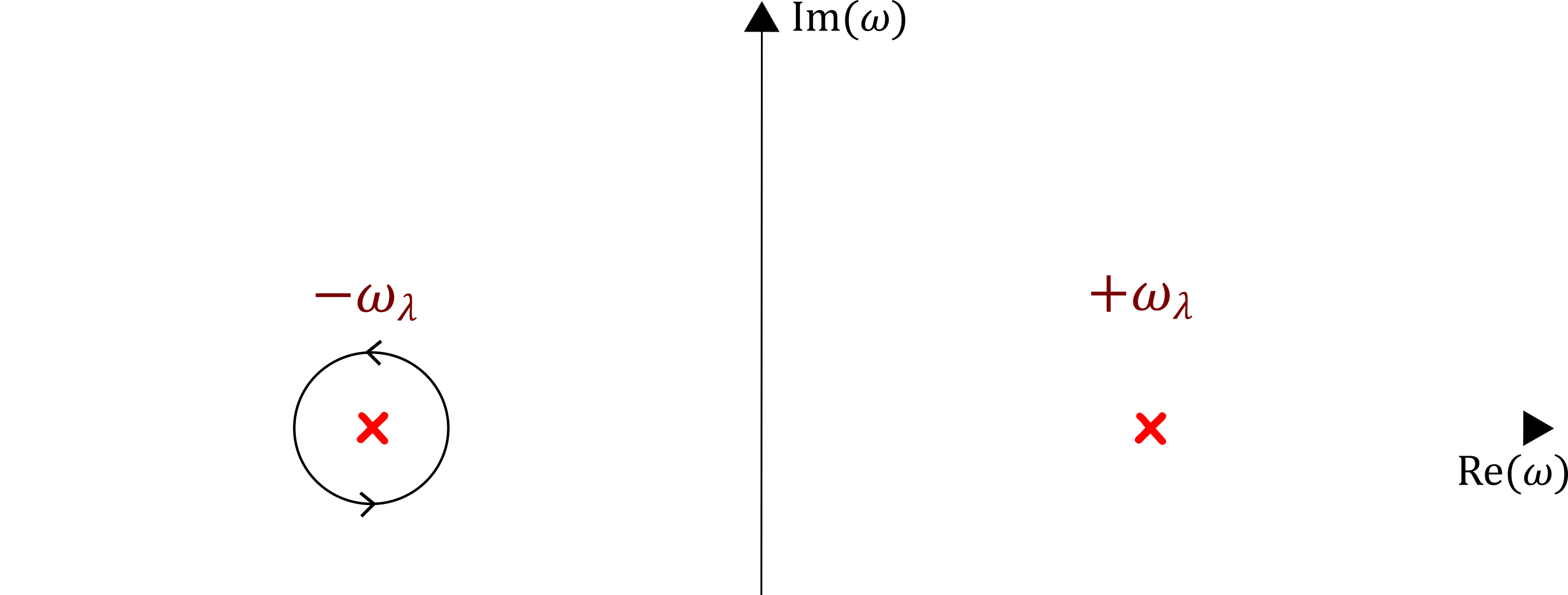}
\caption{Closed contour corresponding $iG^-$. Here, one goes around the left pole once, counterclockwise. \\  Source: By the author. }
\label{GWIGHTMAN-}
\end{figure}

\textbf{7-The Commutator $G$:} go around both poles counterclockwise. This can be obtained subtracting $G_{adv}$ from $G_{ret}$:

\begin{align}
G = G_{adv} - G_{ret} = -i(G_+ - G_-) = 2 \operatorname{Im}(G_+). \label{Gcomm0}
\end{align}

(The reason why we call this function the commutator will be clear briefly, when we analyze the connection with field operators.) We have its functional form from \eqref{Gcomm0}:

\begin{align}
G(x,x') = -\int \frac{d\mu(\lambda)}{\omega_\lambda} \sin\bigl( \omega_\lambda(t-t') \bigl) \psi^*_\lambda(\mathbf{x'})\, \psi_\lambda(\mathbf{x}).
\end{align}

\begin{figure}[H]
\centering
\includegraphics[width=0.6\linewidth]{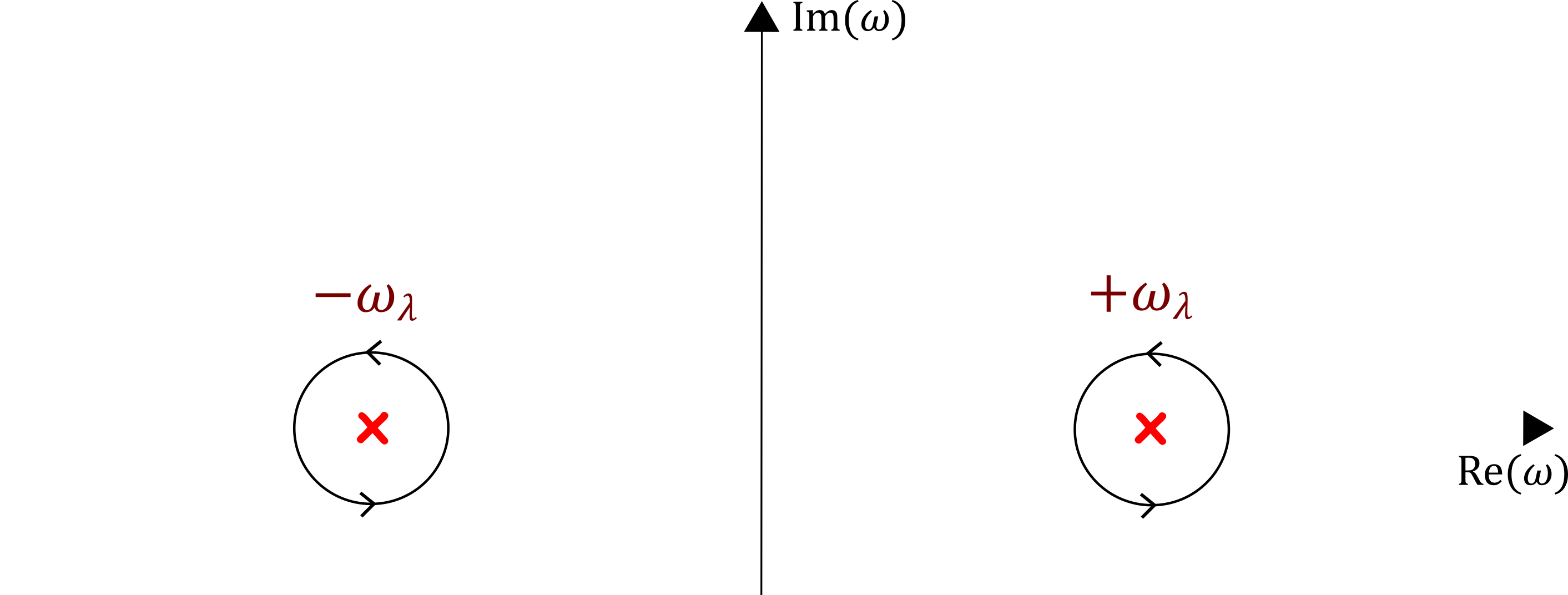}
\caption{Closed contour corresponding $G$. Here, one goes around both poles, counterclockwise. \\  Source: By the author. }
\label{GCOMM}
\end{figure}

\textbf{8- The Anticommutator (also known as Hadamart elementary function or Schwinger function) $G^{(1)}$:} go around the right pole clockwise, and the left one counterclockwise.

\begin{align}
G^{(1)} &= G^+ + G^- = 2 \operatorname{Re}(G_+) \nonumber \\
        &= -i[(G_F-G_{adv}) + (G_F-G_{ret})]  = 2i(\bar{G}-G_F) . \label{Gantcomm0}
\end{align}

Thus

\begin{align}
G^{(1)}(x,x') = \int \frac{d\mu(\lambda)}{\omega_\lambda} \cos\bigl( \omega_\lambda(t-t') \bigl) \psi^*_\lambda(\mathbf{x'})\, \psi_\lambda(\mathbf{x}).
\end{align}

\begin{figure}[H]
\centering
\includegraphics[width=0.6\linewidth]{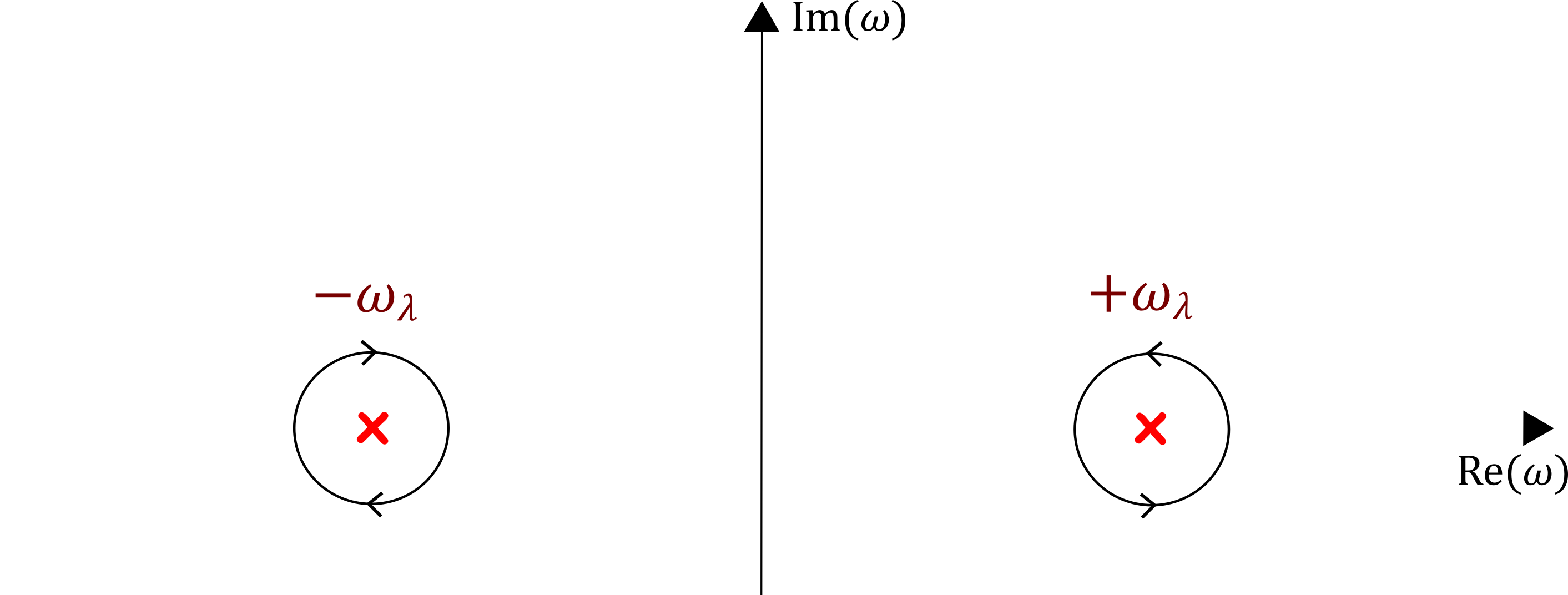}
\caption{Closed contour corresponding $G^{(1)}$. One goes around the right pole clockwise and the left one counterclockwise. \\  Source: By the author. }
\label{GANTCOMM}
\end{figure}

One important feature to notice in these Green functions concerns their spacetime support. Taking a fixed $x$, we have already seen that $G_{ret}$ and $G_{adv}$ are null for $t<t'$ and $t>t'$, respectively. To take the analysis further, it is particularly enlightening to consider $t=t'$. We see that, at equal times, $G$, $G_{ret}$, $G_{adv}$, $\bar{G}$ all vanish (whereas $G^\pm$, $G_F$ and $G^{(1)}$ do not, even for $\mathbf{x} \neq \mathbf{x'}$). Further, by analyzing first time derivative of $G$, for example, we find

\begin{align}
\partial_t G(x,x')\bigl|_{t=t'} \; = -\int d\mu_\lambda \cos\bigl( \omega_\lambda \times 0 \bigl) \psi^*_\lambda(\mathbf{x'})\, \psi_\lambda(\mathbf{x}) = \delta(\mathbf{x}-\mathbf{x'}).
\end{align}

Due to this localized initial data, $G$ (as a function of $x'$) will be only supported inside the light cone of $x$. Consequently (see eq. \eqref{Gcomm0}), $G_{adv}$ will be only supported in the future light cone of $x$, and $G_{ret}$ in its past light cone. In contrast, $G^\pm$ and $G^{(1)}$ spread through all spacetime, even for spacelike separated events.

\begin{figure}[H]
\centering
\includegraphics[width=0.6\linewidth]{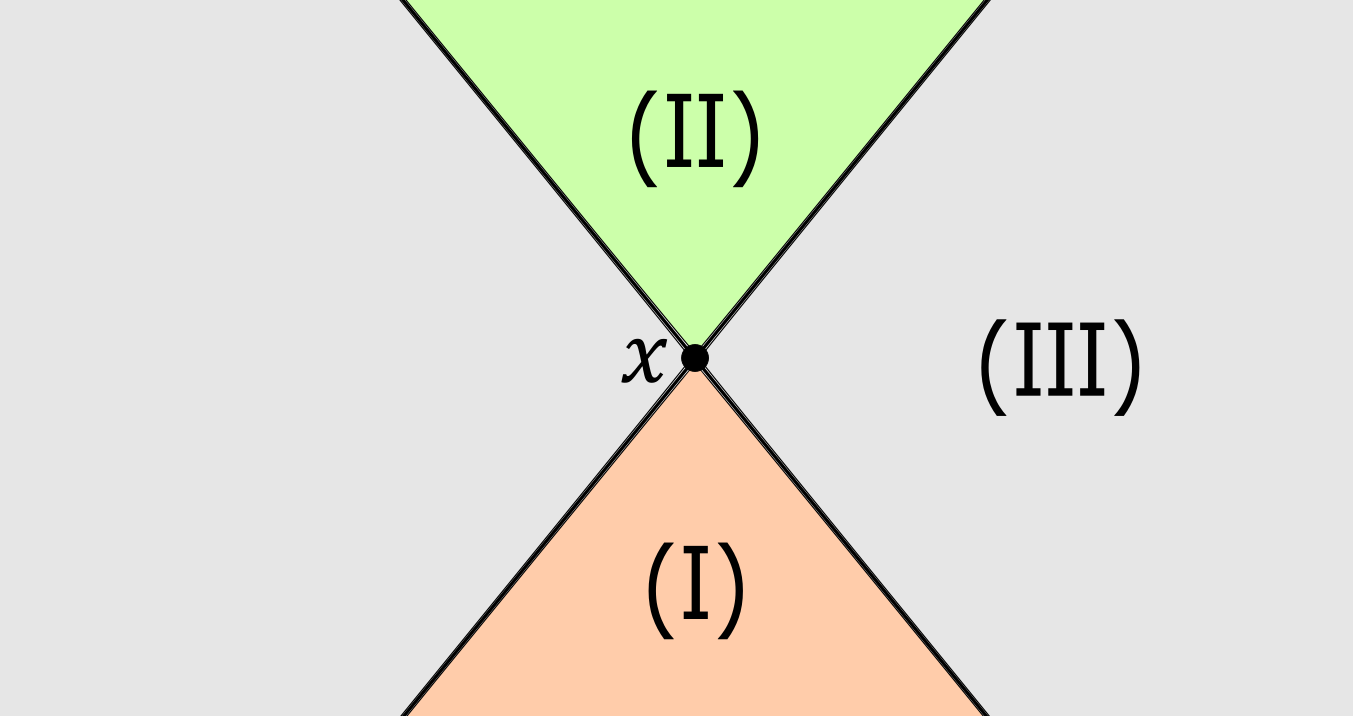}
\caption{Green functions and their various supporting regions. $G_{ret}$ will only be nonzero inside the past light cone (region (I)), $G_{adv}$, inside the future light cone, and $G$, $\bar{G}$ inside both the entire light cone (regions (I) and (II)). On the other hand, $G^\pm$ and $G^{(1)}$ will also be nonvanishing in the spacelike-separated region (III), spreading through all of spacetime. \\  Source: By the author. }
\label{GREEN_V3}
\end{figure}

Now that we have constructed various Green functions as solutions of the field equation, we shall identify them with the correspondent vacuum expectation values of bilinear field operators. When we make the formal expansion of a few of the vacuum expectation values mentioned above, we see that we can immediately identify them with some of the above Green functions. For example, let us evaluate the product $\phi(x)\phi(x')$:

\begin{align}
\braket{0|\phi(x)\phi(x')|0} &= \iint \! \frac{d\mu(\lambda)}{\sqrt{2\omega_\lambda}} \frac{d\mu(\lambda')}{\sqrt{2\omega_{\lambda'}}} \braket{0\bigl| \bigl(a_\lambda u_\lambda(x) + a^\dagger_\lambda u^*_\lambda\!(x) \bigl)  \bigl(a_{\lambda'} u_\lambda(x') + a^\dagger_{\lambda' }u^*_{\lambda'}\!(x') \bigl) \bigl|0} \nonumber \\[6pt]
  &= \iint \! \frac{d\mu(\lambda)}{\sqrt{2\omega_\lambda}} \frac{d\mu(\lambda')}{\sqrt{2\omega_{\lambda'}}}  u_\lambda(x) u^*_{\lambda'}\!(x') \delta(\lambda, \lambda') \nonumber \\[6pt] 
  &= \int \! \frac{d\mu(\lambda)}{2\omega_\lambda} e^{-i\omega_\lambda (t-t')} \psi_\lambda(\mathbf{x}) \psi^*_{\lambda}\!(\mathbf{x'}).
\end{align}

Thus, we immediately identify it with the Wightman function

\begin{align}
G^+(x,x') = \braket{0|\phi(x)\phi(x')|0} .\label{fWhightmann+}
\end{align}

Similarly, $\phi(x')\phi(x)$ yields

\begin{align}
G^-(x,x') = \braket{0|\phi(x')\phi(x)|0}. \label{fWhightmann-}
\end{align}

Then, from eqs \eqref{Gcomm0} and \eqref{Gantcomm0}, we immediately identify the commutator and anticommutator:

\begin{align}
iG(x,x') &= \braket{0|[\phi(x)\phi(x')]|0}, \label{fGComm+} \\
G^{(1)}(x,x') &= \braket{0|\{\phi(x)\phi(x')\}|0}. \label{fHadamart}
\end{align}

Note that this last function symmetrizes the two-point field product before evaluating its expectation value. For this reason, one often computes $G^{(1)}$, instead of working with $G^+$ and/or $G^-$ directly.

Finally, we identify Feynman's propagator with the time-ordered product:

\begin{align}
-iG_F(x,x') = \braket{0|\mathcal{T}(\phi(x)\phi(x'))|0}, \label{fFeynmann}
\end{align}
where we have defined

\begin{align}
\mathcal{T}(\phi(x)\phi(x')) = \phi(x) \phi(x') \Theta(t-t') + \phi(x') \phi(x) \Theta(t'-t) = 
\begin{cases} \phi(x) \phi(x') , \;\;t>t' \\ \phi(x') \phi(x), \;\;t'>t \end{cases} .
\end{align}

Having reconstructed all of these operators as expectation values, we shall not rederive that they obey the (homogeneous or inhomogeneous) wave equations in all cases. We just note that, taking in consideration that $\phi(x)$ obeys the homogeneous field equation \eqref{elliptic} (and $[\partial^2_t+H^2_\mathbf{x}]$ does not act on $x'$), and that

\begin{align}
\partial_t \Theta(t-t') = \delta(t-t'), \qquad \qquad \int d\mu(\lambda) \psi(\mathbf{x}) \psi^*(\mathbf{x'}) = \delta(\mathbf{x}-\mathbf{x'}),
\end{align}
it is straightforward to recover them from the field-operator definition.

%% file: chap3.tex
\chapter{Quantum Field Theory in Curved Spacetime} \label{QFTCS}

In this chapter, we shall generalize in a straightforward manner the basic formalism of quantization of noninteracting fields to curved spacetimes, explicitly developed through the paradigmatic example of a real scalar field. In this particular quantization procedure, we promptly use the existence of a decomposition of the solutions of the classical field equations in orthonormal modes to impose the usual commutation relations $[a_i,a^\dagger_j] = \delta_{ij}$ and follow in general lines some of its consequences.

To concretely carry this procedure, we start by defining the classical prerequisites for our theory in curved spacetime. In section \ref{Geometric Fundamentals}, we give the basic outline of theory of General Relativity and how to formulate the joint dynamics of matter and spacetime in a Lagrangian formalism. We also generalize the notion of `equal-time' surfaces to (\emph{globally hyperbolic}) curved spacetimes, defining the notion of \emph{Cauchy Surfaces}. 

Then, in section \ref{curved quantization}, we explicitly develop the quantization procedure for a noninteracting scalar field in curved spaces, and discuss some basic aspects of QFTCS, such as the absence of a physically priviledged vacuum state, as well as how it requires bosonic statistics (commutation relations) to be internally consistent under general mode transformations.

After that, in section \ref{particle detectors}, we give particular emphasis to the construction of an operational definition of particles, based on the response of actual particle detectors, as well as to the nontrivial relation between the different vacua associated with different mode decompositions, which prepares the ground for a more meaningful discussion of the processes of particle creation in dynamical spacetimes in section \ref{particle creation}.

Finally, in section \ref{adiabatic}, we analyze the limits in which descriptions in terms of particle modes are meaningful to define what is called the adiabatic vacuum in dynamical spacetimes. Along with the corresponding asymptotic expansions of the field modes (the so-called adiabatic expansions), it will play a key role in the discussion of renormalization in the following chapter.

\section{General Relativity and the Structure of Spacetime} \label{Geometric Fundamentals}

In the present section, we give a brief overview of General Relativity (GR), discussing some of its geometrical and dynamical features. After laying key aspects in the interplay between matter (fields) and spacetime geometry given by the Einstein Equations, we show how these can be derived through a minimal action principle, extending the formalism of section \ref{CFT} that will allow for Lagrangian formulation of GR permeated by matter fields.
Given the overwhelming challenges in obtaining a fully quantum theory of gravity \cite{QGreview} (either in the vacuum or in the presence of matter), we turn to the well-established and fruitful approach of quantizing matter fields in a classical curved spacetime.

theory of General Relativity is without doubt a major revolution in the way we conceive space, time, and gravity. Rather than a static immutable stage through which matter propagates passively, spacetime comes to be conceived as a dynamic entity, curved by the matter within it. In a formal perspective, this step is achieved by letting go the assumption that spacetime is decribed as a flat space armed with a \emph{given} flat (Minkowski) metric $(\mathbb{R}^4,\eta_{ab})$, as in Special Relativity, and allowing for the more general structure of a 4-dimensional manifold with a (generally curved) \emph{dynamic} metric $(\mathcal{M},g_{ab})$, which is not \textit{a priori} defined, but rather must be determined jointly with the matter evolving under its influence.

This dynamical content of GR may be very elegantly summarized through the Einstein Equations, which govern how the matter content in spacetime acts as a source for its curvature (for the definitions of curvature tensors and covariant derivatives, see Appendix B):

\begin{align}
R_{ab} - \tfrac{1}{2}R g_{ab} = -8\pi G T_{ab}, \label{EE}
\end{align}
where $T_{ab}$ is the matter stress tensor and we have kept Newton's constant $G$ for later convenience in chapter \ref{renormalization}, where we show that it can be renormalized as a coupling constant between matter and spacetime. Equation \eqref{EE} is almost the most general covariant second-order equation which automatically leads to the covariant quantization of the stress tensor\footnote{As we have derived in appendix \ref{geometry}, the Bianchi identity implies that the covariant derivative of the LHS of Einstein equations should be null (\textit{i.e.}, $\nabla_{\!a}G^{ab} = 0$). Thus the same must be true for the RHS. } that one can write for $g_{ab}$; the most general form is achieved by simply adding a term proportional to $g_{ab}$, introducing a cosmological constant $\Lambda$,

\begin{align}
R_{ab} - \tfrac{1}{2}R g_{ab} +\Lambda g_{ab} = -8\pi G T_{ab}. \label{EEwL}
\end{align}

These equations, eventually supplemented by the equations of motion of matter, and initial/boundary conditions, will allow us to predict the geometry of all spacetime and of the matter propagating within it.

\subsection{Lagrangian Formulation of General Relativity}

Both in classical particle mechanics and field theories in flat spacetime, one can express their entire dynamical content through their equations of motion (such as Newton's or Maxwell's equations). It allows one to tell the evolution of a system from given initial conditions and thus to make any possible physical predictions on it. Similarly, in General Relativity, its dynamical content can be fully expressed through Einstein's equations (\ref{EE}).

However, just as it happens with the former theories, it is desirable to present GR with a Lagrangian (or Hamiltonian) formulation for a number of reasons. Besides aesthetical and simplicity considerations, our known methods of quantization employ either of these formulations. Thus, not only do they prove central if one attemps to quantize gravity through a recognizable approach, but also they are necessary for QFTCS (and semiclassical gravity), so that one has a formulation of field theory in curved space liable to quantization (and is eventually able to connect quantum fields as a source of curvature for classical spacetimes).

For the purposes of this work, it will suffice for us to present only a Lagrangian Formulation\footnote{ 
 For a Hamiltonian formulation of GR, we refer the reader to Appendix E of \cite{wald}, on which much of the presentation of the Lagrangian formulation in the present section is based.}.
 It has the advantages of providing a manifesly covariant description of our theories (whereas a Hamiltonian relies on a split between space and time), and of allowing us to very simply obtain our dynamical equations. The mere existence of the correspondence with a Hamiltonian formulation will allow us to directly implement the scheme of canonical quantization, but rather than applying it to configuration and momenta variables, we impose the commutation relations directly to field mode operators.

With that said, we turn our attention to the construction of a Lagrangian formulation of GR. Before discussing a full dynamic theory of spacetime and matter, and elaborating on how we may adapt the latter to include gravity, let us begin by showing how we can encompass the spacetime geometry alone in a Lagrangian formulation, and obtain the vacuum Einstein Equations through an action principle. 

Generally, for field theories, we have been considering an action functional $S$ which only depends on its field variables \emph{locally}, in the form of a spacetime integral of a scalar Lagrangian\footnote{Technically, this is what we called a \emph{Lagrangian density} in Chapter 2, where we reserved the term `Lagrangian' to spatial integrals of $\mathscr{L}$. From this point onwards we shall refer to $\mathscr{L}$ only as the Lagrangian; the term \emph{Lagrangian density} will be assigned with a different meaning below.}
 function $\mathscr{L}$ (see eq. \ref{action}). Here, we want to build a purely geometrical action, which will likewise be constructed from a local scalar function of the metric, $S_G[g_{ab}]$, to be written in the form

\begin{align}
S_G = \int_\mathcal{M} d\mu_g(x) \mathscr{L}_G\bigl(g_{ab}(x)\bigl),
\end{align}
where $\mathscr{L}_g\bigl(g_{ab}(x)\bigl)$ depends only on the metric and its spacetime derivatives (which shall appear through curvature terms) at the event $x$, and $d\mu_g(x)$ is the natural volume element in the spacetime manifold $\mathcal{M}$\footnote{For more details on integrations in manifolds, and volume elements see appendix B of \cite{wald}. Further reference on the subject can be found in chapter I of \cite{yvone}.} \emph{induced by the metric $g_{ab}$}.
Thus, compared with theories analyzed in the last chapter, GR presents us with a difficulty. If we attempt to look at variations of $S_G$ with respect to the metric, we are faced with the awkward convolution that not only $\mathscr{L}_G$ but also the volume element itself depend on $g_{ab}$. To circumvent this, we begin by noting that the covariant (coordinate-independent) volume element can be expressed in any coordinate system as $d\mu_g(x) = d^4\!x|g(x)|^{1/2}$, being $d^4x$ a (coordinate-dependent) coordinate volume element in $\mathbb{R}^4$ and $|g(x)|^{\frac{1}{2}}$ the Jacobian associated to it; its dependence on the metric can be simply codified as a determinant of its components in the coordinate basis\footnote{In fact, this was already true for flat spacetimes, but there are two key differences: (i) there, one can always find globally inertial coordinates, making Jacobian $|\eta(x)|^\frac{1}{2}$ trivially $1$, and (ii) while there the metric was merely a background structure, here it is a dynamical variable and we must compute variations with respect to it.}\cite{wald, yvone}: $|g(x)| = |\det(g_{\mu \nu}(x))|$. 

Now that we have properly isolated the metric dependence in the volume element, one particularly convenient way to handle it is to absorb this dependence in the integrand and perform the integrations in the (metric-independent) \emph{coordinate volume}. To do so, we define \textit{tensor densities} as follows: given any \emph{tensor} field $T_{abc...}^{def...}$, whose definition does not make reference to any particular coordinate system, \emph{we construct an associated tensor density field $\tilde{T}_{abc...}^{def...}$ in a given coordinate system} by defining its value in each point as: $\tilde{T}_{abc...}^{def...}(x) \equiv |g(x)|^{1/2}T_{abc...}^{def...}(x)$. Particularly, for a scalar field $\mathscr{L}$ we will have an associated \textit{scalar density} $\tilde{\mathscr{L}}=|g|^\frac{1}{2}\mathscr{L}$. 

With these considerations, let us show how the vacuum Einstein equations may be very elegantly obtained from what is arguably the most simple nontrivial action one can build from purely geometrical scalars. Postulating the Lagrangian $\mathscr{L}=R$, or equivalently, the \emph{Lagrangian density} $\tilde{\mathscr{L}}= |g|^{1/2}R$, we obtain the famous Einstein-Hilbert action:

\begin{align}
S_G[g^{ab}] = \int d^4x |g(x)|^{1/2}R(x), \label{EinsteinHilbert}
\end{align}
where, for a matter of convenience, we are regarding $S_G$ as a function of the \textit{inverse metric} $g^{ab}$, rather than of $g_{ab}$. Now, using the same apparatus as in section \ref{CFT}, let us explicitly show how to compute its functional derivatives and obtain the associated dynamical equations. It proves convenient to evaluate them in the form of infinitesimal variations:

\begin{align}
\delta(|g|^{1/2}g^{ab}R_{ab}) = \delta(|g|^{1/2})g^{ab}R_{ab} + |g|^{1/2}\delta g^{ab} \,R_{ab} + |g|^{1/2}g^{ab}\delta \!R_{ab}.  \label{varEH}
\end{align}

The second term is already proprortional to a variation in the argument $g^{ab}$. The first is also relatively straightforward to compute in terms of it, as it is a direct function of $g^{ab}$:

\begin{align}
\delta|g|^{1/2} = -\tfrac{1}{2}|g|^{-1/2} \delta g = +\tfrac{1}{2}|g|^{1/2} \Bigl[g^{-1} \delta g \Bigl] = \tfrac{1}{2}|g|^{1/2} g^{ab}\delta g_{ab} = -\tfrac{1}{2}|g|^{1/2} g_{ab}\delta g^{ab}. \label{varDet}
\end{align}
(Here we stress that, in the middle equality, we have rewritten a product involving the variations of the metric \emph{determinant} in terms of the trace of product involving variations of the metric \emph{tensor}.)

The third term, however, involves a variation in curvature. This makes it somewhat more convoluted to compute, since its relation to the metric is only indirectly defined through covariant derivatives. We make a more complete discussion on how to compute these variations in appendix \ref{geometry}, from which we merely quote the result

\begin{align}
\delta R_{ab} = \tfrac{1}{2} g^{cd} [ \nabla_{\!a} \nabla_{\!b} \delta g_{cd} + \nabla_{\!c} \nabla_{\!d} \delta g_{ab} - 2 \nabla_{\!c} \nabla_{\!(b} \delta g_{a)d} ] .
\end{align}

Thus, we have that the third term in \eqref{varEH} is proportional to

\begin{align}
g^{ab}\delta R_{ab} = \nabla^a \nabla_{\!a} (g^{cd}\delta g_{cd}) - \nabla^a \nabla^b \delta g_{ab} = \nabla^a v_a, \label{varR}
\end{align}
where we have defined $v_a \equiv \nabla_{\!a}(g^{cd}\delta g_{cd}) - \nabla^b \delta g_{ab} $.

Thus, we see this term takes the form of a perfect divergence, making only a boundary contribution to the variations in $S_g$. Since boundary terms do not make any contribution to the local degrees of freedom in the bulk (and thus to the dynamic equations), we temporarily just ignore this term and obtain the following variation:

\begin{align}
\frac{\delta S_G}{\delta g^{ab}} = |g|^{1/2} \bigl( R_{ab} - \tfrac{1}{2}R g_{ab} \bigl).
\end{align}

Thus, extremizing the Einstein-Hilbert action, 

\begin{align}
\frac{\delta S_G}{\delta g^{ab}}=0,
\end{align}
we obtain precisely Einstein's equations in the vacuum: 

\begin{align}
R_{ab} - \tfrac{1}{2}R g_{ab} = 0. \label{vEE}
\end{align}

Before we proceed, we briefly comment on the matter of boundary terms. Usually, such terms make no contribution to $\delta S$ whatsoever, provided that one forces the variations of the relevant field (in our case $\delta g^{ab}$) to vanish at the boundary. \emph{However, this is actually not the case for the Einstein-Hilbert action}; due to the fact that $R$ involves second derivatives of the metric, one must \emph{also require} that the derivativatives of the variations $\nabla_{\!c} \delta g_{ab}$ vanish at the boundary, or else define a boundary \emph{counterterm} to subtract in the action. In the scope of this work, we shall not occupy ourselves with these boundary terms. The interested reader can find a few comments on the subject in the aforementioned Appendix E of \cite{wald}, and a quite thorough discussion in \cite{sumanta}.

At this point, one can very simply incorporate a cosmological constant $\Lambda$ to the Einstein equations simply by adding a constant term $\Lambda$ in the Einstein-Hilbert Lagrangian. More precisely, by making: $\tilde{\mathscr{L}}_G = |g|^{1/2}(R-2\Lambda)$. The last term only yields a variation due to \eqref{varDet}, whereupon one can easily verify that its addition brings us from \eqref{vEE} to

\begin{align}
R_{ab} - \tfrac{1}{2}R g_{ab} +\Lambda g_{ab} = 0. \label{vEEwL}
\end{align}

At this point, we take the chance to note that Einstein equations (with or without a cosmological constant) are nontrivial in 4 (or more) dimensions. In this case, spacetime alone turns out to have local degrees of freedom, which counting the metric symmetries and all its nondynamical components related to gauge symmetries, the number of degrees of freedom amount to 2 per point in space, which will correspond to two independent polarizations of gravitational waves.

Now, we will show how we can incorporate matter in this formalism, and provide a full general-relativistic theory of matter and curved spacetime. Our previous requirements that the (matter) Lagrangian must be a scalar and that it takes a covariant form can be quite directly transported to curved spacetime through the prescription known as ``minimal substitution''. It goes as follows: given a special-covariant theory, defined in Minkowski spacetime by a Lagrangian involving the metric $\eta_{\mu\nu}$ and spacetime derivatives $\partial_\mu$, one shall everywhere substitute $\eta_{\mu\nu} \rightarrow g_{\mu\nu}$ and $\partial_\mu \rightarrow \nabla_{\!\mu}$, making it generally covariant in curved spaces\footnote{For a more detailed discussion on special and general covariance, as well as the notions of covariance in prerelativistic physics, see chapter 4 of \cite{wald}.}. Thus, for instance, the Klein-Gordon field \eqref{KGlagrangian} minimally substituted in curved space would be

\begin{align}
\mathscr{L}_M = \frac{1}{2}g^{\mu\nu}(\nabla_{\!\mu} \phi) (\nabla_{\!\nu} \phi) - \frac{m^2}{2}\phi^2. \label{KGMSlagrangian}
\end{align}

We stress that this procedure is by no means the only possible generalization of special-covariant theories to curved spacetimes (one could, for instance, add covariant terms proportional to curvature, which will vanish as $g_{\mu\nu} \!\rightarrow\! \eta_{\mu\nu}$, recovering \eqref{KGlagrangian} in flat space), nor is it always free of ambiguities (as when one has 2 equivalent formulations in flat space, in terms of fields or of potentials, and these do not necessarily remain equivalent in curved space after minimal substitution \cite{wald}). Nevertheless, it is a consistent and practical prescription, and often the first one has at hand when trying to generalize a theory to curved spacetimes.

Now, as the geometrical portion of the action $S_G$ does not depend on any matter fields, the dynamical equations of the latter (\textit{i.e.} the Euler-Lagrange equations) will spring solely from the matter portion $S_M$:

\begin{align}
S_M[\phi_a,g^{bc}] = \int_\mathcal{M} d^4x |g(x)|^\frac{1}{2} \mathscr{L}_M \bigl( \phi_a, g^{bc} \bigl).
\end{align}

Note that, although $S_G$ does not carry any dependence on the matter fields $\phi_a$, $S_M$ necessarily depends on the metric. This codifies the fact that our fields are propagating through curved spacetimes, and will necessarily be influenced by its geometry. We obtain their Euler-Lagrange equations by extremizing $S_M$ with respect to the fields:

\begin{align}
\frac{\delta S_M}{\delta \phi_a} = 0.
\end{align}
For instance, for our minimally substituted scalar field \eqref{KGMSlagrangian}:

\begin{align}
g^{\mu\nu} \nabla_{\!\mu} \nabla_{\!\nu} \phi + m^2\phi \equiv \bigl[ \Box + m^2 \bigl] \phi = 0, \label{ckge}
\end{align}
which turns out formally identical to \eqref{fkge}, as we defined the general-covariant D'Alembertian \linebreak $\Box \equiv g^{\mu\nu} \nabla_{\!\mu} \nabla_{\!\nu}$, although \eqref{fkge} and \eqref{ckge} are different (nonequivalent) equations!

On the other hand, to look at the effect that matter has on spacetime, acting as a source of curvature, we must consider the entire action:

\begin{align}
S[\phi_a,g^{ab}] = S_G[g^{ab}] + S_M[g^{ab},\phi_a].
\end{align}

When we first defined $S_G$, we were not worried about its normalization, as it turned out superfluous for the vacuum equations. To reobtain Einstein's equations with a source, however, one must adjust a relative normalization between $S_G$ and $S_M$ (and, of course, one must assure both terms dimensionally consistent, although this matter is entirely hidden in Planck Units, and partially hidden in natural units). This can be achieved by readjusting $\mathscr{L}_G$ as

\begin{align}
\mathscr{L}_G \equiv \frac{R-2\Lambda}{16\pi G}.
\end{align}

Then, by extremizing the total action with respect to the metric,

\begin{align}
\frac{\delta S}{\delta g^{\mu\nu}} = \frac{|g|^{\frac{1}{2}}}{16\pi G} \bigl( R_{\mu \nu} - \tfrac{1}{2}R g_{\mu \nu} +\Lambda g_{\mu\nu} \bigl) + \frac{\delta S_M}{\delta g^{\mu\nu}} = 0,
\end{align}
we obtain a natural definition for the stress tensor of the matter fields, so that we recover the full Einstein equation \eqref{EEwL}:

\begin{align}
T_{ab} \equiv \frac{2}{|g|^\frac{1}{2}} \frac{\delta S_M}{\delta g^{ab}}.
\end{align}

Finally, we note that, the imposition that both $S_G$ and $S_M$ must be scalars, and thus invariant under any spacetime transformations (or, equivalently, any coordinate transformations) will imply that $G_{ab}$ and $T_{ab}$ must be both covariantly conserved, regardless of the Einstein equations.

In direct analogy to what we did in the past chapter, it would seem like a very natural next step to try to quantize this full theory of gravity and matter (or perhaps gravity alone, for a start), imposing some procedure of quantization to the fields $\phi_a$ and $g_{ab}$. However, the attempts to carry out a quantization for spacetime itself, be it through a metric field or more profound changes in the whole spacetime structure classically described by $(\mathcal{M},g_{ab})$, have met enormous challenges in the past decades, so that we are still far from a satisfactory solution for such a theory\footnote{
 Again, for a thorough review on the state-of-the-art of many contemporary approaches to quantum gravity, see \cite{QGreview} and references therein. }.
 A far more manageable approach is to quantize matter fields alone in a classical curved background geometry, which gives rise to what we call \emph{quantum field theory in curved spaces}. This approach consists of finding a way to generalize some of the procedures and basic definitions originally carried in flat, Minkowski background space to generally curved background spaces; it has proved quite successful in describing phenomena for quantum matter fields in which curvature plays a relevant role, but is not too extreme so that it itself does not need to be considered quantized. Furthermore, this curved space theory has raised many relevant questions to QFT in Minkowski spaces, which were previously unnoticed due to the fact that its traditional approaches relied heavily on the Poincaré group of symmetries.

Then, it is this approach of quantum field theory in curved space that we shall develop throughout the rest of this chapter. Before we can proceed to it, however, we shall take a moment in the next section to impose appropriate restrictions in our curved spaces, so that we can make meaningful extensions of many of the concepts defined in Minkowski space and use them for quantization in curved ones.

\subsection{Spacetime Geometry and Quantum Field Theory}

We have just seen how General Relativity provides a quite natural framework to analyze the mutual dynamic of matter and spacetime in a classical context. Indeed, it gives us well-defined local dynamical equations, which should allow us to predict the behaviour of matter and geometry from an appropriate set of initial conditions\footnote{This is generally known as the Initial Value Problem (IVP), or Boundary and Initial Value Problem (BIVP), when spatial boundary conditions are also required. In-depth discussions of the IVP in GR can be found in chapter 10 of \cite{wald} and in chapter 7 of \cite{hawkellis}.}.
 However, the curved nature of spacetime in GR confers it with a few subtleties and complications for the initial value formulation, when compared to flat space. Although classical field theories are not indifferent to these subtleties -- particularly in terms of predictability and a well-posed initial value formulation--, they manifest more acutely in the quantum case, where nonlocal features play a more proeminent role in theory and, particularly, a notion of `equal-time' surfaces is required to postulate the canonical commutation relations in the Hamiltonian formalism. With those matters in mind, we give a brief account of the necessary structure of spacetime to our present formulation of QFTCS, with particular emphasis on its causal structure. This is intended to be just an overview on the topic, sufficient to situate the unfamiliar reader in the subsequential discussion; for a more complete account of the subject, we refer the reader to \cite{wald, hawkellis, penrose, parrado}\footnote{We warn, however, that they require basic notions of topology for a fluid reading.}, which are the direct sources of the present exposition.

In a pregravitational context, thoroughly discussed in the last chapter, we have seen that a crucial structure to the initial value formulation (\textit{i.e.} to obtain a unique solution from the field equations with a given initial condition), as well as to the field mode decomposition and to the postulation of canonical commutation relation, was \emph{equal-time surfaces}. These surfaces allowed us to speak meaningfully of field configurations (`at a given time') and perform ``complete'' spatial integrations, for example for the Poisson brackets \eqref{bfipobra} and the inner product \eqref{fsp}. There, since we were handling either Galilean or Minkowskian spacetimes, where we have either absolute time or a very simple notion of flat equal-time surfaces attached to congruences of inertial worldlines (\textit{i.e.}, to families of inertial observers), we have restricted our analysis to these simple surfaces.

In GR \footnote{Or in any modified-gravity theories that share the basic spacetime structure as a manifold with a pseudo-Riemannian metric $(\mathcal{M},g_{ab})$.}, the curved nature of spacetime does not generally allow for such a distinct and simple construction of `equal-time' surfaces. Notwithstanding, we shall see that for a quite general class of spacetimes, the so-called \emph{globally hyperbolic} spacetimes, one has a generalized notion of simultaneity surfaces, whose causal domains extend to the entire spacetime: \emph{Cauchy Surfaces}. In order to properly define the latter, and provide a little physical intuition on them, we go over a few basic concepts on the causal structure of spacetime.

It follows immediately from the equivalence principle -- which states that, \emph{locally}, any curved spacetime `looks like' flat (Minkowski) spacetime: that is, one can always construct a \emph{local inertial frame} such that the metric components $g_{\mu\nu}$ at an event $x$ are equal to $\eta_{\mu\nu}=(+1;-1,-1,-1)$ and its first derivatives vanish -- that the \emph{local} causal structure of general-relativistic spacetimes is the same as in Minkowski spacetime. Its \emph{global} structure, however, may differ radically, for instance due to nontrivial topologies, to the ``tipping of light cones'', or even to singularities. Let us then classify causal structures and point out some desirable features for a `well-behaved' spacetime (our counterexamples may seem particularly artificial at times, but that is in part recourse to pedagogical examples).

\textbf{- Time Orientability:} A very basic property we would like for physically plausible spacetimes is the possibility to determine, for every event $x$, its past and future directions, and unambiguosly distiguish them. Locally, this is done by constructing the light cones around each event, which divides all events with a positive, timelike separation in two disconnected regions (``above'' and ``below'' the light cone); one can then identify one of them with the (chronological) past and the other with the (chronological) future of that event. Trouble may arise, however, when we try to extend this identification globally. In Minkowski, this can be trivially achievable through the affine structure of space: if one chooses a fiducial event, traces its light cone, and identifies its past and future, one needs simply to translate this rigidly through all spacetime to obtain a unique and consistent identification; equivalently, one may identify future-directed (past-directed) timelike vectors in any two events directly\footnote{ A timelike vector is said to be future-directed (past-directed) if it is in the future (past) section of the interior of the light cone.}.

\begin{figure}[h]
\centering
\includegraphics[width=0.5\linewidth]{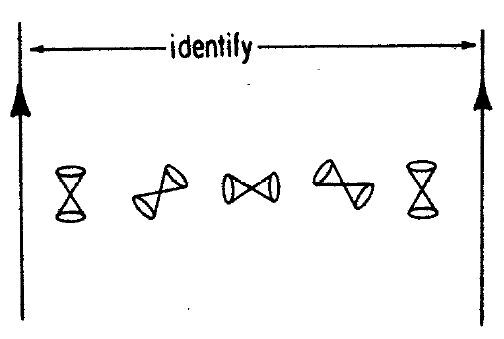} 
\caption{ Pictoric example of a non time-orientable manifold. Here the light cones tipp over 180 degrees as one transcurs a full period from left to right, making it impossible to consistently identify past and future directions. \\ Source: WALD \cite{wald}} \label{WALD}
\end{figure}

In curved spacetimes, however, one cannot automatically identify the tangent vectors in distinct events. The best one can do is to identify them continuously through parallel transport. However, due to spacetime curvature, there will generally be a ``tipping'' of the light cones throughout spacetime. In extreme cases, this tipping may result in a loss of global orientability of the space (something like in Moebius strip kind of spacetime), such that, along a closed curve one may ``tip the light cone upside-down'' and be unable to obtain a globally consistent time orientation (see figure \ref{WALD}).

Then, the first and most basic requirement that we shall make for our spacetimes is that they are time-orientable.

\textbf{-Chronal/Causal past and future:} For any time-orientable spacetime $(\mathcal{M},g_{ab})$, one can identify the \emph{chronological future} of each event $x$, $I^+(x)$, the set of all events that can be reached by a future-oriented timelike geodesic\footnote{ A geodesic $\gamma(t)$ is said to be future-directed as a function of the parameter $t$ if its tangent vector $(\frac{\partial}{\partial t})^a$ is everywhere future-directed.}
 at a strictly positive proper-time interval $|\Delta \tau|>0$ (we require that $|\Delta \tau| \neq 0$ to leave out a curve of null arclength taking $x$ into $x$). Likewise, we define its \emph{cronological past}, $I^-(x)$, as all events that can be reached by a past-directed timelike geodesic at a strictly positive proper-time interval. Then, for any two events $x,y \in \mathcal{M}$, it is obvious that $y \in I^+(x) \Leftrightarrow x \in I^-(y)$. 
Similarly, we define the \emph{causal future} (\emph{causal past}) of $x$,  $J^+(x)$ ( $J^-(x)$)  as the set of all events that can be reached by future-directed (past-directed) timelike \emph{or null} geodesics (these are collectively called \emph{causal geodesics}). Note that, unlike its chronological future (past), this encompasses the possibility of null length curves, so that we always have that $x \in J^+(x)$ and $x \in J^-(x)$.

Now, one can see that if in a spacetime $(\mathcal{M}, g_{ab})$, there are events such that $x \in I^+(x)$, this will mean that $\mathcal{M}$ possesses nontrivial closed timelike curves. A quite straightforward example of a spacetime that does possess closed timelike curves is a flat ``timelike-torus'', which can be obtained from Minkowski spacetime by identifying two equal-time surfaces $t=0$ and $t_0>0$\footnote{
 One may argue that such a spacetime is too ``artificial'', being produced merely by strange topological identifications. However, one can more generally build solutions with closed timelike curves without such topological identifications, such as Gödel's Universe (see section 7.7 of \cite{hawkellis}).}.
 Such spaces are generally regarded as unphysical, and may lead to paradoxes as events may lie in their own chronological future. Thus, generally, we shall require that the spacetimes we are considering do not possess any closed timelike curves, such that $x \not\in I^+(x), \; \forall x\in \mathcal{M}$.

For any subset $A \subset \mathcal{M}$ we can define its chronal and causal pasts and futures, $I^\pm(A)$, $J^\pm(A)$ as the union of the respective regions for each of their events, that is:

\begin{align}
I^\pm(A) \equiv \bigcup_{x \in A} I^\pm (x), \qquad J^\pm(A) \equiv \bigcup_{x \in A} J^\pm (x).
\end{align}

\textbf{-Achronal sets}: An important definition to start to encompass the notion of an equal time surface is that of achronal sets. A subset $S \subset \mathcal{M}$ is said to be achronal if no two events $x$ and $y$ belonging to it are chronologically related (\textit{i.e.}, if $y \not\in I^\pm(x)$, $\forall x,y \in S$), that is:

\begin{align}
I^+(S) \cap S = \emptyset.
\end{align}

This definition prevents one from obtaining an inconsistent notion of simultaneity, as any useful notion of simultaneous events will certainly exclude events that are in the chronological past or future of one another (note that this will only be possible throughout spacetime if it does not have closed timelike curves, such that no event can lie in its own causal past/future).

A particular class of achronal sets that will be of interest to us is that of spacelike differentiable surfaces in $\mathcal{M}$. Although these generally allow one to identify a notion of `simultaneity' in spacetime, they still do not emcompass all we need for a well-posed initial value formulation. For that, we must still require that they are, in a sense, complete. We shall give that a more precise meaning through the definition of domains of dependence. 

\textbf{-Domains of dependence and Cauchy Surfaces}: given any achronal set $S$, we define its future domain of dependence $D^+(S)$ as the set of all events $y$, for which \emph{any} past-inextendible\footnote{
 A causal curve $\gamma$ in $\mathcal{M}$ is said to be past inextendible if it has no past endpoints. That means it will either run off to infinity or `fall in an edge' of spacetime (such as singularity). One can similarly define future-inextendible curves.}
  causal geodesics intersecting $y$ will intersect $S$. Likewise, one defines its past domain of dependence, $D^-(S)$ as the set of all events $x$, for which \emph{any} future-inextendible causal geodesics intersecting $x$ will intersect $S$. One then defines its total domain of dependence $D(S)$ as the union $D^+(s) \cup D^-(s)$.
  
\begin{figure}[H]
\centering
\includegraphics[width=0.45\linewidth]{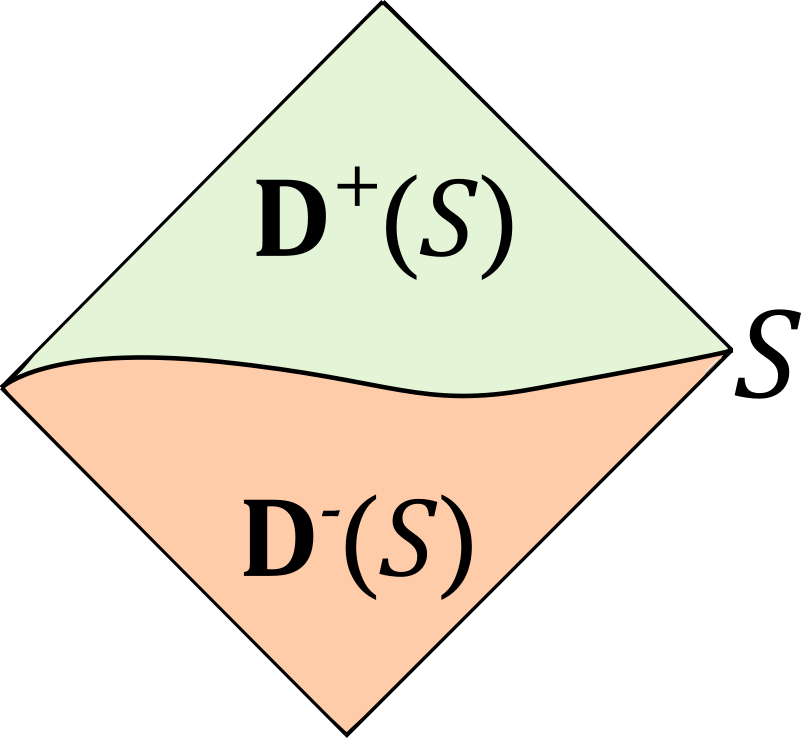}
\caption{ Domains of dependence of a compact achronal set $S$ (represented as a curvy line in the middle). Its future domain of dependence, $D^+(S)$, is represented in green and its past domain of dependence $D^-(S)$ is represented in orange. In the 4-dimensional compact region formed by $D(S)$, one can predict the configurations of a matter field solely from knowledge from it on $S$. \\ Source: By the author. }
\label{Domain}
\end{figure}

This notion is very important for the initial formulation of any causal field theory because, if information from a field can only be transported along causal curves, then knowledge from the field (and its independent derivatives) at an achronal surface $S$ will allow us to determine the field throughout all $D(S)$ (see Figure \ref{Domain}).

\begin{figure}[H]
\centering
\includegraphics[width=0.65\linewidth]{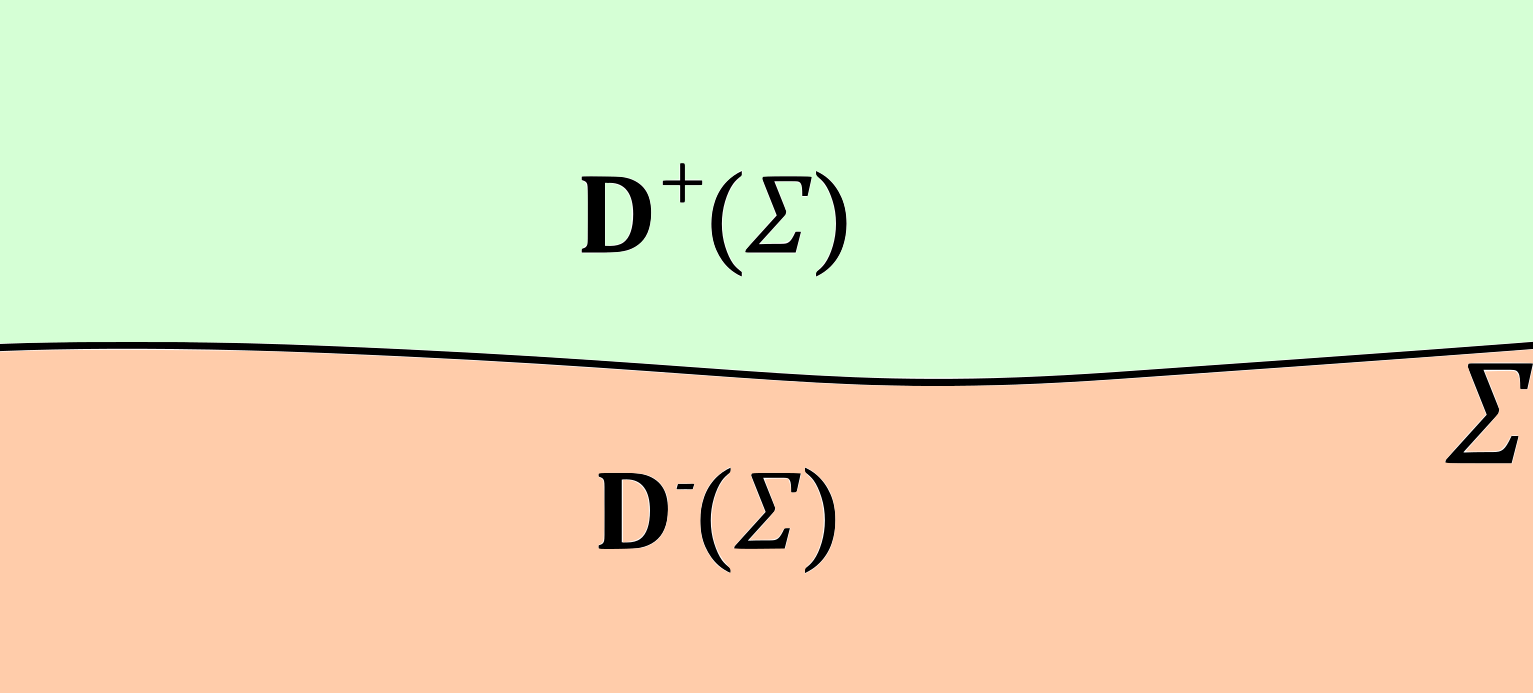}
\caption{ A Cauchy surface $\Sigma$ and its domains of dependence, $D^+(S)$ (green) and $D^-(S)$ (orange). In this case we see that $D(\Sigma)= \mathcal{M}$, such that one can predict dynamical information in the entire spacetime if one has appropriate initial conditions at $\Sigma$. (Here, one should imagine $\Sigma$ and $D^\pm(\Sigma)$ as extending all the way to infinity.) \\ Source: By the author. }
\label{DomainSigma}
\end{figure}

Finally, we are at a place to define a useful generalization of equal-time surfaces in curved spaces, in terms of which we can have a well-posed initial value formulation: \emph{Cauchy surfaces}. A closed achronal surface $\Sigma$ is said to be a Cauchy surface if its domain of dependence extends to the entire spacetime, that is: $D(\Sigma) = \mathcal{M}$. Spacetimes which possess Cauchy surfaces are called \emph{globally hyperbolic}. The most important property of globally hyperbolic spacetimes is that they will allow us to predict the state of a field at any event if we have completely determined its state (configurations and derivatives) at a given `instant of time', that is, at a given Cauchy surface (see Figure \ref{DomainSigma}).

Then, in the following sections, we will require our background spacetimes to be always smooth (so that we may define derivatives to any order) and globally hyperbolic pseudo-Riemannian manifolds; these will be sufficient conditions for us to define a well-posed quantized theory in them.

\section{Quantization of a Scalar Field} \label{curved quantization}

With these classical foundations at hand, we are in position to extend the formalism of chapter \ref{QFTMS} and carry out the quantization of noninteracting fields \emph{in a curved background spacetime} $(\mathcal{M},g_{ab})$ -- \textit{i.e.} in a classical spacetime with a \textit{given} curved metric $g_{ab}$. We shall implement this procedure by appealing to the existence of complete sets of normal modes $\{u_i, u_i^*\}$ to our linear field equations, and promoting the classical amplitudes $\alpha_i$ of these modes to linear operators $a_i$ (in a suitable Hilbert Space $\mathcal{H}$), upon which we impose the mode commutation relations analogous to \eqref{mocorepw}. In summary:

\begin{align*}
\bullet \text{ Classical field:} \qquad &\phi(x) = \sum_i \alpha_i u_i(x) + \alpha^*_i u_i^*(x) \\
\bullet \qquad \alpha_i,\alpha^*_i \in \mathbb{C} \longrightarrow& \;a_i,a^\dagger_i \in GL(\mathcal{H}), \qquad [a_i,a^\dagger_j] = \delta_{ij}, \;\; [a_i,a_j] = 0 = [a^\dagger_i, a^\dagger_j] \\[4pt]
\bullet \text{ Quantized field:} \qquad\!\!\! &\phi(x) = \sum_i a_i u_i(x) + a^\dagger_i u_i^*(x)
\end{align*}

With this basic prescription in mind, let us develop such a process more explicitly. As in the last chapter, we take a real scalar field as a working model. Consider a field $\phi(x)$ with a Lagrangian density:

\begin{align}
\tilde{\mathscr{L}} = |g|^{\frac{1}{2}}\bigl( \tfrac{1}{2}g^{ab}(\nabla_a \phi)(\nabla_b \phi) - \tfrac{1}{2}[m^2 + \xi R]\phi^2 \bigl). \label{lagcurv}
\end{align}

This has the usual kinetic and mass terms, obtained directly from the flat space theory via ``minimal substitution'' (see \eqref{KGMSlagrangian}), as well as a local `nonminimal' covariant coupling with the spacetime curvature, expressed in the term $\tfrac{1}{2}\xi R\phi^2$. Special interest attaches to the values $\xi=0$ (minimal coupling) and $\xi=1/6$ (conformal coupling, in 4 spacetime dimensions; see appendix \ref{geometry}). From the Lagrangian density (\ref{lagcurv}), we immediately obtain the action and derive the dynamic equations for $\phi$:

\begin{align}
S = \int d^4x\, \tilde{\mathscr{L}}(x) = \int d^4x\, |g|^{\frac{1}{2}}\bigl( \tfrac{1}{2}g^{ab}(\nabla_a \phi)(\nabla_b \phi) - \tfrac{1}{2}[m^2 + \xi R]\phi^2 \bigl), \label{actioncurv}
\end{align}

\begin{align}
\frac{\delta S}{\delta \phi(x)} = 0 \quad \Rightarrow \quad \bigl[ \Box_x + m^2 + \xi R(x) \bigl]\phi(x) = 0, \label{ELcurv}
\end{align}
where we have defined the D'Alembertian operator \emph{in curved spacetime} as $\Box = g^{ab}\nabla_{\!a}\nabla_{\!b}$. When acting on a scalar field, it may be written in terms of mere partial derivatives in the form $\Box \phi = |g|^{-1/2} \partial_\mu(|g|^{1/2}g^{\mu\nu} \partial_\nu \phi)$ (see eq. \eqref{Dalembertianphi} ).

Since we are working with a fixed background geometry -- that is, we are ignoring  the gravitational effects of $\phi$ in the metric --, eq. (\ref{ELcurv}) will indeed be a linear second-order PDE, such that any of its solutions can be expanded in a given basis of modes. 

Similarly to the case of flat spacetime, we define an inner product which will allow us to compute projections and decompose any solutions in a given set of modes. These projections will allow us to extract the maximal information of our field from some set of initial conditions in a Cauchy surface $\Sigma$. We make our generalization as follows: we foliate our (globally hyperbolic) spacetime by an arbitrary family of Cauchy surfaces, and we pick \emph{any} surface $\Sigma$ from this family to compute

\begin{align}
(\phi,\psi) &\equiv i \int_\Sigma d^3x |g_\Sigma(x)|^{\frac{1}{2}} \,n^\mu\!(x) \phi^*(x)\overleftrightarrow{\partial_\mu} \psi(x) \nonumber\\
 &= i \int_\Sigma d^3x |g_\Sigma(x)|^\frac{1}{2} \,n^\mu\!(x) \bigl( \phi^*(x)\partial_\mu \psi(x) - (\partial_\mu \phi^*(x)) \psi(x) \bigl), \label{scalarproduct}
\end{align}
where $n^\mu(x)$ is the unitary, future-directed vector normal to $\Sigma$ at $x\in\Sigma$, and $d^3x$ the coordinate 3-volume element. $(-g_\Sigma)_{ij}$ is the (positive-definite) metric induced on $\Sigma$ by $g_{\mu \nu}$, so that the (coordinate independent) induced volume element on $\Sigma$ is $d\mu_{g_\Sigma}(x) = |g_\Sigma(x)|^{\frac{1}{2}} d^3x$.

Just as with (\ref{fsp}), the definition above allows us to immediately verify the elementary properties \eqref{spp} of sesquilinearity. In spite of these useful properties, since our definition relies on an arbitrary choice of integration surface (which, furthermore, is not related to any `special' family of observers), it may not be \textit{a priori} obvious that this product bears similar physical significance to \eqref{fsp}, and whether arbitrary (surface-dependent) elements might appear in it. Indeed, as it happened in flat space, for a completely arbitrary pair of scalar functions $u,w$ in $\mathcal{M}$, the result of (\ref{scalarproduct}) will obviously depend on the choice of $\Sigma$. We shall show, however, that if these functions are solutions of the field equations (\ref{ELcurv}), then their scalar product is independent of $\Sigma$. 

\begin{figure}[H]
\centering
\includegraphics[width=0.6\linewidth]{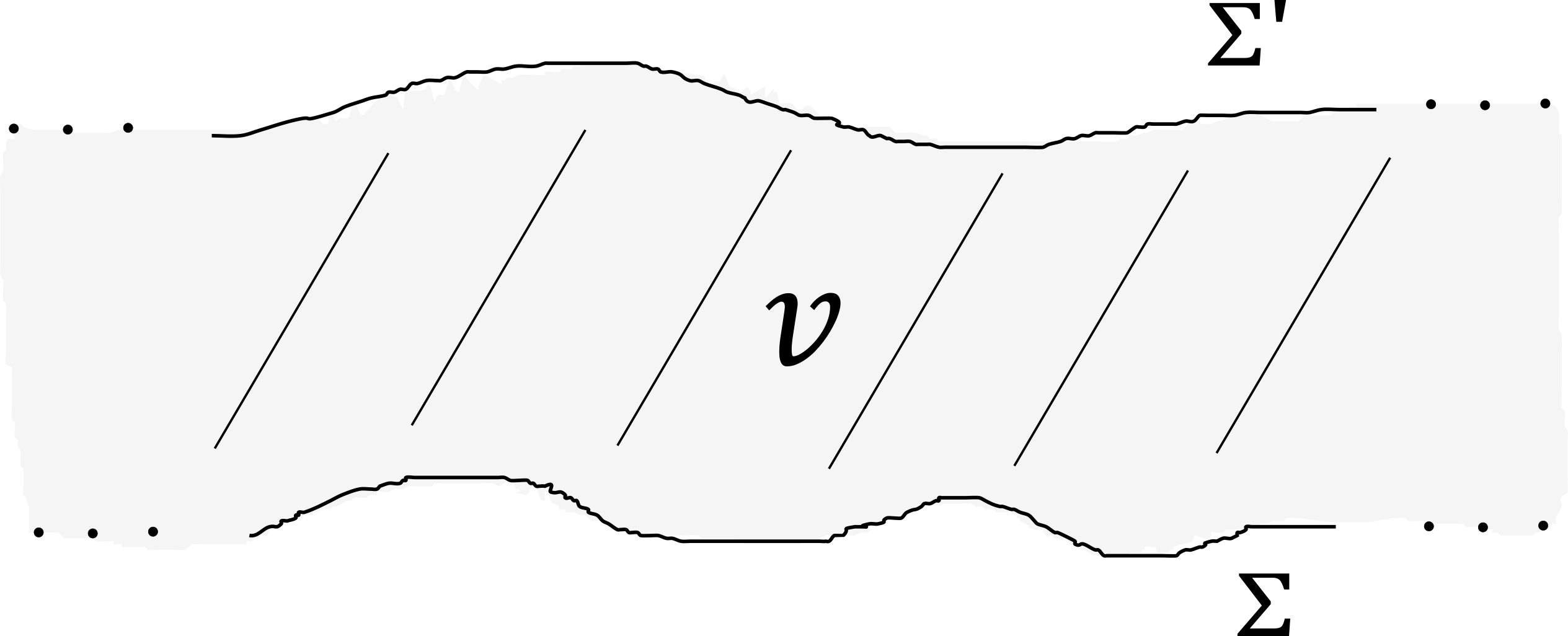}
\caption{Spacetime volume $v$ (grey hatched area) whose bondary is composed by the two Cauchy surfaces $\Sigma$ and $\Sigma' \!\subset\! I^+\!(\Sigma)$. \\ Source: By the author}
\label{CauchyVolume}
\end{figure}

The proof is very similar as in flat space: again, we consider the difference of the product evaluated in two surfaces $\Sigma$ and $\Sigma' \subset I^+(\Sigma)$ and write them as a boundary term; then, using Gauss's theorem, we express it as a volume integral, which will be identically vanishing for any two functions obeying the field equations (see Figure \ref{CauchyVolume}):

\begin{align}
(u,w)_{\Sigma'} - (u,w)_\Sigma & = \int_{\Sigma'} \!d\mu_{g_\Sigma}(x) \;n^\mu\!(x) \, u^*\!(x)\!\overset\leftrightarrow{\partial}_{\!\!\mu} w(x)
 -\int_\Sigma \!d\mu_{g_{\Sigma'}}(x) \;n^\mu\!(x) \, u^*\!(x)\!\overset\leftrightarrow{\partial}_{\!\!\mu} w(x) \nonumber \\
&= \int_v d\mu_g(x) \;\nabla^\mu \!(u^*\!(x)\!\overset\leftrightarrow{\partial}_{\!\!\mu} w(x)) \nonumber \\
&= \int_v d\mu_g(x)  (u^*\!(x)\!\nabla^\mu\nabla_{\!\mu} w(x) - w(x)\nabla^\mu\nabla_{\!\mu} u^*\!(x)) \nonumber \\
&= 0.
\end{align}

Armed with this inner product, it is possible to find an orthonormal basis of solutions to the field equations $\{u_i(x),u_i^*(x)\}$:

\begin{align}
(u_i,u_j) = \delta_{ij} = -(u_i^*,u_j^*), \qquad \qquad (u_i,u^*_j) = 0, \label{orthonormal}
\end{align}
so that we can then expand the classical field in the form

\begin{align}
\phi(x) = \sum_i \alpha_i u_i(x) + \alpha_i^* u^*_i(x) \,. \label{clacurfi}
\end{align}

Now, we can proceed to quantization in an entirely analogous manner to \eqref{mocorepw}, by promoting the classical mode amplitudes $\alpha_i, \alpha^*_i$ to quantum operators $a_i, a^\dagger_i$ with the usual commutation relations\footnote{We stress that these will be \emph{equivalent} to the canonical commutation relations, which can be defined for a given choice of foliation $\Sigma_t$ in $\mathcal{M}$.}:

\begin{subequations} \label{mocore}
\begin{align}
[a_i,a_j] = [a^\dagger_i,a^\dagger_j] &= 0, \\
[a_i,a^\dagger_j] &= \delta_{ij},
\end{align}
\end{subequations}
such that the \emph{quantized field operator} reads

\begin{align}
\phi(x) = \sum_i a_i u_i(x) + a_i^\dagger u^*_i(x) \,. \label{quacurfi}
\end{align}

It then follows, just as in the case of flat spacetime, that we can build a C.S.C.O. in the Hilbert space of our theory from the number operators $N_i = a^\dagger_i a_i$ from an infinite collection of decoupled harmonic oscillators, and so define a Fock Space as usual. We define the vacuum state $\ket{0}$ as the one which is annihilated by all destruction operators $a_i$:

\begin{align}
a_i \ket{0} = 0, \qquad \forall i.
\end{align}

And we can construct the \textit{n}-particle states through successive applications of the creation operators $a^\dagger_i$:

\begin{align}
\ket{n_1,n_2...} = \frac{1}{\sqrt{n_1!n_2!...}}(a_1^\dagger)^{n_1}(a_2^\dagger)^{n_2}... \ket{0}.
\end{align}

However, unlike in Minkowski spacetime, there are in general no ``natural'' sets of modes $\{u_i,u^*_i\}$ in terms of which to define a vacuum state. More precisely, there is no natural way to divide the space of solutions of the field equations in positive- and negative-frequency subspaces (spanned by modes $\{u_i\}$ and $\{u^*_i\}$, respectively). In the Minkowski case, we had a natural choice of coordinates (namely, globally inertial coordinates) and family of modes (plane waves) given by the Poincaré group of all isometries of the Minkowski spacetime. In particular $t^a$ is a Killing field generating time translations, of which the plane waves $u_\mathbf{k}$ are eigenfunctions:

\begin{subequations} \label{mink2posneg}
 \begin{empheq}[left=\empheqlbrace]{align}
 i\partial_t u_\mathbf{k} = +\omega_k u_\mathbf{k}, \label{minkpos} \\
 i\partial_t u^*_\mathbf{k} = -\omega_k u^*_\mathbf{k}, \label{minkneg}
 \end{empheq}
\end{subequations}
where $\omega_k>0, \, \forall \mathbf{k}$.

In a general curved spacetime, with no such symmetry to distinguish particular sets of modes, we could on equal footing consider a distinct set of normal modes, $\{\bar{u}_i,\bar{u}^*_i\}$, obeying the same orthonormality conditions (\ref{orthonormal}), and write the classical field expansion as

\begin{align}
\phi(x) = \sum_i \bar{\alpha}_i \bar{u}_i(x) + \bar{\alpha}^*_i \bar{u}^*_i(x) .
\end{align}
Then we could quantize $\phi$ by promoting these new mode amplitudes $\bar{\alpha}_i,\bar{\alpha}^*_i$ to operators $\bar{a}_i,\bar{a}^\dagger_i$ obeying the same commutation relations as (\ref{mocore}):

\begin{align}
\phi(x) = \sum_i \bar{a}_i \bar{u}_i(x) + \bar{a}^\dagger_i \bar{u}^*_i(x) , \label{quacurfi2}
\end{align}

\begin{subequations} \label{mocore2}
\begin{align}
[\bar{a}_i,\bar{a}_j] = [\bar{a}^\dagger_i,\bar{a}^\dagger_j] &= 0 , \\
[\bar{a}_i,\bar{a}^\dagger_j] &= \delta_{ij}.
\end{align}
\end{subequations}

Just as in the previous case, the number operators $\bar{N}_i =\bar{a}^\dagger_i \bar{a}_i$ make a C.S.C.O. in our Hilbert space, and we may once again define a Fock Space as usual, starting from a vaccum state $\ket{\bar{0}}$, annihilated by all $\bar{a}_i$:

\begin{align}
\bar{a}_i \ket{\bar{0}} = 0, \qquad \forall i,
\end{align}
and similarly defining all particle states through successive applications of $\bar{a}^\dagger_i$ upon it.

Comparing the field expansions \eqref{quacurfi} and \eqref{quacurfi2}, each associated to their respective commutation relations and Fock Spaces, a few relevant questions arise. First of all, are both quantization procedures necessarily equivalent? (For instance, do they lead to mutually consistent field operator commutators? Are their Fock Spaces equivalent, and their physical predictons the same?) Second, do they share a common notion of vacuum? (That is, are $\ket{0}$ and $\ket{\bar{0}}$ always the same, perhaps up to a phase factor?) These questions turn out to reveal deep and interesting features of QFTCS, such as the nature (and inherent ambiguity) of the concept of particles, as well as surprising perspective on the connection between spin and statistics.

In order to address them, we must first specify the relations between these two sets of modes. Being both sets complete, they can each be expanded in terms of one another. For example, we could write $\bar{u}_i$ as

\begin{align}
\bar{u}_i = \sum_j (u_j,\bar{u}_i) u_j - (u_j^*,\bar{u}_i)u_j^* \;= \sum_j \alpha_{ij} u_j + \beta_{ij} u_j^* \,,
\end{align}
where the minus sign on the second term comes from the normalization (\ref{orthonormal}). The coefficients $\alpha_{ij}$ and $\beta_{ij}$ are known as \textit{Bogolubov coefficients}, and are defined above as the projections

\begin{subequations} \label{bogolubov}
\begin{empheq}[left=\empheqlbrace, right = {\;\;.}]{align}
\alpha_{ij} &\equiv (u_j,\bar{u}_i) = (\bar{u}_i,u_j)^* \\ 
\beta_{ij} &\equiv - (u_j^*,\bar{u}_i) = -(\bar{u}_i,u_j^*)^*
\end{empheq}
\end{subequations}
(Here, we warn that the exact conventions may vary somewhat in the literature.)

Conversely, we could expand the $u_i$ modes in terms of $\bar{u}_j$ modes, as well as write similar expansions for the creation and annihilation operators. The latter expansions may be easily computed from the former (or vice-versa) by using the relations $a_i = (u_i,\phi)$ (or $u_i = [\phi,a^\dagger_i]$). We summarize all these expansions in terms of the Bogolubov coefficients \eqref{bogolubov}:

\vspace{10pt}

\begin{minipage}[0.0\linewidth]{0.45\linewidth}
 \begin{subequations}
  \label{modes}
  \begin{align}
  u_i &= \sum_j \alpha_{ji}^* \bar{u}_j - \beta_{ji} \bar{u}_j^* \\
  \bar{u}_i &= \sum_j \alpha_{ij} u_j + \beta_{ij} u_j^*
  \end{align}
 \end{subequations}
\end{minipage} \hfill \vline \hfill
\begin{minipage}[0.0\linewidth]{0.45\linewidth}
 \begin{subequations}
  \label{operators}
  \begin{align}
  a_i &= \sum_j \alpha_{ji} \bar{a}_j + \beta_{ji}^* \bar{a}_j^\dagger \\
  \bar{a}_i &= \sum_j \alpha_{ij}^* a_j - \beta_{ij}^* a_j^\dagger 
  \end{align}
 \end{subequations}
\end{minipage} \vspace{6pt}

Also, the orthonormality and completeness of both sets of modes will imply in self-consistency properties for the Bogolubov coefficients. It is easy to deduce them by performing a back and forth transformation for any fixed mode $\bar{u}_i$:

\begin{align}
\bar{u}_i &= \sum_j \alpha_{ij} u_j + \beta_{ij} u_j^* \nonumber \\
 &= \sum_k \biggl \{ \Bigl[\sum_j \alpha_{ij} \alpha_{kj}^* - \beta_{ij} \beta_{kj}^* \Bigl] \bar{u}_k - \Bigl[\sum_j \alpha_{ij} \beta_{kj} - \beta_{ij} \alpha_{kj} \Bigl] \bar{u}_k^* \biggl\},
\end{align}
which immediately implies

\begin{subequations} \label{Bogolubov OCness}
\begin{empheq}[left = \empheqlbrace, right={\;\;.}] {align}
\sum_j \alpha_{ij} \alpha_{kj}^* - \beta_{ij} \beta_{kj}^* = \delta_{ik} \\
\sum_j \alpha_{ij} \beta_{kj} - \beta_{ij} \alpha_{kj} = 0
\end{empheq}
\end{subequations}

Let us then begin to investigate our first question. With the above relations and properties at hand, it is straightforward to verify whether the commutation relations (\ref{mocore}) and \eqref{mocore2} are mutually consistent. If we analize the covariant commutator $[\phi(x),\phi(y)]$ in both expansions, such consistency must imply the equality

\begin{align}
\sum_i u_i(x)u^*_i(y) - u_i(y)u_i^*(x) = [\phi(x),\phi(y)] = \sum_i \bar{u}_i(x)\bar{u}^*_i(y) - \bar{u}_i(y)\bar{u}_i^*(x). \label{equality}
\end{align}

Expanding the RHS of (\ref{equality}) by means of (\ref{modes}), we then obtain

\begin{align}
 &\sum_i \Bigl\{ \bigl({\textstyle\sum_j} \alpha_{ij} u_j(x) \!+\! \beta_{ij} u_j^*(x) \bigl) \bigl( {\textstyle\sum_k} \alpha_{ik}^* u_k^*(y) \!+\! \beta_{ik}^* u_k(y) \bigl) \nonumber \\[-6pt]
   & \qquad \; -\bigl({\textstyle\sum_j} \alpha_{ij} u_j(y) \!+\! \beta_{ij} u_j^*(y) \bigl) \bigl( {\textstyle\sum_k} \alpha_{ik}^* u_k^*(x) \!+\! \beta_{ik}^* u_k(x) \bigl) \Bigl\} \nonumber \\[6pt]
 =& \sum_{j,k} \Bigl[ \bigl({\textstyle\sum_i}\alpha_{ij}\alpha_{ik}^* \!-\!\beta_{ik}\beta_{ij}^* \bigl)u_j(x)u_k^*(y) - \bigl({\textstyle\sum_i}\alpha_{ik}\alpha_{ij}^* \!-\!\beta_{ij}\beta_{ik}^* \bigl) u_k(y)u_j^*(x) \nonumber \\[-6pt]
 & \qquad \; + \bigl({\textstyle\sum_i}\alpha_{ij}\beta_{ik}^* \!-\!\alpha_{ik}\beta_{ij}^* \bigl)u_j(x)u_k(y) + \bigl({\textstyle\sum_i}\beta_{ij}\alpha_{ik}^* \!-\!\beta_{ik}\alpha_{ij}^* \bigl)u_j^*(x)u_k^*(y) \Bigl].  \label{bigequality}
\end{align}

Then, comparing this to the LHS of (\ref{equality}), we get the following conditions for the Bogolubov coefficients:

\begin{subequations} \label{bogolubov conditions}
\begin{align}
\sum_i \alpha_{ij}\alpha_{ik}^* - \beta_{ik}\beta_{ij}^* &= \delta_{jk}, \\
\sum_i \alpha_{ij}\beta_{ik}^* - \alpha_{ik}\beta_{ij}^* &= 0.
\end{align}
\end{subequations}

But these conditions merely express the orthonormality and completeness of both bases (being equivalent to \eqref{Bogolubov OCness}), thus showing that \eqref{mocore} and \eqref{mocore2} are indeed equivalent.

On the other hand, had we tried to quantize our fields by imposing anticommutation relations to these operators:

\begin{subequations} \label{anticore}
\begin{align}
\{a_i,a_j\} = \{a^\dagger_i,a^\dagger_j\} = &\; 0 = \{\bar{a}_i,\bar{a}_j\} = \{\bar{a}^\dagger_i,\bar{a}^\dagger_j\},   \\
\{a_i,a^\dagger_j\} = &\; \delta_{ij} = \{\bar{a}_i,\bar{a}^\dagger_j\},
\end{align}
\end{subequations}
we immediately see that the corresponding consistency check would yield

\begin{align}
\sum_i u_i(x)u^*_i(y) + u_i(y)u_i^*(x) = \{\phi(x),\phi(y)\} = \sum_i \bar{u}_i(x)\bar{u}^*_i(y) + \bar{u}_i(y)\bar{u}_i^*(x), \label{Aequality}
\end{align}
such that the corresponding conditions for the Bogolubov coefficients, namely

\begin{subequations} \label{bogolubov anticonditions}
\begin{align}
\sum_i \alpha_{ij}\alpha_{ik}^* + \beta_{ik}\beta_{ij}^* &= \delta_{jk}, \\
\sum_i \alpha_{ij}\beta_{ik}^* + \alpha_{ik}\beta_{ij}^* &= 0,
\end{align}
\end{subequations}
are generally not satisfied for $\beta_{ij} \neq 0$. (In section \ref{particle creation}, we shall reinterpret this result in terms of particle creation.)

Thus, we see that we can achieve consistency for quantization in two arbitrary families of modes (for a free scalar field) \emph{only if we impose commutation relations} (rather than anticommutation relations) for our field mode operators. These, on their turn, will imply that $\phi$ must obey Bose-Einstein statistics.

Then, having assured mutual consistency between the commutation relations defined for any two bases of orthonormal modes, we would like to know the relation between their respective Fock spaces, particularly, how one may relate states $\ket{n_1, n_2\hdots}$ defined with the occupation numbers of the modes $\{u_i\}$ in terms of $\ket{\bar{n}_1, \bar{n}_2, \hdots}$, defined with the occupation numbers of the modes $\{\bar{u}_i\}$.

To draw these relations, all one needs to determine is the general form of the projections $\braket{\bar{n}_1,\bar{n}_2 ...|n_1, n_2...}$ ($\propto \braket{\bar{0}|0}$). We shall not deduce the general form of those projections here\footnote{ Although we will explicitly calculate them in a special case in section \ref{particle statistics}.}, 
 but we note that they can be computed with some algebraic effort employing the expansions \eqref{operators} for the creation and annihilation operators (explicit expressions for them in terms of the Bogolubov coefficients can be found in \cite{birrell}, eqs. (3.45)-(3.47)). However, there are two important features of these amplitudes that we would like to point out: (i) the vacuum to many-particles transitions are only nonzero when the number of particles is even, as the creation (annihilation) operators $\bar{a}_i$ ($\bar{a}^\dagger_i$) are always linear in the operators $\bar{a}_i$ and $\bar{a}^\dagger_i$ and one needs an even number of such operators to match the number of created and annihilated particles and produce nonorthogonal states; (ii) these amplitude transitions are \emph{always} proportional to the vacuum to vacuum amplitudes $\braket{\bar{0}|0}$, and thus it is necessary that $\braket{\bar{0}|0} \neq 0$ for both Fock spaces to represent the same Hilbert space.
 
When $\braket{\bar{0}|0}=0$, the modes $\{u_i\}$ and $\{ \bar{u}_i \}$ are said to yield unitarily inequivalent quantized theories. In the scope of the present work, we shall not go into detail of unitary (in)equivalence. For an in-depth discussion of the topic, see \cite{redwald}; also, the reader may find a quite simple example of unitarily inequivalent mode choices in Minkowski spacetime in the section II of \cite{parkerIR}, where one considers modes of the form:

\begin{align}
\bar{u}_{\mathbf{k}} \propto (\alpha(k) e^{-i\omega_k t} + \beta(k)e^{i\omega_k t} )e^{i \mathbf{k}\!\cdot\!\mathbf{x}}.
\end{align}

Then, restricting ourselves to the cases where our theories are unitarily equivalent, what can we say about the relation between their vacua $\ket{0}$ and $\ket{\bar{0}}$ other than they are nonorthogonal? First, by inspecting the operator expansions (\ref{operators}), one immediately sees that the vacuum states $\ket{0}$ and $\ket{\bar{0}}$ will not in general coincide. For example, we see that

\begin{align}
a_i \ket{\bar{0}} = \sum_j \beta_{ji}^* \ket{\bar{1}_j} \neq 0 , \label{bogolubovparticles}
\end{align}
which yields a nonzero expectation value for particle numbers in a `mismatched' vacuum, such as

\begin{align}
\bra{\bar{0}} N_i \ket{\bar{0}} = \sum_j |\beta_{ji}|^2 . \label{bogolubovParticlesEV}
\end{align}

Such nonzero expectation values come from the fact that generally the annihilation operators from one family mixes creation and annihilation ones from the other, which, in its turn, can be traced back to the $\beta$ coefficients, which mix the `not-conjugated' modes (associated with annihilation operators on quantization) with the `conjugated' modes (associated with creation operators on quantization). In Minkowski spacetime, time-translation symmetry gave a distinct meaning to both subspaces of modes: they were associated with positive and negative frequencies, respectively (see \eqref{minkpos}).

A more general class of spacetimes where this priviledged distinction between positive and negative frequency modes arises are stationary spacetimes. These spacetimes will possess (at least one) timelike Killing field $\xi^a$ (see Appendix \ref{geometry}) such that, analogously to \eqref{minkpos}, we can define positive frequency modes $u_j$ as:

\begin{align}
i\pounds_{\!\xi} u_j(x) = \omega_j u_j(x)\,, \;\;\omega_j>0 \label{stationary}
\end{align}
(where $\pounds_{\!\xi}$ denotes the Lie derivative with respect to $\xi^a$).

However, for general curved spacetimes with no such symmetries, there will be no physically priviledged modes in terms of which to define positive-frequency solutions. In the next section, we shall show that this reflects the fact that the concept of particle as occupation numbers of some field modes has generally no direct physical interpretation in terms of what observers would measure with particle detectors.

\subsection{Relating mode and Canonical Commutation Relations}

Before we proceed to the next section, we shall explicitly show how one may reobtain the canonical commutation relations from \eqref{mocore}. As in the case of flat space (where we conversely derived \eqref{mocorepw} from \eqref{fcacore} ), the essential factor to this equivalence is that the (classical) maximal information of the field can be extracted both from $\phi$ and $\pi$ in a Cauchy surface, or by the (space)time-independent amplitudes $\alpha_i$ and $\alpha^*_i$ for a complete set of modes; correspondingly, in a quantum description one may write the infinite collections of operators $\{\phi(x),\pi(x)\}, x\in\Sigma$ and $\{a_i,a^\dagger_i\}, i\in I$ in terms of one another.

The demonstration will be more convenient if we pick a time coordinate $t$ (in order to define the momentum $\pi$) and a foliation $\Sigma_t$ such that the timelike vector field $t^\mu$ whose integral lines generate the evolution in $t$ coincides with $n^\mu$, the unit vectors orthogonal to $\Sigma_t$ at each event. In this case, we write the metric components in the form

\begin{align}
g_{\mu\nu} = n_\mu n_\nu - h_{\mu\nu},
\end{align}
with $n_\mu n^\mu = 1$, and being $h_{\mu\nu}$ tangent to each $\Sigma_t$ (its action restricted to these surfaces defines a positive-definite metric on them), such that $h_{\mu\nu}n^\nu=0$.

Then, we have the velocity $\dot{\phi} \equiv n^\mu \nabla_{\!\mu} \phi \equiv n^\mu \partial_\mu \phi = \partial_0 \phi $. From the Lagrangian \eqref{lagcurv}, we obtain the canonically conjugated momentum

\begin{align}
\pi \equiv \frac{\partial \mathscr{L}}{\partial \dot{\phi}} = g^{\mu 0} \partial_\mu \phi = n^\mu \partial_\mu \phi.
\end{align}

Then using the field mode expansions \eqref{quacurfi}, and the commutation relations \eqref{mocore}, we immediately obtain:

\begin{subequations} \label{curvedcanonical}
\begin{align}
[ \phi(x), \phi(x')] &= \sum_i u_i(x)u_i^*(x') - u_i^*(x)u_i(x') , \\
[ \pi(x), \pi(x')] &= \sum_i n^\mu n^{\nu} \bigl( \partial_\mu u_i(x) \partial'_{\nu} u_i^*(x') - \partial_\mu u_i^*(x) \partial'_{\nu} u_i(x') \bigl) , \\
[ \phi(x), \pi(x')] &= \sum_i n^{\mu} \bigl(  u_i(x) \partial'_{\mu} u_i^*(x') -  u_i^*(x) \partial'_{\mu} u_i(x') \bigl) 
\end{align}
\end{subequations}
(where $\partial'_\mu = \frac{\partial}{\partial {x'}^\mu}$).

Then, if we wish to analyze these relations for equal times, we must only restrict $x$ and $x'$ to belong to the same Cauchy surface $\Sigma_t$, for which we denote $x = (\mathbf{x},t)$ and $x' = (\mathbf{x'},t)$. Now, by virtue of the completeness of $\{u_i,u_i^*\}$, we should be able to expand any arbitrary solutions $v(x)$ to the field equations \eqref{ELcurv} in terms of it:

\begin{align}
v(x) &= \sum_i (u_i,v)u_i(x) - (u^*_i,v)u^*_i(x) \nonumber \\
 &= \sum_i \int_{\Sigma_t} d\mu_h(\mathbf{x'}) \, n^\mu \bigl( u_i^*(x') \partial'_\mu v(x') - \partial'_\mu u^*_i(x') v(x') \bigl)u_i(x) \nonumber \\[-6pt] & \qquad \qquad \qquad \qquad - n^\mu \bigl( u_i(x') \partial'_\mu v(x') - \partial'_\mu u_i(x') v(x') \bigl)u_i^*(x) \nonumber \\[6pt]
 &= \int_{\Sigma_t} d\mu_h(\mathbf{x'})\, \Bigl[n^\mu \textstyle{\sum_i} \bigl( u_i(x) \partial'_\mu u_i^*(x') - u_i^*(x) \partial'_\mu u_i(x') \bigl)\Bigl] v(x') \nonumber \\ & \qquad \qquad \qquad \; -  \Bigl[\textstyle{\sum_i}\bigl( u_i(x) u^*_i(x') - u^*_i(x) u_i(x') \bigl) \Bigl] n^\mu\partial'_\mu v(x'), \label{functionexpansion}
\end{align}
whence we conclude that

\begin{align}
\sum_i n^\mu\Bigl( u_i(x) \partial'_\mu u_i^*(x') - u_i^*(x) \partial'_\mu u_i(x') \Bigl) = \delta(\mathbf{x},\mathbf{x'}), \label{s1} \\
\sum_i \Bigl( u_i(x) u^*_i(x') - u^*_i(x) u_i(x') \Bigl) = 0. \label{s2}
\end{align}

Additionally, taking the time derivative of \eqref{functionexpansion}, we have

\begin{align}
n^\mu \partial_\mu v(x) &= \sum_i n^\mu \bigl( (u_i,v) \partial_\mu u_i(x) - (u^*_i,v)\partial_\mu u^*_i(x) \bigl), \label{flowfree}
\end{align}
as the inner product is time(Cauchy surface)-invariant. From \eqref{flowfree}, one can analogously derive that

\begin{align}
\sum_i n^\mu n^\nu \Bigl( \partial_\mu u_i(x) \partial'_\nu u_i^*(x') - \partial_\mu u_i^*(x) \partial'_\nu u_i(x') \Bigl) = 0. \label{s3}
\end{align}

Thus, applying \eqref{s1}, \eqref{s2} and \eqref{s3} to \eqref{curvedcanonical}, we immediately recover the canonical commutation relations:

\begin{subequations}\label{cacore}
\begin{align}
[\phi(\mathbf{x},t), \phi(\mathbf{x'},t)] = 0 = [\pi(\mathbf{x},t), \pi(\mathbf{x'},t)], \\
[\phi(\mathbf{x},t), \pi(\mathbf{x'},t)] = i \delta(\mathbf{x},\mathbf{x'}).
\end{align}
\end{subequations}

\section{Particle Detectors: an empirical notion of particles} \label{particle detectors}

From the fact that there are different sets of modes associated with different vacuum states, the question arises of which of these should yield the ``most physical vacuum''; that is, loosely speaking, the ``most empty'' vacuum, or the vacuum that better corresponds to the ``experience of no particles''. As stated above, this question is notably ill posed, since any empirical notion of ``emptiness'', or the ``experience of no particles'', cannot depend on the state of the field alone; at the very least it also requires an observer interacting with it.

And indeed (as we shall see ahead), for a fixed field state, the number of particles measured by an observer will be highly nonunique; among other factors, it will depend on the observer's state of motion. This is true even in Minkowiski spacetime; what is special about the latter is not the existence of a unique vacuum state, but rather that its high degree of symmetry assures that there is a common vacuum state for \emph{all inertial observers}. In general globally hyperbolic spacetimes, no such state will exist (even if we confine ourselves to inertial observers), and there will be an inherent ambiguity in the number of particles measured by different observers, or even by the same observer at different times (while the field remaining in a fixed state).

A great deal of this ambiguity in the concept of particles springs from the fact that they are defined as excitations (occupation numbers) of field modes, which are defined \emph{globally, in the entire spacetime} (indeed, we have for example that a particle with momentum $\mathbf{k}$ will be completely spatially delocalized). This global nature makes it impossible to generally draw simple relations between the expected values $\braket{N_i}$ and the (statistical) results of mesurements carried by spatially localized observers (and much less to write simple transformation laws between the results of measurements of 2 distinct observers). In contrast, local observables such as $\braket{T_{\mu\nu}(x)}$ allow a more direct interpretation in terms of measurements carried by localized observers. Furthermore, they are subject to simple transformation laws relating what is measured by two different observers; in the specific case of a tensorial quantity, such as the stress tensor $\braket{T_{\mu\nu}(x)}$, this relation should be a simple coordinate transformation relating two reference frames. Particularly, if $\braket{T_{\mu\nu}(x)} = 0$ for one observer, the same should be true for \emph{all} observers.

Still, in a few highly symmetric spacetimes, a privileged notion of particles may arise, which will be associated with special modes (following spacetimes symmetries and being related to special families of observers). In such cases, simple relations will emerge between the expected values $\braket{N_i}$ and the particles measured by these special observers, recovering, for example, the well-known particle notion in Minkowski spacetime.
One particular case of interest is that of spacetimes which are assimptotically Minkowskian in the remote past and remote future: in this case, both regions will have special vacuum states, which we respectively denote as $\ket{0_p}$ and $\ket{0_f}$.

We remind that, as we are working in the Heisenberg picture, the vector states remain unchanged under time evolution, and the same is true for any number operators $N_{\mathbf{k}}$, as they are defined in respect to global modes, defined through all of spacetime. However, the \emph{physical} notion of particles, as measured by particle observables (and particularly \emph{which} set of number operators may be of direct physical significance) will generally change with time.

We shall illustrate all of the above considerations by exploring an idealized model of a particle detector\footnote{Here, we follow (and extend a little) the exposition in \cite{birrell}. See section 3.3 on it, and references therein. }. This model consists of a point-like physical system, whose internal degrees of freedom correspond to a discrete set of energies $\{E\}$, whose internal dynamics are given by the Hamiltonian $H_0$: $H_0 \ket{E} = E\ket{E}$. For simplicity, we take its energy levels to be nondegenerate (\textit{i.e.} we take $H_0$ to be a C.S.C.O. for the system). 

Now, this probe system (the `detector') shall be weakly coupled to our scalar field by a local monopole interaction, given by the Lagrangian

\begin{align}
\mathscr{L}_I = c \,m(\tau) \phi(x^\mu\!(\tau)), \label{detecoupling}
\end{align}
where $\tau$ denotes the proper time of the detector, $x^\mu\! (\tau)$ its (classical) trajectory in spacetime, $m(\tau)$ its monopole moment, and $c$ a ``small'' coupling constant.

Within this framework, we are interested in deriving the probabilities that the interaction with the field will promote a ``detection'', that is, an excitation of our probe system from its ground state $\ket{E_0}$ to an excited state $\ket{E}$, $E\!>\!E_0$. 
We are demanding the coupling to be weak, so that the interactions between the field and detector may be treated perturbatively. Formally, we shall derive the probabilities of excitation of the detector like a scattering problem\footnote{Note that this entails the assumption that interactions are transient. However, as one can easily see from \eqref{detecoupling}, the interactions are generally persistent. This will lead to a few incongruences below, which will be addressed in due time.},
 through the S matrix formalism\footnote{The reader unfamiliar with this formalism is referred to chapter 6 of \cite{mandlshawn} for further details.}. We thus switch from the Heisenberg to the Dirac picture, where the field and detector observables evolve through their free Hamiltonians, whereas states evolve through the interaction Hamiltonian $\mathcal{H}_i = -\mathscr{L}_I$. To first perturbative order, this entails the transition amplitudes $\mathcal{A}$ between two states $\ket{E,\Psi}$ and $\ket{E',\Psi'}$:
 
\begin{align}
\mathcal{A}\bigl(\ket{E,\Psi} \rightarrow \ket{E',\Psi'}\!\bigl) = ic \bra{E',\Psi'} \int_{-\infty}^{+\infty} \!\!\! m(\tau)\phi(x^\mu(\tau))d\tau \ket{E,\Psi}. \label{amp0}
\end{align}
 
Particularly, we are interested in the possibility of making a transition to make a detection in the vacuum state. That is, of starting with our field in a vacuum $\ket{0}$ and our detector at ground state $\ket{E_0}$, and ending up with an excited detector $\ket{E}$ and some final state for the field $\ket{\Psi}$ (the precise state $\ket{\Psi}$ of the field \textit{after} the measurement is of little importance to us; the relevant question here is if we can make a detection). Let us then analyze the probabilities of detection in Minkowski space, in the usual Minkowski vacuum $\ket{0_M}$:

\begin{align}
\mathcal{A}\bigl(\ket{E_0,0_M} \rightarrow \ket{E,\Psi}\!\bigl) = ic \bra{E,\Psi} \int_{-\infty}^{+\infty} \!\!\! m(\tau)\phi(x^\mu(\tau))d\tau \ket{E_0,0_M}. \label{amp1}
\end{align}

Since we are working in the Dirac Picture, $m(\tau)$ simply evolves through the free Hamiltonian $H_0$:

\begin{align}
m(\tau) = e^{i H_0 \tau} m(0) e^{-i H_0 \tau}.
\end{align}

Substituting in \eqref{amp1}, we obtain

\begin{align}
\mathcal{A}(\ket{E_0,0_M} \rightarrow \ket{E,\Psi}) = ic \bra{E} m(0) \ket{E_0} \int_{-\infty}^{+\infty} \!\!\! e^{i(E-E_0)\tau} \bra{\Psi} \phi(x^\mu(\tau)) \ket{0_M} d\tau. \label{amp2} 
\end{align}

Since $\phi$ is linear in creation and annihilation operators, the only transitions that may occur (\emph{in first perturbative order}) are those to one-particle states: $\ket{\Psi} = \ket{1_{\mathbf{k}}}$. If we consider the continuum normalization \eqref{C plane waves}, we get the amplitudes

\begin{align}
\bra{1_{\mathbf{k}}} \phi(x) \ket{0} &= \int d^3 \mathbf{k'} (16\pi^3 \omega_{k'})^{-1/2} \bra{1_{\mathbf{k}}} a^\dagger_{\mathbf{k'}} \ket{0} e^{i \omega't - i \mathbf{k'}\!\cdot\! \mathbf{x}} \nonumber \\
 &= (16\pi^3 \omega_{k})^{-1/2} e^{i \omega t - i \mathbf{k}\!\cdot\! \mathbf{x}}. \label{amp3}
\end{align}

Inserting the result in (\ref{amp2}), we see that we must indeed specify a spacetime trajectory $x^\mu(\tau)$ to the detector to compute well-defined transition amplitudes. Let us first consider an inertial world-line:

\begin{align}
\mathbf{x} &= \mathbf{x_0} + \mathbf{v}t = \mathbf{x_0} + \mathbf{v} \gamma_v \tau, \label{inertialdetector}
\end{align}
where $\gamma_v$ is the Lorentz factor $\gamma_v = (1-v^2)^{-\frac{1}{2}}$.
In this case, we have

\begin{align}
\mathcal{A}(\ket{E_0,0_M} \rightarrow \ket{E,1_\mathbf{k}}) &= \frac{ic \bra{E} m(0) \ket{E_0} }{16\pi^3 \omega} e^{-i \mathbf{k}\!\cdot\!\mathbf{x_0}} \int_{-\infty}^{+\infty} \!\!\! e^{i(E-E_0)\tau} e^{i(\omega - \mathbf{k}\!\cdot\!\mathbf{v})\gamma_v\tau} d\tau \nonumber \\
 &= \frac{ic \bra{E} m(0) \ket{E_0} }{4\pi \omega} e^{-i \mathbf{k}\!\cdot\!\mathbf{x_0}} \delta( E-E_0 + [\omega - \mathbf{k}\!\cdot\!\mathbf{v}]\gamma_v). \label{ampinm}
\end{align}

But since $E>E_0$ and $\omega > |\mathbf{k}\!\cdot\!\mathbf{v}|$ (as $v<1$ for any timelike trajectory and $\omega \!=\! \sqrt{k^2+m^2} \geq k$) there are no roots in the arguments of the $\delta$ distribution in (\ref{ampinm}), and the transition amplitude is always zero, as dictated by energy conservation -- a direct consequence of time translation symmetry (as energy is the global Noether charge associated to this symmetry).

For more complicated trajectories, however, the transition amplitudes (\ref{amp2}) do not generally yield $\delta$'s, and nonzero transition probabilities may emerge from the vacuum! (As we shall demonstrate briefly.) In such cases, we will be interested in summing the transition probabilities over all possible final states $\ket{\Psi}$ and $\ket{E}$ ($\neq\ket{E_0}$) to obtain the total probability that \textit{any} transition (detection) may occur:

\begin{align}
\sum_{E,\Psi} \bigl| \mathcal{A}(\ket{E_0,0_M} \rightarrow \ket{E,\Psi}) \bigl|^2 = c^2 & \sum_E  \Bigl\{ |\!\braket{E|m(0)|E_0}\!|^2 \times \nonumber \\ 
 &\iint d\tau \, d\tau' e^{i(E-E_0)(\tau-\tau')} \braket{0_M| \phi(\tau') [{\textstyle\sum_\Psi} \!\ket{\Psi}\!\!\bra{\Psi}] \phi(\tau) |0_M} \Bigl\}.
\end{align}

Using the completeness relation $\sum_\Psi \!\ket{\Psi}\!\!\bra{\Psi} = \mathbb{1} $, and recognizing the vacuum two-point correlation as the Wightman function (\ref{fWhightmann+}), we have

\begin{align}
P &= c^2 \sum_E |\!\braket{E|m(0)|E_0}\!|^2 \iint d\tau d\tau' e^{-i(E-E_0)(\tau-\tau')} G^+\!(x(\tau),x(\tau')) \nonumber \\
  &= c^2 \sum_E |\!\braket{E|m(0)|E_0}\!|^2 \mathscr{F}(E-E_0), \label{probtotal}
\end{align}
where we defined the response function of the detector $\mathscr{F}(E)$: 

\begin{align}
\mathscr{F}(E) \equiv \iint d\tau \, d\tau' e^{-iE (\tau-\tau')} G^+\!(\tau,\tau'). \label{responsefunction}
\end{align}
(Here, we simplified the notation of $G^+$, leaving implicit the dependence on the detector trajectory $x(\tau)$.)

Taking a closer look at expression \eqref{probtotal}, we see that the details regarding the inner structure of the detector enter only in the prefactor $c^2|\braket{E|m|E_0}|^2$, whereas the response function carries the dependence on the field variables (of course, it will also depend on the detector energy differences, just like the response of an atom interacting with radiation will depend on its spectrum). If we are not particularly interested in this inner structure, but rather in the field-related response, we may just focus on the latter.

Then, to evaluate \eqref{responsefunction} more closely, it is convenient to perform change of variables in this double integral, analyzing it in terms of the time average $\bar{\tau} = \tfrac{1}{2}(\tau+\tau')$ and time difference $\Delta\tau = \tau - \tau'$. Since the transformation $(\tau,\tau') \rightarrow (\bar{\tau},\Delta \tau)$ has unit Jacobian, we have

\begin{align}
\mathscr{F}(E) = \iint d\bar{\tau} \, d(\Delta\tau) e^{-iE \Delta\tau} \tilde{G}^+(\bar{\tau},\Delta\tau),
\end{align}
where $\tilde{G}^+(\bar{\tau},\Delta\tau) \equiv G(\tau,\tau')$.

Particularly, if we analyze a \emph{stationary} trajectory, that is, one for which the correlations $G^+(\tau,\tau')$ only depend on the proper-time \emph{differences}, $G^+(\tau,\tau') = G^+(\Delta \tau)$ (here, we drop the tilde in our notation since there is no risk of ambiguity), we obtain trivially separable integrals:

\begin{align}
\mathscr{F}(E) = \biggl( \int_{-\infty}^{\infty} \!\!\! d \bar{\tau} \biggl) \biggl( \int_{-\infty}^{\infty} \!\! d(\Delta \tau) e^{-iE \Delta \tau} G^+(\Delta \tau) \biggl),
\end{align}
which can be immediately interpreted as a (constant) \emph{transition rate} multiplied by the (infinite) time interval of the interactions $T \equiv \int \!d\bar{\tau}$. It is clear that, whenever we have non null transition rates, such a case will yield divergent transition probabilities. This evidently points to a break in our perturbative approximation for indefinitely long time scales with persistent interactions, as anticipated earlier; this occurs because a (first-order) perturbative approach fails to account for the possibility that the system already transitioned in a past instant, acumulating an (unboundedly) increasing transition probability.

Nonetheless, this approach should render a good approximation if we restrict our analysis to sufficiently short time intervals $T$, for which $\mathscr{F}(E) \ll 1$, that is

\begin{align}
T \ll \biggl( \int_{-T}^{T} \!\! d(\Delta \tau) e^{-iE \Delta \tau} G^+(\Delta \tau) \biggl)^{-1},
\end{align}
but sufficiently long so that there will not be a great difference in setting the integration limits at this finite $T$, rather than at infinity -- note that the faster the vacuum correlations $G^+\!(\Delta\tau)$ decay, and the higher the energy jump $E$ in the detector is, the smaller this lower bound will be (and the greater the upper bound will be). (Physically, one can think of this restriction as considering a detector that does not eternally interact with the ``background'' field, but rather that is set to interact with it for a finite time interval $T$.\footnote{In this case, one requires the ``offswitch'' (decoupling) of the detector to occur adiabatically (in a sufficiently smooth and slow manner so that no particles are created by the process).})

As long as we remain in these consistency intervals, it is actually quite more convenient to work directly with \emph{transition rates}. Thus, we define the \emph{response function per unit time}:

\begin{align}
\mathscr{F}'(E) = \frac{\mathscr{F}(E)}{T} = \int_{-\infty}^{\infty} d(\Delta \tau) e^{-iE \Delta \tau} G^+(\Delta \tau). \label{responserate}
\end{align}

Let us then attempt to evaluate this function explicitly. Even with all the simplifications so far, Green function $G^+$ is still a little convoluted to resolve analytically in the massive case, $m>0$. Thus, we restrict our attention to the simpler case of a massless field, $m\!=\!0$, and analyze it in further detail. In this case, for an arbitrary pair of events $(x,x')$, $G^+$ reads

\begin{align}
G^+(x,x') &= \frac{-i}{(2\pi)^4} \int d^4\!k \frac{e^{-ik(x-x')}}{(k^0)^2-\mathbf{k}^2} \nonumber \\
  &= \frac{1}{(2\pi)^3} \int \frac{d^3\!\mathbf{k}}{2|\mathbf{k}|} e^{-i|\mathbf{k}|\Delta t + i \mathbf{k}\!\cdot\!\Delta\mathbf{x}} \nonumber \\
  &= \frac{1}{(2\pi)^3} \int_0^\infty \frac{d|\mathbf{k}|}{2|\mathbf{k}|} |\mathbf{k}|^2 e^{-i|\mathbf{k}| \Delta t} \int_{-1}^{1} d(\cos \theta) e^{i |\mathbf{k}||\Delta\mathbf{x}| \cos \theta} \Bigl( \int_0^{2\pi} \!\!\! d\phi \Bigl) \nonumber \\
  &= \frac{1}{4\pi^2} \frac{1}{2i|\Delta \mathbf{x}| } \int_0^\infty d |\mathbf{k}| (e^{-i |\mathbf{k}|(\Delta t - \Delta x)} - e^{-i |\mathbf{k}|(\Delta t + \Delta x)} ). \label{m=0 correlation}
\end{align}

This integral obviously does not converge in the usual functional sense. As we have seen in Section \ref{2-PFs} and Appendix \ref{distributions}, we must generally interpret two-point functions in integrals such as \eqref{probtotal} in the distributional sense.

However, a convenient trick to work directly with $G^+$ (\textit{i.e.} to get a closed expression for $G^+$, carrying the $k$-integral \eqref{m=0 correlation} \emph{before} the $\Delta \tau$ integral in \eqref{responserate}) is to introduce the regularizer $e^{-\epsilon |\mathbf{k}|}$ ($\epsilon > 0$), making \eqref{m=0 correlation} absolutely convergent. In the end of \emph{all} integrations, we may relax the regularization and take the limit $\epsilon \rightarrow 0^+$. Denoting this regularized function by $G^+_\epsilon$, we have

\begin{align}
G^+_\epsilon(x,x') &= \frac{1}{4\pi^2} \frac{1}{2|\Delta \mathbf{x}| } \Bigl( \frac{1}{\Delta t - i\epsilon - |\Delta \mathbf{x}|} - \frac{1}{\Delta t - i\epsilon + |\Delta \mathbf{x}|} \Bigl) \nonumber \\
 &= \frac{1}{4\pi^2} \frac{1}{ (\Delta t - i\epsilon)^2 - |\Delta \mathbf{x}|^2}.
\end{align}

In the case of an inertial detector \eqref{inertialdetector}, we have

\begin{align}
\frac{1}{ (\Delta t - i\epsilon)^2 - |\Delta \mathbf{x}|^2} = \frac{1}{ (\gamma_v\Delta \tau - i\epsilon)^2 - (\gamma_v v\Delta \tau)^2} = \frac{1}{ \Delta \tau^2 - 2i \Delta \tau \gamma_v\epsilon +\mathcal{O}(\epsilon^2)}. \nonumber
\end{align}

We then absorb the positive factor $\gamma$ into $\epsilon$ and ignore any higher order ($\mathcal{O}(\epsilon^2)$) corrections to write

\begin{align}
G^+_\epsilon(x,x') = \frac{1}{4\pi^2 (\Delta \tau - i \epsilon)^2}. \label{invaccor}
\end{align}

Substituting this in the integral \eqref{responserate}, we can easily compute it as a contour integral, invoking Cauchy theorem. For $E>0$, we should close the integration contour at the lower half of the complex plane. Then, since the only pole of the integrand lies in the upper plane, at $\Delta \tau = +i\epsilon$, we have that the response rate of the detector is null. Transporting this result to \eqref{probtotal}, with $E\!-\!E_0>0$, obtain a null detection probability, in perfect accordance with our previous result \eqref{ampinm}.

However, even in the simple situation of stationary response in Minkowski spacetime, we can still find nontrivial examples of particle detection. A case of particular interest is a uniformly accelerated detector, with constant proper acceleration $a=\alpha^{-1}$. Such a detector describes a hyperbolic trajectory in spacetime, which may be conveniently described by \textit{inertial coordinates} in the $xt$-plane as

\begin{subequations} \label{hipercoord}
\begin{empheq}[left=\empheqlbrace, right = \quad .]{align}
x(\tau) &= \alpha \cosh(\tau/\alpha) \\
t(\tau) &= \alpha \sinh(\tau/\alpha)
\end{empheq}
\end{subequations}

By substituting \eqref{hipercoord} in \eqref{m=0 correlation}, one finds (with some algebraic effort) that

\begin{align}
G_\epsilon^+\!(\Delta \tau) = \Bigl[ 16\pi^2 \alpha^2 \sinh^2 \Bigl( \frac{\Delta \tau - 2i\epsilon}{2\alpha} \Bigl) \Bigl]^{-1}, \label{UnruhGreen}
\end{align}
where we have once again absorbed a finite positive factor, $f(\tau,\tau')$, into $\epsilon$, given by

\begin{align}
f(\tau,\tau') \equiv \frac{\sinh(\tau/\alpha) - \sinh(\tau'/\alpha)}{\sinh((\tau-\tau')/\alpha)} > 0, \quad \forall \tau, \tau'.
\end{align}

Then, substituting (\ref{UnruhGreen}) in \eqref{responserate}, we can once again compute the integral through Cauchy Theorem, in a conveniently chosen contour (see Figure \ref{UnruhResidue}). Note that $G_\epsilon^+$ is periodic along the imaginary axis, and its poles lie regularly at $z=2i(\epsilon + n\pi\alpha) $. Then, by closing the contour rectangularly after one period, as illustrated in the figure, and denoting the \emph{regularized} integral in the real axis as $I_\epsilon$, we get:

\begin{figure}[H]
\centering
\includegraphics[width=0.8\linewidth]{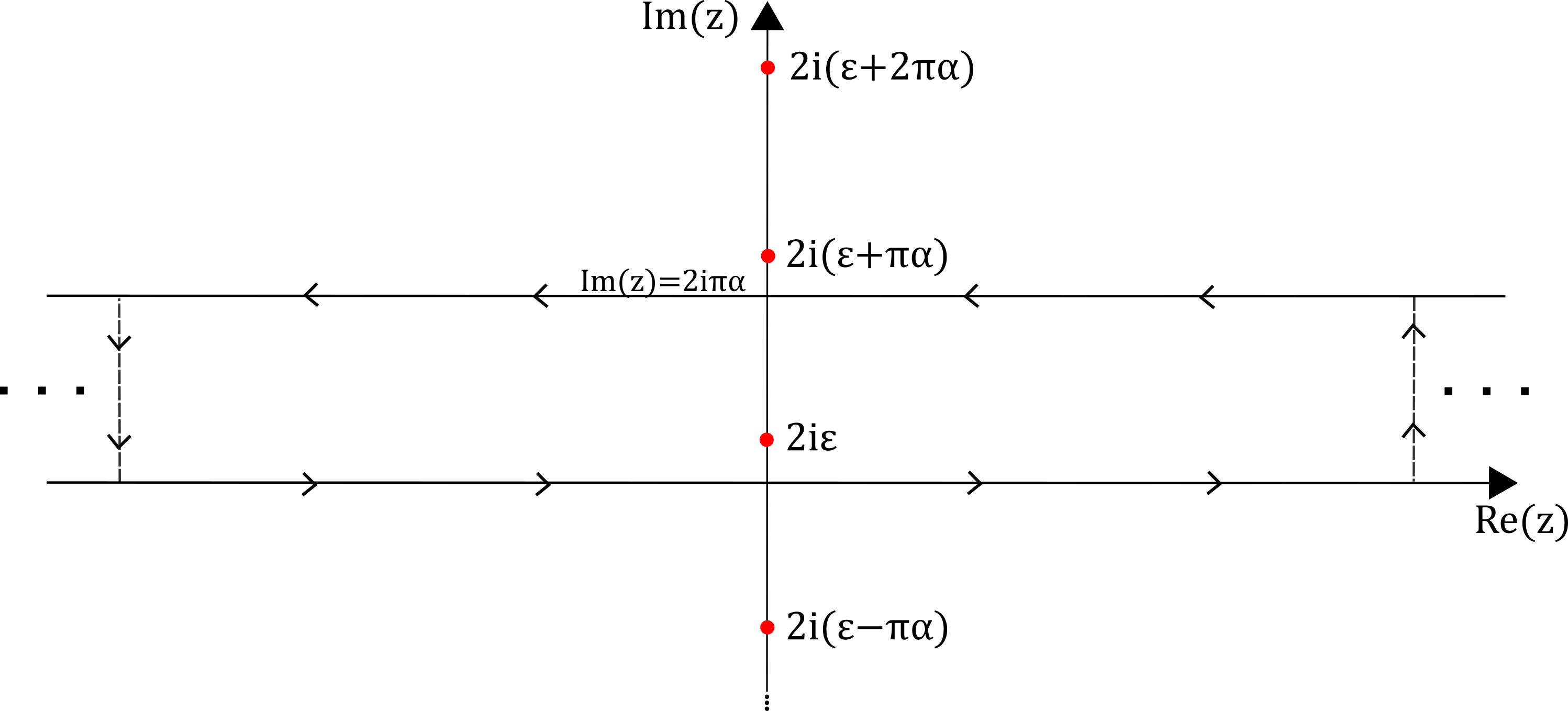}
\caption{Illustration of the contour integral used to calculate the reponse rate for the accelerated detector. The poles of the integrand lie along the imaginary axis, and are represented by red dots. The chosen contour is shown in thick black lines (the dashed lines that close the rectangle should be considered at infinity, where they will give no contribution to the integral), and it only encompasses the pole at $z=2i\epsilon$.  \\  Source: By the author. }
\label{UnruhResidue}
\end{figure}

\begin{align}
(1-e^{2\pi\alpha(E-E_0)})I_\epsilon = 2\pi i \underset{\; z= 2i \epsilon}{\operatorname{Res}}\bigl(e^{-iEz}G^+\!(z) \bigl).
\end{align}

This residue may be calculated for this order 2 pole by:

\begin{align}
\underset{\quad z=2i \epsilon}{\operatorname{Res}}(g) = - \frac{1}{4\pi^2} \lim_{z \rightarrow i\epsilon} \frac{d}{dz} \biggl[ \, \frac{ (\tfrac{z - 2i\epsilon}{2\alpha})^2 } {\sinh^{\!2}\!(\tfrac{z - 2i\epsilon}{2\alpha})} e^{-i(E-E_0)z} \biggl] =  \frac{i}{4\pi^2} (E-E_0)e^{2\epsilon(E-E_0)},
\end{align}
so that we finally obtain the transition probability rate per unit time, $P' = P/T$, in \eqref{probtotal}:

\begin{align}
P' = \lim_{\epsilon \rightarrow 0^+} c^2 \sum_E |\braket{E|m(0)|E_0}|^2 I_\epsilon = \frac{c^2}{2\pi} \sum_E \frac{ |\braket{E|m(0)|E_0}|^2}{e^{2\pi(E-E_0)\alpha}-1} . \label{P'}
\end{align}

Upon immediate inspection of this transition rate, we identify a Planck factor $({e^{2\pi(E-E_0)\alpha}-1})^{-1}$, showing that this detection rate corresponds to a thermal distribution of particles, with an effective temperature proportional to the proper acceleration of the detector: $T= (2\pi\alpha)^{-1} = a/2\pi$.

Note, moreover, that a transition will generally excite \emph{both the detector and the field}, being the final state $\ket{\Psi}$ of the latter generally a 1-particle state. At first sight, this seems very strange on energy grounds, since both the field and the detector will raise their energy (and, in this example, we remain in Minkowski spacetime, whose time translation symmetry should enforce energy conservation). How, then, can we reconcile these nonvanishing transition rates with energy conservation? As it turns out, we ought to attribute the injection of energy in our system to the agent that is imprinting acceleration in our detector: as the latter is coupled to $\phi$, it should indeed cause the emission of particles whenever it is under accelerated motion, imposing upon it a `breaking' force in the opposite direction of the acceleration (this is analogous to electromagnetic \textit{Brehmstrahlung}, where an accelerated charge emits radiation). Thus, the external agent that maintains an acceleration on the coupled (``charged'') detector must do work on it against this breaking force, providing energy for both the excitation of the detector and the emission of particles.

Now that we have seen a nontrivial application of particle detection in the vacuum, let us show next that our detector model indeed reproduces the expected results of particle detection for inertial observers in Minkowski space. These will \emph{exceptionally} bear a simple relation with the expected values $\braket{N_\mathbf{k}}$ for occupation numbers in plane-wave modes. We begin by analyzing the response rate \eqref{responserate} for general many-particle states $\ket{\Psi} = \ket{n_{\mathbf{k_1}}, n_{\mathbf{k_2}},...}$. In this case, we must substitute the Wightman function $G^+$ by

\begin{align}
G^+\!(x,x') \equiv \braket{0_M|\phi(x)\phi(x')|0_M} \longrightarrow & \braket{\Psi|\phi(x)\phi(x')|\Psi} \nonumber \\
 & = G^+\!(x,x') + \sum_\mathbf{k} n_\mathbf{k} \bigl( u_\mathbf{k}(x)u^*_\mathbf{k}\!(x') + u^*_\mathbf{k}\!(x)u_\mathbf{k}(x') \bigl),
\end{align}
or, taking the continuum limit,

\begin{align}
\braket{\Psi|\phi(x)\phi(x')|\Psi} = G^+\!(x,x') + \int \!\! d^3\mathbf{k} \, n(\mathbf{k}) u_\mathbf{k}(x)u^*_\mathbf{k}\!(x') + \int \!\! d^3\mathbf{k} \, n(\mathbf{k}) u^*_\mathbf{k}\!(x)u_\mathbf{k}(x').  \label{MPcorrelation}
\end{align}

This expression gives us three contributions for the response rate. However, we already know that the first term corresponds to the vacuum contribution, which yields a null response for an inertial observer. Let us then look at the contributions from the second and third terms of \eqref{MPcorrelation} in \eqref{responserate}:

\begin{subequations} \hspace{-4pt}
\begin{empheq}[left={\hspace{-4pt}\empheqlbrace}, right= .]{align}
 \frac{1}{(2\pi)^3} \int \frac{d^3\mathbf{k}}{2\omega} n(\mathbf{k}) \!\int\limits_{-\infty}^{\infty}\!\! d(\Delta \tau) e^{-i[E+\gamma_v(\omega-\mathbf{k} \!\cdot\! \mathbf{v})]\Delta\tau} = \frac{1}{(2\pi)^3} \int \frac{d^3\mathbf{k}}{2\omega} n(\mathbf{k}) \cancelto{0}{\delta\bigl(E+\gamma_v(\omega-\mathbf{k}\! \!\cdot\!\! \mathbf{v}) \bigl)} \label{term2} \\[6pt]
 \frac{1}{(2\pi)^3} \int \frac{d^3\mathbf{k}}{2\omega} n(\mathbf{k}) \!\int\limits_{-\infty}^{\infty}\!\! d(\Delta \tau) e^{-i[E-\gamma_v(\omega-\mathbf{k} \!\cdot\! \mathbf{v})] \Delta\tau} = \frac{1}{(2\pi)^3} \int \frac{d^3\mathbf{k}}{2\omega} n(\mathbf{k}) \delta\bigl(E-\gamma_v(\omega-\mathbf{k} \!\!\cdot\!\! \mathbf{v}) \bigl) \label{term3}
\end{empheq}
\end{subequations}

Note that the $\delta$ in \eqref{term2} has no roots for $E>0$. The one in \eqref{term3}, however, has roots in the domain of integration, and will yield a nonnull contribution to the response rate. We can already see from this expression that, generally, this contribution for an excitation $\Delta E$ in our detector will come precisely from particles that have energy equal to $\Delta E$ \emph{as seen in the detector reference frame}. Let us compute this response rate explicitly for an isotropic particle distribution (\textit{i.e.} $n(\mathbf{k}) = n(k)$) and a detector at rest in the isotropic frame ($\mathbf{v}=0$):

\begin{align}
\mathscr{F}'(E) &= \frac{1}{(2\pi)^3} \Bigl( \int \!\!d\Omega \Bigl)\int\limits_0^\infty \frac{dk}{2\omega}k^2n(k) \delta(E-\omega) \nonumber \\
 &= \frac{1}{4\pi^2} \int\limits_m^\infty d\omega \sqrt{\omega^2-m^2} \,\bar{n}(\omega) \delta(E-\omega) \nonumber \\
 &= \frac{1}{4\pi^2} \sqrt{E^2-m^2} \,\bar{n}(E) \Theta(E-m), \label{isotropicresponse}
\end{align}
where we have defined the energy particle distribution $\bar{n}(\omega)\equiv n(\sqrt{\omega^2-m^2}) = n(k)$.

The results in \eqref{isotropicresponse} are quite straightforward to interpret. The detection rates at an energy $E$ are proportional to the particle density at that energy (and a factor accounting for the spectral surface area $k^2$ divided by the $k$-dependent normalization), and the Heaviside $\Theta$ (re)assures that one can make detections only above the minimum threshold energy, given by $\omega = m$.

For an anysotropic observer ($\mathbf{v} \neq 0$), the response function does not look as simple due to a Doppler spreading, but the results are also easy to account for:

\begin{align}
\mathscr{F}(E) &= \frac{1}{(2\pi)^3} \Bigl( \int\limits_0^{2\pi} \!\!d\varphi \Bigl)\int\limits_0^\infty \frac{dk}{2\omega}k^2n(k) \int\limits_{-1}^{+1} d(\cos(\theta))  \cancelto{\hspace{-140pt}\text{Has roots if } \omega-kv < \frac{E}{\gamma_v} < \omega + kv}{\delta\bigl(E- \gamma_v(\omega-kv\cos(\theta) \bigl)} \nonumber \\
 &= \frac{1}{4\pi^2} \frac{1}{\gamma_v v} \int\limits_0^\infty \frac{dk}{2\omega}kn(k) \bigl[ \Theta(\omega-E_-) - \Theta(\omega-E_+) \bigl] \nonumber \\
 &= \frac{1}{4\pi^2} \biggl( \frac{1-v^2}{v^2} \biggl)^{1/2} \int_{E_-}^{E_+} d\omega \; \bar{n}(\omega),
\end{align}
where

\begin{align}
E_{\pm} = \frac{E \pm \sqrt{(E^2-m^2)v^2}}{\sqrt{1-v^2}}.
\end{align}

Again, we find that nonnull transition rates are only possible for $E>m$. This formula is particularly simple in the massless case, where all the energy of the particles comes from a kinetic term. In this case, the Doppler-shifted energies are just

\begin{align}
E_\pm = E\frac{1 \pm v}{(1-v^2)^{\frac{1}{2}}} = E \biggl[ \frac{1\pm v}{1 \mp v} \biggl]^{\frac{1}{2}}.
\end{align}

To wrap-up the discussion in this section: we have taken a closer look at the ambiguities in the concept of particles and, with the aid of simple models of particle detectors, we have seen how such ambiguities reflect in highly nontrivial observation relations for particles, even in (what should be arguably the most trivial and devoid of all states:) the Minkowski vacuum. Already through these examples, we may glimpse that \emph{there is in general no simple relation between the expected value $n_i \equiv \braket{\Psi|N_i|\Psi}$ and the number of particles measured by an actual detector, even for an inertial (free-falling) detector}\footnote{In some instances, as in the case of the accelerated observer in Minkowski, one can actually build appropriate accelerated modes, in terms of which the Minkowski vacuum is a thermal distribution of accelerated particles. This construction, however, relies on an entire family of accelerated observers covering (a wedge of) spacetime, whereas we are interested in the response of just individual localized detectors; this brings us back to the matter that \emph{modes are defined globally}, and, generally, they will not bear a simple relation to locally measured quantities.}.

However, we have shown that \emph{in the very particular case} of free-falling detectors in Minkowski spacetime, this simple relation does exist and the operational definition of particles constructed from detectors coincides with that given by the populations of normal (plane-wave) modes. In the next section, we will be interested in a less trivial context for which the normal-mode definition of particles is still useful -- namely, spacetimes with transient dynamics --, and will allow us to define the phenomenon of particle creation in dynamical spacetimes.

\section{Particle Creation in Asymptotically Flat Spacetimes} \label{particle creation}

In the last section, we have seen that only in very special cases we will find a simple correspondence between the idealized concept of particles as occupation numbers of normal modes, and the empirical one of particles as what particle detectors measure. This correspondence will be possible only when there is a high degree of symmetry in the spacetime under consideration, which picks out both special families of normal modes and special families of observers. Particularly, \emph{time translation symmetry} (and thus stationary spacetimes) plays a proeminent role, since it allows one to define positive-frequency modes (see eq. \ref{stationary}) and gives a very simple class of special observers, namely, \emph{stationary observers} (\textit{i.e.} the ones whose worldlines coincide with the orbits of the time translation Killing field $\xi^\mu$).

Now, in order to analyze some effects in dynamical (nonstationary) spacetimes but still keep things simple enough so that the mode particle definition is still useful to draw simple predictions, we turn our attention to the slightly more general case of spacetimes which go through a dynamical period, but that are asymptoptically stationary in the remote past and future; we denote such asymptotic regions $\Omega_{p}$ and $\Omega_{f}$, respectively. In these spacetimes, both regions will have special sets of normal modes associated with them (which we denote $\{u_j^{(p)}\}$ and $\{u_j^{(f)}\}$, respectively) whose asymptotic behaviour will be of the form\footnote{
 From this point onward, we shall always denote the set of spatial coordinates $\{x^j\}_{j=1,2,3}$ by boldface letters, as we do for ordinary spatial vectors in $\mathbb{R}^3$, even though we are not necessarily considering spatially flat Cauchy surfaces. We do so to compactly distinguish it from 4-dimensional spacetime coordinates/events, which we shall denote just as $x=(t, \mathbf{x})$.}:

\begin{subequations} \label{uin_uout}
  \begin{empheq}[left= \empheqlbrace, right = {\qquad ,}]{align}
  u_i^{(p)}(x) &\simeq \frac{e^{-i \omega_i t}}{\sqrt{2\omega_i}} \psi_i^{(p)}(\mathbf{x}), \qquad x \in \Omega_{p} \label{uin} \\
  u_j^{(f)}(x) &\simeq \frac{e^{-i \omega_j t}}{\sqrt{2\omega_j}} \psi_j^{(f)}(\mathbf{x}) , \qquad x \in \Omega_{f} \label{uout}
\end{empheq}
\end{subequations}
\textit{i.e.} they approximate positive-frequency modes in their respective asymptotic regions. We stress that both (sets of) modes are defined \emph{in the entire spacetime}, as they are \emph{exact} solutions to the field equations everywhere. However their form outside of their respective asymptotic regions is generally quite complicated and will depend heavily on the spacetime evolution.

As discussed before, we may write field expansions in both mode sets:

\begin{align}
\phi(x) = \sum_i a_i^{(p)} u_i^{(p)}\!(x) + a_i^{\dagger(p)} u_i^{*(p)}\!(x) = \sum_j a_j^{(f)} u_j^{(f)}\!(x) + a_j^{\dagger(f)} u_j^{*(f)}\!(x),
\end{align}
and define number operators for each of them, $N_i^{(p)} \equiv a_i^{\dagger(p)}a_i^{(p)}$ and $N_j^{(f)} \equiv a_j^{\dagger(f)}a_j^{(f)}$, as well as their respective vacuum states $\ket{0_p}$ and $\ket{0_f}$.

It will be of special interest to us when the regions $\Omega_{p}$ and $\Omega_{f}$ are asymptotically \emph{Minkowskian}, in which case the modes $\psi^{(p)}$ and $\psi^{(f)}$ will be just ordinary plane waves, which are particularly simple to operate with.

Since in this simple case one can ascribe a very clear physical meaning to the expected values $\braket{\Psi| N_i^{(p)} | \Psi}$ and $\braket{\Psi| N_j^{(f)} | \Psi}$ in terms of particles measured by inertial detectors in either the far past or future, one can then refer to the phenomenon of particle creation (or annihilation) between these two regions by means of simple Bogolubov transformations. For example, if we consider our field to be in the vacuum state $\ket{0_p}$, inertial observers in the far past ($x\in\Omega_p$) would indeed measure no particles $\braket{ 0_p| N_i^{(p)} | 0_p} = 0, \; \forall i $. However, \emph{after the time evolution through the dynamical region}\footnote{Recall that this evolution leaves the state $\ket{0_p}$ unchanged in the Heinsenberg picture.}, inertial observers will generally measure a nontrivial particle content at late times ($x\in\Omega_f$), given by eq (\ref{bogolubovparticles}):

\begin{align}
\braket{0_p| N_j^{(f)} | 0_p } = \sum_i |\beta_{ji}|^2,
\end{align}
being $\beta_{ji} = -(u_j^{(p)},u_i^{*(f)}) $ the usual $\beta$ Bogolubov coefficients between past and future modes.

Of course, one could in principle also have a symmetrical situation of particle annihilation, starting from the \emph{final} vacuum state $\ket{0_f}$, for which one would measure particles in the past,

\begin{align}
\braket{0_f| N_j^{(p)} | 0_f } = \sum_i |\beta_{ij}|^2,
\end{align}
but none in the future. However, although this situation is perfectly compatible with an idealized time evolution of a pure state, \emph{it corresponds to a diminishing in enytropy}. As we have commented in section \ref{curved quantization} (and we shall show more explicitly for FLRW spaces in this section), particles are always created in correlated pairs. This means that a state that evolves from many particles into a vacuum would correspond to an initial state of highly correlated particles, that are perfectly adjusted to be annihilated in pairs (this would be analogous, for example, to postulating an extremely fine-tuned choice of initial conditions for molecules of gas in a box, allowing one to evolve from a state in which the gas is filling the entire box to one in which it spontaneouly concentrates in a fraction of its volume).

So far, the analysis seems quite simple. In practice, however, it is generally quite complicated to actually solve the field equations \emph{exactly} in such generic spacetimes and properly combine a basis of \emph{exact} solutions to obtain the modes whose asymptotic behaviour is that of plane waves in either remote region, as well as to further calculate the Bogolubov coefficients $\alpha_{ij}$ and $\beta_{ij}$ to \emph{every} pair of modes $u_i^{(p)}$ and $u_j^{(f)}$. Notwithstanding, there is a particular class of spacetimes for which these calculations are greatly simplified: spatially homogeneous and isotropic universes. Throughout this section, we shall explore them as a tractable case of study, and analyze the phenomenon of particle creation in more detail.

\subsection{Particle creation in FLRW spacetimes}

A very distinguished class of spacetimes, which is of special interest in the context of cosmology, are the ones which possess maximally symmetrical space sections, \textit{i.e.} which are spatially homogeneous and isotropic (but which may still have a nontrivial time evolution). For historical reasons, they are also known as \textit{Friedmann-Lemaitre-Robertson-Walker} (FLRW) spaces or universes (for a more complete account of the development and properties of FLRW spaces, as well as their use in cosmology, see section \ref{FLRW}). All spacetimes in this class may be described by a metric of the form

\begin{align}
ds^2 = dt^2 - a^2(t) d\Sigma^2,
\end{align}
where $t$ represents the proper time of observers whose worldlines are orthogonal to the isotropic space sections $\Sigma_t$ (which foliate the entire spacetime). Such observers, commonly called \textit{comoving observers}, for reasons to be made apparent, comprise a special family in FLRW spaces, as they are the ones who will perceive space (\textit{i.e.} their spatial sections $\Sigma_t$) as homogeneous and isotropic. $d\Sigma^2$ represents a static spatial metric (common to all surfaces $\Sigma_t$) and $a(t)$ is called the \emph{scale factor}; it dictates how spatial distances expand or shrink with time (\textit{e.g.} $a(t')/a(t)$ gives the ratio of the distances between 2 comoving observers measured along the surfaces $\Sigma_{t'}$ and $\Sigma_t$). Particularly, for a FLRW spacetime that is asymptotically static, we must have that:

\begin{subequations}
 \begin{empheq}[left=\empheqlbrace]{align}
 a(t) &\rightarrow a_1, \qquad t \rightarrow -\infty \\
 a(t) &\rightarrow a_2, \qquad t \rightarrow +\infty
 \end{empheq}
\end{subequations}
(being $a_1$ and $a_2$ constants).

What makes these spacetimes special in the context of particle creation is that they bear separable field equations \textit{at all times}, so that one may always find a complete set of field solutions of the form:

\begin{align}
u_i(x) = \chi_i(t)\psi_i(\mathbf{x}).
\end{align}

Presently, we shall not go into detail for the dynamical equations (these will be further developed for a conformal time coordinate in section \ref{adiabatic}, and in proper-time in section \ref{adiabatic subtraction} ). We just note here that, by defining $h_i(t) \equiv a^{-\frac{3}{2}}(t)\chi(t) $ they will take the general form:

\begin{subequations}
  \begin{empheq}[left=\empheqlbrace, right = {\quad ,}]{align}
  \frac{d^2}{dt^2}h_i(t) &= -\Omega_i^2(t)h_i(t) \\
  H^2_x \psi_i(\mathbf{x}) &= \Omega^2_i(t)\psi_i(\mathbf{x})
  \end{empheq}
\end{subequations}
where $H^2_x = \bigl[ -a^{\!-2}(t)\nabla^2_{\!\mathbf{x} } + m^2 \bigl]$ (being $\nabla^2_{\!\mathbf{x}}$ the Laplacian operator corresponding to the metric $d\Sigma^2$), and $\Omega_i(t)$ a \emph{time-dependent frequency}. Since we are particularly interested in asymptotically Minkowskian spaces, we shall restrict ourselves to the case of spatially flat homonegeous surfaces $\Sigma_t = \mathbb{R}^3$. In this case, $\nabla^2_{\!\mathbf{x}}$ is an ordinary 3D Laplacian and we have simple exponential solutions, labeled by a wave-vector $\mathbf{k}$:

\begin{align}
\psi_\mathbf{k}(\mathbf{x}) = \frac{1}{\sqrt{V}} e^{i \mathbf{k} \!\cdot\! \mathbf{x}} \qquad \text{and} \qquad \psi^*_\mathbf{k}(\mathbf{x}) = \frac{1}{\sqrt{V}} e^{-i \mathbf{k} \!\cdot\! \mathbf{x}}, \label{spacePW}
\end{align}
where we have $\psi^*_\mathbf{k} \!=\! \psi_{-\mathbf{k}}$ and $\nabla^2_{\!\mathbf{x}}\psi_{\pm\mathbf{k}}(\mathbf{x}) \!=\! \mathbf{k}^2\psi_{\pm\mathbf{k}}(\mathbf{x})$. To each pair of spatial solutions with wave vector $\pm\mathbf{k}$ corresponds a pair of linearly independent temporal solutions $h_k(t)$, whose quadratic frequencies $\Omega^2_k(t)$ are given by (see section \ref{adiabatic subtraction}):

\begin{align}
\Omega^2_k(t) = \omega^2_k(t) + \sigma(t), \!\!\!& \\[9pt]
\omega_k(t) = \sqrt{\frac{k^2}{a^2(t)} + m^2}, \qquad \qquad \quad \sigma(t) = &\Bigl(6\xi- \frac{3}{4}\Bigl) \frac{\dot{a}^2}{a^2} + \Bigl(6\xi- \frac{3}{2}\Bigl)\frac{\ddot{a}}{a}. \nonumber
\end{align}

Then, \textit{in the asymptotic regions}, these pairs of \emph{exact solutions} can be decomposed in positive and negative frequency solutions $\{h_k^{(p)},h_k^{*(p)}\}$ and $\{h_k^{(f)},h_k^{*(f)}\}$, such that

\begin{subequations} \label{asymtotic FLWR frequencies}
  \begin{empheq}[left= { \hspace{-9pt} \empheqlbrace}]{align}
  i\frac{d}{dt}h_k^{(p)}(t) &\simeq \omega_{k1}h_k^{(p)}(t), \qquad i\frac{d}{dt}h_k^{*(p)}(t) \simeq -\omega_{k1}h_k^{*(p)}(t), \qquad x \in \Omega_p \; (t\rightarrow -\infty) \\[13pt]
  i\frac{d}{dt}h_k^{(f)}(t) &\simeq \omega_{k2}h_k^{(f)}(t), \qquad i\frac{d}{dt}h_k^{*(f)}(t) \simeq -\omega_{k2}h_k^{*(f)}(t), \qquad x \in \Omega_f \; (t\rightarrow +\infty)
\end{empheq}
\end{subequations}
where we have defined the past and future asymptotic frequencies:

\begin{align}
\Omega_{k(1,2)} = \omega_{k(1,2)} \equiv \sqrt{\frac{k^2}{a^2_{(1,2)}} + m^2}.
\end{align}

Here, we stress once again that \emph{solutions belonging to different $\{\mathbf{k},-\mathbf{k}\}$ pairs remain orthogonal at all times}. Thus, to evaluate particle creation, we just have to consider (2x2) block diagonal Bogolubov transformations among the pairs $\{u^{(p)}_\mathbf{k},u^{*(p)}_\mathbf{k}\}$ and $\{u^{(f)}_\mathbf{k},u^{*(f)}_\mathbf{k}\}$, from which we find

\begin{subequations} \label{diagonalBogolubov}
  \begin{empheq}[left= \empheqlbrace]{align}
\alpha_{\mathbf{k}\mathbf{k'}} = \alpha_k \delta_{\mathbf{k},\mathbf{k'}}, \\
\beta_{\mathbf{k}\mathbf{k'}} = \beta_k \delta_{\mathbf{k},-\mathbf{k'}},
  \end{empheq}
\end{subequations}
where (i) the coefficients $\alpha_k$ and $\beta_k$ only depend on the magnitude of $\mathbf{k}$ as a consequence of spatial isotropy and (ii) we have enforced that the $\alpha$ coefficients must be strictly diagonal $(\mathbf{k},\mathbf{k})$, whereas the $\beta$ ones must be crossed $(\mathbf{k},-\mathbf{k})$, since the spatial ($\mathbf{x}$) dependence for $u_\mathbf{k}$ ($u^*_\mathbf{k}$) is given by $\psi_\mathbf{k}$ ($\psi^*_\mathbf{k}$) \emph{at all times}. More explicitly:

\begin{align}
u_\mathbf{k}^{(f)}(x) = h^{(f)}_k(t)\psi_\mathbf{k}(\mathbf{x})  &= \alpha_k u_\mathbf{k}^{(p)}(x) + \beta_k u^{*(p)}_{-\mathbf{k}}(x) \nonumber \\
 &= \bigl( \alpha_k h^{(p)}_k(t) + \beta_k h^{*(p)}_k(t) \bigl) \psi_\mathbf{k}(\mathbf{x})
\end{align}
(had we had contributions from $u^{(f)}_{-\mathbf{k}}$ or $u^{*(f)}_{\mathbf{k}}$, we would end up with terms proportional to $\psi^*_\mathbf{k} = \psi_{-\mathbf{k}}$, and the equality with the LHS could not match).

In the special case \eqref{diagonalBogolubov}, there are great simplifications in the relations between both modes and results for particle creation. For example, the expected value \eqref{bogolubovParticlesEV} for the total number of particles measured in the asymptotic future, starting from a vacuum state in the past, will be just

\begin{align}
\braket{0_p| N^{(f)} |0_p} = \sum_\mathbf{k} |\beta_k|^2. \label{FLRW bogolubov particles}
\end{align}

Further, the consistency condition \eqref{bogolubov conditions} for the Bogolubov coefficients greatly simplify to

\begin{align}
|\alpha_k|^2 - |\beta_k|^2 = 1. \label{a-b}
\end{align}

As in the general case, these will be compatible with commutation relations, while anticommutation relations would yield \eqref{bogolubov anticonditions}:

\begin{align}
|\alpha_k|^2 + |\beta_k|^2 = 1 \label{a+b}
\end{align}
which are only compatible with \eqref{a-b} when \emph{all} $\beta_k$'s are null, that is, when there are no created particles whatsoever. Thus, in this particular context, where the Bogolubov coefficients can be interpreted dynamically in terms of particle creation, one could argue (as in \cite{parker}) \emph{that the scalar (spin 0) field statistics must be bosonic in curved spacetimes by virtue of its dynamics}. We stress that, generally, a bosonic statistic is enforced \emph{as a consistency condition} (so that one may perform the quantization on equal footing for any orthonormal mode expansion), whether or not one may interpret it dynamically. Nonetheless, it is interesting that in some special contexts, one can make such dynamical interpretation of the spin-statistics relation.

Finally, we note that the mode operators can be written in terms of one another as (eqs \ref{operators}):

\begin{subequations} \label{FLRW operator}
\begin{align}
a^{(f)}_\mathbf{k} &= \alpha_k a^{(p)}_\mathbf{k} + \beta^*_k \,a^{\!\dagger(p)}_{-\mathbf{k}}, \\
a^{(p)}_\mathbf{k} &= \alpha_k^* a^{(f)}_\mathbf{k} - \beta^*_k \,a^{\!\dagger(f)}_{-\mathbf{k}} .
\end{align}
\end{subequations}

Then, since these transformations are `quasidiagonal', they are extremely simpler to invert than in the general case. These will allow us to compute vacuum to many-particle state projections with considerable ease, which we shall use to analyze the statistics and correlations for created particles in the next subsection.

\subsection{Correlations and statistics of created particles} \label{particle statistics}

We have seen that asymptotically flat FLRW spaces make a very convenient stage to analyze particle creation, and so far we have found that one can find the total expectation values for particles in the asymptotic future by \eqref{FLRW bogolubov particles} (or for particles of each type, $\braket{N_\mathbf{k}^{(f)}}$, by withholding the sum and just looking at a particular $\mathbf{k}$ value). However, these expectation values alone do not tell us all about the statistics of the created particles; they just convey information about its \emph{averages}. Indeed, it is easy to see that, for instance, the states $\ket{\Psi_1} = \ket{1_\mathbf{k}}$ and $\ket{\Psi_2} = \frac{1}{\sqrt{2}}(\ket{0} + \ket{2_\mathbf{k}})$ both yield the same expected values:

\begin{align}
\braket{\Psi_1 | N_\mathbf{k'} | \Psi_1} = \braket{\Psi_2 | N_\mathbf{k'} | \Psi_2} = 1 \times \delta_{\mathbf{k}, \mathbf{k'}},
\end{align}
even though $\braket{\Psi_1|\Psi_2} = 0$. To obtain a more detailed statistical information of the created particles, we must analyze general transition amplitudes of the form $\braket{n^{(f)}_\mathbf{k_1},n^{(f)}_\mathbf{k_2},...|0_p}$. Before we analyze these in full generality, it is constructive that we look at more simple particle states. The fact that the relation \eqref{FLRW operator} mixes $\mathbf{k}$ and $-\mathbf{k}$ modes is suggestive that it will be useful to start with amplitudes of the form

\begin{align}
\mathcal{A}_n(\mathbf{k}) \equiv \braket{n^{(f)}_\mathbf{k},n^{(f)}_{-\mathbf{k}},|0_p}, \label{Tramp-Pairs}
\end{align}
\textit{i.e.} the probability amplitude that $n$ pairs of particles were created in the modes $u^{(f)}_\mathbf{k}$ and $u^{(f)}_{-\mathbf{k}}$ (and no others). We have that

\begin{align}
\mathcal{A}_n(\mathbf{k}) &= \frac{1}{n!} \braket{0_f | (a^{(f)}_\mathbf{k})^n (a^{(f)}_{-\mathbf{k}})^n |0_p} \nonumber \\[4pt]
  &= \frac{1}{n!} \braket{0_f | (a^{(f)}_\mathbf{k})^n (\alpha_k a^{(p)}_{-\mathbf{k}} + \beta^*_k a^{\!\dagger(p)}_{\mathbf{k}} )^n |0_p} \nonumber \\
  &= \frac{1}{n!} (\beta^*_k)^n \braket{0_f | (a^{(f)}_\mathbf{k})^n ( a^{\!\dagger(p)}_{\mathbf{k}} )^n |0_p} \nonumber \\[4pt]
  &= \frac{1}{n!} (\beta^*_k)^n \braket{0_f \bigl| (a^{(f)}_\mathbf{k})^n \Bigl( \frac{ a^{\!\dagger(f)}_{\mathbf{k}} - \beta_k a^{(p)}_{-\mathbf{k}} }{\alpha_k^*} \Bigl)^{\!\!n} \bigl|0_p} \nonumber \\[4pt]
  &= \frac{1}{n!} \left( \frac{\beta^*_k}{\alpha_k^*} \right)^{\!\!n} \braket{0_f \bigl| (a^{(f)}_\mathbf{k})^n (a^{\!\dagger(f)}_{\mathbf{k}} )^{n} \bigl|0_p} \nonumber \\[4pt]
  &= \frac{1}{(n-1)!} \left( \frac{\beta^*_k}{\alpha_k^*} \right)^{\!\!n} \braket{0_f \bigl| (a^{(f)}_\mathbf{k})^{n-1} (a^{\!\dagger(f)}_{\mathbf{k}} )^{n-1} \bigl|0_p} \nonumber \\[4pt]
  & \qquad \;\; \scalebox{1.3}{$\vdots$} \nonumber \\[4pt]
  &= \left( \frac{\beta^*_k}{\alpha_k^*} \right)^{\!\!n} \braket{0_f|0_p},
\end{align}
where, in the last lines, we have recursively applied the commutation relations $[(a_\mathbf{k})^n,a^\dagger_\mathbf{k}] = n(a_\mathbf{k})^{n-1} $. From these same lines it is also easy to see that, for $m\neq n$:

\begin{align}
\braket{m^{(f)}_\mathbf{k},n^{(f)}_{-\mathbf{k}},|0_p} = 0.
\end{align}

Therefore, we conclude that particles are always produced in pairs with the same energy and opposite momenta. Indeed, this is to be expected in FLRW spacetimes, since spatial homogeneity implies the conservation of 3-momentum. Note, however, that we have deduced a stronger restriction, since conservation of momentum alone could still allow for created particles in sets like $\ket{1_\mathbf{k}, 2_{-\mathbf{k}/2}}$, $\ket{1_\mathbf{k}, 3_{-\mathbf{k}/3}}$ and other similar combinations. The restriction we have just deduced means that \emph{the only states in the Fock Space built on $\ket{0_f}$ that are not orthogonal to $\ket{0_p}$ are those built with pairs of particles in the modes $u^{(f)}_\mathbf{k}$ and $u^{(f)}_{-\mathbf{k}}$}. For brevity, we drop the superscripts $(f)$ of the future modes and denote these states as

\begin{align}
\ket{ \{ {n_j}(\mathbf{k}_j) \} } = \ket{ \prescript{1}{}{n}_\mathbf{k_1}, \prescript{1}{}{n}_{-\mathbf{k_1}}; \prescript{2}{}{n}_\mathbf{k_2}, \prescript{2}{}{n}_{-\mathbf{k_2}};...}.
\end{align}

We can then write a completeness relation for $\ket{0_p}$:

\begin{align}
\ket{0_p} = \sum_{ \{ {n_j}(\mathbf{k}_j) \} } \ket{ \{ {n_j}(\mathbf{k}_j) \} } \! \braket{ \{ {n_j}(\mathbf{k}_j) \} |0_p } = \sum_{ \{ {n_j}(\mathbf{k}_j) \} } \ket{ \{ {n_j}(\mathbf{k}_j) \} } 
\biggl[ \prod_j \Bigl( \frac{\beta^*_k}{\alpha_k^*} \Bigl)^{\!\!n_j} \biggl] \braket{0_f|0_p}  .
\end{align}

From this equation, we can compute the norm of the vacuum to vacuum transition $|\braket{0_f|0_p}|$ using the normalization condition

\begin{align}
1 = |\braket{0_p|0_p}|^2 = \Biggl( \sum_{ \{ {n_j}(\mathbf{k}_j) \} }  
\prod_j \Bigl| \frac{\beta_k}{\alpha_k} \Bigl|^{2n_j}  \Biggl)|\braket{0_f|0_p}|^2 .
\end{align}

Assuming all summations and products converge appropriately, we may commute them, by noting that

\begin{align}
\sum_{ \{ n_j \} }  \prod_j x_j^{n_j} &= \sum_{n_1=0}^\infty \sum_{n_2=0}^\infty \hdots  (x_1)^{n_1}(x_2)^{n_2} \hdots \nonumber \\[4pt]
 &= \Bigl( \sum_{n_1=0}^\infty x_1^{n_1} \Bigl) \Bigl( \sum_{n_2=0}^\infty x_2^{n_2} \Bigl) \hdots \nonumber \\[4pt]
 &= \prod_j \Bigl( \sum_{n_j=0}^\infty x_j^{n_j} \Bigl),
\end{align}
where these summations are just familiar geometric series. Thus, we have

\begin{align}
1 &= |\braket{0_f|0_p}|^2 \prod_j \biggl( \sum_{n_j=0}^\infty \Bigl| \frac{\beta_{k_j}}{\alpha_{k_j}} \Bigl|^{2n_j} \biggl) \nonumber \\
  &= |\braket{0_f|0_p}|^2 \prod_j \biggl[ 1 - \Bigl|\frac{\beta_{k_j}}{\alpha_{k_j}} \Bigl|^{2}  \biggl]^{-1} \nonumber \\
  &= |\braket{0_f|0_p}|^2 \prod_j |\alpha_{k_j}|^2, \label{hahahaha}
\end{align}
where we have used \eqref{a-b}. Eq \eqref{hahahaha} immediately entails

\begin{align}
|\braket{0_f|0_p}|^2 = \prod_j |\alpha_{k_j}|^{-2}.
\end{align}

Finally, we obtain the explicit transition probabilities:

\begin{align}
P\bigl( \{ {n_j}(\mathbf{k}_j) \} \bigl) \equiv \bigl| \braket{ \{ {n_j}(\mathbf{k}_j) \} |0_p } \bigl|^2 = \Biggl[ \prod_j |\alpha_{k_j}|^{-2} \biggl| \frac{\beta_{k_j}}{\alpha_{k_j}} \biggl|^{2n_j} \Biggl]. \label{FLRW TransitionAmplitudes}
\end{align}

Using this expression, it is particularly interesting to note the marginal probabilities that emerge for the creation of $n$ pairs of just one type. If we fix only one of the $n_j$ (setting $n_j=n$) in \eqref{FLRW TransitionAmplitudes}, and sum over all the possibilities for the remaining modes (with $j' \neq j$), we obtain the probability that $n$ pairs will be created in mode $\mathbf{k}$:

\begin{align}
P_n(\mathbf{k}) \equiv \bigl| \mathcal{A}_n(\mathbf{k}) \bigl|^2 = |\alpha_{k}|^{-2} \biggl| \frac{\beta_{k}}{\alpha_{k}} \biggl|^{2n}. \label{n-Single_Type}
\end{align}

We could also compute a marginal probability for creating $n_1$ particles in a mode $\mathbf{k_1}$ \textit{and} $n_2$ particles in a mode $\mathbf{k_2}$. From eq. \eqref{FLRW TransitionAmplitudes}, we see immediately that pair production for distinct modes ($j \neq j'$) are independent events, since their joint probability is just the product of the individual marginal probabilities:

\begin{align}
P\bigl({n_1}(\mathbf{k}_1), {n_2}(\mathbf{k}_2) \bigl) = P_{n_1}(\mathbf{k}_1) P_{n_2}(\mathbf{k}_2).
\end{align}

However, \emph{the production of multiple pairs in the same mode $\mathbf{k}$ are not independent events}. One can see directly from \eqref{n-Single_Type} that

\begin{align}
P_2(\mathbf{k}) = \frac{P_1^2(\mathbf{k})}{P_0(\mathbf{k})} \geq P^2_1(\mathbf{k}),
\end{align}
where the equality will only occur for $\beta_k = 0$, when particle creation in mode $\mathbf{k}$ is trivial ($P_0(\mathbf{k})=1$, $P_{n\geq1}(\mathbf{k})=0$ ). More generally, we have

\begin{align}
P_n(\mathbf{k}) \frac{P^n_1(\mathbf{k})}{P^{n-1}_0(\mathbf{k})} = \left( \frac{P_1(\mathbf{k})}{P_0(\mathbf{k})} \right)^{\!\!n}P_0(\mathbf{k}) \geq P_1^n(\mathbf{k}) .
\end{align}

Thus, the probability of creating $n$ pairs in the same mode $\mathbf{k}$ is generally greater than the probability of creating all of these pairs independently. This is analogous to the phenomena of spontaneous and stimulated emission (\emph{e.g.} for atoms interacting with radiation), where the probability of emitting one more photon increases as there are more photons present.

From eq. \eqref{n-Single_Type} it is easy to recover the known \emph{average} results for particle creation. In fact, it is not hard to compute any statistical moments; in zeroth order, we reobtain the normalization of probability:

\begin{align}
\sum_{n=0}^{\infty} P_n(\mathbf{k}) = 1.
\end{align}

Here one must just sum geometric series, as in \eqref{hahahaha}. Then, the first order moment recovers the average/expected value:

\begin{align}
\braket{ 0_p | N_\mathbf{k} | 0_p } &= \sum_{n=0}^\infty \braket{0_p|n_\mathbf{k},n_{-\mathbf{k}} } \braket{n_\mathbf{k},n_{-\mathbf{k}} | a_\mathbf{k}^\dagger a_\mathbf{k} |   n_\mathbf{k},n_{-\mathbf{k}} } \braket{ n_\mathbf{k},n_{-\mathbf{k}} |0_p} \nonumber \\
  &= \sum_{n=0}^\infty \braket{0_p|0_f } \left( \frac{\beta_k}{\alpha_k} \right)^{\!\!n} n \left( \frac{\beta^*_k}{\alpha^*_k} \right)^{\!\!n}   \braket{ 0_f |0_p} \nonumber \\
  &= \sum_{n=0}^\infty n P_n(\mathbf{k}) \nonumber \\
  &= |\beta_k|^2,
\end{align}
where we already know the last equality to be true from the Bogolubov transformations \eqref{FLRW bogolubov particles}. Still, it is not difficult to compute it directly through the summation $\sum_n n P_n$ by employing a little trick of taking partial derivatives with respect to $\beta_k$:

\begin{align}
\sum_{n=0}^\infty n P_n(\mathbf{k}) &= \sum_{n=0}^\infty n \biggl| \frac{\beta_{k}}{\alpha_{k}} \biggl|^{2n} |\alpha_{k}|^{-2} \nonumber \\
 &= |\beta_k|^2 \frac{\partial}{\partial |\beta_k|^2} \sum_{n=0}^\infty \biggl| \frac{\beta_{k}}{\alpha_{k}} \biggl|^{2n} |\alpha_{k}|^{-2} \nonumber \\
 &= \biggl|\frac{\beta_k}{\alpha_k}\biggl|^2 \frac{\partial}{\partial |\beta_k|^2} \Bigl( 1 - \Bigl|\frac{\beta_k}{\alpha_k}\Bigl|^2 \Bigl)^{\!\!-1} \nonumber \\
 &= \biggl|\frac{\beta_k}{\alpha_k}\biggl|^2 |\alpha_k|^{-2} \Bigl( 1 - \Bigl|\frac{\beta_k}{\alpha_k}\Bigl|^2 \Bigl)^{\!\!-2} \nonumber \\
 &= |\beta_k|^2. \label{deldelbeta}
\end{align}
(In implementing this trick, however, one must be careful to \emph{only impose eq \eqref{a-b} after taking the derivatives with respect to $|\beta_k|$}, as treated $\alpha_k$ and $\beta_k$ as independent variables to write the second equality.)

Then, if one wishes, it is possible to carry analogous calculations for higher statistical moments (such as the variance).

With the above results, we can recover the expected value for the total particle density in the asymptotic future, due created particles. Particularly, taking the continuum limit, we obtain

\begin{align}
\braket{0_p|N|0_p} &= \lim_{L \rightarrow \infty} \frac{1}{L a_2^3} \sum_\mathbf{k} |\beta_k|^2 \nonumber \\
 &= \frac{1}{2\pi^2 a_2^3} \int\limits_0^\infty \! dk \, k^2 | \beta(k)|^2 \nonumber \\
 &= \frac{1}{2\pi^2} \int\limits_{0}^{\infty} dk' (k')^2 |\beta(a_2k')|^2,
\end{align}
where we have absorbed the scale factor $a_2$ in the definition of the physical momentum in the asymptotic future $k' \equiv k/a_2$.

\subsection{A simple model for particle creation}

Now that we have developed many features of particle creation in a model-independent way\footnote{That is, we have not assumed a particular metric. Even when we specialized to FLRW metrics, we have not assumed a specific form for $a(t)$.}, we would like to better grasp this phenomenon through a simple, tractable model, for which we can explicitly compute the Bogolubov coefficients. This shall serve both to illustrate the general (dynamic-independent) features presented so far, and to give a glimpse of how particle creation ultimately depends on the \textit{dynamics} of spacetime in its nonstationary phase, preparing the ground for how we may define suitable extensions of (approximate) concepts of vacuum and particles \emph{to fully dynamical spacetimes} (letting go the hyphothesis of asymptotic flatness). Here, we shall explore a simple model presented in section 3.4 of \cite{birrell}.

\begin{figure}[H]
\centering
\includegraphics[width=0.6\linewidth]{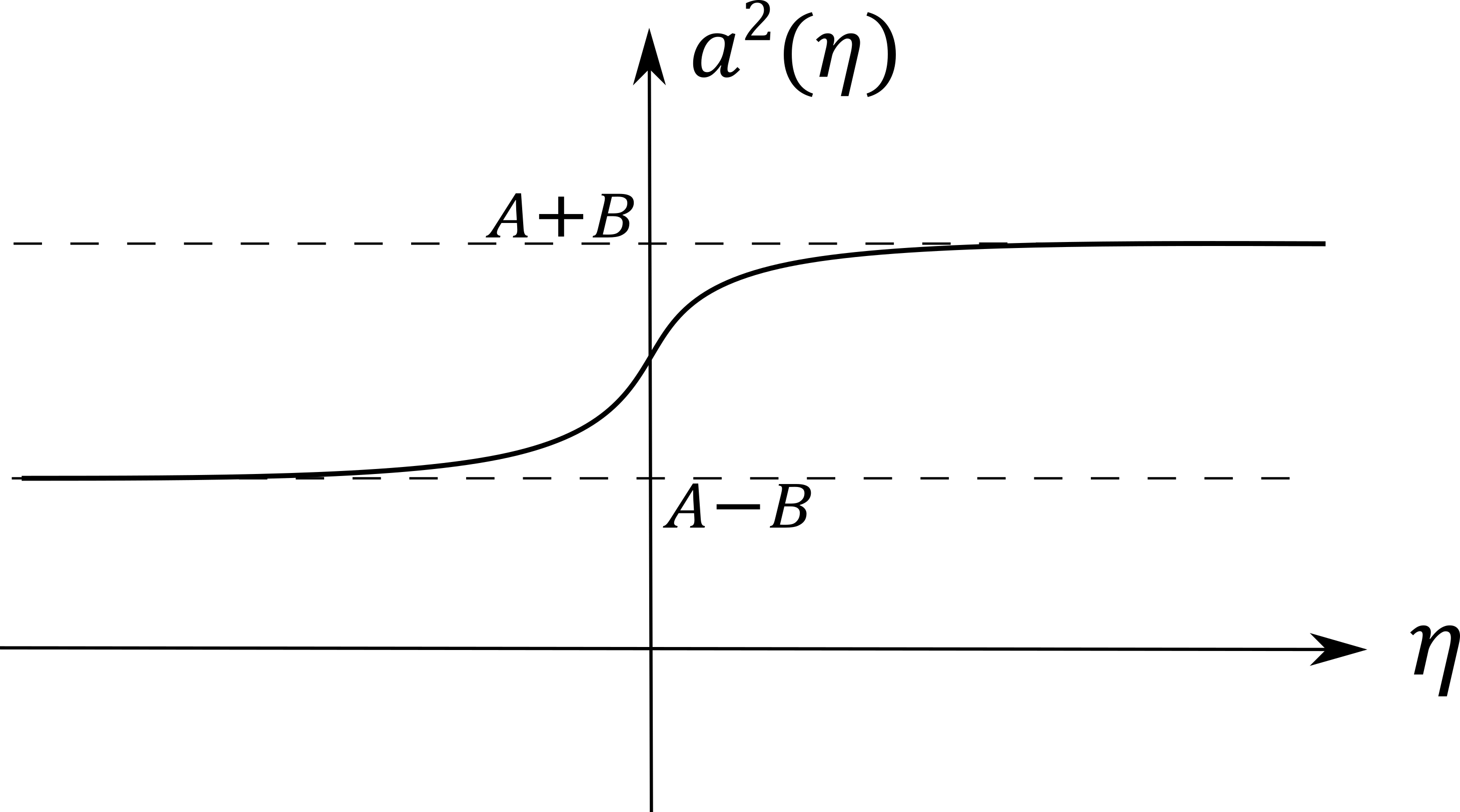}
\caption{ Scale factor for a simple model of expansion displayed as a function of conformal time in the asymptotic regions $\eta \rightarrow \pm \infty$ it becomes asymptotically flat, with $a \rightarrow (A \pm B)$. \\ Source: By the author. }
\label{TANHCONF}
\end{figure}

For simplicity, this model is built in $1+1$ spacetime dimensions in a FLRW metric. Here we make explicit use of the conformally flat form of the metric (see appendix \ref{geometry}), writing it in conformal coordinates $(\eta,x)$:

\begin{align}
ds^2 = dt^2 - a^2(t)dx^2 = a^2(\eta) \bigl[ d\eta^2 - dx^2 \bigl],
\end{align}
where $\eta$ is called the \emph{conformal time}, defined by: $\eta = \int^t \frac{dt'}{a(t')}$. We then define the scale factor as a function of $\eta$ to be (see Figure \ref{TANHCONF}) :

\begin{align}
a^2(\eta) = A + B \tanh(\rho \eta), \label{tanh}
\end{align}
where $A$, $B$ and $\rho$ are constant parameters. Note that $a^2(\eta) \rightarrow A \pm B$ as $\eta \rightarrow \pm \infty$.

We leave to the next section a more thorough discussion of the form and solutions to the massive field equations for a conformally flat spacetime. For the time being, we note that, analogously to when we employed proper-time coordinates, we shall obtain separable solutions, with simple exponential dependences in space, and time-dependent harmonic oscillators in time, whose frequencies $\omega_k(\eta)$ are given by \eqref{conformal frequencies}. This model then yields the asymptotic frequencies for each wave vector $\mathbf{k}$ (we omit the $k$ subscript for cleaness) in the far past and future:

\begin{subequations}
 \begin{empheq}[left= \empheqlbrace, right = {\qquad .}]{align}
 \omega_1 &\equiv \sqrt{k^2+m^2(A-B)} \\
 \omega_2 &\equiv \sqrt{k^2+m^2(A+B)}
 \end{empheq}
\end{subequations}

For later convenience, we also define the frequencies

\begin{align}
\omega_{\pm} \equiv \frac{1}{2}\bigl(\omega_2\pm\omega_1\bigl).
\end{align}
 
The exact field equations will be given by \eqref{TDHO} (with the scale factor \eqref{tanh}), for which it is possible to obtain the (normalized) exact mode solutions $u^{(p)}_{\mathbf{k}}$ and $u^{(f)}_{\mathbf{k}}$, which behave as positive frequencies in the asymptotic past and  the asymptotic future, respectively. They read (see \cite{birrell})

\begin{align}
u_k^{(p)}(\eta,x) &= \frac{1}{\sqrt{4\pi\omega_1}}e^{ikx -i\omega_+\eta -i\frac{\omega_-}{\rho}\ln[2\cosh(\rho\eta)]} \times \prescript{}{2}{F}_1 \bigl( 1+ i\tfrac{\omega_-}{\rho}, i\tfrac{\omega_-}{\rho}; 1 - i\tfrac{\omega_1}{\rho}; \tfrac{1}{2}(1+\tanh \rho\eta) \bigl) \nonumber \\
  & \longrightarrow \frac{1}{\sqrt{4\pi\omega_1}}e^{ikx - i\omega_1 \eta}, \quad \text{as} \; \eta \rightarrow -\infty, \\
u_k^{(f)}(\eta,x) &= \frac{1}{\sqrt{4\pi\omega_2}}e^{ikx -i\omega_+\eta -i\frac{\omega_-}{\rho}\ln[2\cosh(\rho\eta)]} \times \prescript{}{2}{F}_1 \bigl( 1+ i\tfrac{\omega_-}{\rho}, i\tfrac{\omega_-}{\rho}; 1 - i\tfrac{\omega_2}{\rho}; \tfrac{1}{2}(1-\tanh \rho\eta) \bigl) \nonumber \\
  & \longrightarrow \frac{1}{\sqrt{4\pi\omega_2}}e^{ikx - i\omega_2 \eta}, \quad  \text{as} \; \eta \rightarrow +\infty,
\end{align}
where $\prescript{}{2}{F}_1$ is a hypergeometric function. Here, we need not to worry about the details in obtaining these solutions; it is not difficult to verify those indeed satisfy the field equations and have the appropriate asymptotic limits (see section 9.1 of \cite{gradstein}). Also, it is easy to see that those solutions do not coincide, such that one will generally have nonzero $\beta_k$ coefficients and there will be particle creation. One may verify (see \cite{birrell} and references therein, or section 7.5 of \cite{gradstein}) that the Bogolubov transformations take the form

\begin{align}
u_k^{(p)} (\eta,x) = \alpha_k u_k^{(f)}(\eta,x) + \beta_k u_{-k}^{(f)}(\eta,x),
\end{align}
with

\begin{subequations}
 \begin{empheq}[left= \empheqlbrace, right = {\qquad .}]{align}
 \alpha_k &= \biggl( \frac{\omega_2}{\omega_1} \biggl)^{\!1/2} \frac{\Gamma(1\!-\!i\omega_1/\rho)\Gamma(-i\omega_2/\rho)}{\Gamma(1\!-\!i\omega_+/\rho)\Gamma(-i\omega_+/\rho)} \\[2mm]
 \beta_k  &= \biggl( \frac{\omega_2}{\omega_1} \biggl)^{\!1/2} \frac{\Gamma(1\!-\!i\omega_1/\rho)\Gamma(i\omega_2/\rho)}{\Gamma(1\!+\!i\omega_-/\rho)\Gamma(i\omega_-/\rho)}
 \end{empheq}
\end{subequations}

Then, using the properties of the gamma function

\begin{align*}
\Gamma(1+x) &= x\Gamma(x), \\
|\Gamma(iy)|^2 &= \frac{\pi}{y\sinh(\pi y)},
\end{align*}
one can immediately obtain the quadratic Bogolubov coefficients:

\begin{subequations}
 \begin{empheq}[left= \empheqlbrace, right = {\qquad .}]{align}
 |\alpha_k|^2 &= \frac{\sinh^2(\pi\omega_+/\rho)}{\sinh(\pi\omega_1/\rho)\sinh(\pi\omega_2/\rho)} \\[2mm]
 |\beta_k|^2  &= \frac{\sinh^2(\pi\omega_-/\rho)}{\sinh(\pi\omega_1/\rho)\sinh(\pi\omega_2/\rho)} \label{betak2}
 \end{empheq}
\end{subequations}

In this form, it is easy to verify the Bogolubov condition:

\begin{align}
|\alpha_k|^2 - |\beta_k|^2 = 1,
\end{align}
and to verify a few consistency checks. For example, this formula gives us a quite intuitive dependence on the frequencies; particularly, the $|\beta_k|^2$, which accounts from particle creation is proportional to $\sinh^2(\pi \omega_- /\rho)$, so that it increases the more $\omega_2$ differs from $\omega_1$, and vanishes in the limit of no expansion ($B \rightarrow 0$), when $\omega_2 \rightarrow \omega_1$.
 Note that this will always be the case for a conformally coupled massless field, whose frequencies remain in the form $\omega_k = k$. One may then interpret that the mass, which breaks conformal invariance, couples the field nontrivially to gravity, allowing the spacetime expansion to inject it with the energy necessary for particle creation. Furthermore, even in the massive case, note that this frequency difference becomes progressively smaller for higher values of $k$, such that creation of particles will be suppressed for arbitrary high-energy, short-wavelength modes.
 We shall discuss these features in more detail in the next section, where we will try to circumscribe an appropriate extension to the concepts of vacuum and particles in more general dynamic spacetimes.

\section{Adiabatic vacuum} \label{adiabatic}

As we have seen in the last section, dynamical spacetimes will generally not possess a distinguished notion of vacuum, even if we restrict ourselves to inertial (free falling) observers. Particularly, when there were asymptotically flat regions of our spacetime, this phenomenon could be better grasped in terms of particle creation, which could be analyzed simply in terms of asymptotically positive-frequency modes in the far past and future.

In such context, given that there are particles present \emph{after} the expansion, but not \emph{before} (as measured by any inertial observers in the asymptotic regions $\Omega_f$ and $\Omega_p$, respectively), one may be tempted to infer that the particle creation must have ocurred \emph{during} the expansion, and, thus, that measurements performed between these regions would yield an intermediate number of particles. However, these claims do not survive upon closer inspection. As we have thoroughly discussed in section \ref{particle detectors}, detectors in a dynamical region will generally respond in a quite complicated way to their interactions with the field, and one should not \textit{a priori} expect them to measure an intermediate particle content between $\Omega_p$ and $\Omega_f$.

As discussed above, there are no physically privileged definitions of vacuum and particles for general spacetimes, so that it is not always possible to define particle number without ambiguity away from asymptotically flat regions.
We have also seen that a special class of \emph{dynamical} spacetimes for which a privileged (nonstationary) family of inertial observers does exist, are the FLRW spaces, with their comoving observers. In this case, we could try to identify the presence of particles \emph{throughout dynamical regions} according to the detection rates for particle detectors carried by these observers.

However, even in such highly symmetric cases, for which a preferred physical definition of particles is possible, particle numbers are not conserved quantities in nonstationary spaces, which makes their measurement inherently uncertain. If, say, the rate of particle creation is $A$, then a precise measurement of particle numbers in a given time must be carried in a sufficiently short time window $\Delta t$, such that $|A| \Delta t \ll 1$. However there is a fundamental limitation on how short $\Delta t$ can be not to violate the (time-energy) uncertainty principle\footnote{
 For a thorough and pedagocical exposition of time-energy uncertainty princliple, see chapter 3 of \cite{matheus} and references therein.} :
  if one is to make a detection of an excitation within a precision $\Delta E$, we must have that $\Delta t \gtrsim \Delta E^{-1}$. Since any single particle will cause an excitation of \emph{at least} $\Delta E_1 = m$, a precise detection of $N$ particles will be associated with a minimal time interval $\Delta t \sim \Delta E_N^{-1} \sim (m \Delta N)^{-1}$. Taking into account both sources of uncertainty, we have a rough estimate on the limits on the precision for measuring $N$:

\begin{align}
\Delta N \gtrsim (m \Delta t)^{-1} + |A| \Delta t,
\end{align}
so that we have a minimal uncertainty $\Delta N_{min} \sim 2 (|A|/m)^{1/2}$ for $\Delta t = (m|A|)^{-1/2}$.

Thus we see that, for a nonzero particle creation rate $A \neq 0$, and a field of finite mass, there is a fundamental limitation in the precision of particle measurements for any modes in a given time\footnote{ 
Note that, in the massless case, this uncertainty diverges. This is due to the fact that one can have particles of arbitrarily low energies, which would require arbitrarily high precision to be accounted for. This problem is part of a more general hall of infrared divergencies that occur for massless theories, even in Minkowski space. For an introductory account of these divergencies in the simple context of Minkowski spacetimes, see chapter 7 of \cite{mandlshawn}.}.
 Nevertheless, note that the number uncertainty for arbitrarily high-energy (short-wavelenghth) particles -- for which $\Delta E_1 \simeq \sqrt{\mathbf{k}^2 + m^2} \gg m  $ -- will be vanishingly small. This will be true even if the particle creation rate happens to be the same for all modes, since the higher energy modes are associated with smaller time uncertainty in measurements. But besides that, particle production itself is generally suppressed at high energies, as we have seen in last section. 

Having pointed out these fundamental limitations, we know nonetheless that there must be appropriate limits for which particle numbers must be meaningful observables. Particularly, given the astounding success of QFT in Minkowski spacetime to describe our terrestrial experiments, as well as high energy astrophysical observations, one should expect to reobtain this theory as a sufficiently good approximation for QFTCS in our own expanding universe. Furthermore, this approximation should be increasingly better for a correspondingly slower rate of expansion.

To investigate these considerations more concretely, we turn to the simple model that was presented at the end of the previous section. There, we found that particle creation was supressed when $\omega_- \rightarrow 0$, and that would indeed occur in the limit of no expansion; more precisely, if we take $B \rightarrow 0$ and expand $|\beta_k|^2$ to the lowest order in $B$, we find that it decays as $|\beta_k|^2 \propto B^2 \rightarrow 0$.

Moreover, upon closer inspection of \eqref{betak2}, we find that, for a fixed value of \emph{total expansion} $B$, if we take the expansion \emph{rate} $\rho$ to vanishingly small values, particle creation will be exponentially suppressed for all modes, \textit{i.e.}:

\begin{align}
|\beta_k|^2 \propto e^{-2\pi \omega_p/\rho} \rightarrow 0, \qquad \rho \ll |\mathbf{k}|, m\sqrt{B}.
\end{align}

For any finite $\rho$, this exponential suppression will hold approximately for $\rho \ll \omega_p$, that is, always that $\rho \ll k$ or $\rho \ll B^\frac{1}{2} m$. Physically, we can interpret this suppression as a limitation in the production of particles for modes whose frequencies are much larger than the relative rate of expansion $\omega \gg \dot{a}/a \equiv H$ of the universe, so that particle creation should be negligible for high-energy modes of all fields (and, particularly, for any modes of a very massive field); we only expect there to be an appreciable amount of created particles for modes of frequencies $\omega \lesssim H$, comparable to the fractional rate of expansion of the universe or lower. (For our current universe, we see that this rate is extremely low: $H_0 \approx 2 \times 10^{-18}s$, corresponding to energies of about $\hbar H_0 \approx 8 \times 10^{-33}eV $.)

Although we have only deduced these conditions in the context of a simple model, they turn out to be valid in general \cite{birrell}. Then, to generalize an approximate notion of particles \textit{during} the expansion, it will be particularly useful to refer to the limits of very slow expansions, which will allow us to construct a corresponding approximation for positive-frequency modes. Such an approximation should become increasingly precise as the rate of expansion becomes arbitrarily slower; in the limit of an infinitely slow expansion -- which we will baptize as the \emph{adiabatic limit} ahead --, they should be exact, matching the fact that there will be no particles created.

As in the discussion of particle creation, it may be convenient to work with either conformal time or proper time, depending on the application at hand. The former is somewhat easier to handle the dynamical equations (especially in the massless, conformally invariant limit), as well as to present adiabatic expansions and discuss adiabatic orders in an algebraic manner, and we shall employ it in the present section. The latter, on its turn, has a more direct physical interpretation and can be worked without much difficulty with a conveniently chosen decomposition of field modes; we shall discuss it in more detail in the next chapter, in the context of adiabatic subtractions.

In order to keep the discussion simple in this first exposition, we restrict the treatment in this section to FLRW spacetimes conformally related to Minkowski spacetime, and to fields conformally coupled to gravity, \textit{i.e.} with $\xi = \xi(n)$ in $n$ spacetime dimensions (see Apendix \ref{liekillcon}). In this case, we write the line element as

\begin{align}
ds^2 = a^2(\eta) [ d\eta^2 -  d\mathbf{x}^2  ].
\end{align}

Implementing the conformal transformation to the field equations in this case (again, see appendix \ref{liekillcon}), and accounting for the presence of a mass term, which adds a conformally noninvariant contribution, but which can be accounted for with a simple term $m^2\tilde{\phi}=m^2 \Omega^s \phi$ in the transformed equation, we obtain

\begin{align}
\bigl[ \tilde{\Box}_x + \xi(n)\tilde{R}(x) + m^2 \bigl] \tilde{\phi}(x) = \Omega^{s-2}\bigl[\Box_x + \xi(n)R(x) + m^2\Omega^2(x)\bigl]\phi(x) = 0. \label{CFE}
\end{align}

For a conformally flat metric $\tilde{g}_{ab}= \Omega^2\eta_{ab}$, we have simply $R(x)=0$ and $\Box_x = \frac{\partial^2}{\partial \eta^2} - \mathbf{\nabla_{\!x}}^{\!\!\!2}$. Furthermore, the conformal factor $\Omega$ is seen to be simply the scale factor of the universe, $\Omega^2(\eta,\mathbf{x})=a^2(\eta)$, so that we can simplify the right equation in (\ref{CFE}) to

\begin{align}
\Bigl[ \frac{\partial^2}{\partial \eta^2} - \mathbf{\nabla_{\!x}}^{\!\!\!2} + m^2 a^2(\eta) \Bigl]\phi(x) = 0. \label{FLWRCFE}
\end{align}

This is a manifestly separable equation, so that we can decompose normal modes $v_\mathbf{k}$ (we save the notation $u_\mathbf{k}$ for the solutions of the original field equation $u_k = \Omega^{s-2}v_\mathbf{k}$) in the form

\begin{align}
v_\mathbf{k}(\eta,\mathbf{x}) = \frac{e^{i\mathbf{k}\!\cdot\!\mathbf{x}} }{(2\pi)^{3/2}} \chi_k (\eta). \label{FLRWseparation}
\end{align}

Substituting them in \eqref{CFE}, we find that $\chi_k$ will be just a time-dependent harmonic oscillator:

\begin{align}
\frac{d^2}{d \eta^2}\chi_k + \omega_k^2(\eta) \chi_k = 0, \label{TDHO}
\end{align}
where the positive frequencies $\omega_k(\eta)$ are defined by:

\begin{align}
\omega_k(\eta) \equiv + \sqrt{ \mathbf{k}^2 + m^2a^2(\eta) }. \label{conformal frequencies}
\end{align}
 
Once again, in the case of static universe $a=cte$, we trivially recover plane-wave modes, $\chi_k = (2\omega_k)^{-1/2}e^{-i\omega_k \eta}$. In the general case, however, not only are these equations hard to solve, but also their solution space cannot be globally separated in positive and negative frequency subspaces. To try to make sense of such a separation \emph{locally}, we write formal WKB solutions, which take the form

\begin{align}
\chi_k = \frac{1}{\sqrt{W_k(\eta)}}e^{-i\int^\eta d\eta' W_k(\eta')}.
\end{align}

Substituting these in (\ref{TDHO}), we obtain a nonlinear equation for $W_k$:

\begin{align}
W_k^2(\eta) &= \omega_k^2(\eta) -\frac{1}{2} \Bigl(\frac{\ddot{W_k}}{W_k} -\frac{3}{2}\frac{\dot{W}_k^2}{W_k^2} \Bigl) \nonumber \\[4pt]
 &= \omega_k^2(\eta) + W_k^{\frac{1}{2}} \frac{d^2}{d\eta^2} W_k^{-\frac{1}{2}} . \label{Wk}
\end{align}

At this point, the reader may wonder why one would choose to work with this rather complicated nonlinear equation (\ref{Wk}) instead of (\ref{TDHO}) directly. The advantage here lies not in obtaining exact field solutions to these equations, but rather in analyzing their behaviour in the limit of a very slow expansion, when the time derivatives of $W_k$ ($\omega_k$) become negligible. Particularly in the limit of an infinitely slow expansion, we will have that $\dot{\omega}_k \rightarrow 0$, and (\ref{Wk}) will yield a purely algebraic relation

\begin{align}
W_k(\eta) = \omega_k(\eta), \label{zerothsol}
\end{align}
where we have identified the positive roots of $W_k$ and $\omega_k$. For a finitely slow expansion -- where it still holds that $\dot{\omega} \!\ll\! \omega^2$, as well as the inequalities obtained through further time derivatives: $\ddot{\omega} \ll 2\omega\dot{\omega} \ll 2\omega^3$, etc. --, equation (\ref{zerothsol}) can be regarded as a zeroth order approximation for $W_k$.

In order to refine that approximation beyond lowest order in a systematic manner, and quantify a more precise notion of `slowness', we introduce the so-called \emph{adiabatic parameter} $T$, transforming the time variable as $\eta \rightarrow \eta_1 \equiv \eta/T$:

\begin{align}
\eta \rightarrow \eta_1 \equiv \frac{\eta}{T};
\end{align}
$T$ will play the role of stretching $\Delta \eta$ time intervals into the corresponding $T\Delta \eta_1$ transformed time intervals, making time variations go slower for larger values of $T$. A more adequate way to implement this transformation is to consider a 1-parameter family of (FLRW) metrics, with a scale factor given by $a_T(\eta) \equiv a(\eta/T)$. Then, any metric-dependent functions $f(\eta)= f(a(\eta))$, such as $\omega_k(\eta)$, will accordingly transform as

\begin{align}
f_T(\eta) \equiv f(\eta/T) = f(\eta_1).
\end{align}

In practice, this transformation will ``make the spacetime expansion go slower'' as we take larger values of $T$, which will modify our dynamic equations (we can recover the original equations taking $T=1$) by diminishing the relative magnitude of terms associated with time variations (time derivatives) of metric-dependent quantities. More precisely:

\begin{align} 
\frac{d}{d\eta}f\left( \frac{\eta}{T} \right) = \frac{1}{T}\frac{d}{d\eta_1}f(\eta_1) \equiv \frac{1}{T}f'(\eta)
 \qquad \quad \Rightarrow \qquad \quad
\frac{d^n}{d\eta^n}f\left( \frac{\eta}{T} \right) = \frac{1}{T^n}f^{(n)}(\eta_1). \label{adiabatic functions}
\end{align}

Particularly, as $T \rightarrow \infty$ all terms that contain any time derivatives of the metric vanish, producing the so called adiabatic limit. Terms with different powers of $T^{-1}$ will decay at different rates, which we can use to hierarchize different contributions in function of slowness. Thus, we refer to terms proportional to $T^{-n}$ as \textit{nth} adiabatic order terms; in practice, the adiabatic order will be simply a count of time derivatives, as we can see in \eqref{adiabatic functions}. With this hierarchy in mind, we can recursively compute an asymptotic series for $W_k$ in equation (\ref{Wk}), starting from the 0th adiabatic order solution $(W_k)^{(0)}$:

\begin{align}
\bigl( (W_k)^{(0)}(\eta_1) \bigl)^2 = \omega_k^2(\eta_1).
\end{align}

Iterating this at \eqref{Wk}, we obtain the 2nd order solution for $W_k$:

\begin{align}
\bigl( W_k^{(2)}(\eta_1) \bigl)^2 = \omega_k^2(\eta_1) -\frac{1}{2T^2} \Bigl(\frac{\ddot{\omega}_k(\eta_1)}{\omega_k(\eta_1)} -\frac{3}{2}\frac{\dot{\omega}_k^2(\eta_1)}{\omega_k^2(\eta_1)} \Bigl),
\end{align}
such that $W_k$ will differ from $W^{(2)}_k$ only by terms of 3rd adiabatic order, or higher. In fact, as we can see from eq. (\ref{Wk}), successive iterations produce only terms of even adiabatic order, all the odd-order terms vanishing identically. We can then write $W_k = W_k^{(2)} + \mathcal{O}(T^{-4})$.

Illustrating the procedure a little further, we write the calculation results to the 4th order term:

\begin{align}
\bigl( W_k^{(4)} \bigl)^2 &= (W_k^{(2)})^2 -\frac{1}{2T^2} \Bigl(\frac{\ddot{W}^{(2)}_k}{W^{(2)}_k} -\frac{3}{2}\frac{(\dot{W}^{(2)}_k)^2}{(W_k^{(2)})^2} \Bigl) \nonumber \\[6pt]
 &= \omega_k^2 -\frac{1}{2T^2} \Bigl(\frac{\ddot{\omega}_k}{\omega_k} -\frac{3}{2}\frac{\dot{\omega}_k^2}{\omega_k^2} \Bigl) 
 + \frac{1}{8T^4} \biggl[ \frac{\ddddot{\omega}_k}{\omega_k^3} - 10 \frac{\dot{\omega_k}\dddot{\omega_k}}{\omega_k^4} - \frac{11}{2}  \frac{\ddot{\omega_k}^2}{\omega_k^2} + \frac{93}{2} \frac{\dot{\omega_k}^2 \ddot{\omega_k}}{\omega_k^5} - \frac{279}{8} \frac{\dot{\omega_k}^4}{\omega_k^6} \biggl] .
\end{align}

These adiabatic expansions then provide us with a natural (approximate) generalization for the concept of vacuum and particles to fully dynamic spacetimes. The case where we had asymptocally flat regions is seen to be a case where the zeroth order adiabatic approximation becomes asymptotically exact. For other, dynamical regions of spacetime, there will always be \emph{exact} solutions of the field equations which we can \emph{locally} identify as positive frequency by matching them with the positive-frequency adiabatic expansion at a given time\footnote{
 Although these expansions are only asymptotic (meaning they will generally diverge, rather than converge to a solution), the asymptotic expansion of a function (in our case, of an exact solution) will be unique. The converse, however, is not generally true; a particular asymptotic expansion may represent more than one function (more than one exact solutions). This will reflect in the fact that the determination of a vacuum state will not be unique in dynamical spacetimes, even if we constrain our exact modes up to arbitrarily high adiabatic orders. }.
 Particularly, in our separable FLRW case, for which each value of $\mathbf{k}$ will be only associated with two linearly independent solutions, $\{u_\mathbf{k},(u_\mathbf{k})^*\}$, we can identify a positive frequency mode with an $Ath$-order adiabatic approximation $u_\mathbf{k}^{(A)}$ as

\begin{align}
u_\mathbf{k}(\mathbf{x},\eta) = \alpha_\mathbf{k}^{(A)}(\eta)u_\mathbf{k}^{(A)}(\mathbf{x},\eta) + \beta_\mathbf{k}^{(A)}(\eta) (u_\mathbf{k}^{(A)})^*(\mathbf{x},\eta) ,
\end{align}
where:

\begin{subequations}
  \label{AdiabaticBogolubov}
  \begin{align}
  \alpha_\mathbf{k}^{(A)}(\eta) &= 1 + \mathcal{O}(T^{-(A+1)}), \\
  \beta_\mathbf{k}^{(A)}(\eta) &= 0 + \mathcal{O}(T^{-(A+1)}).
  \end{align}
\end{subequations}

Here, we can match the solutions for a given time $\eta_0$ by identifying, $u_\mathbf{k}(\mathbf{x},\eta_0) = u_\mathbf{k}^{(A)}(\mathbf{x},\eta_0)$. However, making the identifications at different times will generally yield different exact modes $u_\mathbf{k}$.

This is entirely analogous to when we write the expansions of the positive frequency modes in asymptotically flat regions in terms of one another (particularly, in the `quasidiagonal' FLRW case):

\begin{align}
u_\mathbf{k}^{(f)} = \alpha_\mathbf{k}u_\mathbf{k}^{(p)} + \beta_\mathbf{k} (u_\mathbf{k}^{(p)})^* ,
\end{align}
only in this latter case all nonzero adiabatic orders asymptotically vanish in the remote regions, making $u_\mathbf{k}^{(p)}$ and $u_\mathbf{k}^{(f)}$ asymptotic approximations up to infinite order.

In the general, dynamical regime, we can use the exact modes matched to an adiabatic approximation $u_\mathbf{k}^{(A)}$ to define a vacuum state $\ket{0^{(A)}}$. Although this state is highly nonunique, and inertial observers will generally measure particles in them, the particle content for \emph{any} adiabatic vacuum states will be suppressed at high energies (at least as fast as $k^{-(A+1)}$ or $m^{-(A+1)}$ \cite{birrell}). Thus, these states allow for an approximate generalization of the concept of vacuum in dynamical spacetimes, which will display a consistent behaviour in the limit of arbitrarily high frequencies. In the next chapter, we shall see that these adiabatic expansions will play a key role in the renormalization of UV divergences for our theory in curved spaces, particularly in FLRW spaces.

%% file: chap4.tex
\chapter{Regularization and Renormalization} \label{renormalization}

After going through the basic procedures of quantization for free fields in a classical curved background spacetime, and exploring some of its most direct physical aspects and physical consequences, we now turn our attention to the more delicate and intricate problem of handling formally divergent quantities in our quantized theory, and making physical sense out of them.

As we can already anticipate from the energy divergences in flat spacetime -- which had to be properly circumvented to calculate the Casimir Effect (see section \ref{Casimireff}) -- certain key observables in our theory will be plagued by divergences. In fact, as we shall demonstrate briefly, these divergences are generally worse for fields quantized in a curved spacetime background than their flat counterparts, even for free (noninteracting) fields; it turns out that the implicit interactions with gravity give rise to extra divergent terms.

As we have discussed in Appendix \ref{distributions}, the appearence of divergences is not surprising whenever we are dealing with observables quadratic in field amplitudes, such as $\phi^2(x)$ or $T_{\mu\nu}(x)$. Nonetheless, such observables are a vital portion of the dynamical elements of our theory, and if we are to make full sense of them and derive physical predictions, we must find a suitable way to modify these formally divergent expressions in order to obtain finite physical results.

In this chapter, we attack the intricate problem of renormalization as follows: first, in section \ref{renormalizationfundamentals}, we underline some fundamental remarks in the nature of this problem, illustrating how divergences in curved space are generally worse than in flat spaces, and briefly mentioning how these may be reabsorbed in the definition of gravitational parameters in semiclassical gravity; this shall be the basis for a more general approach to renormalization in curved spaces. However, due to the more convoluted nature of this approach, we postpone its discussion to section \ref{ARenormalization}\footnote{ 
 This is by no means the most \emph{logical} presentation sequence, but it will allow one to `get to the physics' more quickly and develop some level of intuition and operational experience in this intricate subject, before dwelling into more complicated calculations.}.
 In section \ref{adiabatic subtraction} we present a rather practical and more physically intuitive renormalization scheme: adiabatic subtraction. Then, in section \ref{ARenormalization}, we present a brief introduction to Lagrangian approaches to quantum theory, both in the form of Feynman path integrals and of the Schwinger action principle, and use them to derive the effective action. We then exhibit the divergences of the effective action, and isolate them in just a finite number of geometrical terms in an asymptotic expansion, showing that it can be be rendered finite by a renormalization of geometrical parameters in a semiclassical theory of gravity.

\section{Divergences in Curved Spaces and Semiclassical Gravitation} \label{renormalizationfundamentals}

As we have stated many times before, the values obtained for many formal expressions quadratic in field operators are in general divergent. Even in the simplest example of the `standard' vacuum in Minkowski spacetime, expected values such as $\braket{0|T_{\mu \nu}(x)|0}$ and $\braket{0|\phi^2(x)|0}$ present ultraviolet divergences.

In the more particular case of flat spacetimes, it is typically possible to renormalize the values of vacuum energy (either in nontrivial topologies or in the presence of flat boundaries\footnote{Curved boundaries in flat space turn out to be a more complicated issue. For a further account of that matter see the last section of chapter 5 of \cite{fulling}, and references therein.})
 by subtracting the ``Minkowski vacuum corresponding value'', which is taken as a reference for null energy density. In curved spacetimes, however, this procedure is generally not possible. A first reason is that, while in flat spacetimes only \emph{energy differences} are directly observable, in General Relativity (and, to an appropriate extent, in QFTCS) \emph{absolute energy values} appear as sources for spacetimes curvature. Furthermore, the implicit gravitational interactions may cause additional divergences; qualitatively, this is much similar to the case of free vs. interacting field theories in Minkowski spacetime where there are fundamental differences between the asymptotic limits of a weakly interacting theory\footnote{
 For more complete accounts on that point, see \textit{e.g.} \cite{mandlshawn} for a textbook introduction on interacting fields or \cite{schwinger} for a critical collection of founding papers in the field.}
 (\textit{e.g.}, in the gravitational case, taking $G \rightarrow 0$) and its ``free'' counterpart (\textit{e.g.}, taking $G=0$ to start with). 
 
Following \cite{birrell}, let us then show a simple example to illustrate the extra divergences that arise due to spacetime curvature. We consider a conformally flat FLRW space, whose scale factor $a$ is defined by

\begin{align}
a(t) = \sqrt{1 - A^2t^2}, \qquad A = cte, \qquad |t|<A^{-1},
\end{align}
and a massless scalar field $\phi$, minimally coupled to gravity, whose equations of motion are simply $\Box \phi = 0$.

Using its differential definition, $d\eta = \pm a(t)dt$, it is easy to obtain conformal time $\eta$ as a function of proper time $t$, and vice-versa; the two are related by

\begin{align}
\qquad A \eta = \arcsin(At) \quad \Leftrightarrow \quad At = \sin(A\eta) \quad \Rightarrow |\eta| < -\frac{\pi}{2A}.
\end{align}

We then have the metric components in conformal coordinates:

\begin{align}
g^{00} &= - g^{ii} = a^{-2}(\eta) = \cos^{-2}(A\eta), \label{gii} \\
|g| &= a^8(\eta) = \cos^8(A\eta),
\end{align}
from which it is simple to compute the D'Alembertian: $\Box \phi = |g|^{1/2} \partial_\mu \Bigl( |g|^{1/2} g^{\mu \nu} \partial_\nu \phi \Bigl) $. Using our well-known ansatz, $u_\mathbf{k}(\mathbf{x},\eta) = e^{i\mathbf{k}\!\cdot\!\mathbf{x}} \chi_k(\eta)$, we obtain the equation

\begin{align}
\ddot{\chi}_k + 2 H \dot{\chi}_k + k^2 \chi_k = 0,
\end{align}
where $H$ denotes the fractional expansion rate $\dot{a}/a$. In conformal time, this yields simply $H = -A\tan(A\eta)$.

Then, defining $h_k = a^{-1} \chi_k$, we arrive at a simple (\emph{time-independent}) harmonic oscillator:

\begin{align}
\ddot{h}_k + (A^2 + k^2)h_k = 0.
\end{align}

Thus, implementing a proper normalization, we arrive at the complete field modes:

\begin{align}
u_\mathbf{k} = (16 \pi^3)^{-1/2} a^{-1}(\eta)(A^2+k^2)^{-1/2} e^{ i\mathbf{k}\!\cdot\!\mathbf{x} -i(A^2+k^2)^{1/2}\eta}, \label{RTUmodes}
\end{align}
in terms of which we can write the field expansion (\ref{quacurfi}).

Having obtained a quantized field expansion, the main goal of our analysis is to compute the expected values of energy-momentum observables and identify their divergent terms. In this simple case of a massless, minimally coupled scalar field, the stress tensor is given simply by\footnote{We shall analyze the stress tensor in more detail in section \ref{VECS}. There is a subtlety in taking the massless limit for a quantized field, but we shall ignore it at this point, as it is irrelevant to the structure of the UV divergences in which we are interested at this point.}

\begin{align}
T_{\mu\nu} = \nabla_{\!\mu}\phi\, \nabla_{\!\nu}\phi - \tfrac{1}{2}g_{\mu\nu}\nabla_{\!\alpha}\phi\nabla^{\alpha}\phi.
\end{align}
Further, we have that $\nabla_{\!\mu}\phi = \partial_\mu \phi$. Then, particularly, we have the energy density operator

\begin{align}
T_0^{\;0} = \tfrac{1}{2} \bigl(\partial_{0}\phi\, \partial^{0}\!\phi - \partial_i\phi\, \partial^i\!\phi \bigl) = \tfrac{1}{2} g^{00} \bigl(\dot{\phi}^2 + (\mathbf{\nabla}\phi)^2 \bigl),
\end{align}
where we have used (\ref{gii}) in the last equality.

Let then $\ket{0}$ be the vacuum state associated to the modes (\ref{RTUmodes}). We may compute its corresponding vacuum energy density

\begin{align}
\rho(x) = \braket{0|T_0^{\;0}(x)|0} &= \tfrac{1}{2} g^{00}(x) \int d^3\mathbf{k} \Bigl[ \dot{u}_\mathbf{k}(x) \dot{u}^*_\mathbf{k}(x) + (\mathbf{\nabla} u_\mathbf{k}(x)) \!\!\cdot\!\! (\mathbf{\nabla} u^*_\mathbf{k}(x)) \Bigl] \nonumber \\
 &= \frac{1}{32\pi^3a^4(\eta)} \int d^3\mathbf{k} \bigl[ (k^2+A^2)^{1/2} + (k^2+H^2)(k^2+A^2)^{-1/2} \bigl]. \label{RTUT00}
\end{align}

From the asymptotic form of the integrand in (\ref{RTUT00}) as $k\rightarrow\infty$, one can easily see that this energy density diverges quartically in the UV. Similarly to what we did in the Casimir effect, we may keep track of the divergent terms by introducing a regularizer $e^{-\alpha(k^2+A^2)^{1/2}}$. For convenience, we also multiply both sides of \eqref{RTUT00} by $a^4(\eta)$\footnote{
 One may think of the observable in the LHS of this equation as the (classically conserved) quantity associated to the energy in a coexpanding unitary volume $1 \times a^3$ corrected by the cosmological redshift factor $\times a$. See the next chapter (and subsection \ref{VECS}) for more details on this point.}, 
 obtaning the following result:

\begin{align}
\rho a^4 &= \frac{1}{32\pi^3}\,4\pi\! \int_0^\infty dk \,e^{-\alpha(k^2+A^2)^{1/2}} k^2\bigl[ (k^2+A^2)^{1/2} + (k^2+H^2)(k^2+A^2)^{-1/2} \bigl] \nonumber \\
  &= \frac{1}{32\pi^2} \,4\! \int_A^\infty d\omega \,e^{-\alpha\omega} \omega(\omega^2-A^2)^{1/2} \bigl[ \omega + (\omega^2 - A^2 + H^2)\omega^{-1} \bigl] \nonumber \\
  &= \frac{1}{32\pi^2} \biggl\{ 4\! \int_A^\infty d\omega \,e^{-\alpha\omega}  \Bigl[ 2\omega^3 + (H^2 - 2A^2)\omega + \tfrac{1}{2}A^2(H^2-A^2)\omega^{-1} + \mathcal{O}(\omega^{-3}) \Bigl] \biggl\}.
\end{align}

The regularizer has allowed us to temporarily tame the divergences in the first 3 terms, which are quartic, quadratic and logarithmic, respectively. Carrying integrations by parts for the first two terms and making a change in variables in the third one ($\omega \rightarrow \alpha\omega$), one may then arrive at the expansion

\begin{align}
\rho a^4 = \frac{e^{-\alpha A}}{32\pi^2} \Bigl[ \frac{48}{\alpha^4} + \frac{4H^2-8A^2}{\alpha^2} + 2A^2(H^2-A^2)\ln(\alpha) + \mathcal{O}(\alpha^0) \Bigl]. \label{toyvacuumenergy}
\end{align}

Of course, if we relax the regularization, letting $\alpha \rightarrow 0^+$, we reobtain a quartic, a quadratic and a logarithmic divergent terms. Note that, in the limit of a Minkowski spacetime $A,H \rightarrow 0$, this vacuum energy is still divergent, due to the quartic term. Nevertheless, we see explicitly that spacetime curvature has led to the emergence of \emph{additional} quadratic and logarighimic divergences. Thus, in general, \emph{one cannot obtain a finite energy in curved spaces just by subtracting a Minkowski-vacuum contribution}.

An alternative strategy to handle the infinities in $\braket{T_{\mu\nu}}$ in curved spacetimes is presented when we consider not only a theory of quantized fields propagating in a \emph{fixed} background geometry, but rather a wider dinamical theory which couples quantized fields to a \emph{classical, but dynamical} spacetime: semiclassical gravitation. In this approach, one attempts to incorporate the gravitational backreaction of the quantum fields under consideration, by coupling the \emph{expected values} of energy-momentum currents as the source of spacetime curvature.

In the purely classical case, we had Einstein's equations \eqref{EEwL} coupling $T_{\mu\nu}$ to spacetime curvature:

\begin{align}
R_{\mu\nu} - \tfrac{1}{2}g_{\mu\nu} R + \Lambda g_{\mu\nu} = -8\pi G T_{\mu\nu}. \label{EEwL2}
\end{align}

If we quantize the matter fields alone, the proposed anologue in this semiclassical scheme is to substitute $T_{\mu\nu}$ by $\braket{T_{\mu\nu}}$, so that we have

\begin{align}
R_{\mu\nu} - \tfrac{1}{2}g_{\mu\nu} R + \Lambda_B g_{\mu\nu} = -8\pi G_B \braket{T_{\mu\nu}}. \label{scEE}
\end{align}

For the time being, \eqref{scEE} is merely a formal equation, since $\braket{T_{\mu\nu}}$ is generally divergent. The rough idea of renormalization in this wider theoretical framework is to \emph{absorb} its infinities by redefining the theory's so-called \emph{bare parameters} (such as $\Lambda_B$ or $G_B$) into new, renormalized ones, which will appear in the equations with the finite `physical' part of $\braket{T_{\mu\nu}}$:

\begin{align}
R_{\mu\nu} - \tfrac{1}{2}g_{\mu\nu} R + \Lambda g_{\mu\nu} = -8\pi G \braket{T_{\mu\nu}}_{phys}.
\end{align}

In order to achieve this renormalization in a systematic way, one first turns to a more fundamental object in the theory, the action, in terms of which $T_{\mu\nu}$ in classically defined. Recall that we define the classical action $S$ with a purely geometrical term and a matter one:

\begin{align*}
S = S_G + S_M, 
\end{align*}
with $\frac{\delta G}{\delta g^{\mu\nu}} = 0$ implying the Einstein equation above, such that the matter stress tensor is defined as

\begin{align}
\frac{2}{\sqrt{-g}} \frac{\delta S_M}{\delta g^{\mu\nu}} = T_{\mu\nu}.
\end{align}

Then, in the semiclassical theory, we seek for an object $W$, which we will call the \emph{effective action}, whose functional derivative with respect to the metric yields:

\begin{align}
\frac{2}{\sqrt{-g}} \frac{\delta W}{\delta g^{\mu\nu}} = \braket{T_{\mu\nu}}, \label{treaction}
\end{align}
where a more precise definition of the braket $\braket{...}$ will be given in section \ref{ARenormalization}.

To find an object $W$ that satifies the relation \eqref{treaction}, we will present in section \ref{ARenormalization} the Schwinger action principle formulation to quantum mechanics, which is based on a Lagrangian formalism and is intimately related to Feynman path integrals. Then, in the context of field theory, we will find that $W$ is generally divergent, and that one way to obtain finite quantities for observables of the quantized matter fields will be to renormalize $W$; this can be achieved by absorbing its divergent portion $W_{div}$ \emph{in the geometrical action} $S_G$, redefining (renormalizing) its basic geometrical parameters.

At this point, however, we shall postpone the treatment of a great such an intricate approach to renormalization and explore a simpler and subtraction scheme, working directly with the spectral representation of the stress tensor: adiabatic subtraction. In the next section, we will develop this aproach comprehensively, and use it both to obtain some operational intuition with renormalization in curved spaces, and to derive physical results of interest, such as the renormalized stress tensor in FLRW spacetimes.

\section{Adiabatic Subtraction} \label{adiabatic subtraction}

The method of renormalization that we shall concretely develop in this dissertation is the one called \emph{adiabatic subtraction}. It makes thorough use of the adiabatic expansions presented in section \ref{adiabatic}, which are employed both conceptually, insofar as the adiabatic condition allows us to separate positive- and negative-frequency modes and obtain a useful notion of a vacuum state, and operationally, as they are used to subtract UV-divergent terms and yield a finite result for the expectation values of various observables.

In general lines, the procedure consists of a mode-by-mode subtraction of divergent terms in the formal expression of several observables, by representing this expression in an asymptotic expansion, identifying in this expansion the terms which contain UV-divergent terms when summed (or integrated) up to arbitrarily high momenta, and then subtract such terms \emph{inside the integration}.

In order to carry that procedure, we begin by exposing a slightly different construction for adiabatic expansions as in section \ref{adiabatic}, by making these expansions in proper time, as well as explicitly computing the $nth$ adiabatic order solutions $h_n(t)$ before turning to WKB frequencies $W^{(n)}$. We then show how successive adiabatic orders follow a well defined hierarchy in the divergences of observables, whereupon we can design a prescription to systematically eliminate these divergences by subtracting a finite number of terms in the adiabatic expansion. Finally, we shall apply this prescription to compute the power spectrum, as well as the stress tensor for a scalar field, and interpret the obtained results.

\subsection{Adiabatic Expansion in proper time}

As in the previous approach, our starting point is to write \emph{exact} field mode solutions (conveniently separated in this spatially flat FLRW spacetime) in the form:

\begin{align}
f_{\mathbf{k}}(x) = \frac{1}{\sqrt{Va^3(t)}}e^{i\mathbf{k}\!\cdot\!\mathbf{x}}h_{\mathbf{k}}(t), \label{FLRWptModes}
\end{align}
for which the dynamic equations $[\Box + m^2 + \xi R]f_{\mathbf{k}}  = 0$ also yield a time-dependent harmonic oscillator for the temporal amplitude $h_{\mathbf{k}}$:

\begin{align}
\frac{d^2h_\mathbf{k}}{dt^2} + \Omega^2_k(t) h_{\mathbf{k}} = 0. \label{tdho}
\end{align}

Only in this form, it has different expression for the frequency $\Omega_k$, which we write in the form:

\begin{align}
\Omega^2_k = \omega^2_k + \sigma, \label{fatomega}
\end{align}
where we define:

\begin{align}
\omega_k \equiv \sqrt{\frac{k^2}{a^2}+m^2} \, , 
\qquad \sigma \equiv \Bigl(6\xi- \frac{3}{4}\Bigl)\frac{\dot{a}^2}{a^2} + \Bigl(6\xi- \frac{3}{2}\Bigl)\frac{\ddot{a}}{a}, \label{omegasigma}
\end{align}
splitting $\Omega_k$ in an ``instantaneous frequency'', $\omega_k$, and an extra contribution due to $\sigma$, associated with the time variation of the scale factor of the universe $a(t)$ (note that $\sigma$ has contributions both from the direct coupling with the curvature, which can be found in terms proportional to $\xi$, and from time derivatives of the frequency $\dot{\omega}_k$).

Of course, in the limit of a static Minkowski spacetime, $a \rightarrow cte$ (more precisely, $\dot{a}/a \rightarrow 0$, and all expressions obtained by subsequent time derivatives of it), we recover the familiar plane-wave solutions: $h_k = (2\omega_k)^{-1/2}e^{-i\omega_k t}$. But in general, for a non-static spacetime, we have the more complicated time-dependent harmonic oscillator (\ref{tdho}).

Now, just as we did in section \ref{adiabatic}, we want to analyze (\ref{tdho}) in the limit of an arbitrarily slow expansion/variation, in order to build an asymptotic series to $h_k$, starting from the zeroth order plane-wave approximation. Once again, we introduce the adiabatic parameter $T$, substituting any metric-dependent functions $f(t)$ by $f_T(t) = f(t/T)$, which can be thought of as a time rescaling:

\begin{align}
t \rightarrow t' = \frac{t}{T} \qquad \Rightarrow \qquad \frac{d^nf}{dt^n} \Bigl( \frac{t}{T} \Bigl) = \frac{1}{T^n} f^{(n)}(t') ,
\end{align}
(such that, in all metric dependent functions, the coordinate time shifts will be rescaled as $t \rightarrow tT$).

In practice, all these changes are mediated by the transformation in the scale factor $a(t) \rightarrow a_T(t) = a(t/T)$. We then note that one could generalize this approach to arbitrary metrics by reescaling \emph{all} spacetime distances as $x \rightarrow xT$ (\emph{e.g.}, rescaling geodesic distances by means of Riemann normal coordinates), or, as we do above, by transforming the metric (and any metric-dependent functions) as $g_{\mu\nu}(x) \rightarrow \prescript{T}{}{g}_{\mu\nu}(x) = g_{\mu\nu}(x/T)$. Then, as we take $T\rightarrow\infty$ we are effectively stretching spacetime distances in all directions, and diluting any effects of curvature.

Returning to our discussion in FLRW spaces, a quite direct way to implement successive adiabatic approximations is to define iterative variable changes, starting with:

\begin{align}
t_1 &= \int^t dt' \Omega(t'), \\
h_1 &= \Omega^{1/2}h.
\end{align}

Then, it is relatively straightforward to find the form of our equations (\ref{tdho}) in the transformed variables. From the differential relation $dt_1 = \Omega(t)dt$, we have that

\begin{align}
\frac{df}{dt} = \Omega(t)\frac{df}{dt_1}.
\end{align}

Thus

\begin{align}
0 &= \frac{d^2}{dt^2}h + \Omega^2h \nonumber \\
  &= \Omega\frac{d}{dt_1}\Bigl( \Omega\frac{d}{dt_1}(\Omega^{-1/2}h_1) \Bigl) + \Omega^{3/2}h_1 \nonumber \\
  &= \Omega^{3/2} \Bigl\{ h_1'' + \Bigl( \frac{1}{4}\frac{\Omega'^2}{\Omega^2} -\frac{1}{2}\frac{\Omega''}{\Omega} + 1 \Bigl)h_1 \Bigl\},
\end{align}
where the primes $'$ denote differentiation with respect to $t_1$. We can then rewrite the last equation as

\begin{align}
\frac{d^2}{dt_1^2}h_1 + \Omega_1^2h_1 = 0, \label{order1}
\end{align}
where we have defined:

\begin{align}
\Omega_1^2 \equiv 1 + \epsilon_2, \qquad \quad \epsilon_2 = \frac{1}{4}\frac{\Omega'^2}{\Omega^2} -\frac{1}{2}\frac{\Omega''}{\Omega} = \Omega^{-1/2}\frac{d^2}{dt_1^2}\Omega^{1/2} .
\end{align}

Since each derivative of any function with respect to $t_1$ is proportional to its derivative with respect to $t$ (or, more loosely speaking, since an infinitesimal variation in $t_1$ is proportional to an infinitesimal variation in $t$), we can immediately assert that $\epsilon_2$ contains only terms of 2nd adiabatic order or higher. In the lowest (zeroth) order, equation (\ref{order1}) is merely a time-independent oscillator, whose linearly independent solutions are

\begin{align}
h_1(t) \propto e^{\mp it_1} + \mathcal{O}(T^{-2}) = e^{\mp i \int^t \Omega(t')dt'} + \mathcal{O}(T^{-2}).
\end{align}

Evaluating the solution in the limit of an infinitely slow expansion ($T \rightarrow \infty$), we take the positive-frequency solutions in a general FLRW spacetime to be those which match the usual Minkowski positive-frequency solutions, namely, the ones with a minus sign on the exponent.

Upon comparison of equations (\ref{tdho}) and (\ref{order1}), we see that they are formally identical in their respective parameters. Such self-similarity allows us to easily define iterations to obtain higher adiabatic orders:

\begin{align}
t_2 &= \int^t_1 dt_1' \Omega_1(t_1'), \\
h_2 &= \Omega_1^{1/2}h_1,
\end{align}
whence it follows immediately that

\begin{align}
\frac{d^2}{dt_2^2}h_2 + \Omega_2^2h_2 = 0, \label{order2}
\end{align}
being

\begin{align}
\Omega_2^2 = 1 + \epsilon_4, \quad \epsilon_4 \equiv \Omega_1^{-1/2}\frac{d^2}{dt_2^2}\Omega_1^{1/2}.
\end{align}

Since the derivatives act only on the higher order terms (cancelling the constant, zeroth order one), we have that $\epsilon_4$ only contains terms of 4th adiabatic order or higher:

\begin{align}
\epsilon_4 &= (1+\epsilon_2)^{-1/2}\frac{d^2}{dt_2^2}(1+\epsilon_2)^{1/2} \nonumber \\ &= \Bigl(1 - \frac{1}{2}\epsilon_2 + \mathcal{O}(T^{-4}) \Bigl)\frac{d^2}{dt_2^2}\Bigl(\cancelto{0}{1} + \frac{1}{2}\epsilon_2 + \mathcal{O}(T^{-4}) \Bigl) \nonumber \\
&= \frac{1}{2}\frac{d^2}{dt_2^2}\epsilon_2 + \mathcal{O}(T^{-6}).
\end{align}

Again, we can immediately write the solution up to 4th adiabatic order:

\begin{align}
h_2 \propto e^{\mp i t_2} + \mathcal{O}(T^{-6}),
\end{align}
and this substitution process can be repeated up to any desired adiabatic order, recursively defining $t_n$, $h_n$, $\Omega_n$, which will obey analogous equations to \eqref{tdho}:

\noindent
\begin{minipage}[t]{0.47\linewidth}
 \begin{align}
 t_n &= \int^t_{n-1} dt_{n-1}' \Omega_{n-1}(t_{n-1}') \\[4pt]
 h_n &= \Omega_{n-1}^{1/2}h_{n-1}
 \end{align}
\end{minipage} \hfill \vline 
\begin{minipage}[t]{0.43\linewidth}
 \begin{align}
 &\frac{d^2}{dt_n^2}h_n + \Omega_n^2h_n = 0 \label{ordern} \\
 &\Omega_n = 1 + \epsilon_{2n}
 \end{align}
\end{minipage}

\strut \vspace{6pt}
\begin{align}
\Rightarrow \qquad h_n \propto e^{-it_n} + \mathcal{O}\bigl(T^{-2(n+1)}\bigl).
\end{align}

To make use of this expansion to perform \emph{adiabatic subtractions}, it will be particularly useful to compute the expansions of the WKB frequency $W_k$, as well as some elementary functions of it. The iterations for $W$ here are entirely analogous to those in section \ref{adiabatic}; the only difference is that, because we are working with proper time $t$, the frequency that appears in eq. \eqref{tdho} is not $\omega(t)$ but rather $\Omega(t)$, \emph{which already carries contributions of second adiabatic order from $\sigma(t)$}, as $\sigma$ has terms proportional to $\dot{a}^2$ and $\ddot{a}$ (see eqs. \eqref{fatomega} and \eqref{omegasigma}). 

Therefore, just as in equation \eqref{Wk}, if we write $h_k = (2W_k)^{-\frac{1}{2}} \exp(-i\int^t dt'W_k(t'))$, eq. \eqref{tdho} yields

\begin{align}
W_k^2 = \Omega_k^2 + \omega_k^\frac{1}{2}\frac{d^2}{dt^2}\omega_k^{-\frac{1}{2}}. \label{ptWKB}
\end{align}

Before we proceed further in computing its adiabatic expansion, let us lay an unambiguous notation for $W_k$ (in what follows, we shall often suppress the subscript $k$, for cleaness). As in section \ref{adiabatic}, we denote its truncation to $nth$ adiabatic order as $W^{(n)}$. To keep track of the adiabatic order of each term, it is useful to write the expansion in the form:

\begin{align}
W_k &\sim \omega_k + \omega_k^{(2)} + \omega_k^{(4)} + \hdots \,,
\end{align}
where we have used the fact that $\omega^{(0)}_k = \omega_k$. Thus, we write the first few truncations as:

\begin{align*}
&W^{(2)} = \omega + \omega^{(2)}, \\
&W^{(4)} = \omega + \omega^{(2)} + \omega^{(4)} = W^{(2)} + \omega^{(4)}, \\
&etc.
\end{align*}

However, for any other functions of $W$, we shall use $\bigl( f(W) \bigl)^{(n)}$ to denote the terms of \emph{exact adiabatic order $n$} in their expansion, rather than the truncation up to $n$th order. Thus, for example, $(W^{(2)})^2 \neq (W^2)^{(2)}$, since

\begin{align*}
(W^2)^{(2)} &\equiv \bigl( (\omega + \omega^{(2)} + \omega^{(4)} +... )^2 \bigl)^{(2)} \\
 &= \bigl( \omega^2 + 2\omega\omega^{(2)} + (\omega^{(2)})^2 + 2\omega\omega^{(4)} +... \bigl)^{(2)} \\
 &\equiv 2\omega \omega^{(2)},
\end{align*}
which is purely of second adiabatic order, while

\begin{align*} 
 (W^{(2)})^2 &\equiv (\omega + \omega^{(2)})^2 = \omega^2 + 2\omega\omega^{(2)} + (\omega^{(2)})^2,
\end{align*}
which has terms of zeroth, second and fourth adiabatic orders.

A particular class of functions $f(W)$ that we will be operating in the next section are simple powers of $W$, $W^\alpha$, for which we have:

\begin{align}
W^\alpha &\sim \bigl[ \omega + \omega^{(2)} + \omega^{(4)} +... \bigl]^\alpha \nonumber \\
 &= \omega^\alpha \biggl[1 + \frac{\omega^{(2)}}{\omega} + \frac{\omega^{(4)}}{\omega} +... \biggl]^\alpha \nonumber \\
 &= \omega^\alpha \biggl[1 + \alpha \Bigl(\frac{\omega^{(2)}}{\omega} + \frac{\omega^{(4)}}{\omega}+...\Bigl) + \frac{\alpha(\alpha-1)}{2}\Bigl(\frac{\omega^{(2)}}{\omega} + \frac{\omega^{(4)}}{\omega}+...\Bigl)^2 + ... \biggl].
\end{align}

From this, it is easy to group the terms of the same adiabatic order. For the 3 lowest orders, we obtain:

\begin{subequations}
 \begin{align}
 (W^\alpha)^{(0)} &= \omega^\alpha, \\ 
 (W^\alpha)^{(2)} &= \alpha \frac{\omega^{(2)}}{\omega} \,\omega^\alpha, \\ 
 (W^\alpha)^{(4)} &= \Bigl[ \alpha \frac{\omega^{(4)}}{\omega} + \tfrac{1}{2}\alpha(\alpha -1) \Bigl( \frac{\omega^{(2)}}{\omega} \Bigl)^{\!2}  \Bigl]\omega^\alpha.
\end{align}
\end{subequations}

 A case of particular interest ahead will be $\alpha=-1$, which results in:

\begin{align}
(W^{-1})^{(0)} = \omega^{-1}, 
\quad (W^{-1})^{(2)} = \frac{-\omega^{(2)}}{\omega^2}, 
\quad (W^{-1})^{(4)} = \frac{-\omega^{(4)}}{\omega^2} + \frac{(\omega^{(2)})^2}{\omega^3} . \label{W-1}
\end{align}

Now that we have shown how to obtain the expansions of functions of $W$ in terms of the basic building blocks $\omega^{(n)}$, let us explicitly compute the first few orders of the latter. The zeroth order term is simply

\begin{align}
\omega^{(0)} = \omega = \sqrt{m^2 + \frac{\mathbf{k}^2}{a^2}}.
\end{align}

Then, recursively, it is not hard to compute the 2nd term from eq \eqref{ptWKB}:

\begin{align}
(W^{(2)})^2 &= \omega^2 + \sigma + \omega^{\frac{1}{2}} \frac{d}{dt^2} \omega^{-\frac{1}{2}} \nonumber \\
 \Rightarrow W^{(2)} &= \omega + \frac{1}{2\omega} \Bigl( \sigma + \omega^{\frac{1}{2}} \frac{d}{dt^2} \omega^{-\frac{1}{2}} \Bigl),
\end{align}
where we have computed the square root discarding 4th or higher order terms. Thus

\begin{align}
\omega^{(2)} &= \frac{1}{2\omega} \Bigl( \sigma + \omega^{\frac{1}{2}} \frac{d}{dt^2} \omega^{-\frac{1}{2}} \Bigl) \nonumber \\
 &= \frac{1}{2\omega} \biggl\{ \Bigl[ 6\xi - \frac{3}{4} - \frac{3}{2} \frac{k^2}{\omega^2a^2} + \frac{5}{4} \Bigl( \frac{k^2}{\omega^2a^2} \Bigl)^{\!2} \Bigl] \left( \frac{\dot{a}}{a} \right)^{\!2} + \Bigl[ 6\xi - \frac{3}{2} + \frac{1}{2}\frac{k^2}{\omega^2a^2} \Bigl] \frac{\ddot{a}}{a} \biggl\}
\end{align}

The 4th order term, although equally straightforward, is quite laborious to compute manually, and prohibitively large to write down here. Nevertheless, its definition in the recursive expression

\begin{align}
(W^{(4)})^2 = \Omega^2 + (W^{(2)})^\frac{1}{2}\frac{d^2}{dt^2}(W^{(2)})^{-\frac{1}{2}}
\end{align}
allows one to easily implement these calculations symbolically in a computer. We will use such results (presently computed in the software Mathematica) in the following sections to evaluate the renormalization of the stress tensor.

Now that we have extensively developed adiabatic expansions, we shall occupy ourselves in the next section in making use of them to systematically subtract infinities, and obtain finite expectation values for observables of physical interest.

\subsection{Structure of the divergences and Adiabatic Subtraction}

We already know that there will be divergences in the expectation values of many relevant physical observables;  particularly, in FLRW spaces, these expectation values can be put in the form of a Fourier expansion in terms of the $\mathbf{k}$, which takes the form of integrals of (almost everywhere) finite functions of $k$ (i.e. finite integrands). The divergences then occur when we integrate these funtions in the UV region, $k \rightarrow \infty$ \footnote{ 
 Sometimes, there may also be divergences in the IR, $k \rightarrow 0$, but these are generally not so pervasive; they usually appear for specific ranges of parameter in the theory and can be highly dependent on the choice of vacuum state. We shall not systematically occupy ourselves with them in the scope of this text. }.
 By means of the adiabatic expansion, these integrands may be split in a sum of integrable (convergent) and non-integrable (divergent) terms, and, furthermore, \emph{this expansion has a well defined hierarchy for the order of the divergences in its terms}. (The meaning of this last sentence should become clearer below.)

The simplest example at hand is the two-point field amplitude $\phi(x)\phi(x')$. If we keep $x$ and $x'$ independent, this is a well defined (operator-valued) distribution (see Appendix \ref{distributions}). If we attempt to evaluate its vacuum expectation value, we get the following expansion:

\begin{align}
\braket{0|\phi(x)\phi(x')|0} = \sum_\mathbf{k} f_\mathbf{k}(x) f^*_\mathbf{k}(x') \rightarrow \frac{1}{2(2\pi)^3} \bigl[ a(t) a(t') \bigl]^{-3/2} \int d^3\mathbf{k} e^{i \mathbf{k} \!\cdot\! (\mathbf{x} - \mathbf{x'})}h_\mathbf{k}(t) h^*_\mathbf{k}(t'). \label{phixphixp}
\end{align} 
(The vacuum state $\ket{0}$ considered here is that determined by the modes $f_\mathbf{k}$. Among other things, it will be approximated by any adiabatic vacuum $\ket{0^{(A)}}$.)

Although this does not properly yield a function of $(x,x')$, since \eqref{phixphixp} does not absolutely converge, it is well defined as a (number-valued) distribution, and can be straightforwardly evaluated inside integrals (as we did in last chapter for obtaining the response function of particle detectors). However, we are in much more serious trouble when trying to evaluate the integral \eqref{phixphixp} by itself as a function of spacetime, making $x=x'$. Formally, we write the expansion

\begin{align}
\braket{0|\phi^2(x)|0} = \frac{1}{2(2\pi)^3 a^3(t)} \int d^3\mathbf{k} |h_k(t)|^2 = \frac{1}{4\pi^2 a^3(t)} \int_0^\infty dk k^2 |h_k(t)|^2. \label{barepowerspectrum}
\end{align}

If we check the ultraviolet limit of this integrand $k \gg m, \sigma$, we obtain

\begin{align}
k^2 |h_k|^2 \sim k^2\omega_k^{-1} \sim k,
\end{align}
(where we are ignoring the $k$-independent scale factor $a(t)$).

Therefore, if we set a very large UV cutoff $K$ for this integral, we will have

\begin{align}
\int_0^K dk k^2 |h_k(t)|^2 \sim \mathcal{O}(K^2).
\end{align}

This means that this observable diverges quadratically. To be more precise, its dominant term diverges quadratically. Had we considered a higher order expansion of $\omega_k^{-1}$, we would obtain

\begin{align}
\omega_k^{-1} &= \left[ \frac{k^2}{a^2} + m^2 \right]^{-\frac{1}{2}} \nonumber \\
 &= \frac{a}{k} - \frac{1}{2}\frac{m^2a^3}{k^3} + \frac{3}{8} \frac{m^4a^5}{k^5}+ \hdots \;\;. \label{0thtermexpansion}
\end{align}

In that case, the integral above yields

\begin{align}
\int_0^K dk k^2 \omega_k^{-1} \sim \mathcal{O}(K^2) + \mathcal{O}(\ln K) + \mathcal{O}(K^{-2}) + \hdots \;,
\end{align}
where we can see that it has quadratic \emph{and} logarithmic divergences, as well as a series of convergent terms. If we now analyze the first few terms in the adiabatic expansion for $|h_k(t)|^2=W_k^{-1}$, we can verify that \emph{the dominant UV contribution decreases in direct corresponce with the adiabatic order}, i.e.:

\begin{subequations}
  \begin{align}
  (W_k^{-1})^{(0)} &\sim k^{-1}, \\
  (W_k^{-1})^{(2)} &\sim k^{-3}, \\
  (W_k^{-1})^{(4)} &\sim k^{-5}, 
  \end{align}
\end{subequations}
and, generally, $(W_k^{-1})^{(2n)} \sim k^{1-2n}$ (although for some very special parameter values $(\xi,m)$, the coefficients for the highest power terms may turn out to be $0$). Thus, if we evaluate the adiabatic expansion for the $\braket{\phi^2}$ integral \eqref{barepowerspectrum}, we obtain

\begin{align}
\int_0^K dk k^2 |h_k(t)|^2 &= \int_0^K dk k^2 W_k^{-1} \nonumber \\
  &\sim \int_0^K dk k^2 \bigl( (W_k^{-1})^{(0)} + (W_k^{-1})^{(2)} + (W_k^{-1})^{(4)} +... \bigl) \nonumber \\
  &= \mathcal{O}(K^2) + \mathcal{O}(\ln K) + \text{convergent terms},
\end{align}
where the divergences have spawned only from the zeroth and the second order adiabatic terms: $(W_k^{-1})^{(0)}$ and $(W_k^{-1})^{(2)}$. More especifically, $(W_k^{-1})^{(0)}$ yields \emph{both} quadratic and logarithmic divergences (+ convergent terms), while $(W_k^{-1})^{(2)}$ yields \emph{only} logarithmic divergences (+ convergent terms).

Thus, one way to get rid of all infinities and obtain a finite expectation value for $\braket{\phi^2}$ -- to which we shall ascribe physical meaning and compare with experiments -- is to \emph{subtract from $|h_k|^2=W_k^{-1}$ all terms up to second adiabatic order inside the integration sign}, and then carry the integration for a convergent integrand. That is, we \emph{define}:

\begin{align}
\braket{0| \phi^2(x) |0}_{phys} \equiv \frac{1}{4\pi^2 a^3(t)} \int_0^\infty dk k^2 \bigl[ |h_k(t)|^2 - (W_k^{-1})^{(0)}(t) - (W_k^{-1})^{(2)}(t) \bigl].
\end{align}

Another very important bilinear observable in field theories is the stress tensor $T_{\mu\nu}$, which, as we have said before, conveys information about the energy and momentum of the field. Besides its bilinear form in field operators, some of its terms also contain second spacetime derives, or are quadratic in first derivatives. These derivatives give rise to two extra powers of $k$ (or $\omega_k$, which behaves as $\sim k$ in the UV), making the divergences on $T_{\mu\nu}$ worse than those in $\phi^2$: besides logarithmic and quadratic divergent terms, it also contains quartic divergences (indeed, we have already come across them in the previous section; see eq \eqref{toyvacuumenergy}). Therefore, in order to eliminate all divergences in $\braket{T_{\mu\nu}}$, one must generally make subtractions up to 4th adiabatic order.

If we are to generalize this procedure to any observable $Q$ in our theory, and we wish to make it sufficiently systematic to obtain a unique physical prediction (less of residual free renormalization parameters, whose values should be experimentally determined), there are a couple of things we should pay attention to. First, different observables may, quite naturally, present different types of divergency and therefore require different orders of adiabatic subtraction to be rendered finite. Second, within one single adiabatic order, one will generally find both divergent and convergent terms (as is well illustrated at eq \eqref{0thtermexpansion}); in principle, one only has to subtract the divergent contributions to get a finite result, but the questions of how to decompose each adiabatic term and whether to subtract its convergent parts leave ambiguities of which finite result we will end up with. Moreover, the leading UV behaviour of each adiabatic term may depend on the parameters of the theory (in the present example of a free scalar field, nonminimally coupled to gravity, these are basically the mass $m$, and the adimensional coupling constant $\xi$); for some special values of parameters, the leading UV coefficients may turn out to be $0$, making a otherwise divergent term convergent (e.g., for $\xi=1/6$, $(W_k^{-1})^{(2)}$ decays as $k^{-5}$, rather than $k^{-3}$).

Therefore, in order to systematically eliminate divergences and obtain a well-defined finite expectation value, we \emph{define} the procedure of adiabatic subtraction as follows \cite{parker}: given an observable $Q$ that, \emph{for general values of parameters in the theory}, has a formal expansion for its expectation value $\braket{ \Psi| Q | \Psi} $ with divergences up to the adiabatic order $Q^{(n)}$, then its physical expectation value $\braket{ \Psi| Q | \Psi}_{phys}$ is defined by subtracting \emph{in the expansion (i.e. under the integration sign) all terms $Q^{(A)}$ of adiabatic order $A \leq n$}, regardless of whether these terms have convergent contributions or whether they are divergent at all for the specific values of parameters under consideration.

For example, if we consider the power spectrum, the last divergent term is generally $(W^{-1})^{(2)}$. We can rewrite it, arranging all terms proportionally to positive powers of $m$ and $\xi'=\xi-1/6$:

\begin{align}
(W^{-1})^{(2)} &= -\frac{\omega^{(2)}}{\omega^2} \nonumber \\
 &= - \frac{1}{2\omega^3} \left[ \Bigl( 6\xi' - \frac{m^2}{\omega^2} + \frac{5}{4}\frac{m^4}{\omega^4} \Bigl) \left( \frac{\dot{a}}{a} \right)^{\!2} + \Bigl( 6\xi' - \frac{1}{2}\frac{m^2}{\omega^2} \Bigl) \frac{\ddot{a}}{a} \right] \nonumber \\
 &= -\frac{3\xi'}{\omega^3}\frac{\dot{a}^2}{a^2} - -\frac{3\xi'}{\omega^3}\frac{\ddot{a}}{a} + \frac{m^2}{2\omega^5}\frac{\dot{a}^2}{a^2} + \frac{m^2}{4\omega^5}\frac{\ddot{a}}{a} - \frac{5m^4}{8\omega^7}\frac{\dot{a}^2}{a^2}. \label{Wk2}
\end{align}

In this form, we immediately see that in the conformally special case $\xi=1/6$ ($\xi'=0$) this would have no divergent terms for the power spectrum. Still, according to our prescription, we should subtract the renormalized expression by subtracting this term as well. In fact, this will be necessary if we want out theory to depend continuously on its parameters.

This prescription also leads to some intriguing consequenses regarding how quantum field observables could differ from our classical expectations. For example, a quantity that is positive-definite such as $\phi^2(x)$ can, due to the subtractions, present negative expectation values. The same is true for the energy density $\mathcal{H}(x)$ -- which we had already seen in renormalization in flat space, for the Casimir Effect. Moreover, as different observables may require different orders of fundamental subtraction, they could in principle \emph{have different physical expectation values even when their classical expressions coincide for some particular value of parameters}. Indeed, there are known examples of this renormalization discrepancy; among them, we shall explore the so-called \emph{trace anomaly} (or \emph{conformal anomaly}) for the stress tensor in the next section.

In the next subsection, we shall explicitly present the application of this method to compute the stress tensor of a scalar field.

\subsection{Vacuum Energy in Curved Space: adiabatic renormalization of the Stress Tensor} \label{VECS}

Two manifest advantages of the adiabatic subtraction procedure are that (i) it is, in a sense, more physically intuitive than other procedures in its execution, since one operates subtractions directly for the observables of interest in terms of field modes (and has a quite extensive interpretation framework for spectral amplitudes in physics), and (ii) it is extremely straightforward to compute predictions, either analytically or with the aid of symbolic/numerical tools, based on its iterative structure.

Even so, for a number of observables of interest, the necessary computations may be analytically impractical and the results, prohibitively large to even display, obscuring their physical meaning and interpretation. Particularly, this is true for the expectation values of the stress tensor, which is an essential observable for the dynamical predictions of a theory, and crucially so if we wish to explore its gravitational (and cosmological) effects. Thus, in order to properly grasp the results of adiabatic renormalization for the stress tensor, and interpret qualitative and quantitative features of vacuum energy in curved spacetimes, we start by restricting our attention to the conformally special case $m=0, \, \xi = 1/6$; this will greatly simplify our computations, and allow us to explore fully analytical calculations. After those results have been calculated and discussed, we will explore a more general range of parameters with the aid of symbolical and numerical calculations in the next section.

We begin by computing the classical expression for the stress tensor of the scalar field \eqref{lagcurv}, which is found by extremizing its action \eqref{actioncurv} with respect to the metric:

\begin{align}
T_{\mu\nu} &= \frac{2}{\sqrt{-g}} \frac{\delta S}{\delta g^{\mu\nu}} \nonumber \\[6pt]
 &= \frac{1}{2} \frac{\partial \sqrt{-g}}{\partial g^{\mu\nu}} \bigl( (\nabla^\alpha \phi)(\nabla_{\!\alpha} \phi) - (m^2+\xi R)\phi^2 \bigl) + \frac{\sqrt{-g}}{2} \Bigl[ \nabla_{\!\mu} \phi \nabla_{\!\nu} \phi - \xi\phi^2 \Bigl( R_{\mu\nu} +  g^{\alpha\beta}\frac{\partial R_{\alpha\beta}}{\partial g^{\mu\nu}} \Bigl) \Bigl] \nonumber \\[6pt]
 &= \nabla_{\!\mu} \phi \nabla_{\!\nu} \phi - \tfrac{1}{2} \bigl[ \nabla^{\alpha} \phi \nabla_{\!\alpha} \phi - m^2 \phi^2 \bigl] - \xi \bigl[ ( R_{\mu\nu}- \tfrac{1}{2}g_{\mu\nu}R ) \phi^2 + \nabla_{\!\mu} \nabla_{\!\nu} \phi^2 - g_{\mu\nu} \Box \phi^2 \bigl], \label{scalarTmunu}
\end{align}
where we have used equations \eqref{varDet} and \eqref{varR}. We also stress that there will be no variations associated to first derivative terms, as $\nabla_{\mu}$ has an invariant action on scalars (\textit{i.e.}, $\nabla_{\!\mu}\phi = \partial_\mu \phi$).

Equation \eqref{scalarTmunu} then yields the trace

\begin{align}
T_{\mu}^{\;\mu} &= -(\nabla_{\!\mu} \phi)(\nabla^{\mu} \phi) + 2m^2\phi^2 + \xi R \phi^2 + 3\xi \Box \phi^2 \nonumber \\
 &= (6\xi-1) \nabla_{\!\mu} \phi \nabla^{\mu} \phi + m^2\phi^2, \label{TOper trace}
\end{align}
where we have used the dynamic equations for $\phi$ in the last line.

In this expression, if we make $m\!=\!0,\,\xi\!=\!1/6$, we recover the classical result that the trace of a conformally trivial field is null\footnote{See, \textit{e.g.}, appendix D of \cite{wald}.}: $T_\mu^{\;\,\mu}\!=\!0$. For the quantized field, however, this equation must be treated with greater care. Generally, it can be problematic to impose classical field equalities, especially constraints, directly in terms of \emph{operator identities}\footnote{For an illustrative example of this assertion in the context of gauge theories, see for instance chapter 5 of \cite{mandlshawn} and the implementation of the Lorenz gauge condition in the Gupta-Bleuler quantization method \emph{as a constraint in the Hilbert space of the theory, rather than an operator identity}.} (at the very least, there will be no \emph{a priori} guarantee that they coincide with the limits $m \rightarrow 0$, $\xi \rightarrow 1/6$ of the nontrivial theory). Instead, we expect them to be implemented for the quantized theory at the level of expected values:

\begin{align}
\braket{ \Psi \bigl| \,T_\mu^{\;\mu}(x) \bigl| \Psi}\!\!\bigl|_{\xi=\frac{1}{6},m=0} \;=\, 0. \label{Qnulltrace}
\end{align}

Indeed, equation \eqref{Qnulltrace} will hold for the formal, nonrenormalized expression of the trace. However, according to our prescription for adiabatic subtraction, we should determine the physical value of $\braket{T_\mu^{\;\mu}}$ by computing the expected value \emph{with the appropriate (4th order) adiabatic subtraction, regardless of whether the formal expectation value is nondivergent for these specific parameters}. In fact, as $T_{\mu\nu}$ is a tensorial observable, one should actually consider $g^{\mu\nu}\!\braket{T_{\mu\nu}}$ for the expected value of the trace, and the nonrenormalized expression of $\braket{T_{\mu\nu}}$ carries divergences even in this conformally special case.

With some computational effort, one may verify that, indeed, even when the appropriate adiabatic subtractions are performed, the contribution to the expected value $\braket{T_\mu^{\;\mu}}$ from the term with derivatives on \eqref{TOper trace} vanishes when one makes $\xi\rightarrow/6$. \emph{Thus, we take the informal liberty of writing}

\begin{align}
T_\mu^{\;\mu} \bigl|_{\xi=1/6} \;=\, m^2 \phi^2 \label{Ttracce}
\end{align}
\emph{directly at an operator level}.

From \eqref{Ttracce}, we obtain the formal expression for the vacuum expectation value:

\begin{align}
\braket{0| T_\mu^{\;\mu}(x) |0}\!  \bigl|_{\xi=1/6} = m^2 \braket{0|\phi^2(x)|0}.
\end{align}

However, as anticipated in last section, this does not mean that one can automatically identify the physical, renormalized expectation values:

\begin{align}
\braket{0| T_\mu^{\;\mu}(x) |0}_{phys} \not\equiv m^2 \braket{0|\phi^2(x)|0}_{phys}
\end{align}
(where we have omitted the specification $\xi=1/6$).

This is because $\phi^2$ and $T_{\mu\nu}$ generally have different types of divergences. While the former has at most quadratic divergences, and must be subtracted only up to second adiabatic order, the latter also has quartic divergences, so that we must also subtract the fourth adiabatic term. This amounts to

\begin{align}
\braket{0| T_\mu^{\;\mu}(x) |0}_{phys} = m^2 \braket{0|\phi^2(x)|0}_{phys} - \frac{m^2}{4\pi^2 a^3(t)} \int_0^\infty dk k^2 (W_k^{-1})^{(4)}(t), \label{phystrace}
\end{align}
where the last term is the so-called \emph{trace anomaly}; it emerges solely from the process of renormalization and will generally make the trace of a quantized theory differ from its classical counterpart.
 The situation is particularly interesting if we evaluate this result in the limit $m \rightarrow 0$, for which the trace classically vanishes. At first sight, both terms in \eqref{phystrace} seem to vanish in this limit, as they have a $m^2$ prefactor; however, one must look more carefully in the spectral integrals, to see if $\braket{0|\phi^2|0}_{phys}$ and $\int dk k^2 (W_k^{-1})^{(4)}$ do not entail any infrared (IR) divergences that may compensate this factor.

As of the first term, it is easy to verify that both the integrals of $(W_k^{-1})^{(0)}$ and $(W_k^{-1})^{(2)}$ do not yield any IR contributions in that limit. We show that explicitly, beginning with $(W_k^{-1})^{(0)}$:

\begin{align}
\lim_{m \rightarrow 0} \biggl\{ m^2 \!\int_0^K \!\!\! dk\, k^2 \omega_k^{-1}  \biggl\} &= \lim_{m \rightarrow 0} \biggl\{ m^2 \!\int_0^K \!\!\! dk\, k^2 m^{-1}\bigl[1 + \tfrac{k^2}{m^2} \bigl]^{-\frac{1}{2} }  \biggl\} \nonumber \\
 &= \lim_{m \rightarrow 0} \biggl\{ m^4 \!\int_0^K \!\!\! dx\, x^2 [1 + x^2 ]^{-\frac{1}{2} }  \biggl\}, \nonumber \\
 &= 0,
\end{align}
where we have made the substitution $k \rightarrow x \equiv k/m $ to take the $m$ dependence outside of the integral; we have also inserted a UV-cutoff $K$ to tame the UV divergences and focus in the IR behaviour. It is then easy to see that the $x$-integral remains finite in the IR limit, as $x \rightarrow 0$, so that the whole expression vanishes in the massless limit. 

The situation is quite similar for $(W_k^{-1})^{(2)}$. One can see from eq. \eqref{Wk2} that $m$ and $\omega$ always appear in terms with a fixed power proportion $m^i/\omega^{i+3}$, such that the relevant integral takes the form

\begin{align}
\lim_{m \rightarrow 0} \biggl\{ m^2 \!\int_0^K \!\!\! dk\, k^2 (W_k^{-1})^{(2)}  \biggl\} &= \lim_{m \rightarrow 0} \biggl\{ m^2 \!\int_0^K \!\!\! dk\, k^2 \sum_i \alpha^{(2)}_i \frac{m^i}{\omega^{i+3}}  \biggl\} \nonumber \\
 &= \lim_{m \rightarrow 0} \biggl\{ m^2 \!\int_0^K \!\!\! dk\, k^2 m^{-3}\sum_i \alpha^{(2)}_i  \bigl[1 + \tfrac{k^2}{m^2} \bigl]^{-\frac{i+3}{2} }  \biggl\} \nonumber \\
 &= \lim_{m \rightarrow 0} \biggl\{ m^2 \!\int_0^K \!\!\! dx\, x^2 \sum_i \alpha^{(2)}_i  \bigl[1 + x^2 \bigl]^{-\frac{i+3}{2} }  \biggl\} \nonumber \\
 &= 0,
\end{align}
where the $\alpha_i$ are coefficients independent of $k$ and $m$. This similarly vanishes in the massless limit, although as $m^2$ rather than as $m^4$.

Then, it is not hard to anticipate what will happen with $(W_k^{-1})^{(4)}$. Although it is laborious to compute it explicitly, it is not hard to see that all its terms will follow the same tendency, bearing the fixed power proportions $m^i/\omega^{i+5}$. Now, since this term is not UV divergent, we drop the upper cutoff $K \rightarrow \infty$ and directly show the form of its \emph{finite} contribution to the trace:

\begin{align}
\braket{0| T_\mu^{\;\mu} |0}_{phys} &= \frac{1}{4\pi^2 a^3}  \lim_{m \rightarrow 0} \biggl\{ m^2 \int_0^\infty \!\!\! dk \, k^2 (W_k^{-1})^{(4)} \biggl\} \nonumber \\
 &= \frac{1}{4\pi^2 a^3}  \lim_{m \rightarrow 0} \biggl\{ m^2 \int_0^\infty \!\!\! dk \, k^2 \sum_i \alpha^{(4)}_i \frac{m^i}{\omega^{i+5}} \biggl\} \nonumber \\
 &= \frac{1}{4\pi^2 a^3}   \int_0^\infty \!\!\! dx \, x^2 \sum_i \alpha^{(4)}_i (1+x^2)^{- \frac{i+5}{2} } \nonumber \\[4pt]
 &= \frac{1}{4\pi^2 a^3} \sum_i \alpha_i^{(4)} \frac{\sqrt{\pi}\,\Gamma(1 + \tfrac{i}{2})}{4\,\Gamma(2+\tfrac{i}{2})} ,
\end{align}
as all powers of $m$ have cancelled out in the expression in the curly brackets, yielding simply a $m$-independent integral. 

Making use of a symbolically computed expression for $(W^{-1})^{(4)}$, we may find the explicit coefficients $\alpha^{(4)}_i$ in terms of $a(t)$ in this conformally coupled case ($\xi=1/6$):

\begin{subequations}
\begin{empheq}[left=\empheqlbrace]{align}
\alpha_2^{(4)} &= -\frac{\dot{a}^4}{2a^4} - \frac{7\ddot{a}^2}{16a^2} - \frac{\ddddot{a}}{16a} - \frac{33\dot{a}^2\ddot{a}}{16a^3} - \frac{11\dot{a}\dddot{a}}{16a^2}, \\[4pt]
\alpha_4^{(4)} &= \frac{49\dot{a}^4}{8a^4} + \frac{21\ddot{a}^2}{32a^2} + \frac{35\dot{a}^2\ddot{a}}{4a^3} + \frac{7a\dddot{a}}{8a^2}, \\[4pt]
\alpha_6^{(4)} &= - \frac{231\dot{a}^4}{16a^4} - \frac{231\dot{a}^2\ddot{a}}{32a^3}, \\[4pt]
\alpha_8^{(4)} &= \frac{ 1155\dot{a}^4}{128a^4} ,
\end{empheq}
\end{subequations}
and obtain the following expression for the anomalous trace:

\begin{align}
\braket{0| T_\mu^{\;\mu} |0}_{phys} = \frac{1}{480\pi^2} \biggl[ \frac{ \ddot{a}^2}{a^2} + \frac{ \ddddot{a} }{a} -3\frac{ \dot{a}^2 \ddot{a}}{a^3} + 3\frac{ \dot{a}\dddot{a}}{a^2}  \biggl]. \label{traceTab}
\end{align} 

For completeness, we mention that this can be computed in a generally covariant form, in terms of curvature scalars. This yields (see eq. (6.144) of \cite{birrell})

\begin{align}
\braket{0| T_\mu^{\;\mu} |0}_{phys} = -\frac{1}{2880\pi^2} \bigl[ R_{\mu\nu\alpha\beta}R^{\mu\nu\alpha\beta} - R_{\alpha\beta}R^{\alpha\beta} - \Box R  \bigl]. \label{covtraceTab}
\end{align}

Once again, this entails a relatively compact result after a lengthy regularization and subtraction procedure. With \eqref{traceTab}, one can immediately compute the trace expectation value \emph{just from the knowledge of the spacetime expansion $a(t)$}. Note that this purely anomalous trace will actually be state-independent: had we considered a many-particle state $\ket{\Psi} = \ket{n_\mathbf{k_1}, n_\mathbf{k_2}, \hdots}$ with a finite energy difference with respect to the vacuum, we would have just an extra convergent term in $\braket{\phi^2(x)}$:

\begin{align}
\braket{0| \phi^2(x) |0} \longrightarrow \braket{\Psi| \phi^2(x) |\Psi} = \braket{0| \phi^2(x) |0} + \sum_\mathbf{k} 2 N_\mathbf{k} |u_\mathbf{k}(x)|^2,
\end{align}
which would yield a vanishing contribution to $\braket{T_\mu^{\;\,\mu}}$ as $m\rightarrow0$.

In what follows, we will show how one is able to compute the vacuum expectation value for the entire stress tensor $\braket{0|T^{\mu\nu}|0}_{phys}$ in a FLRW spacetime \emph{just from the knowledge of its trace} $\braket{T_\mu^{\!\mu}}_{phys}$. In fact, as this trace is state-independent, the procedure outlined here will actually allow us to compute this expectation value \emph{in any state} $\ket{\Psi}$ that obeys the FLRW symmetries, namely, spatial homogeneity and isotropy.

As a starting point, we note that this renormalized expectation value must obey the covariant conservation law:

\begin{align}
\nabla_{\!\mu} \braket{0| T^{\mu \nu} |0}_{phys} = 0. \label{covconsRenTab}
\end{align}

This is true because (i) this equality holds for the formal, unrenormalized expectation value:

\begin{align}
\nabla_{\!\mu} \braket{0| T^{\mu \nu} |0} = 0, \label{formalcon}
\end{align}
and (ii) it must hold order by order in an adiabatic expansion: 

\begin{align}
\braket{0| T^{\mu \nu} |0} \sim \braket{0| T^{\mu \nu} |0}^{(0)} + \frac{1}{T^2}\braket{0| T^{\mu \nu} |0}^{(2)} + \frac{1}{T^4}\braket{0| T^{\mu \nu} |0}^{(4)} +  \hdots 
\end{align}
so that \eqref{formalcon} can be true for \emph{any} of the spacetimes in the 1-parameter family of FLRW metrics $a_T(t)=a(t/T)$ (more concisely, so that  \eqref{formalcon} can hold for \emph{any} value of $T$). In fact, this condition is a strong motivation to define adiabatic subtractions in the way we did, subtracting \emph{all} contributions from each divergent adiabatic order, even the finite ones; otherwise these subtractions would \emph{not} generally enforce covariant conservation of the renormalized stress tensor.

Then, we resort to a conformal Killing field (see Appendix \ref{liekillcon}) of FLRW spaces, namely, the one that generates time translations:

\begin{align}
\xi^\mu = a(t) \delta^\mu_{\;0}. \label{confkilling}
\end{align}

It is not difficult to verify that this field must indeed take the form \eqref{confkilling} by starting from a generic (homogeneous and isotropic) timelike field $\xi^\mu = f(t) \delta^\mu_{\;0}$ and then solving the Killing equation $\pounds_\xi g_{\mu\nu} = \lambda(t) g_{\mu\nu}$ for the unknown functions $f$ and $\lambda$. Expanding the Lie derivative:

\begin{align}
\lambda g_{\mu\nu} &= \pounds_\xi g_{\mu\nu} \nonumber \\ 
 &\equiv 2\nabla_{\!(\mu} \xi_{\nu)} \nonumber \\ 
 &= \xi^\alpha\partial_\alpha g_{\mu\nu} + 2 g_{\alpha(\mu}\partial_{\nu)}\xi^\alpha \nonumber \\
 &= f \partial_t g_{\mu\nu} + 2 g_{0(\mu}\partial_{\nu)}f, \label{lieconftt}
\end{align}
for which we have the nontrivial time-time and space-space diagonal components $(0,0)$ and $(i,i)$:

\noindent \strut\vspace{14pt} \hspace{14pt}
\begin{minipage}[t]{0.35\linewidth}
 \begin{subequations}
 \begin{empheq}[left=\empheqlbrace]{align}
  2\dot{f} &= \lambda \\
  f.2a\dot{a} &= \lambda a^2
 \end{empheq}
 \end{subequations}
\end{minipage}%
\begin{minipage}[t]{-0.1\linewidth}
  \begin{align*}
  \\[-18pt]
  \qquad \qquad  \scalebox{1.5}{$\Rightarrow$}
  \end{align*} 
\end{minipage}%
\hspace{-4pt}
\begin{minipage}[t]{0.35\linewidth}
 \vspace{0pt}
 \begin{subequations}
 \begin{empheq}[left=\empheqlbrace ]{align}
  f &= a \\
  \lambda &= 2\dot{a} \label{lambda}
 \end{empheq}
 \end{subequations}
\end{minipage}\\
(where we have ignored a free multiplicative constant common to $f$ and $\lambda$, setting it to $1$).

Thus, using eqs. \eqref{covconsRenTab}, \eqref{lieconftt} and \eqref{lambda}, we compute the following perfect divergence:

\begin{align}
\nabla_{\!\mu} \bigl( \braket{T^{\mu \nu}} \xi_\nu \bigl) = \braket{T^{\mu \nu}} \nabla_{\!(\mu}\xi_{\nu)} = \dot{a} g_{\mu\nu} \braket{T^{\mu \nu}} = \dot{a} \braket{ T_\mu^{\;\mu} },
\end{align}
where we have explored the symmetry of the stress tensor, $T^{\mu\nu} =T^{(\mu\nu)}$, to symmetrize the derivative in the second term.

Now, integrating this divergence in a 4-volume delimited by the isotropic Cauchy surfaces $\Sigma_{t_1}$ and $\Sigma_{t_2}$, we may use Gauss's theorem to obtain

\begin{align}
\int d^4\!x \, \sqrt{-g} \dot{a}\braket{T_\mu^{\;\mu}} &= \int d^4\!x \, \sqrt{-g} \, \nabla_{\!\mu} \bigl( \braket{T^{\mu \nu}}\! \xi_\nu \bigl) \nonumber \\
 &= \int_{\Sigma_{t_2}} \!\! d^3\!\mathbf{x} \sqrt{-g_{\Sigma_{t_2}}} \braket{T^{0 \nu}}\! \xi_\nu - \int_{\Sigma_{t_1}} \!\! d^3\!\mathbf{x} \sqrt{-g_{\Sigma_{t_1}}} \braket{T^{0 \nu}}\! \xi_\nu. \label{gausstrace}
\end{align}

In virtue of spatial homogeneity, we need not to evaluate these spatial integrals; we can simply pick any space point $\mathbf{x}$ and evaluate the integrands directly\footnote{
 Or equivalently, we can carry a (trivially homogeneous) integration in a finite coordinate volume $V$ and then divide both sides by $V$. Note that homogeneity will cause the total contribution from the spatial boundaries to be null.}.
  Since the Jacobians $\sqrt{-g}$ and $\sqrt{-g_{\Sigma_t}}$ actually coincide for spatially flat FLRW spaces in proper-time Cartesian coordinates,  $\sqrt{-g}= a^3(t) = \sqrt{-g_{\Sigma_t}}$, we obtain simply

\begin{align}
\int_{t_1}^{t_2}\!\!\! dt \, a^3\!(t) \dot{a}(t) \braket{ T_{\,\mu}^{\:\mu} }\!(t) = a^3\!(t_2)\braket{ T^{00}}\!(t_2)\, a(t_2) - a^3\!(t_1)\braket{ T^{00}}\!(t_1)\, a(t_1) . 
\end{align}

This yields an integral expression for the energy density $\braket{T^{00}}=\braket{T_{00}}=\braket{T_{0}^{\;0}}$ as a function of the trace $\braket{T_{\mu}^{\;\mu}}$:

\begin{align}
\bigl[ a^4\!(t)\langle T_{\,0}^{\:0} \rangle(t) \bigl]^{t_2}_{t_1} = \!\!\int_{t_1}^{t_2}\!\!\! dt \, \dot{a}(t)a^3\!(t) \langle T_{\,\mu}^{\:\mu} \rangle(t) . \label{T00t2t1}
\end{align}

Then, using eq. \eqref{traceTab} for the conformally special case:

\begin{align}
\int_{t_1}^{t_2}\!\!\! dt \, a^3\!(t) \dot{a}(t) \braket{ T_{\,\mu}^{\:\mu} }\!(t) &= \frac{1}{480\pi^2} \int_{t_1}^{t_2} \!\!\! dt \bigl[ a\dot{a}\ddot{a}^2 + a^2\dot{a}a^{(4)} - 3\dot{a}^3\ddot{a} + 3 a\dot{a}^2a^{(3)} \bigl] \nonumber \\
 &= g(t_2) - g(t_1), \label{inta3Tmumu}
\end{align}
where $a^{(3)}\equiv\dddot{a}$ and $a^{(4)}\equiv\ddddot{a}$.

Although it is a little intricate to manually compute a primitive $g(t)$ for the integrand in \eqref{inta3Tmumu}, one can easily verify that one particular solution is given by

\begin{align}
g(t) \equiv \frac{1}{480\pi^2} \bigl( a^2\dot{a} \dddot{a} + a \dot{a}^2 \ddot{a} - \tfrac{1}{2}a^2 \ddot{a}^2 - \dot{a}^4 \bigl).
\end{align}

Equation \eqref{T00t2t1} then entails

\begin{align}
a^4(t) \braket{T_0^{\;0} }\!(t) = g(t) + E \qquad \Rightarrow \qquad  \braket{T_0^{\;0} }\!(t) = \frac{g(t)}{a^4(t)} + \frac{E}{a^4(t)}, \label{ren vac+mat}
\end{align}
where $E$ is simply an integration constant. From the classical behaviour of the energy density of a noninteracting massless field in a FLRW spacetime (which will be discussed in further detail in the next chapter, with emphasis on the electromagnetic field), we can immediately identify this last term as a contribution from an isotropic particle distribution, whose energy density decays as $\rho \propto a^{-4}$. Thus, we identify the remaining term as that due to vacuum energy. It reads

\begin{align}
\rho_0 &= \langle 0| T_{\,0}^{\;\,0} |0\rangle = \frac{1}{480\pi^2} \biggl[ -\frac{ \dot{a}^4}{a^4} + \frac{ \dot{a} \dddot{a}}{a^2} +\frac{ \dot{a}^2 \ddot{a}}{a^3} - \frac{1}{2}\frac{ \ddot{a}^2}{a^2}  \biggl]. \label{CTVrho}
\end{align}

In virtue of spatial isotropy, the spatial components $\braket{T_i^{\;j}}$ must be diagonal and equal (see section \ref{FLRW} in the next chapter for more details on this argument), which allows us to immediately compute the space-space (pressure) components $\braket{T_i^{\;\,i}}$ from $\braket{T_\mu^{\;\,\mu}}$ and $\braket{T_0^{\;\,0}}$:

\begin{align}
\braket{T_\mu^{\;\mu}} = \braket{T_0^{\;0}} + 3\braket{T_i^{\;i}} \quad \Rightarrow \quad \braket{T_i^{\;i}} = \tfrac{1}{3} \bigl( \braket{T_\mu^{\;\mu}} - \braket{T_0^{\;0}} \bigl) 
\end{align}
(where we are not carrying a sum in the spatial index $i$; its repetition just stands for a diagonal component). This gives us the vacuum pressure

\begin{align}
p_0 \equiv -\langle0| T_{\,i}^{\;\,i}|0\rangle &= \tfrac{1}{3} \bigl( \rho - \braket{0|T_\mu^{\;\,\mu}|0}\! \bigl) \nonumber \\
 &=- \frac{1}{3}\frac{1}{480\pi^2} \biggl[ \frac{ \dot{a}^4}{a^4} + 2\frac{ \dot{a} \dddot{a}}{a^2} - 4\frac{ \dot{a}^2 \ddot{a}}{a^3} + \frac{ 3}{2}\frac{ \ddot{a}^2}{a^2} + \frac{ \ddddot{a}}{a} \biggl]. \label{CTVP}
\end{align}

Having obtained these expressions for a conformally trivial case, we now take a moment to interpret them, and address a few pressing questions. What do these results tell us about the properties and dynamical behaviour of vacuum energy? Are they in any way meaningful in respect to the expansion of our own universe? Also, can they be extended beyond the conformally trivial case and into a more general range of parameters $(m,\xi)$? (If so, how?)

First of all, we emphasize that the relations \eqref{CTVrho} and \eqref{CTVP} were computed for a \emph{fixed} background metric, so that they will not be generally compatible with a \emph{dynamical} expansion \emph{driven solely (or mainly) by} vacuum energy. Nevertheless, they should still be useful to consistently calculate the vacuum energy in an expansion dominated by other forms of matter and energy (and eventually even compute its gravitational backreaction perturbatively).

With these caveats duely noted, we proceed to analyze the properties of the vacuum energy we have calculated. From eqs \eqref{CTVrho} and \eqref{CTVP}, we see that both $\rho_0$ and $p_0$ can have either sign, depending on the `kinematics' of the scalar factor $a(t)$. To analyze these more concretely, we evaluate them for a few simple examples of cosmological relevance. First, let us consider a power-law expansion, $a(t) \propto t^\lambda$. In this case, it is easy to see that both $\rho_0$ and $p_0$ will decay as $t^{-4}$ ($\propto a^{-\frac{4}{\lambda}}$); the exact expressions read

\begin{subequations}
 \begin{empheq}[left=\empheqlbrace, right = {\qquad .}]{align}
 \rho_0 &= \frac{\lambda^2(3-6\lambda+\lambda^2)}{960 \pi^2} t^{-4}  \\
 p_0 &= \frac{\lambda(4-11\lambda + \frac{22}{3}\lambda^2 +\lambda^3)}{960\pi^2} t^{-4}
 \end{empheq}
\end{subequations}

These can be either negative or positive, depending on the value of $\lambda$ (since they are both degree 4 polynomials, they will have 4 roots where they may switch in signs). We plot them rescaled by $t^4$ as a function of $\lambda$:

\begin{figure}[H]
\centering
\includegraphics[width=0.45\linewidth]{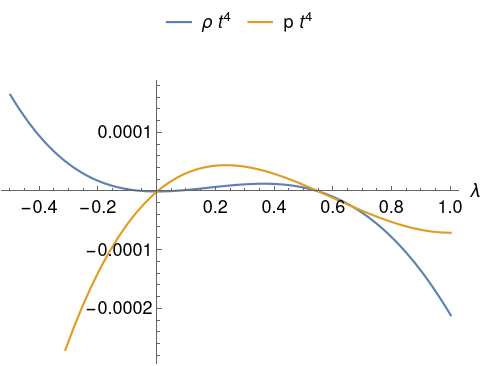}
\caption{Plots for the time independent scaled energy density $\rho_0t^4$ and pressure $p_0t^4$ arising from the vacuum renormalized values in a power-law ($a\propto t^\lambda$) FLRW universe. Note that, in the expanding region $\lambda>0$, $\rho_0$ and $p_0$ can be both positive and negative, and they have matching signs (corresponding to a positive equation of state $w_0$) for $\lambda<\frac{4}{3}$; for $\lambda>\frac{4}{3}$ (region not shown in the plot) $p_0$ turns positive again, yielding a negative $w_0$. \\ Source: By the author. }
\label{CT_rho-p(lambda)}
\end{figure}

In the context of cosmology, we define the equation of state for a given species of matter as the relation between their pressure and their energy density, in the form $p = w\rho$ (see, chapter \ref{cosmology} for more details). We see that the vacuum equation of state here is simply a constant, whose value depends on $\lambda$:

\begin{align}
w_0 = \frac{p_0}{\rho_0} = \frac{4-11\lambda+ \frac{22}{3}\lambda^2 + \lambda^3}{\lambda(3-6\lambda+\lambda^2)}.
\end{align}

Again, we emphasize this is generally not consistent with an expansion driven solely by vacuum energy. As we will see in the next chapter, the relation between $w$ and $\lambda$ that one obtains for a single species source (\textit{i.e.}, a source with a fixed equation of state) in the Friedman Equation \eqref{FrEq1} supplemented by \eqref{conservation} is \eqref{w(l)}: 

\begin{align}
w = \frac{2-3\lambda}{3\lambda}.
\end{align} 

Two particularly important values of $\lambda$ in cosmology are $2/3$ and $1/2$, corresponding to expansions driven by cold matter (``dust'') and radiation, respectively. The corresponding vacuum equation in these backgrounds would be:

\vspace{20pt}\hspace{-0.03\linewidth}
\begin{minipage}{0.45\linewidth}
\begin{subequations}
 \begin{empheq}[left={ \lambda= \displaystyle\frac{1}{2}\; \empheqlbrace } ]{align}
  \rho_0(t) &= \frac{1}{15360\pi^2} \;t^{-4} \\[4pt]
  p_0(t) &= \frac{1}{9216\pi^2} \;t^{-4} \\[4pt]
  w_0(t) &= \frac{5}{3}
 \end{empheq}
\end{subequations}
\end{minipage}\hfill%
\begin{minipage}{0.45\linewidth}
\begin{subequations}
 \begin{empheq}[left={ \lambda= \displaystyle\frac{2}{3} \; \empheqlbrace } ]{align}
  \rho_0(t) &= -\frac{1}{3888\pi^2} \;t^{-4} \\[4pt]
  p_0(t) &= -\frac{1}{3888\pi^2} \;t^{-4} \\[4pt]
  w_0(t) &= 1
 \end{empheq}
\end{subequations}
\end{minipage}

In both cases, we have a positive equation of state, $w_0>0$, although the vacuum energy density (and pressure) do alternate their sign between these two values, being positive for $\lambda=1/2$ and negative for $\lambda=2/3$. Curiously, the latter corresponds to the only nontrivial case $\rho_0, p_0 \neq 0$ for which both quantities coincide: $\rho_0 = p_0$.

Finally, we consider the case of an exponential expansion, $a \propto e^{Ht}$. This is relevant for a universe dominated by a form of energy which behaves like a cosmological constant, such as our current universe, dominated by Dark Energy, or many primordial inflationary scenarios. In this case we obtain simply constant energy densities and pressures, in the form

\begin{subequations} \label{CT vac ener}
 \begin{empheq}[left={ \empheqlbrace } ]{align}
  \rho_0(t) &= \frac{H^4}{960\pi^2} = cte, \\
  p_0(t) &= -\frac{H^4}{960\pi^2 } = cte, \\
  w_0(t) &= -1.
 \end{empheq}
\end{subequations}

This yields precisely a form of energy behaving like a cosmological constant!

\begin{align}
\braket{0|T_{\mu\nu}|0} = \Lambda g_{\mu\nu}, \qquad \Lambda = \frac{H^4}{960\pi^2}. \label{CosConsVE}
\end{align}

Thus, precisely for an exponential expansion, we find a form of vacuum energy which is \emph{qualitatively} self-consistent with the expansion that it would generate. However, we emphasize that the \emph{quantitative} consistence is still not generally satisfied. The value of $\Lambda$ that is obtained in \eqref{CosConsVE} is \emph{generally} not the one that would produce an expansion rate $H$ in Einstein's Equations; the latter would be proportional to $H^2$. Writing it as an energy contribution (i.e., as a term in $T_{\mu\nu}$, as in eq. \eqref{CosConsVE} ) in the RHS of Einstein's equations, and recovering the constants $G,\hbar, c$ in our equations, we can compare both relations between $H$ and $\Lambda$:

\begin{align}
H^2 = \frac{8\pi G}{3c^2}\Lambda; \qquad \qquad H^4 = \frac{960\pi^2 c^3}{\hbar} \Lambda, \label{2 Lambdas}
\end{align}

If we wish to find out for which value of $\Lambda$ (i.e., for which value of vacuum energy density) those would match, we square the first equation and substitute in the second, yielding

\begin{align}
\biggl( \frac{8\pi G}{3c^2} \biggl)^{\!\!2} \Lambda^2 = \frac{960\pi^2 c^2}{\hbar} \qquad \Rightarrow \qquad \Lambda = 45 \frac{c^7}{\hbar G} = 45 \rho_p , \label{self-consistent Lambda}
\end{align}
which is 45 times larger than Planck energy density! Thus, we see that what would be a quantitatively self-consistent case can no longer be described in terms of classical spacetime ( \emph{for these particular values of field parameters}).

As of the last question, of whether (and how) we can extend our results beyond conformally trivial case, the answer is yes, but the calculations will be considerably more complex, and it is very difficult to avoid symbolical and numerical calculations. Although we will no longer have analytical results in these cases, there are a few characteristics that we would like to anticipate: (i) these basic features that the renormalized energy and pressure can have either sign remain valid, and, particularly, for an exponential expansion we \emph{always} obtain a vacuum energy with an equation of state $p_0 = -\rho_0$; (ii) all the deductions we have made from eq \eqref{covconsRenTab} to \eqref{T00t2t1} remain valid for more general $m$ and $\xi$. The fundamental difference is that the trace appearing in the RHS of \eqref{T00t2t1} \emph{will no longer be purely anomalous} -- one must generally compute the full expression of the tensor trace, which will include subtractions of 0th and 2nd adiabatic orders, as well as a contribution stemming from the \emph{exact} field modes $|h_\mathbf{k}|^2$. Thus, in these cases, one must either work in very special FLRW for which known analytical solutions exist\footnote{
 To this author's knowledge, such solutions are still quite scarce in the literature in the massive case. For solutions in the massless, \emph{minimally} coupled case ($m, \xi = 0$), see \cite{parkerIR}. Also, there is a recent analytic treatment for massless nonconformally coupled ($m=0, \, \xi \not\equiv 1/6$) fields in general FLRW spaces given in \cite{yang}. However, in the referred work, the authors set $m=0$ \emph{a priori} for the quantized field, which fails to account for a term $W_k^{(4)}$ that remains finite in the $m \rightarrow 0$ limit and thus results in a null trace anomaly in the conformally special ($\xi=1/6$) case.},
  or work directly with numerical ones. In the next section, we shall take the first approach and carry a detailed analysis of the renormalized stress tensor in exponentially expanding (de Sitter) spacetimes, which not only are more tractable but also happen to be a case of high interest in inflationary cosmology.

\subsection{Renormalization in de Sitter Spacetimes: analyzing the power spectrum and the stress tensor in the Bunch-Davies vacuum} \label{deSitter}

Now that we have developed the basic procedures of adiabatic subtraction and some of its applications in general FLRW spacetimes, we will specialize our approach to a more specific class of spacetimes: de Sitter spaces. De Sitter spaces are a class of curved, yet maximally symmetric spacetimes, corresponding to solutions of the Einstein Equations with a positive cosmological constant (or a spacetime homogeneous form of energy behaving like a positive cosmological constant) and no other types of matter or energy, being therefore of high interest to study inflationary scenarios dominated by this particular type of vacuum energy. These will thus provide us with a nontrivial, yet tractable backgrounds to analyze our quantized fields, and allow us to obtain more results through the procedure of adiabatic subtraction.

Then, before we proceed to our field analysis, we make a brief digression about de Sitter spaces\footnote{ For a more detailed analysis of these spaces, their many different coordinate systems and their role in field theory, see \textit{e.g.} section 5.2 of \cite{hawkellis} and section 5.4 of \cite{birrell}.}.
 A very convenient way to visualize these spaces is by considering a 4-dimensional hyperboloid embedded in a 5-dimensional flat Lorentzian space. If this embedding space is covered with Cartesian coordinates $(T,X,Y,Z,W)$, such that its line element is:

\begin{align}
dS^2 = dT^2 - dX^2 - dY^2 - dZ^2 - dW^2,
\end{align}
the hypersurface that represents a de Sitter space can be written by the equation

\begin{align}
T^2 - X^2 - Y^2 - Z^2 - W^2 = -H^{-2}.
\end{align}

Just like Minkowski spaces, de Sitter spaces have the maximal number of Killing fields (10, in our 4-dimensional case), and there are a large number of convenient choices of coordinates that emphasize different symmetries. Particularly, just as we can cover a portion of Minkowski spacetime (which is obviously stationary) with coordinates that give it a form of a hyperbolic FLRW space\footnote{This is known in the literature as the Milne universe. See \textit{e.g.} section 5.3 of \cite{birrell}.}
 with $a(t) \propto t$, we can cover half of de Sitter space with a coordinate system $(t,x,y,z)$ that makes it look like an exponentially expanding FLRW space:
 
\begin{subequations}
  \begin{empheq}[left=\empheqlbrace, right={\quad ,}]{align}
   T &= H^{-1}\sinh(Ht) + \tfrac{1}{2}H e^{Ht}(x^2+y^2+z^2) \\
   W &= H^{-1}\cosh(Ht) - \tfrac{1}{2}H e^{Ht}(x^2+y^2+z^2) \\
   X &= xe^{Ht}, \qquad Y = ye^{Ht}, \qquad Z = ze^{Ht} 
  \end{empheq}
\end{subequations}
where $(t,x,y,z)$ all range from $-\infty$ to $\infty$, covering half of the hyperboloid with $T+W>0$.

In these coordinates, the line element reads

\begin{align}
ds^2 = dt^2 - e^{2Ht}(dx^2+dy^2+dz^2).
\end{align}

Then, summarizing the spatial coordinates as $\mathbf{x} = (x,y,z)$, we write this compactly as

\begin{align}
ds^2 = dt^2 - e^{2Ht} d\mathbf{x}^2 = (H\eta)^{-2} \bigl( d\eta^2 - d\mathbf{x}^2 \bigl),
\end{align}
where we have given the expression in conformal time $\eta$; in this case, it can be written as a function of $t$ simply as

\begin{align}
\eta = \int^t e^{-Ht'}dt' = -H^{-1}e^{-Ht}.
\end{align}

At this point, we emphasize that although an \emph{eternally} inflating universe (both to the past and to the future) that scales \emph{exactly} exponentially can be analytically extended in a full de Sitter space, a space that is only approximately exponentially expanding for a finite time period \emph{does not have all de Sitter symmetries} (so that it cannot be extended in a full de Sitter space) and is preferrably represented by nonstationary, exponentially expanding coordinates\footnote{
 The situation is somewhat similar to the Schwarzschild spacetime, which allows for the Kruskal extension only for eternal Black Holes; for a Black Hole that forms from the collapse of matter, only a portion of an approximate Kruskal space makes physical sense (see \textit{e.g.} chapter 6 of \cite{wald}).}.

For this spacetime, all curvature tensors are quite simple, as they are highly constrained by symmetry. The only independent quantity is the curvature scalar $R=12H^2=cte$\footnote{
 The remaning curvature tensors are given simply by $R$ and appropriate combinations of $g_{\mu\nu}$. We have: $R_{\mu\nu} = 3H^2g_{\mu\nu}$ and $R_{\mu\nu\alpha\beta} = H^2(g_{\mu\alpha}g_{\nu\beta} - g_{\nu\alpha}g_{\mu\beta})$.}.
 (We give a more complete account of the computation of curvature in FLRW spacetimes in section \ref{FLRW}; in particular, the reader can easily verify that this result follows from \eqref{FLRWscalar}.) The fact that $H$ is a constant turns out to yield relatively simple dynamical equations for our scalar field \eqref{lagcurv}:

\begin{align}
\partial^2_t \phi + 3H\partial_t \phi - e^{-2Ht}\nabla^2 \phi + M^2 \phi = 0, \label{PTdSfe}
\end{align}
where we have defined a new, constant mass parameter $M^2 = m^2 + 12\xi H^2$.

We already know that a particularly convenient mode decomposition in FLRW spaces in given by \eqref{FLRWptModes}. In a de Sitter space, it reads:

\begin{align}
f_{\mathbf{k}}(x) = \frac{e^{i\mathbf{k}\!\cdot\!\mathbf{x}}}{\sqrt{V}} e^{-\frac{3}{2}Ht} h_{\mathbf{k}}(t). \label{dSptModes}
\end{align}

Once again, this results in time-dependent harmonic oscillator \eqref{tdho} for $h_k(t)$. To solve this equation exactly in de Sitter spaces, it is convenient to perform a change in variables of the form $t \rightarrow v \equiv kH^{-1}e^{-Ht} = -k\eta$ \footnote{ We warn here that some authors also use the variable $u=-v$, which is an increasing function of time. This convention leads to switch of roles of the solutions $H_\nu^{(1)}$ and $H^{(2)}_\nu$ below in respect to the adiabatic condition.}.
 Carrying this substitution through, it is straighforward to verify that we arrive at a Bessel equation:

\begin{align}
v^2 \frac{d^2 h_k}{dv^2} + v \frac{dh_k}{dv} + (v^2-\nu^2)h_k =0, \label{besseleq}
\end{align}
where we have defined

\begin{align}
\nu \equiv \left( \frac{9}{4} - \frac{M^2}{H^2} \right)^{\!\!\sfrac{1}{2}} = \left( \frac{9}{4} - \frac{m^2}{H^2} - 12\xi \right)^{\!\!\frac{1}{2}}.
\end{align}

The general solutions to these equations may then be immediatly written in terms of known special functions (see, for example \cite{arfken}). A particularly convenient basis to decompose them is given by the Hankel functions $H^{(1)}_\nu$ and $H^{(2)}_\nu$:

\begin{align}
h_k(t) = \sqrt{\frac{\pi}{2H}} \bigl( E(k)H_\nu^{(2)}(v) + F(k)H_\nu^{(1)}(v) \bigl), \label{generalHankel}
\end{align}
where we have already included a normalization factor for later convenience, and $E(k),F(k)$ are numerical factors  necessary to make a general linear combination of the two solutions for each value of $k$ (we keep a $k$ dependence here because these factors can in principle be fixed to different values for different modes $h_k$ when we impose the adiabatic condition below) \footnote{Bear in mind that the wave vector $\mathbf{k}$ and the (proper-time) variable $t$ in the ODE \eqref{tdho} are independent variables, so that these factors are just constants in the field equations \eqref{PTdSfe}. Similarly, they should not be considered variables in eq \eqref{besseleq}, even though $v$ was defined proportionally to $k$; these factors were just included in the general solutions $h_k$ as \emph{coefficients} of the equation solutions.}. 

Having obtained this general solution, we would like to fix the factors $E(k)$ and $F(k)$ appropriately to obtain a subset of these exact field modes $\{f_\mathbf{k} \}$ \textit{which obey the adiabatic condition (i.e. whose asymptotic behaviour corresponds to positive-frequency modes)}. Since we are dealing with a spacetime that is dynamical \emph{at all times} (i.e. that has no asymptotically static regions) and whose dynamics is governed by a single parameter $H$, a seemingly natural way to investigate the adiabatic limit would be to take $H \rightarrow 0$, for which we approach a static (Minkowski) spacetime. However, it turns out that a much more convenient way to analyze this limit is simply by keeping $H$ fixed and look at the UV ($k \rightarrow \infty$) behaviour of modes. If we take $k \gg m, H$, such that frequency is approximately just $\omega(t)\approx k/a(t)$, the adiabatic condition reads

\begin{align}
h_k \sim (2ke^{-Ht})^{-1/2} \exp\bigl(-i \textstyle{\int^t} ke^{-Ht'} dt'\bigl) = (2Hv)^{-1/2}e^{+iv}, \label{dSadiabatic}
\end{align}
where we have ignored a global phase factor that emerges in the indefinite integral in the exponent.

We then wish to match the adiabatic form \eqref{dSadiabatic} with the UV-limit of \eqref{generalHankel}. To evaluate the latter, we make use of the asymptotic form of the Hankel functions for large $v$ (see \cite{gradstein}, p. 920):

\begin{align}
H_\alpha^{(1)}(v) \longrightarrow \sqrt{\frac{2}{\pi}}v^{-1/2}e^{i(v+\theta(\alpha) )},
\qquad
H_\alpha^{(2)}(v) \longrightarrow \sqrt{\frac{2}{\pi}}v^{-1/2}e^{-i(v+\theta(\alpha) )},
\end{align}
where $\theta(\alpha)= \frac{\pi}{4} + \frac{\pi \alpha}{2}$ is merely a global (spacetime-indepedent) phase factor.

Comparing these with \eqref{dSadiabatic}, and noting the normalization factor that we have included in \eqref{generalHankel}, we immediately find that, in the UV-limit:

\begin{align}
\qquad \qquad E(k) \rightarrow 0, \qquad \quad F(k) \rightarrow 1, \qquad \qquad \text{when}\; k \rightarrow \infty. \label{asymptoticEF}
\end{align}

Now, to extrapolate this condition to any frequencies, we shall make use of the symmetries in de Sitter spaces. These spaces will obviously have the 6 FLRW symmetries corresponding to spatial translations and rotations, of which we have already made use in our mode decomposition. Further, we shall use a symmetry associated with a time translation, which, in our coordinate system, takes the form\footnote{
As with any spacetime transformations, one can either take the ``active'' perspective, conceiving this as an actual spacetime transformation, or the ``passive'' one, conceiving it merely as a change in coordinates covering spacetime.}

\begin{subequations}\label{tIsometry}
   \begin{empheq}[left=\empheqlbrace, right={\quad .}]{align}
   t &\rightarrow t' = t + t_0 \\
   \mathbf{x} & \rightarrow \mathbf{x'} = \mathbf{x}e^{-Ht_0} 
   \end{empheq}
\end{subequations}
We also define a transformed wave vector, $\mathbf{k'}=\mathbf{k}e^{Ht_0}$, such that

\begin{align}
\frac{\mathbf{k'}}{a(t')} = \frac{\mathbf{k}}{a(t)} \qquad \text{and} \qquad \mathbf{k'}\!\cdot\!\mathbf{x'} =\mathbf{k} \!\cdot\! \mathbf{x}.
\end{align}

Then, we write the transformed modes:

\begin{align}
h_{k'}(t') = \sqrt{\frac{\pi}{2H}} \bigl( E(k')H_\nu^{(2)}(v) + F(k')H_\nu^{(1)}(v) \bigl). \label{generalHankelmodified}
\end{align}

Note that \eqref{generalHankelmodified} only differ from \eqref{generalHankel} in the factors $E(k), F(k) \longrightarrow E(k'), F(k')$. We then argue that a priviledged family of modes in de Sitter spacetime, in respect to which we should define a vacuum state, should be one that is invariant under these symmetries as well. That is, it should obey

\begin{align}
f_{k'}(x') = f_k(x) \qquad \Rightarrow \qquad h_{k'}(t') = h_k(t).
\end{align}

In this case, the asymptotic form \eqref{asymptoticEF} will imply that

\begin{align}
\qquad \qquad E(k) = 0, \qquad \quad F(k)=1, \qquad \quad \forall k.
\end{align}

We then have the solutions that match the adiabatic condition (the asymptotically ``positive-frequency'' solutions):

\begin{align}
h_k(t) &= \sqrt{\frac{\pi}{2H}}H^{(1)}_\nu(v) = \sqrt{\frac{\pi}{2H}}H^{(1)}_\nu\Bigl( \frac{k}{H}e^{-Ht} \Bigl) \\[8pt]
 \Rightarrow \qquad f_\mathbf{k}(x) &= \sqrt{\frac{\pi}{2H}} \frac{e^{i\mathbf{k}\!\cdot\!\mathbf{x} }}{\sqrt{2(2\pi)^3 e^{3Ht} }} H^{(1)}_\nu\Bigl( \frac{k}{H}e^{-Ht} \Bigl). \label{de Sitter Adiabatic modes}
\end{align}

Then, quantizing the field through the usual mode expansions \eqref{quacurfi} in $\{f_\mathbf{k}, f_\mathbf{k}^*\}$, we can define a vacuum state $\ket{0}$ associated to them. This is known in the literature as the \emph{Bunch-Davies vacuum}. Besides the usual UV divergences, ubiquitously present in QFT, this vacuum state is known to suffer from infrared divergences in the field amplitudes $\braket{\phi^2}$ and the stress tensor $\braket{T_{\mu\nu}}$ in the minimally coupled, massless case. These divergences emerge due to the higher-order singularities in the Hankel functions $H_\nu(x)$ as $x \rightarrow 0$ for $\nu \geq \frac{3}{2}\,$; for the same reason, it can be troublesome to evaluate parameters for which $M^2<0$ \footnote{
  One could argue that a $M^2<0$ case is in itself pathological, due to vacuum instabilities; however, this will actually be a relevant regime for stable interacting theories with local maxima in their potentials, as we shall see in the next chapter.}
  with this vacuum state. (Surprisingly, it turns out that the adiabatic subtraction procedure, designed to eliminate UV divergences, also cancels the IR divergences in this $M^2=0$ ($\nu=3/2$) case\footnote{
  See section 2.10 \cite{parker}. For a more general treatment of vacuum states and IR divergences in FLRW spaces, including the power-law and exponential cases, see \cite{parkerIR}. See also \cite{allen, allenfollaci} for a rigorous and detailed account of vacuum states in de Sitter spacetime based in its symmetry groups.}; 
  we note, however, that such subtracted divergences can still be a delicate matter in a numerical treatment, both in the IR and in the UV.)$\,$ Notwithstanding, the modes \eqref{de Sitter Adiabatic modes} and the Bunch-Davies vacuum will constitute the basis of our analysis of field theory in de Sitter spaces.

In what follows, we will consider a scalar field with a Lagrangian \eqref{lagcurv} and carry a numerical analysis of two crucial renormalized observables, namely, the quadratic field amplitudes $\braket{\phi^2}$ and the stress tensor varying the parameters $m$ and $\xi$ through a region of parameters.

A good starting point for our renormalization analysis are the expectation values of the field amplitudes $\braket{0| \phi^2(x) |0}$. This observable has a sufficient simple form for us to take a closer look at its spectral expansion and consider adiabatic subtractions to different orders. First, let us consider its formal, unsubtracted expansion:

\begin{align}
\braket{0|\phi^2(x)|0} = \frac{1}{4\pi^2 a^3(t)} \int\limits_0^\infty dk\, k^2 |h_k(t)|^2 \equiv  \int\limits_0^\infty \frac{dk}{k} \mathcal{P}_0(k,t), \label{barepowerspectrum2}
\end{align}
where we have defined the \emph{power spectrum} $\mathcal{P}(k,t)$ (and we use the subscript $0$ in \eqref{barepowerspectrum2} to emphasize that it refers to the unsubtracted value). In terms of our field modes, it reads

\begin{align}
\mathcal{P}_0(k,t) = \; \frac{k^3 |h_k(t)|^2 }{4\pi^2 a^3(t)} \; =  \frac{ k^3e^{-3Ht} }{8\pi H}  \; \biggl| H^{(1)}_\nu\Bigl( \frac{k}{H}e^{-Ht} \Bigl) \biggl|^2.
\end{align}

In terms of the power spectrum, it is easy to see that the integral \eqref{barepowerspectrum2} will diverge in the IR ($k \rightarrow 0$) whenever $\mathcal{P}$ is nonvanishing in this lower limit (and, generally, it will diverge in the UV). We plot the form of this power spectrum at a fixed time $t=0$ for a massive, minimally coupled field (Figure \ref{UnsubPS}) :

\begin{figure}[H]
\centering
\includegraphics[width=0.55\linewidth]{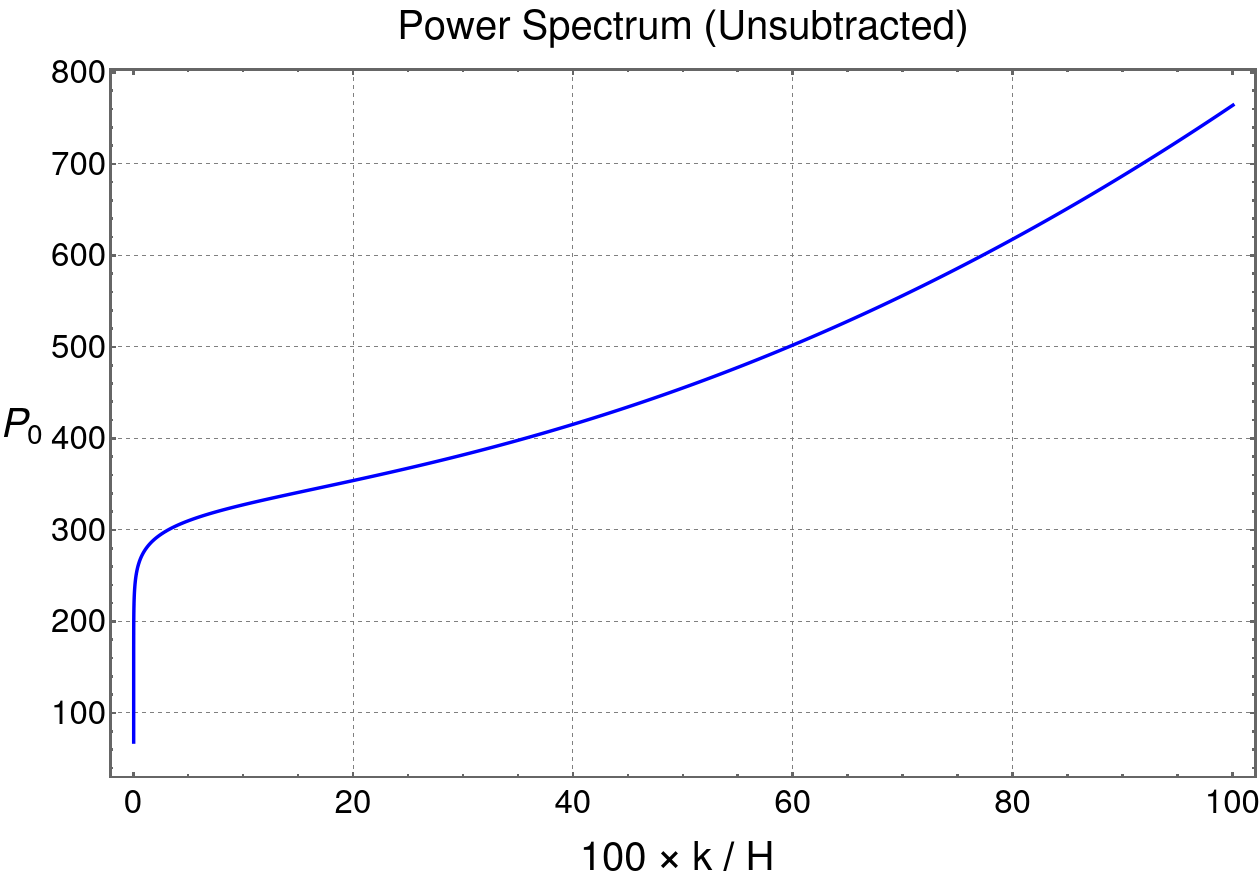}
\caption{Unsubtracted power spectrum for a minimally coupled ($\xi=0$) field with mass $m^2=0.1H^2$. In this case $\nu = [\frac{9}{4}-\frac{1}{10}]^\frac{1}{2} < \frac{3}{2}$, such that $\mathcal{P}_0$ is UV-divergent but it vanishes as $k \rightarrow 0$. \\ Source: By the author. }
\label{UnsubPS}
\end{figure}

This can be seen to yield a divergent expectation value for $\braket{\phi^2}$. According to our adiabatic subtraction prescription, it should be subtracted \emph{only} up to second adiabatic order to render the corresponding finite physical result. By carrying this adiabatic subtraction, we obtain:

\begin{figure}[H]
\centering
\includegraphics[width=0.55\linewidth]{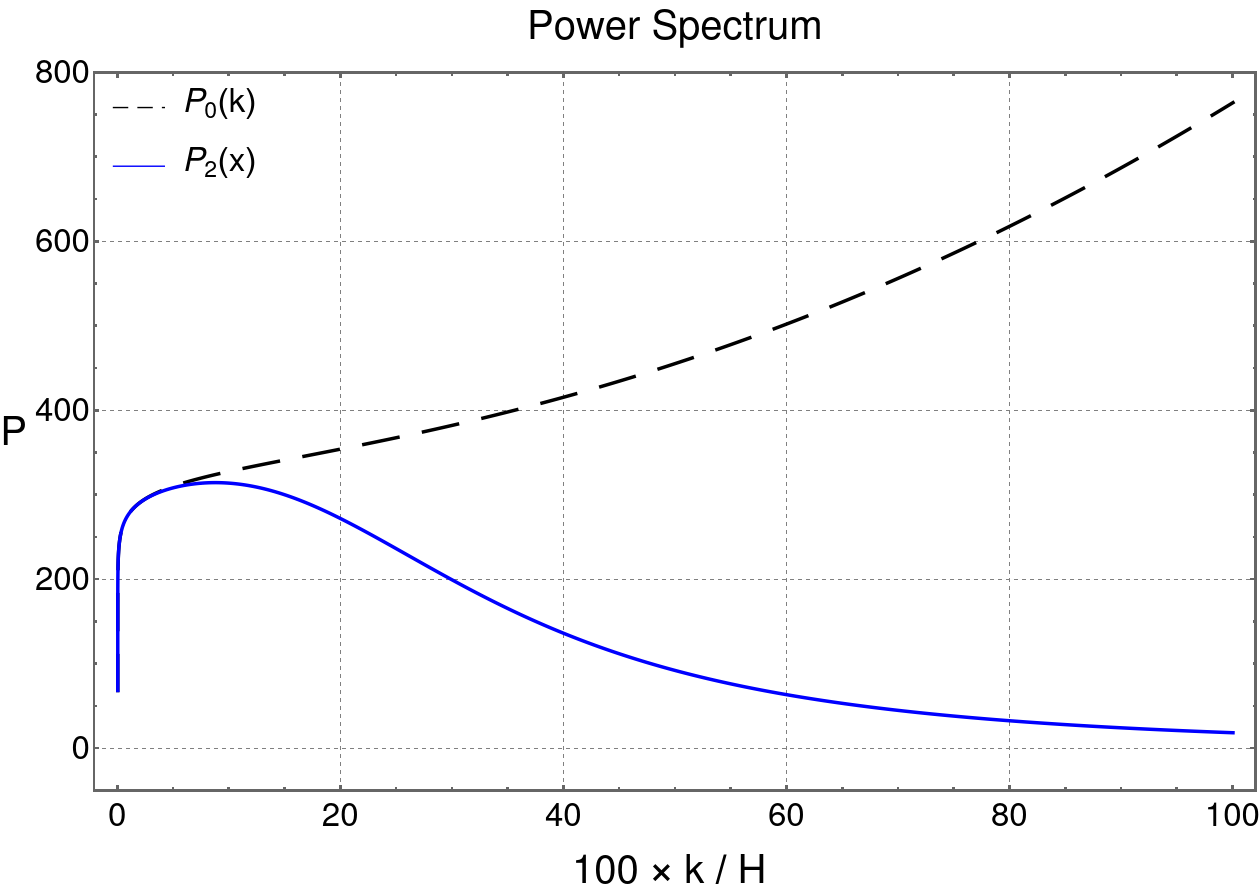}
\caption{Unsubtracted (black dashed line) and 2nd-order-subtracted (Blue filled line) Power Spectra for $\xi=0$ and $m^2=0.1H^2$. While the former is UV-divergent, the latter yields an integrable spectrum, which is associated with the renormalized observable $\braket{\phi^2(x)}_{phys}$. \\ Source: By the author. }
\label{Sub2PS}
\end{figure}

We shall analyze the power spectrum in more detail in chapter 5, when we discuss the potential role of such vacuum fluctuations in the fluctuation spectrum of the Cosmic Microwave Background (CMB) observed today. For completeness, we also show at this point how the Power Spectrum would look like when subtracted up to 4th order, which will be the one relevant in computing the stress tensor (Figure \ref{Sub4PS}):

\begin{figure}[H]
\centering
\includegraphics[width=0.45\linewidth]{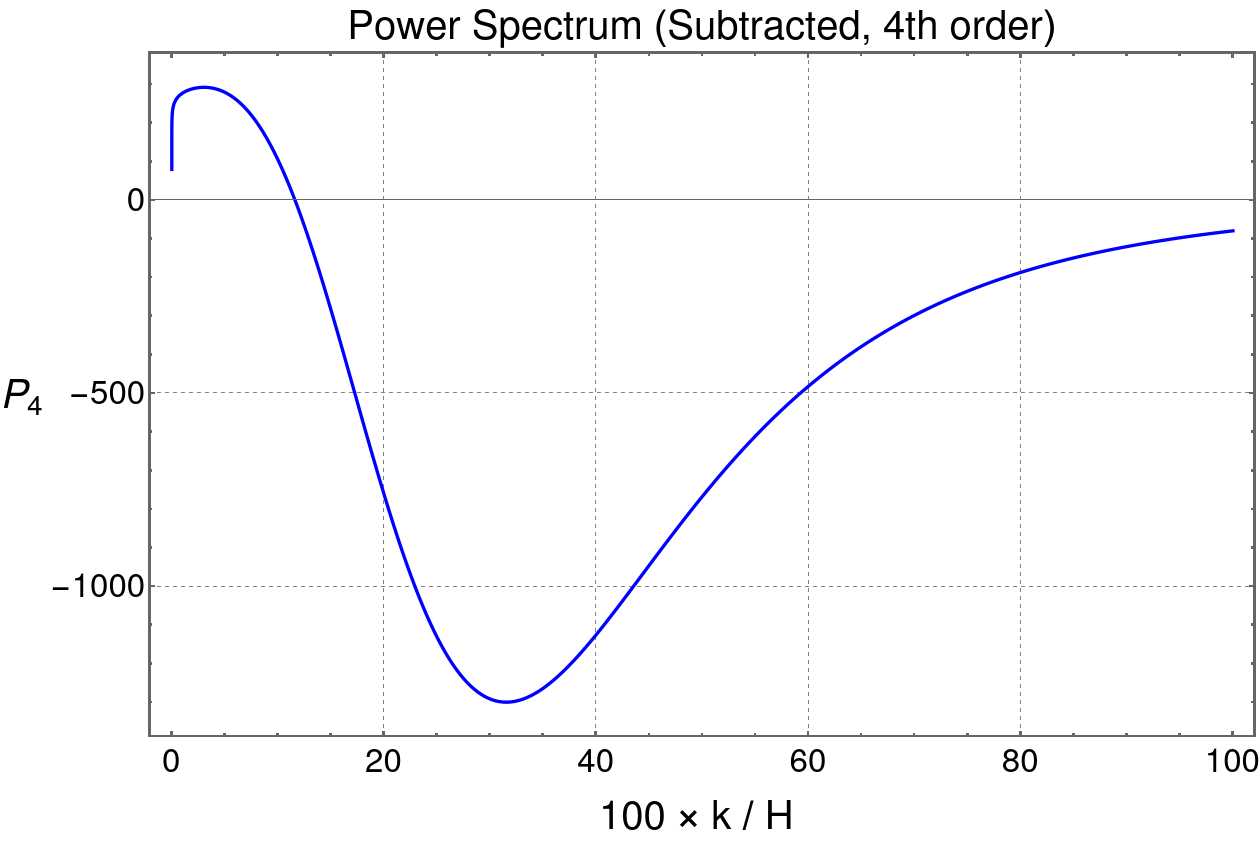}
\caption{Power Spectrum $\mathcal{P}_4$, subtracted up to 4th Adiabatic order for $\xi=0$ and $m^2=0.1H^2$. For these particular parameters, it is only at this order that the loss of positive-definiteness manifests. However, we note once again that this is a general feature of renormalization, and it can manifest in any subtracted orders. \\ Source: By the author. }
\label{Sub4PS}
\end{figure}

Finally, we show the behaviour of the unsubtracted and the subtracted power spectra covering a comprehensive range of parameters:

\begin{figure}[H]
\centering
\hspace{-10pt}\begin{minipage}{0.33\linewidth}
 \includegraphics[width=1\linewidth]{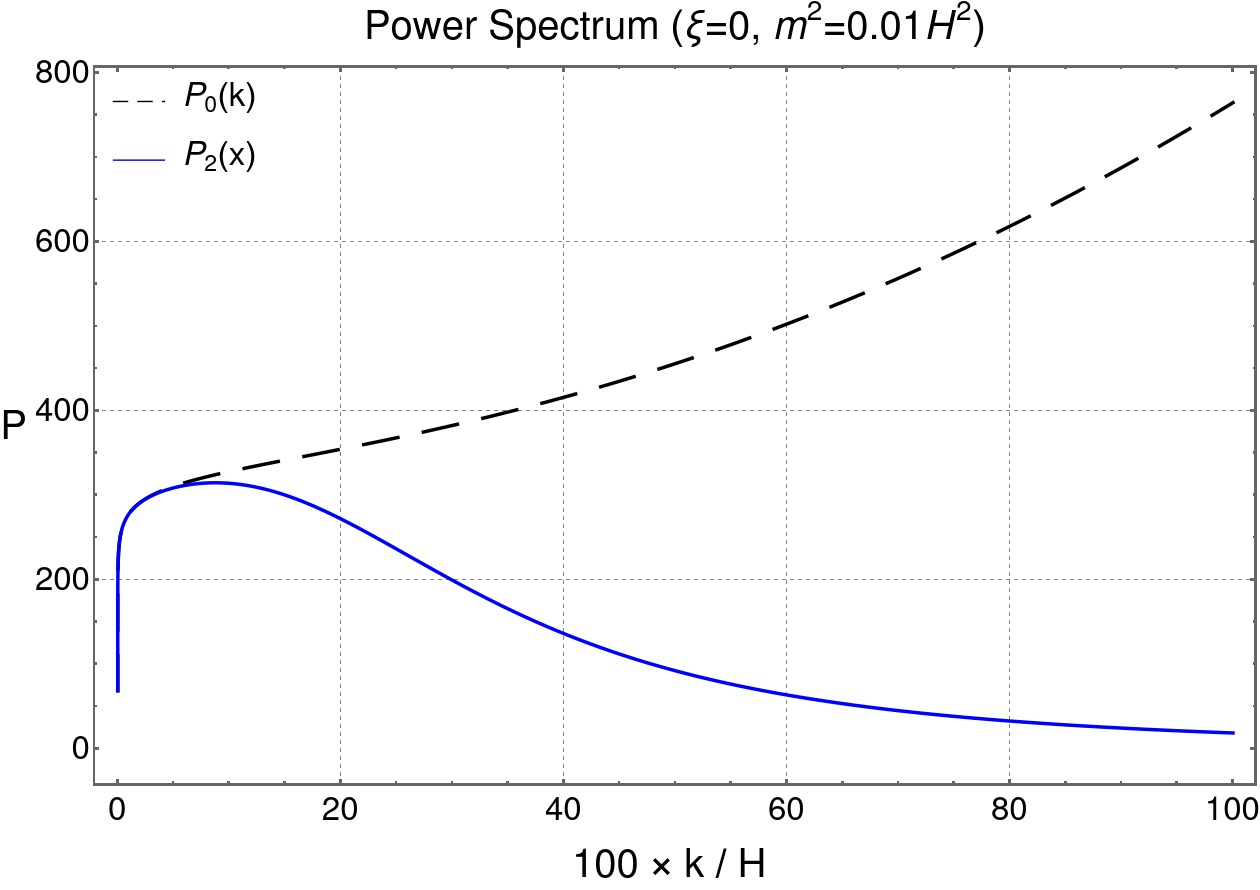}
 \caption{$\xi=0, m^2=0.01H^2$. \\Source: By the author. } \label{PSu+2(0,0;0,01)}
\end{minipage} %
\begin{minipage}{0.33\linewidth}
 \includegraphics[width=1\linewidth]{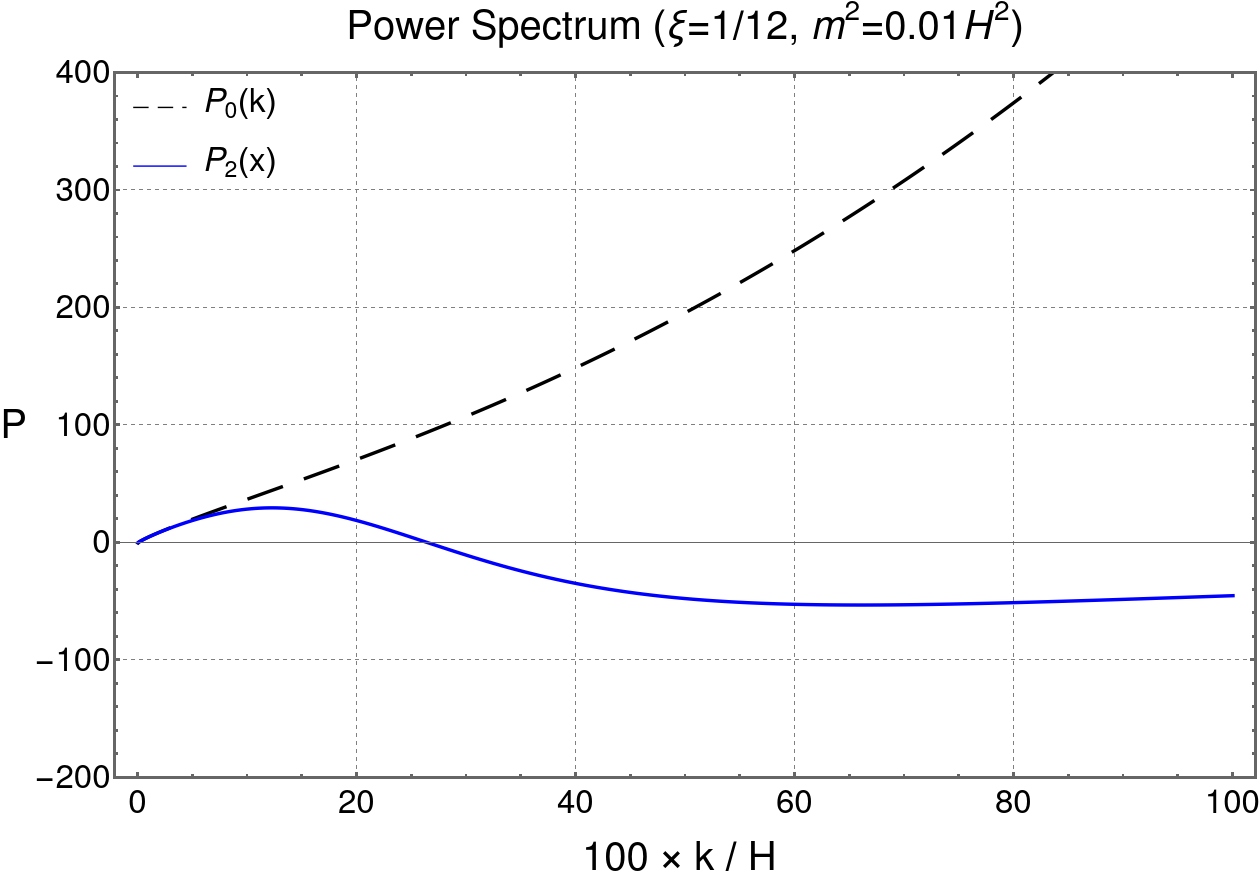}
 {\caption{$\xi\!=\!1/12, m^2=0.01H^2$. \\Source: By the author. }} \label{PSu+2(0,5;0,01)}
\end{minipage} %
\begin{minipage}{0.33\linewidth}
 \includegraphics[width=\linewidth]{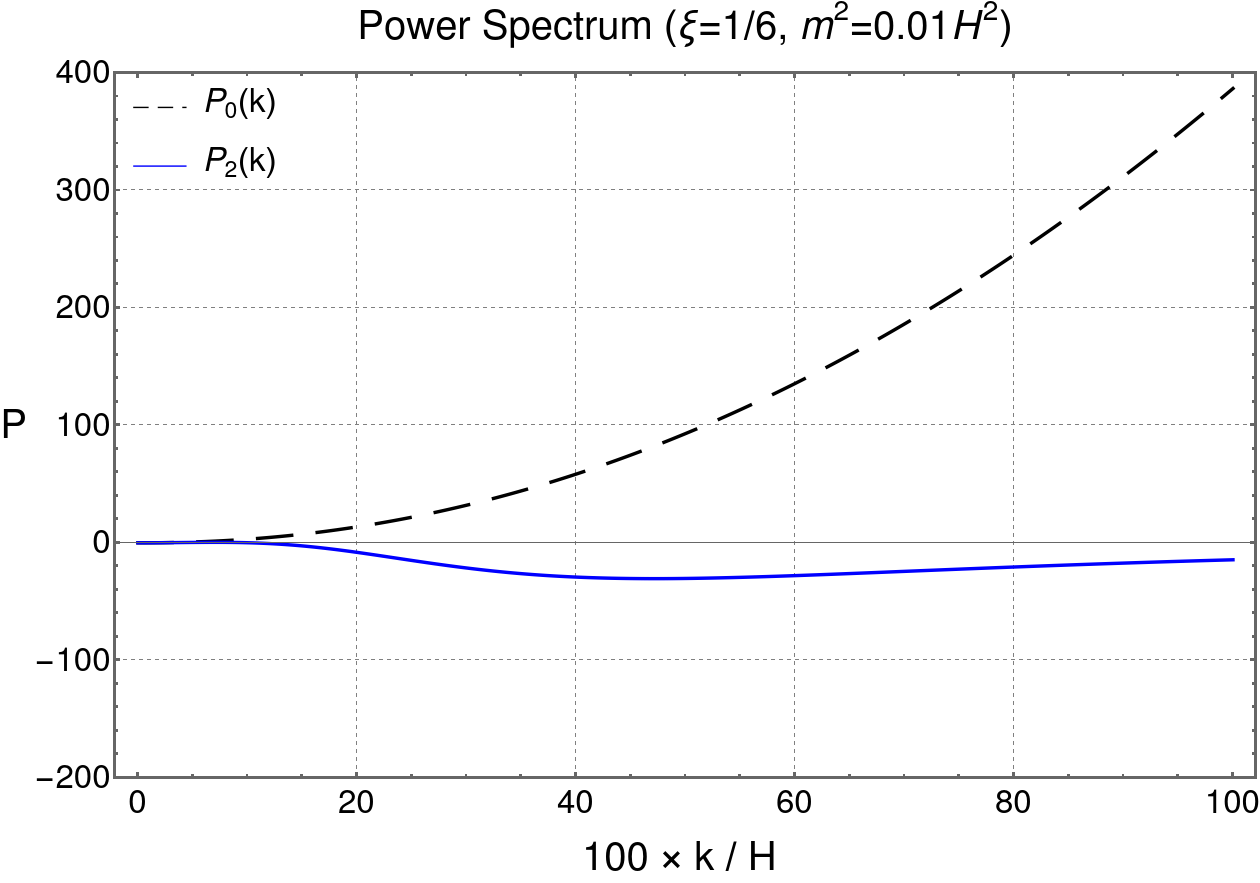}
 \caption{$\xi=1/6, m^2=0.01H^2$. \\Source: By the author. } \label{PSu+2(1,0;0,01)}
\end{minipage} 
\end{figure} 
\begin{figure}[H]
\hspace{-10pt}\begin{minipage}{0.33\linewidth}
 \includegraphics[width=1\linewidth]{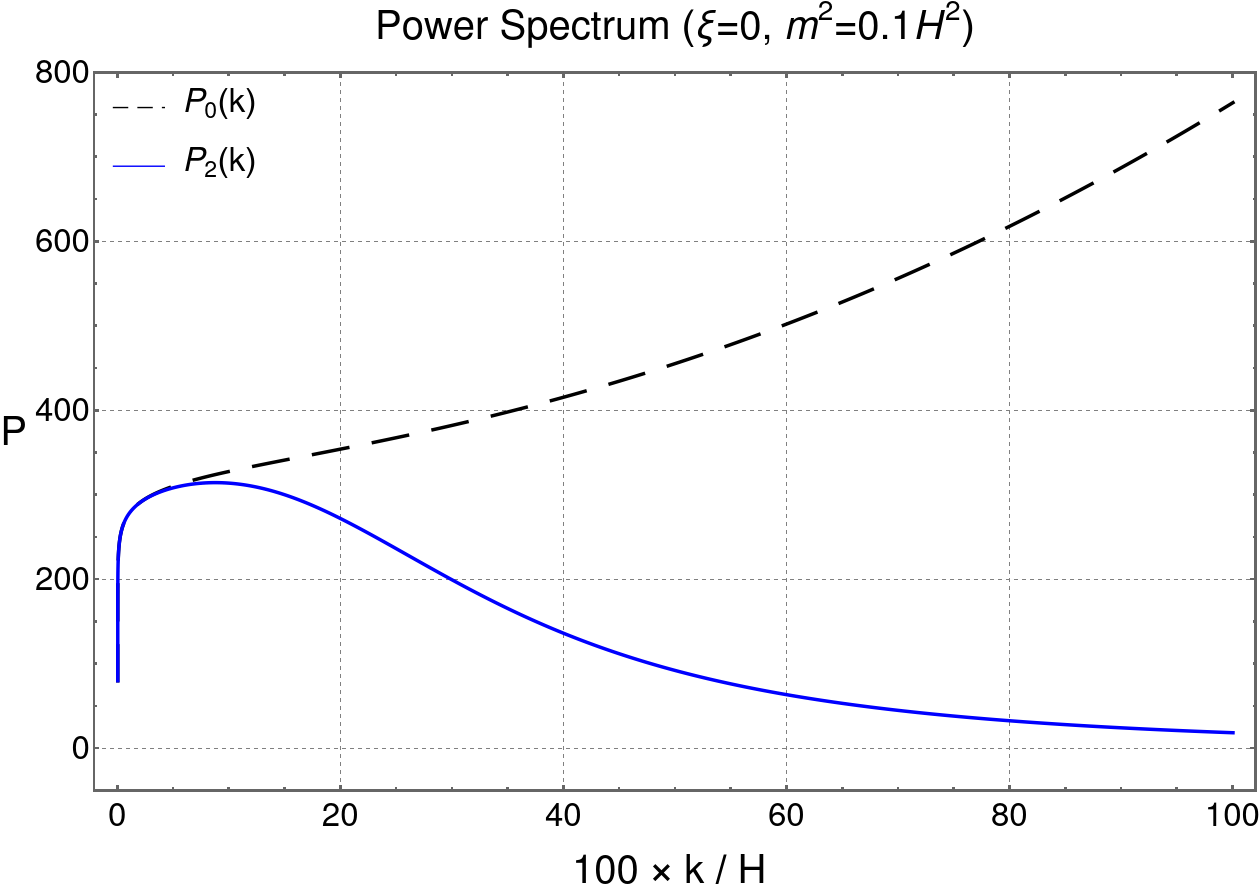}
 \caption{$\xi=0, m^2=0.1H^2$. \\ Source: By the author. } \label{PSu+2(0,0;0,1)}
\end{minipage} %
\begin{minipage}{0.33\linewidth}
 \includegraphics[width=1\linewidth]{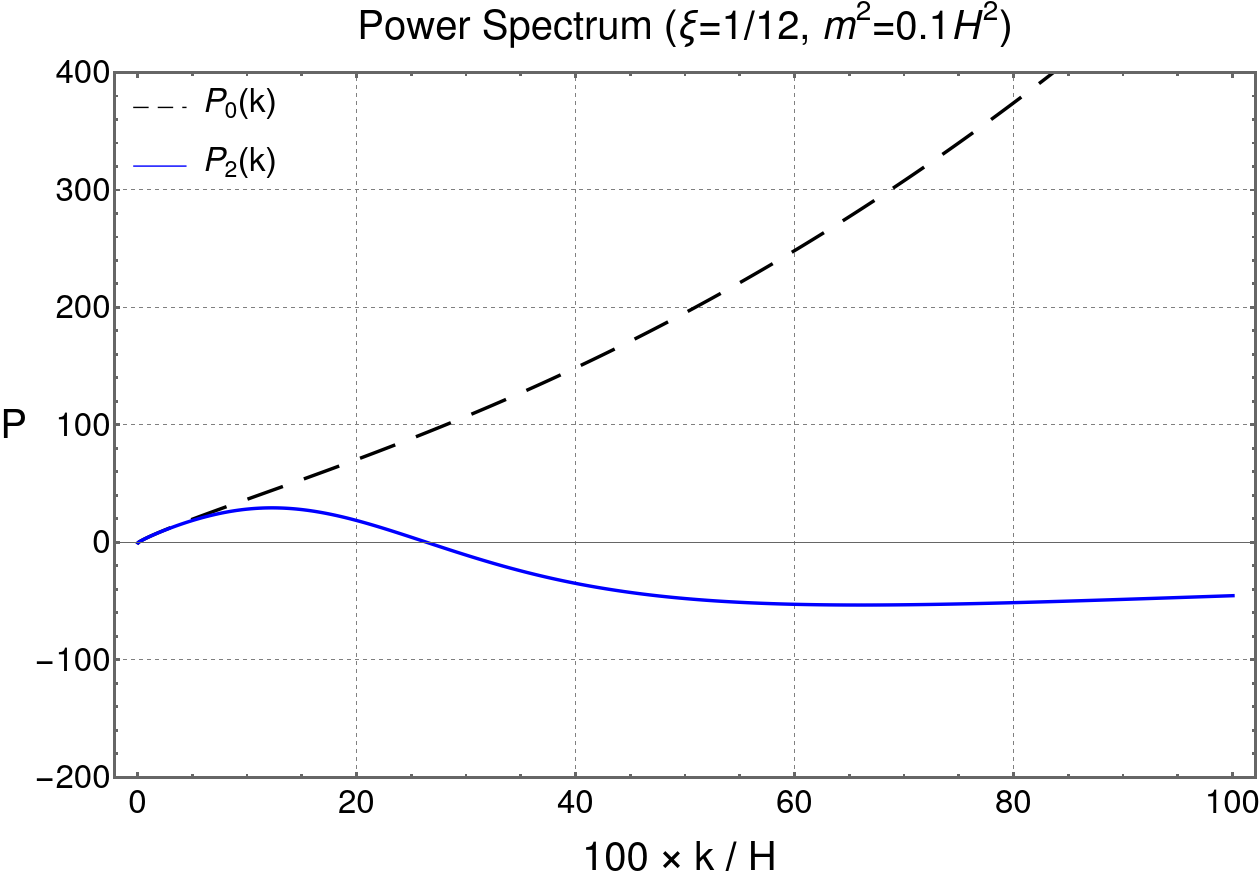}
 \caption{$\xi=1/12, m^2=0.1H^2$. \\ Source: By the author. } \label{PSu+2(0,5;0,1)} 
\end{minipage} %
\begin{minipage}{0.33\linewidth}
 \includegraphics[width=1\linewidth]{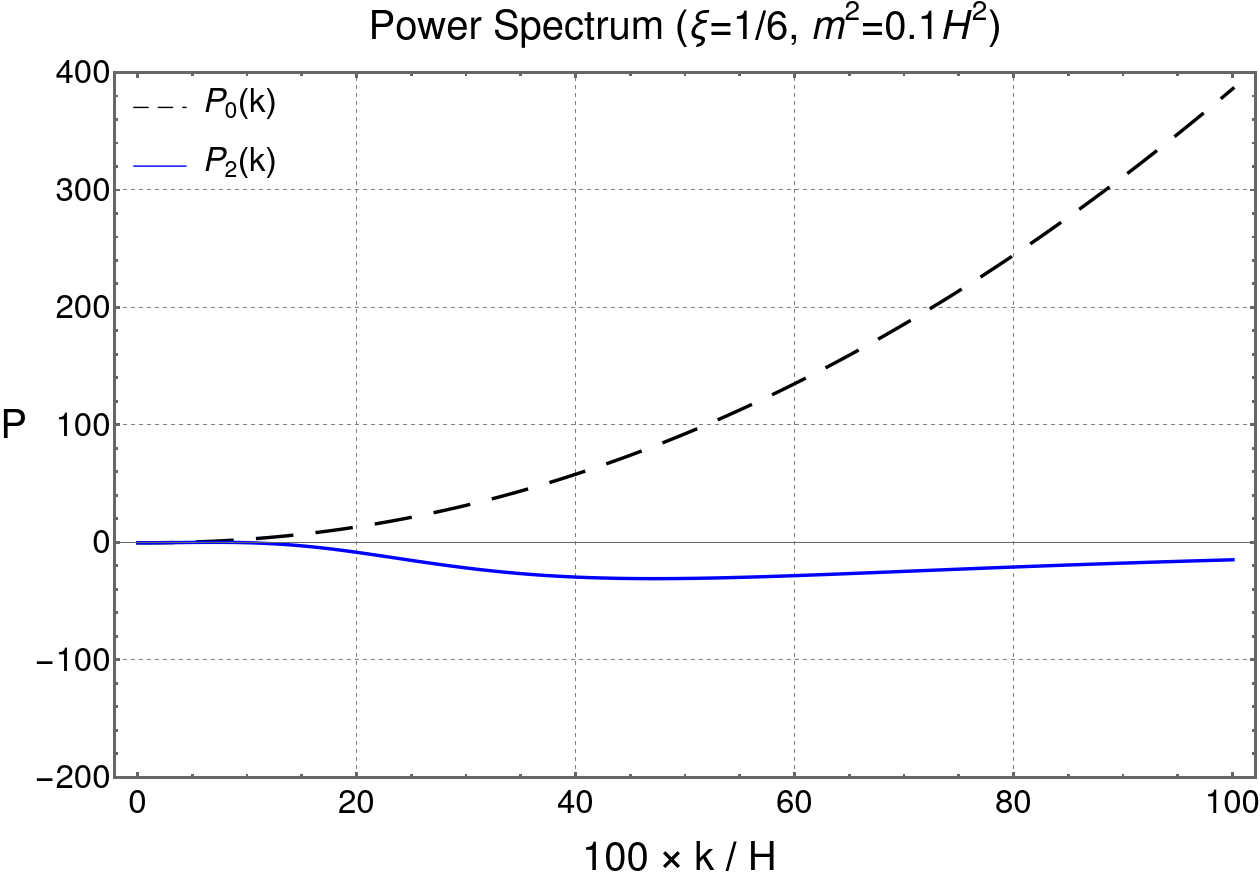}
 \caption{$\xi=1/6, m^2=0.1H^2$. \\ By the author. } \label{PSu+2(1,0;0,1)}
\end{minipage} \\
\end{figure} 
\begin{figure}[H]
\hspace{-10pt}\begin{minipage}{0.33\linewidth}
 \includegraphics[width=1\linewidth]{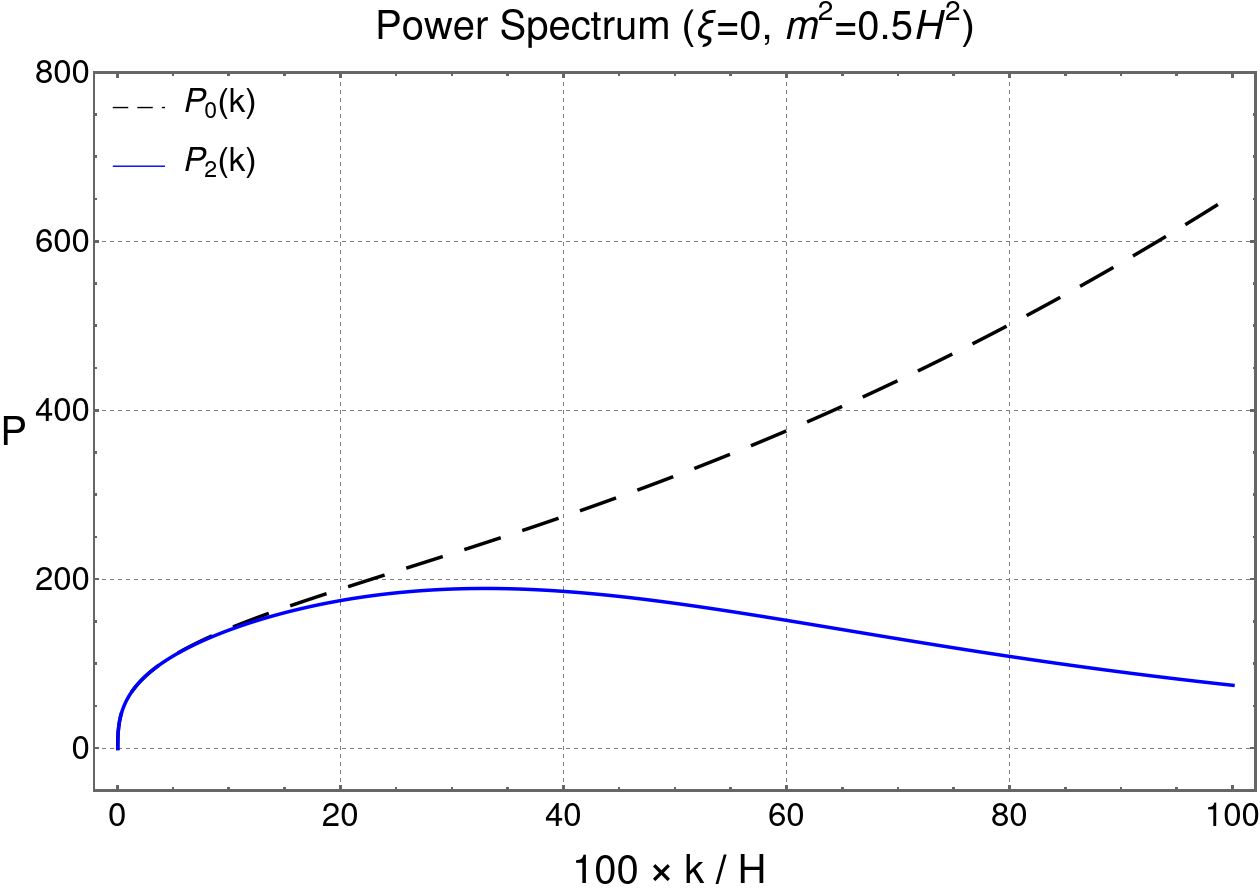}
 \caption{$\xi=0, m^2=0.5H^2$. \\ Source: By the author. } \label{PSu+2(0,0;0,5)}
\end{minipage} %
\begin{minipage}{0.33\linewidth}
 \includegraphics[width=1\linewidth]{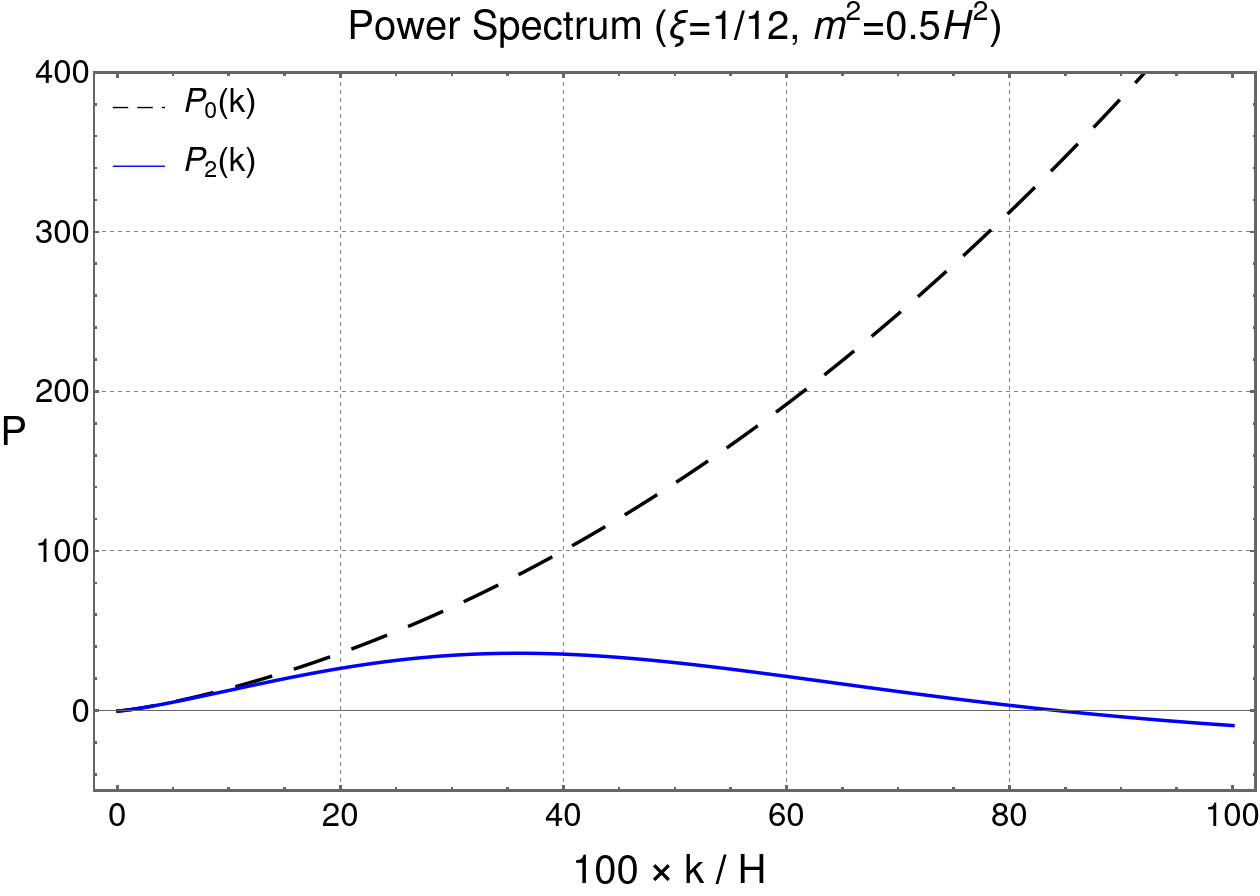}
 \caption{$\xi=1/12, m^2=0.5H^2$. \\ Source: By the author. } \label{PSu+2(0,5;0,5)} 
\end{minipage} %
\begin{minipage}{0.33\linewidth}
 \includegraphics[width=1\linewidth]{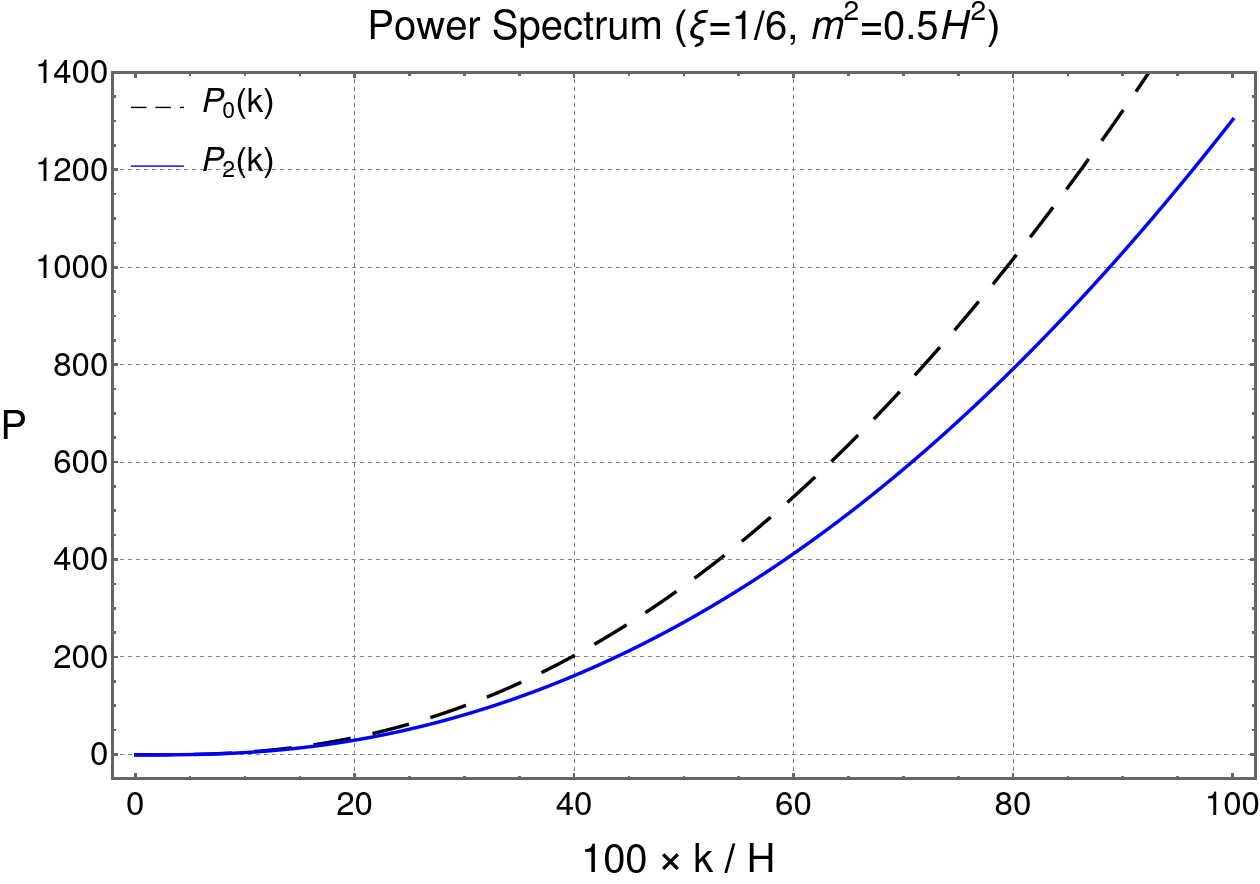}
 \caption{$\xi=1/6, m^2=0.5H^2$. \\Source: By the author. } \label{PSu+2(1,0;0,5)}
\end{minipage} \\
\end{figure}
\begin{figure}[h]
\begin{minipage}{0.45\linewidth}
 \includegraphics[width=1\linewidth]{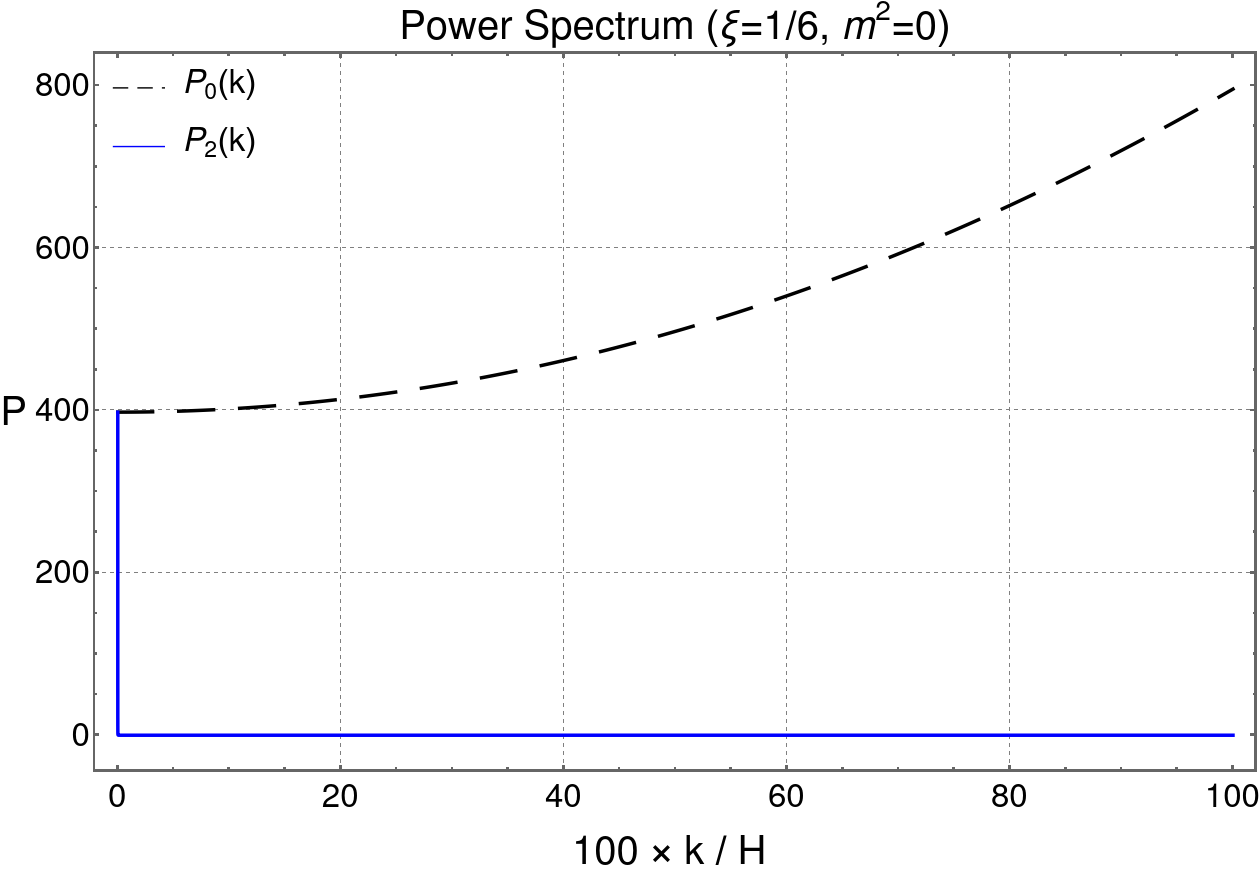}
 \caption{$\xi=1/6, m^2=0$. In this conformally trivial regime, all contributions come from the zeroth order term for any finite $k$, so the subtracted terms yield a trivial spectrum. \\ Source: By the author. } \label{PSu+2(1,0;0,0)_cor}
\end{minipage} \hspace{8mm} %
\begin{minipage}{0.45\linewidth}
 \includegraphics[width=1\linewidth]{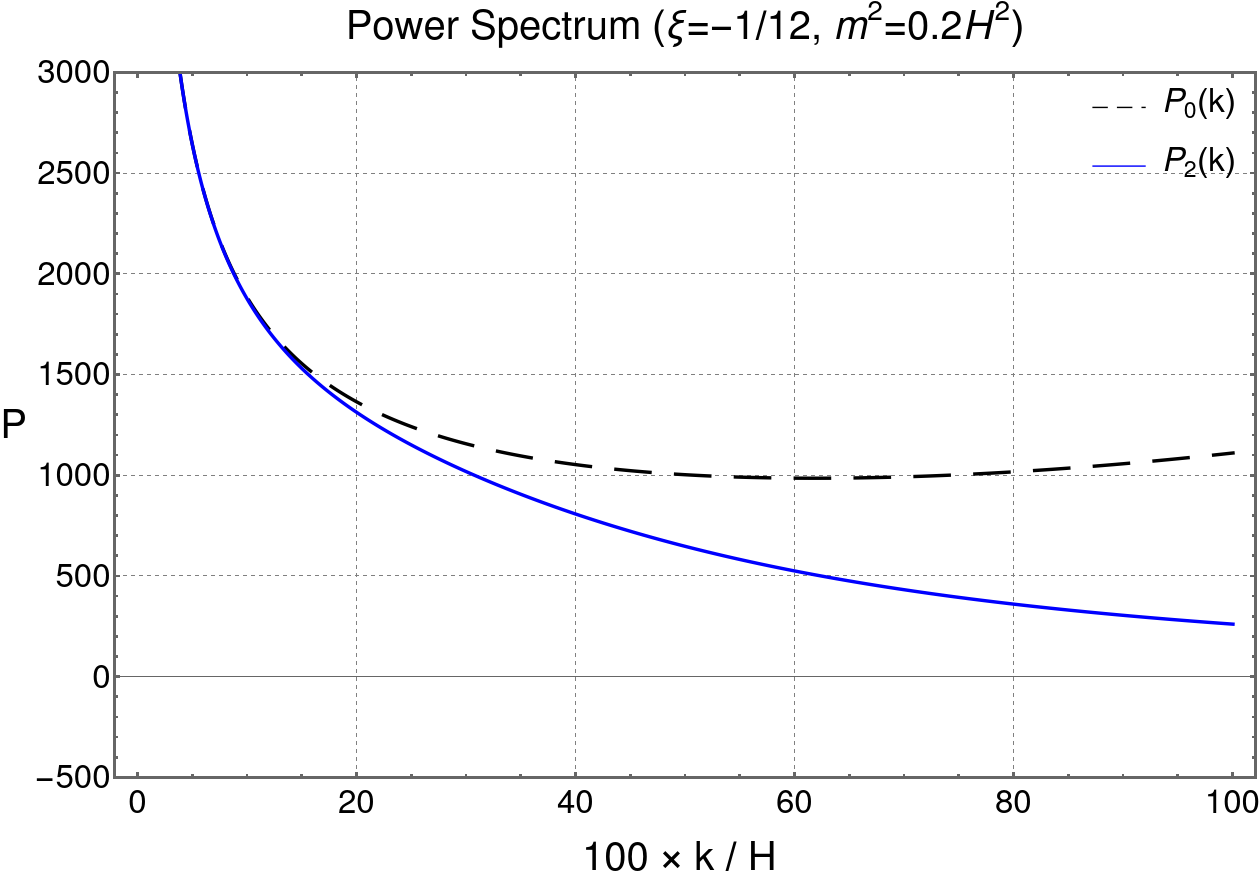}
 \caption{$\xi = -1/12, m^2=0.2H^2$. For this choice of parameters, although $m^2$ is positive, the effective mass $M^2$ is negative, so that $P(k\rightarrow0)$ is finite and there will be IR divergences\\ Source: By the author. } \label{PSu+2(-0,5;0,2)}
\end{minipage} 

\end{figure}

For the parameters that yield $0<M^2<H^2$, we will have well-behaved spectra in both the IR and the UV. For $M^2<0$, however (as we see in Figure \ref{PSu+2(-0,5;0,2)}), we will have IR divergences (even if $m^2>0$, for which $\omega^{-1}_k$ remains bounded) even in the subtracted spectrum. 

Having performed the apropriate adiabatic subtractions for an observable of interest, we must then evaluate the corresponding spectral integral to obtain its renormalized value in position space. Particularly, we are interested in the stress tensor $\braket{T_{\mu\nu}(x)}_{phys}$. As in the previous section, we shall obtain it from its renormalized trace, which will generally have more than just an anomalous contribution. In fact, as this trace is spatially homogeneous and it scales quite simply with time in a de Sitter space (as \emph{all} derivatives of $a(t)$ will have a time dependence proportional to $e^{Ht}$), we can obtain meaningful information from it by analyzing it  at a fixed event (which, for convenience, we shall fix at the origin of our coordinate system). A numerical analysis for a sample of well-behaved parameters $(m,\xi)$ then yields:

\begin{figure}[h]
\centering
\includegraphics[width=0.65\linewidth]{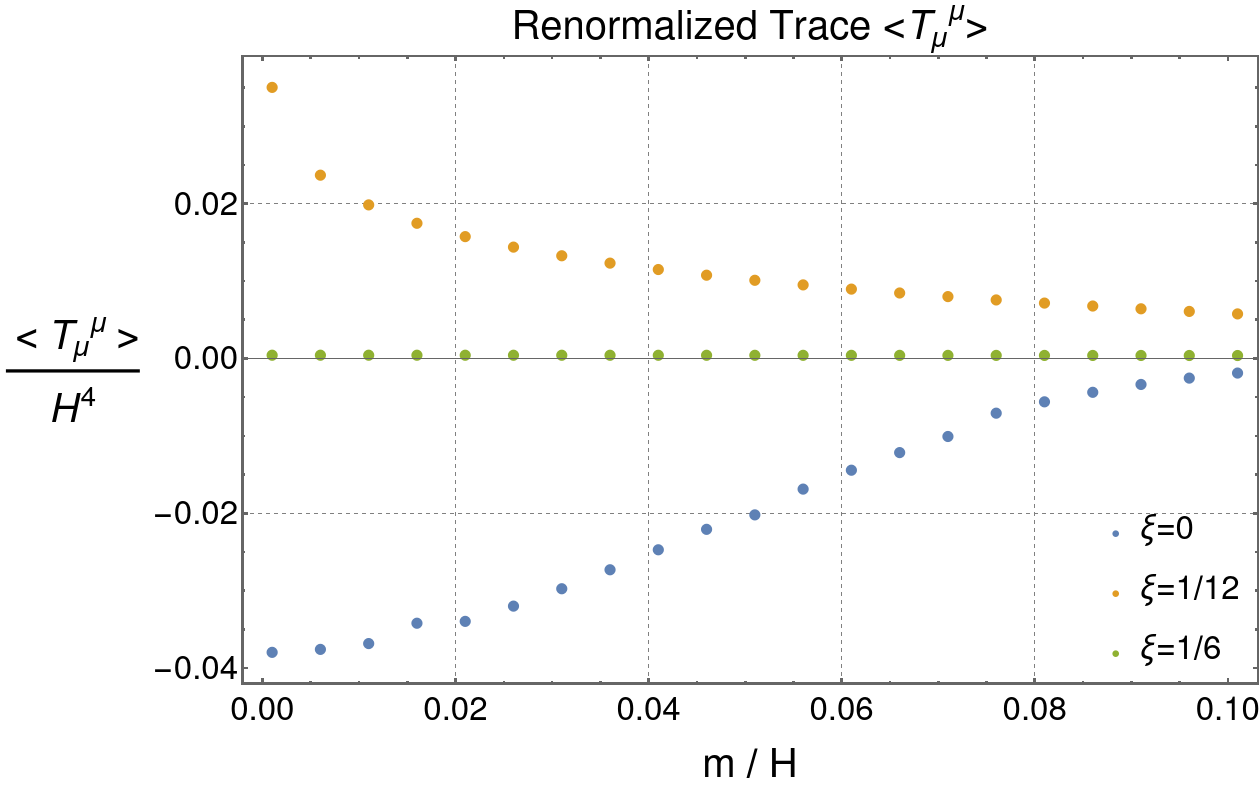}
\caption{ Renormalized value of the stress tensor trace $\braket{T_\mu^{\;\,\mu}}$ at $t=0$ normalized by $H^4$. In the conformally coupled case $\xi=1/6$, its dominant contribution comes from the conformal anomaly (for this range of masses), which turns out to be relatively small compared to its nonanomalous contributions for $\xi \not\simeq 1/6$. \\  Source: By the author.  }
\label{Trace_table3}
\end{figure}

\begin{figure}[H]
\centering
\includegraphics[width=0.65\linewidth]{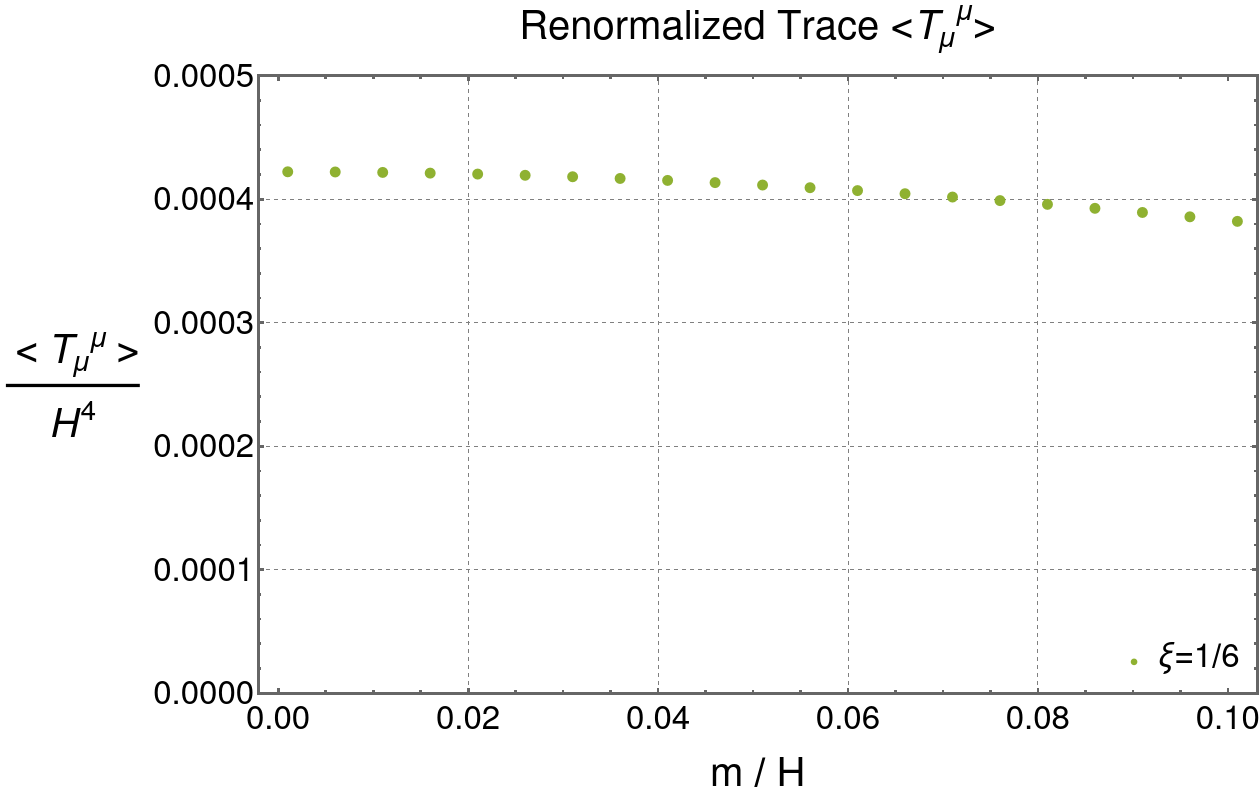}
\caption{ Renormalized value of the stress tensor trace $\braket{T_\mu^{\;\,\mu}}$ at $t=0$, normalized by $H^4$, for the conformally coupled case $\xi=1/6$. In this regime, its dominant contribution comes from the conformal anomaly (for this range of masses); as $m\rightarrow 0$, one can verify its numerical agreement with \eqref{traceTab}. \\ Source: By the author. }
\label{Trace_table1}
\end{figure}

This sample already reveals to us a number of features of the renormalized trace in de Sitter spaces. First, it can have either sign depending on both of the parameter values; a particular consequence of this is that there will be a 1-dimensional region in the $(\xi,m)$ plane for which $\braket{T_\mu^{\;\mu}}$ will vanish. We also stress that, although in Figure \ref{Trace_table3} (\ref{Trace_table5}), the trace may look vanishing in the conformally coupled case, it is actually slightly positive. In fact zooming in this plot a little (see Figure \ref{Trace_table1}), we can verify that it numerically agrees with our analytical result \eqref{traceTab} as $m \rightarrow 0$:

\begin{align}
\frac{\braket{T_\mu^{\;\,\mu}} }{H^4} = \frac{1}{240\pi^2} \approx 0.00042 \,.
\end{align}

For completeness, we plot a similar graphic to \ref{Trace_table3}, with a few more values of parameters, for which we can see the trace actually cross the axis:

\begin{figure}[h]
\centering
\includegraphics[width=0.65\linewidth]{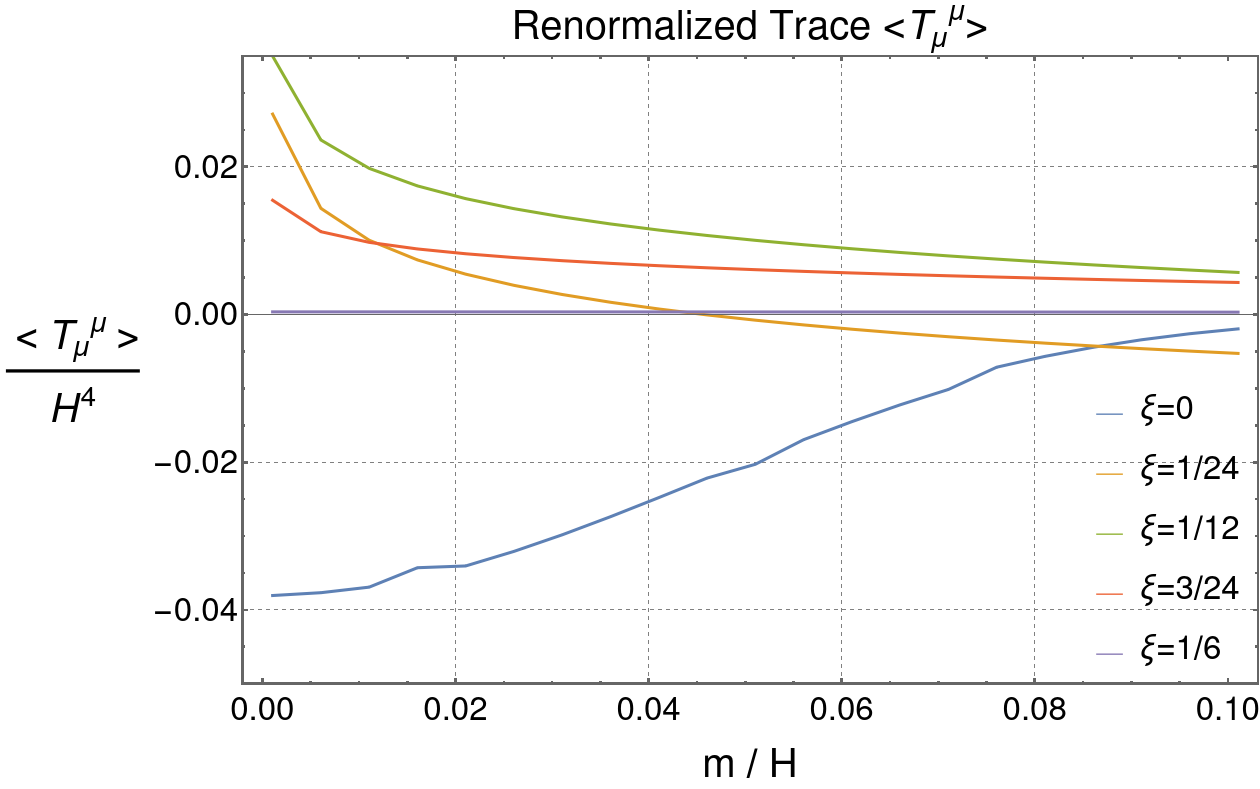}
\caption{ Renormalized value of the stress tensor trace $\braket{T_\mu^{\;\,\mu}}$ at $t=0$ normalized by $H^4$, plotted in lines for better visualization of the crossing values. In this sample, one can see that, for $\xi=1/24$, $\braket{T_\mu^{\;\mu}}$ actually crosses the axis near $m=0.045H$. \\  Source: By the author.  }
\label{Trace_table5}
\end{figure}

Note also that the sign of the trace will be determinant for the sign of the energy density, as we see in eq \eqref{T00t2t1}. By computing numerical integrals both in the spectrum and in time (for the latter we start at $t=0$ and go through a few $e$-folding periods, $H^{-1}$), we may obtain the renormalized values of energy density and pressure, which can in principle be a function of time. We display the typical behaviour for $\rho(t)$ and $p(t)$, exemplified in a minimally coupled massive ($m^2=0.1H^2$) case:

\begin{figure}[H]
\centering
\includegraphics[width=0.65\linewidth]{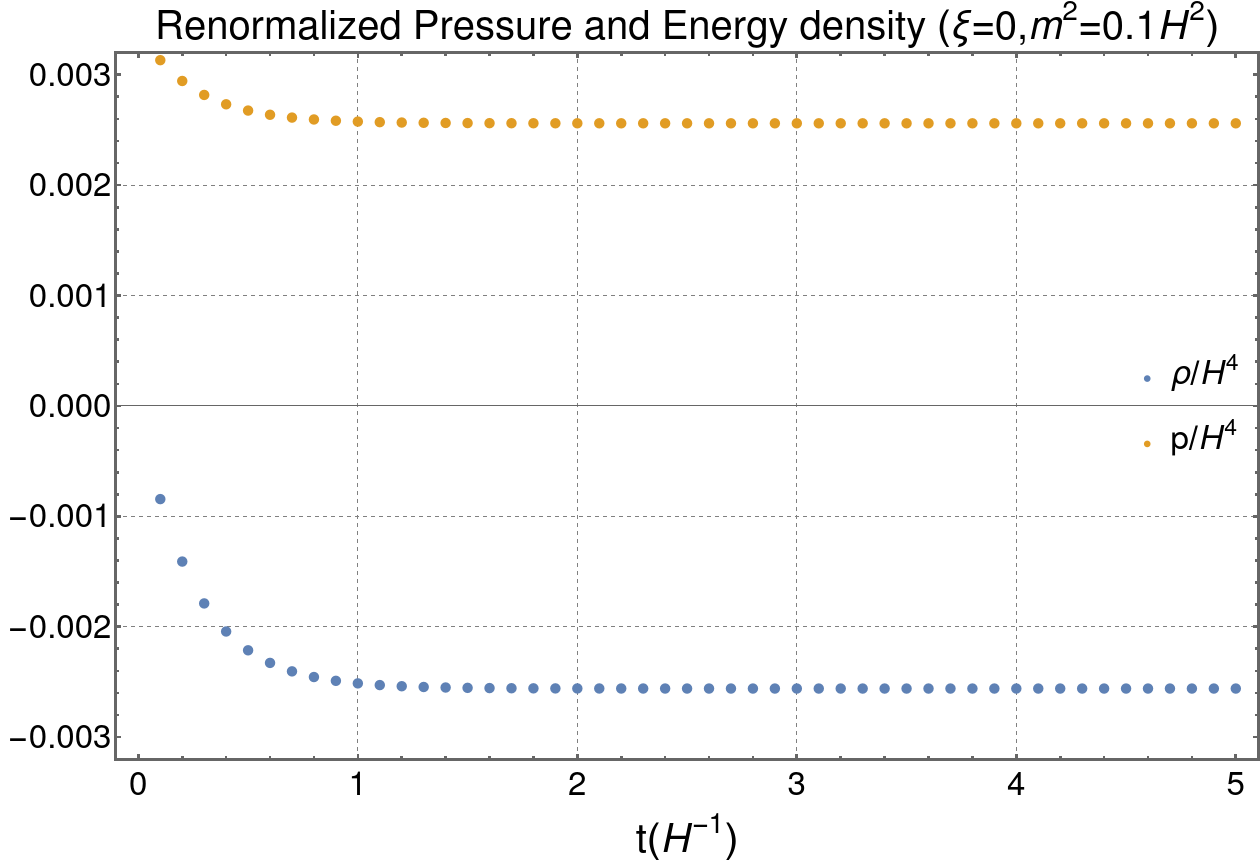}
\caption{ Renormalized energy density and pressure as functions of time, with initial condition $\rho_0=0$. After a few $e$-foldings, these quantities evolve to constant values $\rho_{eq} \approx 0.0025 H^4$ and $p_{eq} \approx -\rho_{eq}$. \\ Source: By the author. }
\label{time_rhop(0,0;0,1)}
\end{figure}

Note that the initial condition imposed in our temporal integral was $\rho(t=0)=0$, as we carried a definite integral starting at $t=0$. Then, after few $e$-foldings, one sees that energy and pressure quickly evolve to constant equilibrium values $\rho_{eq}$ and $p_{eq} \simeq -\rho_{eq}$, obeying the equation of state that is (qualitatively) self-consistent with de Sitter spaces ($\braket{T_{\mu\nu}}=\Lambda g_{\mu\nu}$). Moreover when we analyze the difference term

\begin{align}
\delta \rho(t) \equiv \rho(t) - \rho_eq,
\end{align}
we find that it decays precisely as $\delta \rho(t) \propto e^{-4Ht} = a^{-4}(t)$. Then, in eq \eqref{ren vac+mat} we immediately identify this transient term as the decaying integration constant that was attributed to particle terms, rather than vacuum energy\footnote{The same interpretation could be attained here, as the particle energy density should decay as $a^{-4}(t)$, provided that one generally considers particles with both positive and negative energies, as $\delta \rho$ can have either sign.}, so we identify the renormalized vacuum energy (pressure) as $\rho_0 = \rho_{eq}$ ($p_0 = p_{eq}$).

In this particular case, we have found $\rho_0 < 0$ and $p_0 > 0$, the signs can be reversed, depending on the values of $m$ and $\xi$. In fact, carrying the same analysis for the conformally trivial case, we recover precisely our analytical results \eqref{CT vac ener} (see Figure \ref{time_rhop(1,0;0,0)} ):

\begin{align}
\frac{\rho_{eq}}{H^4} \simeq -\frac{p_{eq}}{H^4} \approx \frac{1}{960\pi^2} \simeq 0.00011 \,.
\end{align}

\begin{figure}[H]
\centering
\includegraphics[width=0.65\linewidth]{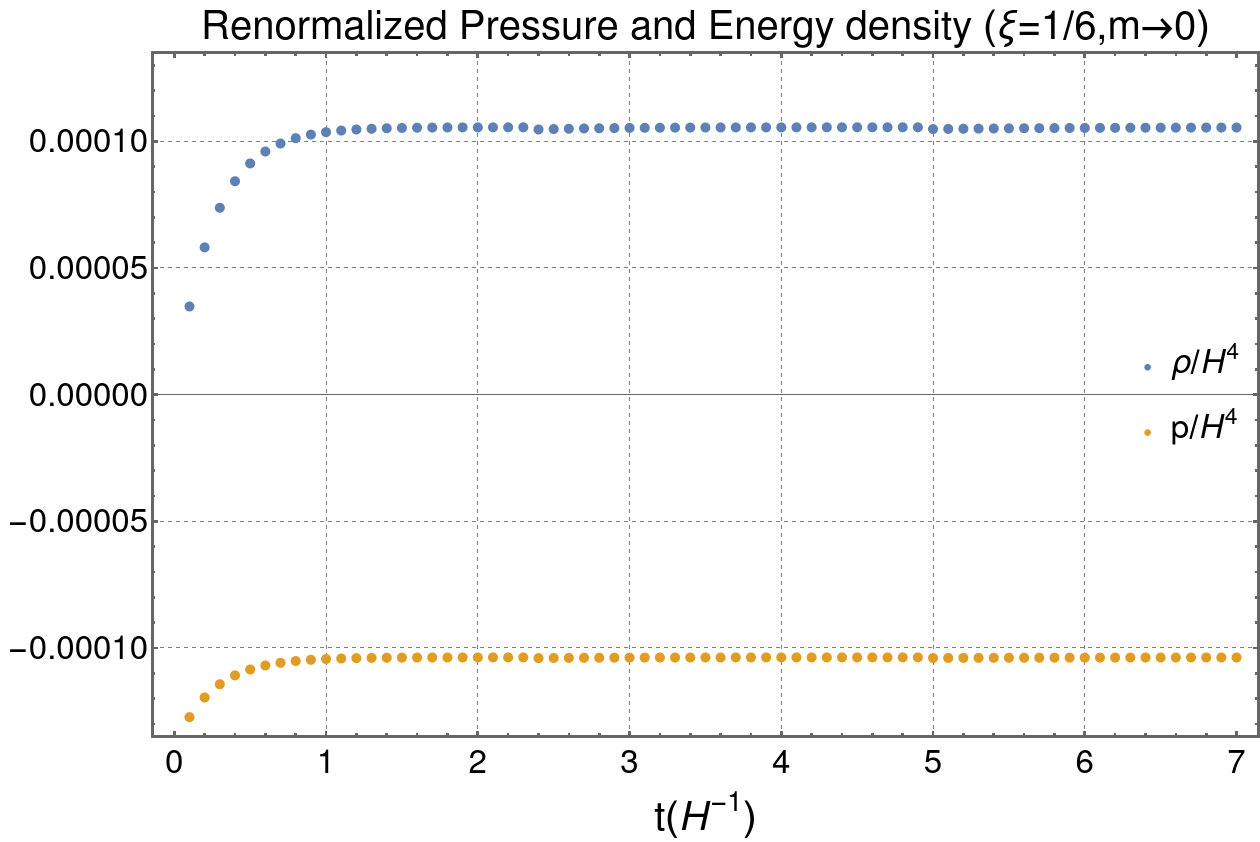}
\caption{ Renormalized energy density and pressure as functions of time in the conformally trivial case. \\ Source: By the author. }
\label{time_rhop(1,0;0,0)}
\end{figure}

Applying the same procedure for various parameter values within a well-behaved range (for which we manage to achieve numerical convergence), we find by inspection that, indeed, the equation of state for our renormalized vacuum energy is always of the form $p_0 \simeq - \rho_0$ in our de Sitter spacetimes, yielding a stress tensor in the form of a cosmological constant $\braket{T_{\mu\nu}}=\Lambda g_{\mu\nu}$. This is actually not surprising, as we have built the Bunch-Davies vacuum to be invariant under the de Sitter symmetries, and, in this maximally symmetrical spacetime, the only possibility for a symmetric rank (0,2) tensor built only from geometrical quantities is $\braket{T_{\mu\nu}} \propto g_{\mu\nu}$. Nonetheless, the present analysis has allowed us to explicitly compute the renormalized values of vacuum energy densities and pressures, for which we not only verify this self-consistency geometrical condition to be satisfied, but also obtain specific values for $\rho_0$ and $p_0$ for sufficiently well-behaved parameters.

Then, to conclude this section, we show our results for the renormalized vacuum energy $\rho_0$ obtained by this procedure for parameters ranging in the intervals $ 0 < \xi \leq 1/6$ and $0 < m^2 \leq 0.1H^2$:

\begin{figure}[H]
\centering
\includegraphics[width=0.7\linewidth]{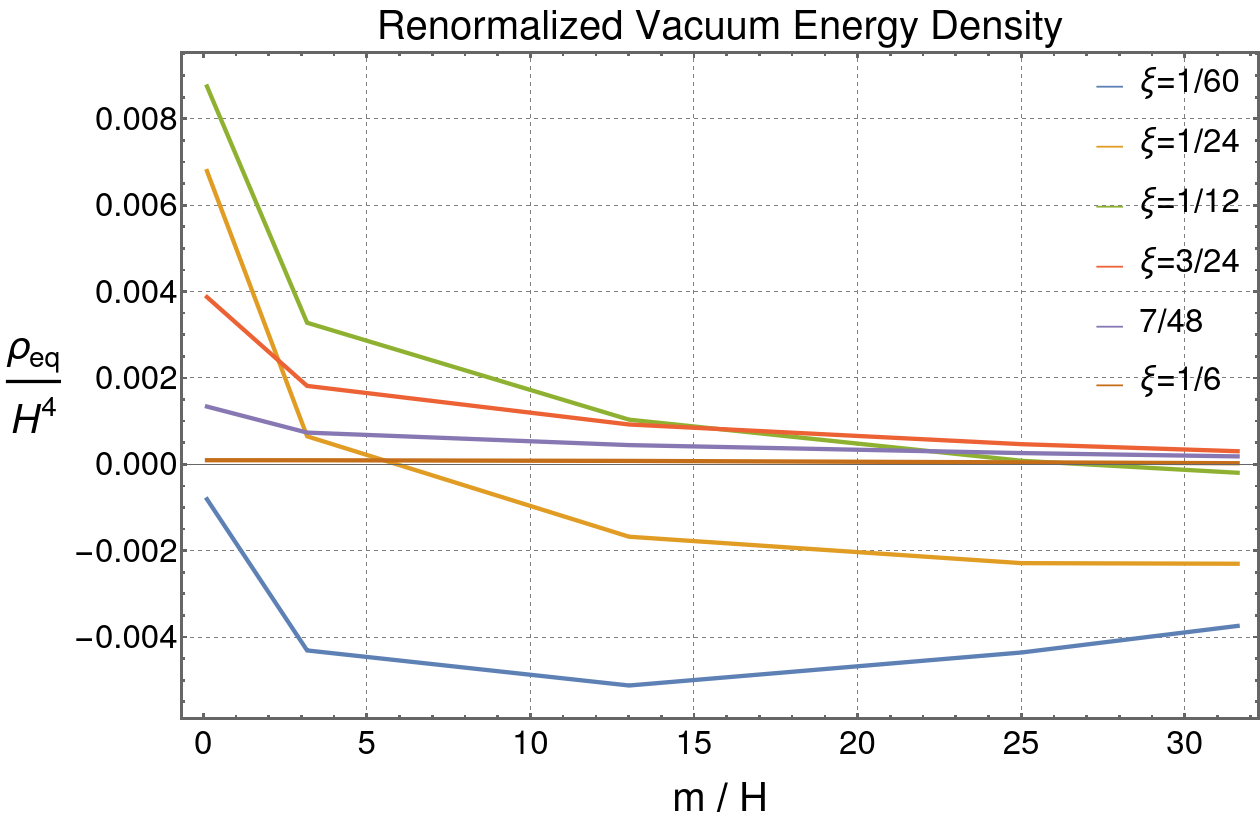}
\caption{ Renormalized energy density for various values of $m$ and $\xi$ within well-behaved intervals. \\ Source: By the author. }
\label{RHO_EQ}
\end{figure}

As we have repeatedly remarked, this renormalized vacuum energy can be found to bear either sign as we sweep the parameters, and, particularly there will be a 1-dimensional region in the $(m,\xi)$ plane for which it will be trivially null; again, we stress that the conformally coupled case actually lies slightly above the axis. We also note that, for $m \approx 0$, the values of $\xi$ that approach $1/6$ successively approach the conformally anomalous vacuum energy for $\xi>1/12$.

Finally, we note that, as the vacuum energy densities $\rho_0$ are significantly higher (for a fixed value of $H$) in the conformally nontrivial cases, or equivalently, the ratio $H^4/\rho_0$ are significantly lower, one finds in eq \eqref{2 Lambdas} that the self-consistency values of $\Lambda$ will be smaller. If we denote the conformally trivial vacuum energy as $\Lambda_t$ and a conformally nontrivial value as $\Lambda = \alpha \Lambda_t$, we find that the corresponding self-consistent values of $H$ and $\Lambda$ would yield (see eq \ref{self-consistent Lambda}):

\begin{align}
\Lambda = \alpha^{-1} \Lambda_t = \frac{45}{\alpha}\rho_p,
\end{align}
which can yield sub-Planckian self-consistent values of $\Lambda$ for a sufficiently high $\alpha$. The energy densities found here, however, although significantly higher than $\Lambda_t$, have only $|\alpha| \lesssim \mathcal{O}(10^2)$, which would still correspond to a Planckian self-consistency regime.

\section{Path integrals, effective action and Renormalization of Gravitational Parameters} \label{ARenormalization}

Now that we have had a first operational contact with the subject of renormalization, in the concrete example of adiabatic subtraction, we would like to better understand it conceptually, and take a glimpse on its links with the wider scheme of renormalization of geometrical parameters in a gravitational context.

To achieve that, we start by giving a brief presentation of the Schwinger Action Principle, which will allow us to construct the effective action $W$. This action principle is intimately related to the path-integral formulation of quantum mechanics, which not only gives a novel conceptual perspective to the theory but also provides us with a very powerful arsenal to operate with the effective action. Not surprisingly, $W$ will present divergences in field theories; to properly handle them and put them in a renormalizable form, we shall write an asymptotic expansion for $W$ in which we can isolate the divergencies in a finite number of terms, and eventually subtract them from the matter action, reabsorbing them in the definition of geometrical parameters.

In Classical Mechanics (be it particle mechanics or classical field theory), one could derive the dynamics from an extreme action principle. This principle could be summarized as ``the path that a system will classically follow to go from a configuration $q_1$ at time $t_1$ (at a Cauchy surface $\Sigma_1$) to a configuration $q_2$ at a time $t_2$ (at a Cauchy surface $\Sigma_2$) will be that which extremizes the action functional $S[q(t)]$''. In Quantum Mechanics, where a precise determination of a classical path for a system is forbidden, one can see this principle in a new light. In Feynman's formulation cite{FeynmanA,FeynmanB}, one can state a quantum version of the action principle as follows: ``the \emph{\underline{conditional} probability amplitude} that a system starting at a configuration $\ket{q'}$ at a time $t_1$ (a Cauchy surface $\Sigma_1$) will be measured at a configuration $\ket{q''}$\footnote{ 
Here, we avoid the most obvious notation $\ket{q_1'}$ and $\ket{q_2'}$ for those configurations to avoid an ambiguity in the following discussion, where we shall consider \emph{different observables} $q_1$ and $q_2$ to be evaluated at times $t_1$ and $t_2$; in this case $q_1'$ and $q_2'$ will denote \emph{eigenvalues of distinct operators}, whereas, for instance, $q_1'$ and $q_1''$ denotes \emph{distinct eigenvalues of the same operator}. }
 at a time $t_2$ (a Cauchy surface $\Sigma_2$) \emph{\underline{having passed through a classical path}}\footnote{
 On first sight, this may sound contradictory to what we just said above, that the \emph{determination} of a classical path is forbidden, but it is not. We are not \emph{determining} a classical path, just associating conditional probability amplitudes to it. This is the same as, \textit{e.g.,} associating amplitudes from the `left' and `right' paths in a double slit experiment; one may then appropriately add these amplitudes to obtain an interference pattern. }
 $q'(t)$ will be proportional to $e^{iS[q'(t)]}$, being $S[q'(t)]$ the classical action associated with that path''. We summarize in a condensed notation:

\begin{align}
\braket{q'',t_2|q',t_1}[q'(t)] \propto e^{\frac{i}{\hbar} S[q'(t)]}, \label{FAmplitudes}
\end{align}
where we temporarily recovered Planck's constant ($\hbar \neq 1$) in order to make a clearer picture of the classical limit ahead. (We also note that these probabilities will be generally nonnormalizable, and one must use sophisticated functional integral techniques to handle them.)

With this physical postulate, one can in principle work out any dynamical predictions of a quantum theory -- which can be put in the form of transition amplitudes $\braket{q'',t_2|q',t_1}$ -- by summing (integrating) the \emph{conditional amplitudes} \eqref{FAmplitudes} over all classical paths $q'(t)$, \emph{\textit{i.e.}, by calculating a path integral}. These predictions turn out to be equivalent to those obtained by the known quantum dynamical equations obtained through canonical quantization; then, conversely, one can deduce these equations from Feynman's basic postulate.

 More generally, one could measure any complete set of commuting observables (C.S.C.O.) to maximally determine a physical state, rather than just its ``position'' configurations $\ket{q'}$; then, denoting eigenstates of two arbitrary C.S.C.O.'s $A$ and $B$ by $\ket{a'}$ and $\ket{b'}$, respectively, the most general probabilities that we could measure are of the form $\braket{b'|a'}$. Usually, as we will be interested in determining transition amplitudes for arbitrary C.S.C.O.'s at times $t_1$ and $t_2$ (Cauchy surfaces $\Sigma_1$ and $\Sigma_2$), we will correspondingly denote these arbitrary observables as $q_1$ and $q_2$ (or $\zeta_1$ and $\zeta_2$) and the amplitudes $\braket{q_2',t_2|q_1',t_1}$ (or $\braket{\zeta_2',\Sigma_2|\zeta_1',\Sigma_1}$).

 Additionally, in this formulation, the classical limit of the theory becomes conceptually quite obvious. Whenever a physical system has an action $S[t]$ that is very large with respect to Planck's constant, such that $S[q(t)]/\hbar \gg 1$ and it varies by a great amount for small path variations, then interference from nearby paths will cancel out almost everywhere, and \emph{nonneglegible probabilities will arise only from the paths around which the action varies the least, i.e., around stationary paths, which extremize the action}. Then, in this limit, the probability that a system goes very nearly around a path that extremizes the action becomes practically unity, so that one recovers the system's classical paths as given by the stationary action principle.
 
Besides, this quantization formalism extends very directly to interacting fields, whereas our approach in chapter \ref{QFTCS} was restricted to noninteracting fields, which could be conveniently put in a form of an infinite collection of decoupled harmonic oscillators. However, path integrals are generally very complicated to calculate. The noninteracting case is one of the few where calculations can be rendered in a computable form, using (infinite-dimensional) Gaussian integrals.

An equivalent and intimately related way to define this manifestly covariant approach to quantum theory is through the Schwinger Action Principle\footnote{For the original works on Schwinger's formulation, see the seminal papers cite{schwingerI,schwingerII}. For a more pedagogical and thorough textbook introduction, see cite{toms}. }.
 In this formulation, one works directly with transition amplitudes and their variations in terms of an action operator: the \emph{effective action}. Thus, it will be more convenient for us to start with this formalism, as it will allow us to directly \emph{define} the effective action as a starting point and then \emph{derive} its relation to path integrals.
 
Before presenting it in the context of field theory, it is constructive to briefly illustrate our considerations in the simpler context of ordinary quantum mechanics. A first point that we stress here is that we are describing the possible \emph{physical states} (configurations) in our system by \emph{complete sets of eigenvalues of time-dependent observables}. Thus, when we write $\ket{a',t}$ as an eigenstate of the C.S.C.O. $A(t)$, we mean that

\begin{align}
\ket{a', t}: \; A(t) \ket{a', t} = a' \ket{a', t},
\end{align}
\emph{where the eigenvalue $a'$ is time-independent.} This representation has the potentially confusing implication that \emph{our \underline{basis} state vectors will be generally \underline{time-dependent in the} \underline{Heinsenberg  picture} and \underline{time-independent in the Schrödinger picture}}. Indeed, consider for instance a free nonrelativistic particle moving in 1 spatial dimension and the C.S.C.O. given by the position operator $x(t)$: in the Schrödinger picture, $x$ is time independent, thus so will be its eigenvectors $\ket{x'}$ (up to an arbitrary phase factor, which we omit), since

\begin{align}
x\ket{x',t} = x'\ket{x',t}, \forall t \quad \Rightarrow \quad \ket{x',t_1} = \ket{x',t_2};
\end{align}
in the Heisenberg picture, on the other hand, $x$ is time-dependent, such that $x(t_1) \not\equiv x(t_2)$, which implies that

\begin{align}
x(t)\ket{x',t} = x'\ket{x',t}, \forall t \quad \Rightarrow \quad \ket{x',t_1} \neq \ket{x',t_2}. \label{heisenbergbasis}
\end{align}

Note, further, that, if Schrödinger ordinary state vectors evolve by the evolution operator $U$ ($\ket{\psi(t)}=U(t)\psi_0$), and thus Heisenberg operators evolve as $U^\dagger(t)A_0U(t)$, then our \emph{Heisenberg basis vectors} will evolve by the \emph{inverse} evolution operator $U^\dagger$, in order to satisfy \eqref{heisenbergbasis} at all times. 

Having clarified our basic notation, we proceed to discuss the dynamics. Let $q_1$ and $q_2$ be two C.S.C.O.'s in our theory, so that our fundamental physical description may be given by the transition rates:

\begin{align}
\braket{q_2',t_2|q_1',t_1}.
\end{align}
Then, the Schwinger Action Principle ascertains that variations on those transitions will take the form

\begin{align}
\delta \!\braket{q_2',t_2|q_1',t_1} = i\braket{q_2',t_2| \,\delta S\, |q_1',t_1}, \label{SAP}
\end{align}
where $S$ is the action functional of the theory. Just like \eqref{1} or \eqref{action}, it is to be regarded as a function of the basic dynamic variables; then, in the quantum case, it will be an operator. We then define the \emph{effective action} $W$ by

\begin{align}
\braket{q_2',t_2|q_1',t_1} = e^{iW}. \label{effective action}
\end{align}

From \eqref{SAP} and \eqref{effective action}, we find that variations in $W$ read

\begin{align}
\delta W = \frac{\braket{q_2',t_2| \,\delta S\, | q_1',t_1}  }{\braket{q_2',t_2|q_1',t_1}}.
\end{align}

In relativistic field theory, we shall have a very similar situation. The most relevant distinction to our discussion so far is that we must trade the simple notion of a time instant for that of a Cauchy surface, so our basic configurations are correspondingly modified $\ket{q',t} \rightarrow \ket{\zeta',\Sigma}$, and our transition amplitudes read

\begin{align}
\braket{\zeta_2', \Sigma_2|\zeta_1', \Sigma_1} \qquad \Rightarrow \qquad \delta W = \frac{\braket{\zeta_2',\Sigma_2| \,\delta S\, | \zeta_1',\Sigma_1}  }{\braket{\zeta_2',\Sigma_2| \zeta_1',\Sigma_1} } . \label{field schwinging}
\end{align}

Now, these variations in the action could come from different sources. The most obvious one is a change in the dynamical variables, of which it is a direct function. But one could also vary the basic \emph{parameters} of the theory (such as masses and coupling constants, or even adding external sources) or Cauchy-Surface boundaries of the action integral \eqref{action}.

A convenient way for us to analyze variations in the action is by modifying it through the addition of an external classical source term. In the case of a single scalar field, we introduce a scalar source $J$, modifying the action $S$ as follows:

\begin{align}
S[\phi] \longrightarrow S_J[\phi] = S[\phi] + \int d\mu_g(x) J(x)\phi(x). \label{action w source}
\end{align}
Thus, the field equations are correspondingly modified to

\begin{align}
\frac{\delta S_J}{\delta \phi(x)} = \frac{\delta S}{\delta \phi(x)} + J(x) = 0. \label{ddelta_S}
\end{align}

Considering the original field equations \eqref{ELcurv}, this yields

\begin{align}
\bigl[ \Box_x + m^2 + \xi R(x) \bigl]\phi(x) = -J(x).
\end{align}

We then consider the basic functional $Z$ of our theory, which takes the form of transition amplitudes:

\begin{align}
Z[J] \equiv \braket{\zeta_2', \Sigma_2|\zeta_1', \Sigma_1}[J] = \braket{2|1}[J],
\end{align}
where we have once a simplified notation in the last equality, denoting our states as $\ket{1}, \ket{2}$. Note that $Z$ relates to the effective action $W$ simply as

\begin{align}
W[J] = -i \ln(Z[J])
\end{align}

Once again, the Schwinger Action Principle states that \emph{general} variations in this modified theory will take the form

\begin{align}
\delta Z[J] = i\braket{2| \,\delta S_J \, |1}.
\end{align} 
Particularly, if we vary \emph{only} the source $J$, leaving the remaining parameters, the dynamical variables and the integration boundaries fixed, we obtain a simple variation in the form

\begin{align}
\delta Z[J] = i \int d\mu_g(x) \Bigl(\braket{2|\phi(x)|1}\![J]\,\Bigl) \delta J(x), \label{delta1_Z}
\end{align}
or, using functional derivatives,

\begin{align}
\frac{\delta Z[J]}{\delta J(x)} = i \braket{2|\phi(x)|1}\![J]. \label{ddelta1_Z}
\end{align}

This analysis can be carried on further, and we can analyze second variations of $Z$ with respect to $J$ -- or, equivalently, variations of \eqref{ddelta1_Z} with respect to it:

\begin{align}
\delta \biggl(\frac{\delta Z[J]}{\delta J(x)} \biggl) = i \delta \bigl(\!\braket{2|\phi(x)|1}\![J] \bigl). \label{delta2_Z}
\end{align}


To evaluate the RHS of \eqref{delta2_Z} it will be convenient to consider an intermediate Cauchy surface $\Sigma$ between $\Sigma_1$ and $\Sigma_2$, containing the event $x$, and decompose the source variation in two parts: $\delta J(x') = \delta J_1(x') + \delta J_2(x')$, such that $\delta J_1$ vanishes identically to the future of $\Sigma$, $I^+(\Sigma)$ and $\delta J_2$ vanishes identically to its past, $I^-(\Sigma$). Correspondingly, we split the total variation in the form

\begin{align}
\delta \braket{2| \phi(x) |1} = \delta_2 \!\braket{2| \phi(x) |1} + \delta_1 \!\braket{2| \phi(x) |1}.
\end{align}

The trick now is to write the completeness relation with the eigenstates of an intermediate C.S.C.O. $\zeta$ in $\Sigma$, $\ket{\zeta',\Sigma}$, to conveniently evaluate each of these variations in terms of \eqref{delta1_Z}. For example, for $\delta_1$, we write

\begin{align}
\delta_1\braket{2|\phi(x)|1} &= \sum_{\zeta'} \delta_1 \Bigl( \braket{2|\phi(x)|\zeta'} \!\braket{\zeta'|1} \Bigl) \nonumber \\
 &= \sum_{\zeta'} \bigl( \delta_1 \!\braket{2|\phi(x)|\zeta'}\! \bigl) \braket{\zeta'|1} + \braket{2|\phi(x)|\zeta'}  \!\bigl( \delta_1\!\braket{\zeta'|1} \bigl).
\end{align}

However, since $\delta_1$ yields null variations between $\Sigma$ and $\Sigma_2$, only the second term will be nonvanishing. Thus

\begin{align}
\delta_1\braket{2|\phi(x)|1} &= \sum_{\zeta'} \braket{2|\phi(x)|\zeta'}  \!\bigl( \delta_1\!\braket{\zeta'|1} \bigl) \nonumber \\
 &= i \int d\mu_g (x') \delta J_1(x') \sum_{\zeta'} \braket{2|\phi(x)|\zeta'}\! \braket{\zeta'|\phi(x')|1} \nonumber \\ 
 &= i \int d\mu_g (x') \delta J_1(x') \braket{2|\phi(x)\phi(x')|1}
\end{align}
(where $\delta J_1(x)$ will vanish identically in $I^+(\Sigma)$).

Similarly, we have the variation $\delta_2$:

\begin{align}
\delta_2\braket{2|\phi(x)|1} &= \sum_{\zeta'} \delta_2 \Bigl( \braket{2|\zeta'} \!\braket{\zeta'|\phi(x)|1} \Bigl) \nonumber \\
 &= \sum_{\zeta'} \bigl( \delta_2 \!\braket{2|\zeta'}\! \bigl) \braket{\zeta'|\phi(x)| 1} \nonumber \\
 &= i \int d\mu_g (x') \delta J_2(x') \braket{2|\phi(x')\phi(x)|1}
\end{align}
(where $\delta J_2(x)$ will vanish identically in $I^-(\Sigma)$).

Then, adding the two variations, we obtain

\begin{align}
\delta \braket{2|\phi(x)|1} &= i \int d\mu_g (x') \braket{2\bigl| \delta J_1(x')\phi(x)\phi(x') + \delta J_2(x')\phi(x')\phi(x) \bigl|1} \nonumber \\
 &= i \int d\mu_g (x') \delta J(x') \braket{2\bigl| \mathcal{T}\bigl(\phi(x)\phi(x')\bigl) |1}, \label{delta21_Z}
\end{align}
where the time ordered product is defined with respect to the (arbitrary) Cauchy surface $\Sigma \ni x$. More generally, one must define a (arbitrary) foliation of spacetime between $\Sigma_1$ and $\Sigma_2$ to concretely define time-ordered products involving any two events. However, note that, since $[\phi(x),\phi(x')]$ vanishes for spacelike separated events, our result will be \emph{foliation-independent}. Finally, we can rewrite \eqref{delta21_Z}  in terms of functional derivatives:

\begin{align}
\frac{\delta^2 Z[J]}{\delta J(x_1) \delta J(x_2)} &= i^2 \braket{2\bigl| \mathcal{T}\bigl(\phi(x_1)\phi(x_2)\bigl) \bigl|1}.
\end{align}

And, by induction, it is not hard to generalize to variations of any order:

\begin{align}
\frac{\delta^n Z[J]}{\delta J(x_1) \hdots \delta J(x_n)} &= i^n \braket{2\bigl| \mathcal{T}\bigl(\phi(x_1) \hdots \phi(x_n)\bigl) \bigl|1} \label{ddeltan_Z}.
\end{align}

Now, we are in position to demonstrate the equivalence between the Schwinger Action Principle and path integrals. To do so, it is convenient to introduce a ``functional index'' notation, so that operations with continuous indexes take a similar form to those with discrete ones; this serves both to compactify our expressions and to promptly recognize operator invariants in the continuum as matrix invariants (such as determinants and traces). In this notation, we write spacetime variables as indices, like $\phi(x) = \phi^i$ or $J(x)=J_i$, integrals as implicit sums, such as

\begin{align}
\int d\mu_g(x) J(x)\phi(x) = J_i\phi^i,
\end{align}
and functional derivatives compactly with indices following commas, such as

\begin{align}
\frac{\delta S}{\delta \phi(x)} = \frac{\delta S}{\delta \phi^i} = S_{,i}
\end{align}
(where the functional variable in respect to which it is being derived is left implicit and to be understood from context, just like when one write partial derivatives compactly as $\frac{\partial}{\partial x^i} = \partial_i$).

Thus, we could rewrite \eqref{ddeltan_Z} as

\begin{align}
Z^{,\,j_1...j_n} = i^n \braket{2| \mathcal{T}( \phi^{j_1}\hdots\phi^{j_n}  ) |1}.
\end{align}

Our goal here will be to write the transition amplitudes as the functional integral of a kernel of a functional differential equation, just like the path integral is a functional integral of one fundamental kernel (given by the imaginary exponential of the action).

To do so, we write $Z$ as a Taylor series in $J$ around $J=0$. In ordinary and in compact notations, it reads

\begin{align}
Z[J] &= \sum_{n=0}^\infty \frac{1}{n!} \int \Bigl( \prod_{j=1}^n d\mu_g(x_j) \Bigl) J(x_1)\hdots J(x_n) \frac{\delta^n Z[J]}{\delta J(x_1) \hdots \delta J(x_n)}\biggl|_{J=0} \nonumber \\
 &= \sum_{n=0}^\infty \frac{1}{n!} J_{i_1}\hdots J_{i_n} Z^{,i_1\hdots i_n}[J=0].
\end{align}
Then, using \eqref{ddeltan_Z}:

\begin{align}
Z[J] &= \sum_{n=0}^\infty \Bigl(\frac{i^n}{n!}\Bigl) J_{i_1}\hdots J_{i_n} \; \braket{2| \mathcal{T}( \phi^{j_1}\hdots\phi^{j_n}  ) |1} \nonumber \\
 &=\braket{2\bigl|\;   \mathcal{T}\Bigl( \scalebox{0.9}{$\displaystyle\sum_{n=0}^\infty$} \Bigl(\frac{i^n}{n!}\Bigl) J_{i_1}\hdots J_{i_n} \phi^{j_1}\hdots\phi^{j_n}  \Bigl) \bigl| 1} \nonumber \\
 &= \braket{2\bigl| \mathcal{T}( e^{i J_i\phi^{i}} \bigl) \bigl| 1}.
\end{align}

Similarly, we can Taylor-expand the action as a function of $\phi$ around $\phi=0$:

\begin{align}
S[\phi] = \sum_{n=0}^\infty \frac{1}{n!} S_{,i_1 \hdots i_n} \phi^{i_1}\hdots \phi^{i_n},
\end{align}
as well as its first derivative:

\begin{align}
S[\phi]_{,i} = \sum_{n=0}^\infty \frac{1}{n!} S_{,i\, i_1 \hdots i_n} \phi^{i_1}\hdots \phi^{i_n}.
\end{align}

This allows us to write the expression

\begin{align}
S_{,i}[\phi]e^{iJ_j\phi^j} &= \sum_{n=0}^\infty \frac{1}{n!} S_{,i\, i_1 \hdots i_n} \phi^{i_1}\hdots \phi^{i_n} e^{iJ_j\phi^j} \nonumber \\
 &= \sum_{n=0}^\infty \frac{1}{n!} S_{,i\, i_1 \hdots i_n} \Bigl[ i^{-n} \frac{\delta^n}{\delta J_{i_1} \hdots \delta J_{i_n}} e^{iJ_i\phi^i} \Bigl] \nonumber \\ 
 &\equiv S_{,i}\Bigl[i^{-1}\frac{\delta}{\delta J}\Bigl]e^{iJ_j\phi^j},
\end{align}
where we have defined the functional differential operator $S[i^{-1} \frac{\delta}{\delta J}]$ in the last line. Then, applying it to our transition amplitudes $Z[J] = \braket{2|1}[J]$, we obtain

\begin{align}
S_{,i}\Bigl[i^{-1}\frac{\delta}{\delta J}\Bigl] Z[J] &= \braket{2 \bigl| S_{,i}\Bigl[i^{-1}\frac{\delta}{\delta J}\Bigl] \mathcal{T}( e^{i J_j\phi^{j}} \bigl) \bigl| 1} \nonumber \\
 &= \braket{2 \bigl| S_{,i}[\phi] \mathcal{T}( e^{i J_j\phi^{j}} \bigl) \bigl| 1} \nonumber \\
 &= -J_i Z[J], \label{FDiff_eq_Z}
\end{align}
where we have used \eqref{ddelta_S} in the last equality.

Then, we want to solve equation \eqref{FDiff_eq_Z} for the transition amplitude $Z$ through a functional integral kernel $F[\phi]$. Of course, a specific solution should not only depend on the differential equation, but also on a given set of boundary conditions. Rigorously, it is far from trivial to properly define functional integrals, as well as providing appropriate boundary conditions. In what follows, however, we shall ignore these subtleties and quite pragmatically assume that it is possible to define appropriate measures on the functional space to which $\phi$ belongs, as well as to, impose boundary conditions that make $F[\phi]$ vanish at infinity ``for every point in space'' (i.e., $F[\phi] \rightarrow 0$, as $\phi(x) \rightarrow \pm \infty$, for any $x$ in $\mathcal{M}$). Then using some measure $\mu[\phi]$ in space function we write the following ``Fourier Transform'' for our functional:

\begin{align}
Z[J] = \int d\mu[\phi] F[\phi] e^{iJ_i\phi^i},
\end{align}
in order to solve for the integral kernel $F$, we enforce equation \eqref{FDiff_eq_Z}:

\begin{align}
0 &= \int d\mu[\phi] \Bigl( S_{,i}[\phi] + J_i \Bigl) F[\phi] e^{iJ_j\phi^j} \nonumber \\
 &= \int d\mu[\phi] \biggl( S_{,i}[\phi]F[\phi] e^{iJ_j\phi^j} -i F[\phi] \frac{\delta}{\delta \phi^i} e^{iJ_j\phi^j}  \biggl) .
\end{align}
Then, integrating the second terms by parts (assuming a boundary condition for which $F[\phi]$ vanishes as $\phi$ reaches infinity at \emph{any} event $x$)\footnote{In the light of the result \eqref{FIntegralKernel} below, one particularly convenient way to enforce this boundary condition, which is intimately associated with the Feynman propagator $G_F$ is to add a small negative imaginary contribution to the mass, $m^2 \rightarrow m^2 - i \epsilon$ so that $S$ gets an infinite imaginary contribution as $\phi \rightarrow \infty$.} , we obtain

\begin{align}
0 &= \int d\mu[\phi] \biggl( S_{,i}[\phi]F[\phi]  + iF_{,i}[\phi]  \biggl) e^{iJ_j\phi^j}.
\end{align}

Since this equation must be valid for any event $x$ (for any index $i$ in our compact notation), and for any source $J$, it implies

\begin{align}
S_{,i}[\phi]F[\phi]  + iF_{,i}[\phi] = 0 \qquad \Rightarrow \qquad F[\phi] = C e^{iS[\phi]}, \label{FIntegralKernel}
\end{align}
being $C$ a normalization constant.

Thus, we finally obtain transition amplitudes in the form of path integrals:

\begin{align}
Z[J] \equiv \braket{2|1}[J] = C \int d\mu[\phi]\, e^{i(S[\phi] + J_j\phi^j)}. \label{ZFeynman}
\end{align}

This recovers Feynman's formulation for our modified action \eqref{action w source}. Particularly, taking $J=0$, we recover Feynman amplitudes for our original action.

From the relation \eqref{ZFeynman}, we can immediately find an expression for the effective action as a function of the source $J$:

\begin{align}
 e^{iW[J]} = Z[J] = \braket{2|1}[J],
\end{align}
which will yield variations with respect to $J$ in the form

\begin{align}
\frac{\delta Z}{\delta J(x)} = i e^{iW} \frac{\delta W}{\delta J} \quad \Rightarrow \quad \frac{\delta W}{\delta J(x)} = \frac{\braket{2|\phi(x)|1}}{\braket{2|1}} \equiv \braket{\phi(x)}.
\end{align}
Note that here we are using the brackets $\braket{\hdots}$ to denote the normalized transition amplitudes between nonorthogonal initial and final states. These will only coincide with ordinary expected values in a fixed state when $\ket{2}=\ket{1}$.

Particularly, we will be interested in the case of a free scalar field, whose action will be bilinear in field operators (and its spacetime derivatives), so that its Taylor expansion on field variables is simply

\begin{align}
S_J[\phi] = \tfrac{1}{2}S_{,ij}\phi^i\phi^j + J_i\phi^i .
\end{align}

This will be a particularly tractable case, because path integrals can then be evaluated as the product of (an infinite number of) Gaussian integrals. Note that we can write an expression for the effective action in the form

\begin{align}
e^{iW[J]} = \int d\mu[\phi] e^{i \frac{1}{2}S_{,ij}\phi^i\phi^j + J_i\phi^i}. \label{GI effective action}
\end{align}

Before we can evaluate this functional integral, let us analyze a simpler analogue of it in a finite-dimensional space. If $A$ is a symmetric operator acting on $\mathbb{R}^n$ -- such that its components can be written as $n\times n$ matrices, with components $A_{ij} = A_{ji}$ -- we define the following integral:

\begin{align}
I(A) = \int d^n\!x \,e^{\frac{i}{2}(x,Ax)} = \int d^n\!x \, e^{\frac{i}{2} A_{ij}x^ix^j}.
\end{align}

Since $A$ is symmetric, it can be diagonalized by an orthogonal matrix $M$: $L = M^T A M$, where $L_{ij} = L_j \delta_{ij}$. Then, we can use $M$ to perform a variable transformation in our space $x \rightarrow y = M^Tx$, with unit Jacobian, such that our integral reads

\begin{align}
I(A) = \int d^n\!y \, \prod_{j=1}^n e^{\frac{i}{2}L_j (y^j)^2} = \prod_{j=1}^n \biggl( \int\limits_{-\infty}^{+\infty} \!dy^j\, e^{\frac{i}{2}L_j(y^j)2} \biggl) = \prod_{j=1}^n \Bigl( \frac{2\pi i}{L_j} \Bigl)^{\!\frac{1}{2}}.
\end{align}

Well, but this is just a product of the inverse eigenvalues of $L$, which are the same as the eigenvalues of $A$; we have that $\prod_j L_j = \det(L) = \det(A)$. Thus, our integral reads

\begin{align}
I(A) = (2\pi i)^{\frac{n}{2}} (\det A)^{-\frac{1}{2}} = (2\pi i)^{\frac{n}{2}} \bigl(\det A^{-1}\bigl)^{\frac{1}{2}}
\end{align}
(where we have written the last equality in terms of the inverse operator $A^{-1}$, $(A^{-1})^{ij}A_{jk} = \delta^i_{\;\,k}$).

Then, if we define a new measure $\mu$ on $\mathbb{R}^n$, by absorbing constant prefactor $(2\pi i)^{1/2}$:

\begin{align}
d\mu(x) = \prod_i \frac{dx^i}{(2\pi i)^{1/2}}, \label{measure safada}
\end{align}
we obtain simply

\begin{align}
\int d\mu(x) e^{i A_{ij}x^ix^j} = (\det A)^{-\frac{1}{2}} = \bigl(\det A^{-1}\bigl)^{\frac{1}{2}}.
\end{align}

Then, back to our infinite-dimensional case, we can take $J=0$ and perform an analogous Gaussian integral, yielding

\begin{align}
e^{iW[0]} = \int d\mu[\phi] e^{i \frac{1}{2}S_{,ij}\phi^i\phi^j} =  \det( S_{,ij} )^{-\frac{1}{2}}. \label{EA deteminant}
\end{align}
Note that, in defining a suitably normalized \eqref{measure safada}, we avoided a divergent constant prefactor in our infinite-dimensional expression. However, generally, any constant multiplicative factor will yield only a constant additive factor to $W$, which will make no contribution to its variations.

Now, we can also write this expression in terms of an inverse operator $G^{ij}$, obeying $G^{ij}S_{,jk} = \delta^i_{\;k}$. Such inverse operators are given exactly by the Green functions to our field equations. In our functional case, however, the specific Green function to be used will also depend on a choice of boundary conditions. We shall not dive in the technical details here about these, but, for our particular choice of boundary condition in our path integrals, the appropriate kernel will be given by the Feynman propagator $-G_F$ (the reversed sign is due to equation \eqref{GF minus equation} ), which obeys

\begin{align}
\int d\mu_g(x') \, \bigl[ (\partial_t^2 + H_\mathbf{x}^2)\delta(x,x') \bigl] G_F(x',x'') = -\delta (x,x'')
\end{align}
(where we take note that $G_F$ appeared in section \ref{2-PFs} as the kernel of particular limit of the modified field equations \eqref{imaginary kernel}). In condensed notation, this reads

\begin{align}
S_{,ij}(G_F)^{jk} = - \delta_i^{\;k} \qquad \Leftrightarrow \qquad S_{,ij}(-G_F)^{jk} =  \delta_i^{\;k}.
\end{align}

Then, we can write the effective action $W \equiv W[J=0]$ as

\begin{align}
W = -\frac{i}{2} \ln(\det(-G_F)) = -\frac{i}{2} \operatorname{Tr}[\ln(-G_F)], \label{WlnGF}
\end{align}
where we evaluate the trace of an operator $K(x,x')$ in the continuum as

\begin{align}
\operatorname{Tr}K = \int d\mu_g(x) K(x,x).
\end{align}

To obtain an operationally useful representation of $G_F$, we use the following integral representation for a regularized inverse operator:

\begin{align}
K^{-1} = \lim_{\epsilon \rightarrow 0^+}(K - i\epsilon)^{-1} = \lim_{\epsilon \rightarrow 0^+} \Bigl\{ -i \int\limits_0^\infty ds\, e^{-i(K-i\epsilon)s} \Bigl\}.
\end{align}

This identity allows one to perform a spectral integral in $G_F$, so that it can be cast in a the form due DeWitt\footnote{ For more details on the derivation of this expression, see section 3.6 of cite{birrell} and references therein.}:

\begin{align}
G^{DS}_F(x,x') = -i(4\pi)^{-\frac{n}{2}} \Delta^{\frac{1}{2}}(x,x') \int\limits_0^\infty ds\, (is)^{-\frac{n}{2}} e^{-im^2s + \frac{\sigma}{2is}}F(x,x';is), \label{DW-S}
\end{align}
where $n$ is the dimension of spacetime, $\sigma(x,x')/2$ is the proper geodesic distance between $x$ and $x'$ and $\Delta(x,x')$ is the so-called Van Vleck determinant:

\begin{align}
\Delta(x,x') = - \det[\partial_\mu \partial'_\nu \sigma(x,x')][g(x)g(x')]^{-\frac{1}{2}}.
\end{align}

The convenient thing about this expression is that $F(x,x';is)$ can be written in the form of an asymptotic expansion:

\begin{align}
F(x,x';is) = \sum_{j=0}^\infty a_j(x,x')(is)^j,
\end{align}
where the coefficients $a_j(x,x')$ depend only on geometrical quantities evaluated at the events $x$ and $x'$. In practice, they are quite complicated to derive in curved spaces, as one must parallel transport various geometric tensors along the geodesics joining $x$ and $x'$. However, when we use this propagator to compute expectation values of \textit{local} observables, taking $x \rightarrow x'$ (going from a well-defined distribution to an ill-defined divergent object) as in the trace \eqref{WlnGF}, they will take a considerably simpler form, depending only on local geometric tensors at $x$. We display the results for the first 3 of them (see eqs. (6.46)-(6.48) of cite{birrell}):

\begin{subequations} \label{a_j}
\begin{align}
a_0(x) &= 1, \\
a_1(x) &= \bigl(\xi-1/6)R(x), \\
a_2(x) &= \frac{1}{180}R_{\alpha\beta\mu\nu}R^{\alpha\beta\mu\nu} - \frac{1}{180}R_{\mu\nu}R^{\mu\nu} + \tfrac{1}{2}(\xi -1/6)R^2 + \tfrac{1}{6}(1-1/5) \Box R . \label{a_2}
\end{align}
\end{subequations}

Then, to cast the effective action in terms of this asymptotic expansion, we note that the logarithm of an operator can similarly be written as

\begin{align}
\ln K = \int_0^\infty \frac{e^{iKs}}{is}ids
\end{align}
(where we ignore an infinite additive constant arising in the lower bound).

Now, identifying $G_F$ with $-K^{-1}$ in the above expressions, we can finally substitute \eqref{DW-S} in \eqref{WlnGF} to obtain

\begin{align}
W = \frac{i}{2} \int d\mu_g(x) \lim_{x' \rightarrow x} \biggl\{ \int\limits_0^\infty ds \, \frac{e^{-i\bigl( m^2s + \sigma/(2s) \bigl)} }{s^3} \Bigl[ \sum_{j=0}^\infty a_j(x,x') (is)^j \Bigl] \biggl\}.
\end{align}

We can then write this expression in terms of an integral of an effective Lagrangian:

\begin{align}
W = \int d\mu_g(x) \mathscr{L}_{eff}(x), \label{W(Leff)}
\end{align}
where we define

\begin{align}
\mathscr{L}_{eff}(x) = - \lim_{x' \rightarrow x} \biggl\{ \frac{\Delta^{\frac{1}{2}}(x,x')}{2(4\pi)^\frac{n}{2}} \int\limits_0^\infty ds \, \frac{e^{-i\bigl( m^2s + \sigma/(2s) \bigl)} }{s^{\frac{n}{2}+1}} \Bigl[ \sum_{j=0}^\infty a_j(x,x') (is)^j \Bigl] \biggl\}. \label{Leff}
\end{align}

From these expressions, it is possible to verify that there will be two types of divergences in $W$. The first one will be associated with taking the integral \eqref{W(Leff)} in an infinite spacetime volume. This one is relatively easy to manage, as we can still derive meaningful local expressions for its integrand. The second type of divergence are those that appear directly in the effective Lagrangian $\mathscr{L}_{eff}$. These appear when we take the limit $x \rightarrow x'$ and will be much more intricate to handle, requiring appropriate procedures of renormalization; in this limit, the damping factor $\sigma(x,x')/(2s)$ will vanish in the integrand (in fact it vanishes in the entire light cone), making the integral divergent in its lower limit.

From our asymptotic expansion, however, we see that, in $n=4$ spacetime dimensions, it will be only the first 3 terms in the integral \eqref{Leff} that will yield divergent contributions as $s\rightarrow 0$, the rest of them being regular. We write this divergent contribution as

\begin{align}
\mathscr{L}_{div}(x) = - \lim_{x' \rightarrow x} \biggl\{ \frac{\Delta^{\frac{1}{2}}(x,x')}{32\pi^2} \int\limits_0^\infty ds \, \frac{e^{-i\bigl( m^2s + \sigma/(2s) \bigl)} }{s^{3}} \Bigl[ a_0(x,x') +  a_1(x,x') (is) + a_2(x,x')(is)^2 \Bigl] \biggl\}. \label{Ldiv}
\end{align}

 Now, although this is an effective Lagrangian associated with the matter fields, the coefficients $a_0, a_1, a_2$ only depend on local geometric tensors as $x \rightarrow x'$ (see eqs. \eqref{a_j}). This fact will allow us to \emph{absorb the divergent terms in the purely geometrical, gravitational action $\mathscr{L}_G$}\footnote{
  Actually, due to the quadratic terms that appear in $a_2$ (eq \eqref{a_2}), one cannot absorb all divergencies in $\Lambda_B$ and $G_B$; it is also necessary to consider additional parameters following quadratic terms in curvature. In practice this could yield quantum corrections to GR. We note, however that it is the value of the \emph{renormalized} parameters that should have physical meaning and these should be ultimately determined by comparison with experiment. For this `corrected' action to be valid in the limits where GR is well tested, though, these quadratic coefficients would have to be relatively small (and, in principle, there is no reason why they could not be zero).
  } , whose ``bare'' (unrenormalized) form reads
 
\begin{align}
\mathscr{L}_G = \frac{R-2 \Lambda_B}{16\pi G_B},
\end{align}
by defining new (renormalized) parameters $\Lambda$, $G$ which will be conceived as a correction of $\Lambda_B$, $G_B$ by the addition of (infinite) contributions from the divergent terms arisinf in $\mathscr{L}_{div}$.

By doing that, one may define the renormalized matter action as the finite remainder:

\begin{align}
\mathscr{L}_{ren} &\equiv \mathscr{L}_{eff} - \mathscr{L}_{div}
 &= - \lim_{x' \rightarrow x} \biggl\{ \frac{\Delta^{\frac{1}{2}}(x,x')}{32\pi^2} \int\limits_0^\infty ds \, \frac{e^{-i\bigl( m^2s + \sigma/(2s) \bigl)} }{s^3} \Bigl[ \sum_{j=3}^\infty a_j(x,x') (is)^j \Bigl] \biggl\}.
\end{align}

Of course, as we can anticipate from the previous sections, actually carrying the required regularizations and subtractions in $\mathscr{L}_{eff}$ takes very cumbersome and tortuous calculations. Unfortunately, it will remain out the scope of this dissertation to actually derive some of them explicitly and illustrate these analytical procedures of renormalization involving the effective action. For those, we refer the reader to the very thorough section 6.2 of cite{birrell}, where the methods of dimensional regularization, zeta function regularization and point-split regularization are explicitly derived, and the renormalization of the geometric parameters is thoroughly discussed.

%% file: chap5.tex
\chapter{Standard and Inflationary Cosmology} \label{cosmology}

In this chapter, we finally pay closer attention to the subject of cosmology, and dwell in some of the ways in which the theoretical framework developed in the preceding chapters may help to elucidate some of the most pressing questions that we have about our own universe.

In section \ref{scosmology}, we discuss at considerable length the foundations and some paradigmatic results of standard cosmology: we start by thoroughly constructing FLRW spaces and some of its relevant cosmological observables, then, making use of these constructions, we give an overview of how they culminate in the standard cosmological model -- the $\operatorname{\Lambda CDM}$ model --, and, finally, we show some of the fundamental issues in it, which motivate the community in the field to posit a primordial inflationary period.

In section \ref{inflationandSSB} we qualitatively discuss the bases of field theory which allow for a dynamical description of an inflationary scenario and briefly comment on the related subject of spontaneous symmetry breaking, within the scope of a few simple models.

Finally, in section \ref{cinflation}, we discuss the bases and some of the developments of inflationary cosmology, within the particular scenario of chaotic inflation. Throughout this section, we consider a simplified model of an interaction scalar field to perform a few concrete computations and draw estimates for some quantities and potential observational predictions of inflation. We begin this analysis with a more thorough discussion of initial conditions, arguing that chaotic inflation should provide a reasonable framework for this matter. We follow by showing how an interacting field $\phi$ may produce a finite quasiexponential inflation phase as $\phi$ slowly decays from its unstable vacuum towards a stable one, and, in the sequence, we quantize its linearized perturbations near its slowly varying equilibrium value and show how the spectrum of this perturbation for very long wavelengths may give rise to a (nearly) scale-invariant spectrum in the CMB. Finally we comment briefly on the evolution of the universe after inflation, and how it can take the form of the hot, radiation-dominated universe that we observe (or draw well-verified predictions from) at later times.

\section{Standard Cosmology: The $\operatorname{\Lambda CDM}$ Model} \label{scosmology}

The origin and development of the universe is something that has raised many questions and speculations throughout the entire history of human civilization. However, our capacity to more closely observe the skyes, as well as our knowledge from laws of nature to systematically analyze our observations has never been nearly as powerful as it has become in the last 100 years. In this section, we shall explore some of the major developments in the field of cosmology that have arisen based on the theory of general relativity, and make for the picture of what is now known as the standard cosmological model.

\subsection{The Cosmological Principle and FLRW metrics} \label{FLRW}

Attempts to apply the General Theory of Relativity to obtain a meaningful description of (some average properties of) the large scale universe date back to the very first years of General Relativity itself. Early attempts were strongly marked by constraints of simplicity and philosophical considerations, such as that we should not occupy a distinguished position in universe, or live in a distinguished time in its history. This led to foundind hypotheses that the universe was (on average) homogeneous and isotropic, and even that it was eternal.
 Einstein himself first introduced a Cosmological Constant in his equations in 1917 \cite{einsteinLambda} to allow for a universe that was spacetime homogeneous (\textit{i.e.}, both spatially homogeneous and eternal). A few years later, in 1922,
Friedmann arrived at the first cosmological solutions for Einsteins equations which contemplated the possibilities of an expanding or contracting universe \cite{friedman}; Lemaître arrived at the same solutions indepently in 1927 \cite{lemaitre}. The possibility of an expanding universe, in opposition to a static one, came to be strongly favored after Hubble's observations in 1929 \cite{hubble} that distant galaxies seemed to be moving apart from us, with receding velocities roughly proportional to their distance. Later, in the 1930's Robertson and Walker rigorously demonstrated that these solutions were indeed unique (up to topological identifications) for a spatially homogenous and isotropic spacetime \cite{robertson,walker}. For all these contributions, these solutions are collectively known as Friedmann-Lemaître-Robertson-Walker (FLRW) spaces.

Let us now detach a little from the historical details, and go through a simple (physically motivated) mathematical construction of the FLRW spaces, starting from the basic assumptions of spatial homogeneity and isotropy. The physical motivation for this simplifying assumption lies in the so-called (modern) cosmological principle, which states that we should not occupy a special position in the universe, or, more generally that there are no distinguished positions in it. We can roughly summarize this as follows: ``for each instant in time, every point in space should look the same''. Similarly, there should be no distinguished directions in space, that is: ``in every point in space at any instant of time, every spatial direction should look the same''. Of course, such considerations do not apply at \emph{any} scales in our universe. It is evident that from subatomic and astrophysical scales the universe is highly inhomogeneous and anysotropic, particularly since the gravitational collapse of matter creates many types of structures at considerably large scales. Nevertheless, these hypotheses turn out to apply very well on \emph{very large, cosmological scales}, and increasingly so as we go backwards in time, as matter becomes less and less gravitationally clumped. Moreover, besides the simplicity, physical appeal and applicability of those hypotheses, there is the further advantage that they result in a cosmological model with very few parameters to be adjusted, as it will be tightly constrained by symmetries. Thus, it is truly remarkable that the  $\operatorname{\Lambda CDM}$ can explain so accurately our most precise cosmological observations up to this date.

Now that we have informally stated intuitive notions of spacial homogeneity and isotropy, let us formulate these in a mathematically precise manner, through geometrical restrictions in our spacetime. 

\begin{figure}[H]
\centering
\includegraphics[width=0.45\linewidth]{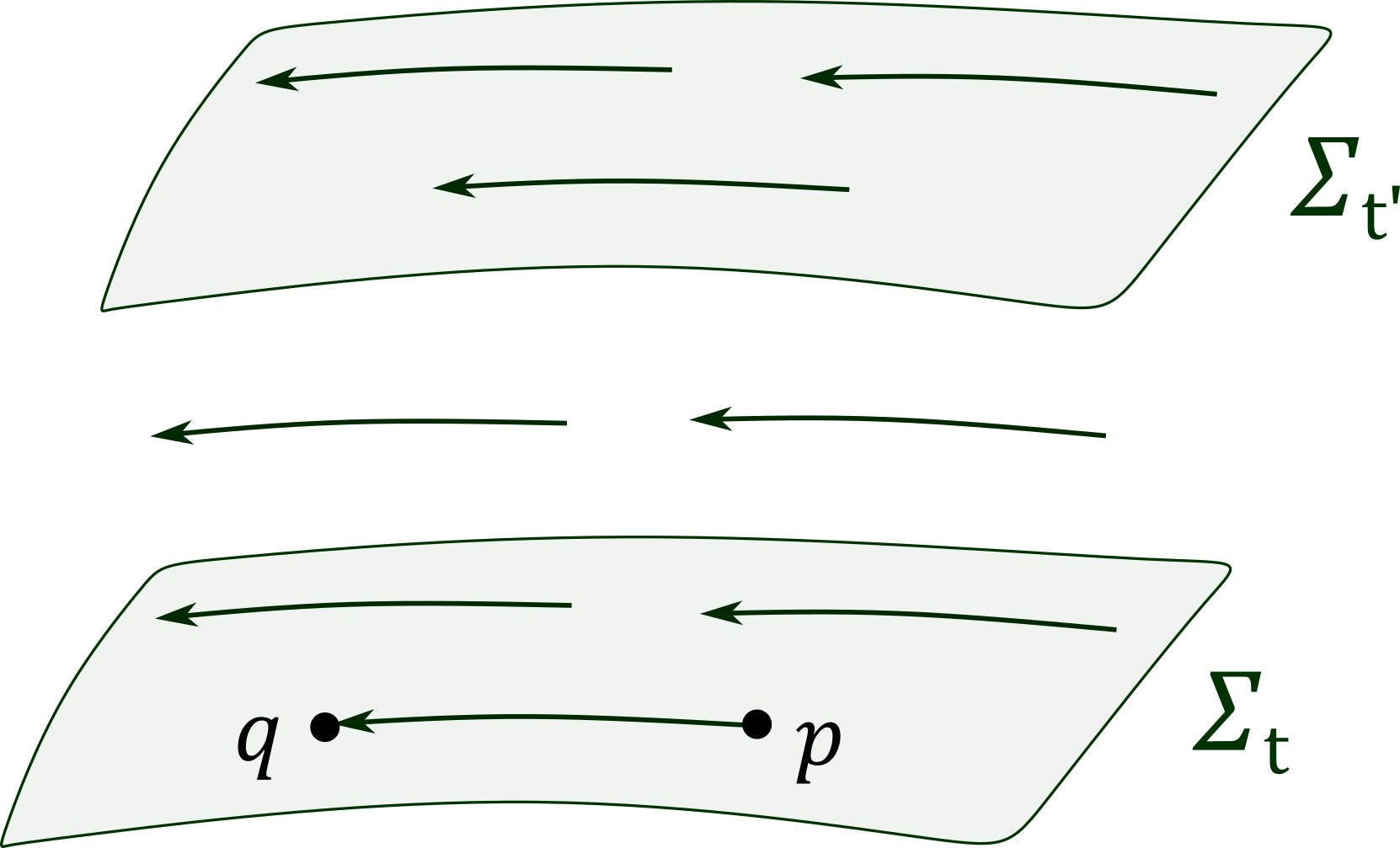}
\caption{Illustration of a translation isometry in a spatially homogeneous spacetime, which takes the point $p \in \Sigma_t$ and maps it into $q \in \Sigma_t$. This picture also tries to convey the fact that a translation isometry acts \emph{in the entire spacetime}, but mapping each homogeneous surface $\Sigma_t$ in itself. \\ Source: By the author. }
\label{Homogenous}
\end{figure}

We start with the notion of spatial homogeneity: a spacetime $(\mathcal{M},g_{ab})$ is said to be spatially homogeneous if there is a 1-parameter-family of spacelike surfaces $\Sigma_t$ foliating $\mathcal{M}$ such that, given a time instant $t$ and any 2 points $p,q \in \Sigma_t$, there is an isometry $\mathcal{I}$ ($\mathcal{I}: (\mathcal{M},g_{ab}) \rightarrow (\mathcal{M},g_{ab})$) that takes $p$ into $q$, $\mathcal{I}(p)=q$ (See Figure \ref{Homogenous}).

As to spatial isotropy, we must first emphasize that, for a general spatially homogeneous spacetime, there can be at most one observer in each event $p$ that `sees space around him as isotropic'; correspondingly, there will be at most one foliation $\Sigma_t$ of $\mathcal{M}$ which will be everywhere spatially isotropic (and homogeneous). This ``isotropic observer'' will be the one whose worldline is orthogonal to the hypersurface $\Sigma_t$ at $p$, such that his `spatial directions' all lie parallel to $\Sigma_t$. Let then $u^a$ be the tangent vector to his worldline at $p$ and let $s_1^a$ and $s_2^a$ be any two normalized purely spatial vectors to him (\textit{i.e.}, tangent to $\Sigma_t$ at $p$, such that $u_as_i^a=0$);
a spacetime will be said spatially isotropic \emph{at a point} $p$ if there is an isometry $\mathcal{I}$ preserving $p$ and $u^a$ and rotating two arbitrary normalized spatial vectors $s_1^a$ and $s_2^a$ into one another, $\mathcal{I}^*(s_1^a) \rightarrow s_2^a$ \footnote{$\mathcal{I}^*$ denotes the \textit{pushforward} map induced by $\mathcal{I}$ in vectors tangent to $M$. For more details on diffeomorphisms between manifolds and their induced maps on tangent tensor fields, see appendix \ref{geometry}.} (see Figure \ref{Isotropic}). 

\begin{figure}[H]
\centering
\includegraphics[width=0.1\linewidth]{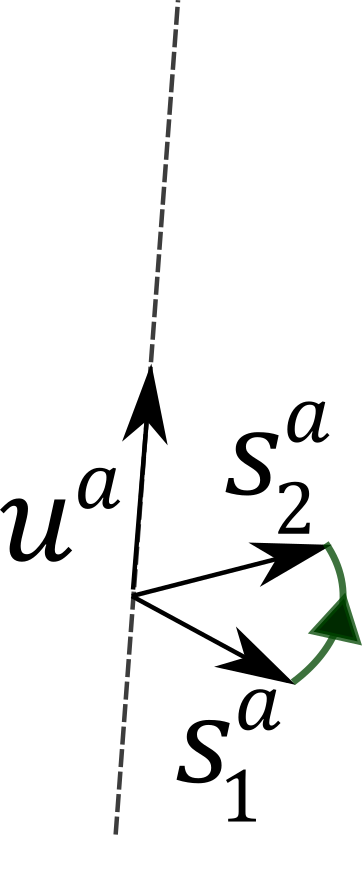}
\caption{Illustration of the action of the pushforward map $I^*$, associated with a spatial rotation isometry. Here $I^*$ rotates the normalized spatial vector $s_1^a$ in another such vector $s_2^a$. \\ Source: By the author. }
\label{Isotropic}
\end{figure}

Now the imposition that we wish to make (on the basis of the cosmological principle) is that $\mathcal{M}$ is spatially isotropic \emph{at all of its points}. As we will demonstrate next, this will be a particularly restrictive condition for spacetime geometry. Let $h_{ab}(t)$ be the (positive-definite) metric induced in $\Sigma_t$ by $g_{ab}$\footnote{
 In the entire spacetime, $h_{ab}$ can be seen as a projector (with an inverse sign for our $(+,-,-,-)$ choice of signature) on the tangent spaces parallel to each $\Sigma_t$: $h_{ab} = -(g_{ab}-u_au_b)$.};
 we may use the covariant derivative $D_a$ associated to it (i.e. $D_a h_{bc}=0$) to construct a spatial curvature tensor on $\Sigma_t$: $K_{abc}^{\;\;\;\;d}$, and then raise its 3rd index with the metric $h_{ab}$: $K_{ab}^{\;\;\;cd} = h^{ce}K_{abe}^{\;\;\;\;d}$. Due to the antissimetry properties of Riemann curvature on the first and second pair of indices ($K_{abcd} = - K_{bacd} = K_{badc}$), $K_{ab}^{\;\;\;cd}$ can be thought of as a map $L$ between 2-forms (antisymmetric rank (0,2) tensors) in this subspace:

\begin{empheq}[]{align*}
  K_{ab}^{\;\;\;cd} \rightarrow L:\;\; &W \rightarrow W \\
  &\omega_{ab} \rightarrow K_{ab}^{\;\;\;cd}\omega_{cd} \;(\equiv L\omega).
\end{empheq}
 
Further, $h^{ab}$ may be used to define an inner-product $H$ between 2-forms.
 
\begin{empheq}[]{align*}
 H: \;\; &W \times W \rightarrow \mathbb{R} \\
 & (\omega_{ab},\mu_{cd}) \rightarrow h^{ac}h^{bd}\omega_{ab}\mu_{cd} \; (\equiv \braket{\omega,\mu}),
\end{empheq}
and it is easy to see that $L$ will be a symmetrical (self-adjoint) map with respect to $H$: $\braket{\omega, L\mu} = \braket{L\omega,\mu}$. Thus, there will be in $W$ a basis of eigenvectors (``eigen-2-forms'') of $L$. The restriction of spatial isotropy will then imply that the eigenvalues of $L$ \emph{must all be the same}, otherwise, one could use this purely geometrical prescription to build distinguished 2-forms, and thus distinguished planes and directions in $\Sigma_t$ (more concretely, we can interpret that different eigenvalues would result in planes with different curvatures tangent to $\Sigma_t$). Thus, $L$ must act as a multiple of the identity operator in $W$ (and annihilate all symmetrical rank (0,2) tensors):

\begin{align}
L = \kappa\, \mathbb{1}_W \quad \Leftrightarrow K_{ab}^{\;\;\;cd} = \kappa\, \delta^c_{\;[a} \delta^d_{\;b]} \quad \Leftrightarrow K_{abcd} = \kappa\, h_{c[a}h_{b]d}.
\end{align}

Further, spatial homogeneity will imply that $\kappa$ must be a constant throughout each $\Sigma_t$. Curiously, this homogeneity actually turns out to be a necessary consequence of isotropy \emph{at all points}, which can be demonstrated by the fact that the curvature tensor $K_{abcd}$ must obey a Bianchi identity:

\begin{align}
0 = D_{[e}K_{ab]cd} = (D_{[e}\kappa)h_{|c|a}h_{b]d},
\end{align}
and, for a manifold $\Sigma$ of dimension 3 (or larger), the rightmost side of this equation will be null if, and only if, $D_e \kappa = 0$ (\textit{i.e.}, if $\kappa$ is a constant in each $\Sigma_t$).

Now, any two isotropic spaces of the same constant curvature $\kappa$ will be locally isometric. The problem of finding instantaneous possible solutions to Einstein equations with such symmetries then reduces to that of classifying all possible 3-dimensional geometries with constant isotropic curvature. As Robertson and Walker first demonstrated in the 1930's \cite{robertson, walker}, there are only 3-possibilities if we assume a usual, simply-connected topology, corresponding to the spatial metrics:

\begin{subequations} \label{FLRWdsc}
  \begin{empheq}[left= {d\Sigma^2 = \empheqlbrace}, right = {\quad,}]{align}
  d\psi^2 &+ \sin^2(\psi)[d\theta^2 + \sin^2\theta d\phi^2] \;\; \;\; (\Sigma = \mathbb{S}^3),\; \quad \kappa>0  \\
  d\psi^2 &+ \psi^2[d\theta^2 + \sin^2\theta d\phi^2] \qquad\;\; \;\; (\Sigma = \mathbb{R}^3),\; \quad \kappa=0  \\
  d\psi^2 &+ \sinh^2(\psi)[d\theta^2 + \sin^2\theta d\phi^2] \;\; (\Sigma = \mathbb{H}^3),\; \quad \kappa<0  
\end{empheq}
\end{subequations}
where we have written the line element in spherical coordinates with a proper-distance radial coordinate $\psi$. 

Along with the orthogonal contribution in the isotropic timelike directions, and accounting for a possible time-dependence in the space metric, this gives us the total spacetime metric

\begin{align}
ds^2 = dt^2 - a^2(t)d\Sigma^2 = dt^2 - d\Sigma_t^2.
\end{align}

Often, it is more convenient to write all three possibilities for the spatial metric compactly in terms of the areal radial coordinate $r$, in terms of the normalized curvature $k=a^2\kappa$, so that $ds^2$ reads

\begin{align}
ds^2 = dt^2 - a^2(t) \biggl[ \frac{dr^2}{1-kr^2} + r^2\bigl( d\theta^2 \sin^2(\theta) d\phi^2 \bigl) \biggl], \qquad 
\begin{cases} k = +1 \; \Rightarrow \Sigma = \mathbb{S}^3  \\ k = 0 \;\;\;\; \Rightarrow \Sigma = \mathbb{R}^3  \\ k = -1 \; \Rightarrow \Sigma = \mathbb{H}^3   \end{cases}\;. \label{FLRW r}
\end{align}

Note that, in the spatially flat case, the instantaneous value of $a(t)$ at any particular time $t_0$ does not have direct physical meaning, and it can always be redefined in a change of spatial coordinates (in \eqref{FLRW r}, this would be $r \rightarrow a(t_0)r$); in this case, the only physically meaningful quantity is its relative time variation $H= \dot{a}/a$. However, in the spatially curved cases (either spheric or hyperbolic) $a(t)$ \emph{directly} provides a physical length scale in the universe, namely, the one associated with the inverse spatial curvature in $\Sigma_t$ (this can be quite intuitively associated with the `radius' of the universe for the spherical case, but, although the hyporbolic case has a noncompact spatial section, \emph{both} curved geometries have intrinsic geometrical observables that reflect their curvature scales. This will become more evident in the Friedmann equations).

Note further that all considerations so far have not specifically assumed GR (except insofar as we are assuming spacetime to have a specific structure of a 4D pseudo-Riemannian manifold with a Lorentzian metric): we have not yet imposed Einstein's equations. Such features will then be common to any gravity theories that share this basic structure, as long as we constrain the analysis with the very strict symmetry hypothesis of perfect spatial homogeneity and isotropy.

Having analyzed some basic geometrical features of a spatially homogeneous and isotropic universe, and arrived in the general form of a FLRW metric, we would now like to substitute the general metric \eqref{FLRW r} in Einstein's equations to derive predictions about the dynamical evolution of our universe -- \textit{i.e.} to obtain an explicit form of $a(t)$ given a distribution of matter and energy. Therefore, to do so, let us begin by making a few considerations about matter and energy content in a homogeneous and isotropic universe.

We begin by noting that the most general matter/energy distribution which is fully consistent with our hypotheses of homogeneity and isotropy will take the form of a perfect fluid (that is, a fluid without viscosity or heat transfer and with null velocity as seen by isotropic observers), less of any terms directly proportional to curvature. One can see why that is by noting that, in that case, the stress tensor for matter can only be built using the metric $g_{ab}$ and the timelike vector field $u^a$ tangent to the isotropic worldlines. Thus, the most general symmetrical rank 2 tensor we can build is of the form\footnote{In principle, if one allows terms proportional to curvature, he/she could also add terms proportional to $R_{ab}$, $R g_{ab}$ and $R u_a u_b$ without violating general covariance or the cosmological principle. However, it is highly unusual to consider such forms of \textit{matter} density that relate directly to curvature. Furthermore, \emph{in the right combination}, these terms could be \textit{partly} shoved to the LHS of Einstein's Equations, redefining $G$.}

\begin{align}
T_{ab} = \alpha u_au_b + \beta g_{ab}, \label{HIfluid}
\end{align}
and, using the standard identifications of energy density $\rho = T_{ab}u^au^b$ and pressure $p=T_{ab}s^as^b$ (where, again, $s^a$ is \emph{any} normalized spatial vector tangent to the isotropic space dections $\Sigma_t$), one can easily cast (\ref{HIfluid}) in the form

\begin{align}
T_{ab} = \rho u_a u_b + p(u_a u_b-g_{ab}), \label{perfectfluid}
\end{align}
which is that of a general perfect fluid. Note that the homogeneity condition further restricts $\rho$ and $p$ to be (at most) functions of time.

In practice, for a cosmological analysis, it will be generally convenient to split $T_{ab}$ in different components to account for different types of matter with different equations of state (that will dictate how $p$ and $\rho$ are related in equilibrium conditions). Such components can be well approximated as noninteracting, for an appreciable part of the history of the universe.

A particularly simple component, which has been dominant for a significant part of the history of our universe, is given by nonrelativistic/cold matter, which is very well modeled as a pressureless fluid $T_{ab}^{dust}= \rho u_au_b$; this component is very commonly known as `dust'. Another significant component is the one given by ultrarelativistic energy contributions -- most proeminently in the form of electromagnetic radiation, although this would equally apply to any massless particles/fields -- whose equation of state in an isotropic distribution is just $p=\rho/3$.

Then, we would like to evaluate Einstein's equation \eqref{EE} (with $\Lambda = 0$) for a FLRW metric (\ref{FLRWdsc}) with a source of the type of a perfect fluid (\ref{perfectfluid}) with a given equation of state, so that we may solve for $a(t)$, $\rho(t)$ and $p(t)$. In this highly symmetric metric, one can show (with similar arguments as those for the spatial curvature) that the purely spatial portion of the Ricci tensor $R_a^{\;\,b}$ must be proportional to the identity operator ($\prescript{(3)}{}{\delta}_a^{\;\,b}$) on the subspaces tangent to $\Sigma_t$, and that its space-time components should vanish, so that we end up with only two independent components: the time-time ($R_t^t$) and the (isotropic) space-space ($R_i^i$) ones. Thus, the independent components of \eqref{EE} read

\begin{subequations} \label{HI_EE}
  \begin{align}
   G_{tt} &\equiv R_{tt} - \tfrac{1}{2}g_{tt}R = -8\pi T_{tt} = 8\pi \rho, \\
   G_{**} &\equiv R_{**} - \tfrac{1}{2}g_{**}R = -8\pi T_{**} = 8\pi p,
  \end{align}
\end{subequations}
where we have denoted the normalized spatial components of the tensors with an asterisk, $T_{**} = T_{ab}s^as^b$. In terms of coordinate components, they are simply $T_{**} = (g_{ii})^{-1}T_{ii}$.

Now, to actually evaluate those equations, we must first obtain the Ricci Tensor explicitly in terms of $a(t)$ to put it in the LHS of the equations. These calculations are somewhat lengthy and quite mechanical, so that it is generally useful to obtain them from a symbolic calculator software. Their results for the Ricci components read

\begin{subequations}
  \begin{align}
   R_{tt} &= 3\frac{\ddot{a}}{a}, \\
   R_{**} &= (g_{ii})^{-1}R_{ii} = - \biggl[ \frac{\ddot{a}}{a} + 2 \Bigl( \frac{\dot{a}}{a} \Bigl)^2 + \frac{2k}{a^2} \biggl].
  \end{align}
\end{subequations}

From those components, it is easy to compute the Ricci scalar:

\begin{align}
R = 6 \left[ \frac{k}{a^2} + \Bigl( \frac{\dot{a}}{a} \Bigl)^{\!2} + \frac{\ddot{a}}{a} \, \right]. \label{FLRWscalar}
\end{align}

Substituting these in \eqref{HI_EE}, we finally obtain the famous Friedmann equations:

\begin{subequations} \label{FrEq}
 \begin{align}
 \frac{\dot{a}^2\!}{a^2\!} &= \frac{8\pi\rho}{3} - \frac{k}{a^2}, \label{FrEq1} \\[3pt]
 \frac{\ddot{a}}{a} \;&= -\frac{4\pi}{3}(\rho+3p). \label{FrEq2}
 \end{align}
\end{subequations}

Before analyzing in further detail the dynamical consequences of these equations, let us see how the matter and energy content should evolve subject to them when we have a simple equation of state. Multiplying equation (\ref{FrEq1}) by $a^2$ and taking a time derivative we get

\begin{align}
2\dot{a}\ddot{a} = \frac{8\pi}{3}(\dot{\rho}a^2 + 2\rho a\dot{a}) \qquad \Rightarrow \qquad \dot{\rho} + \Bigl( 2\rho  - \frac{3}{4\pi} \frac{\ddot{a}}{a} \Bigl) \frac{\dot{a}}{a} = 0
\end{align} 
and, then, using (\ref{FrEq2}):

\begin{align}
\dot{\rho} + 3(\rho+p) \frac{\dot{a}}{a} = 0.
\end{align}

Generally, we must supply additional information regarding the relation between $p$ and $\rho$ (\textit{i.e.} an equation of state) so that we may derive the full joint evolution of spacetime and matter. Since for a great part of the history of the universe we may treat it as dominated by a single component, a very simple and widely applicable class of equations of state will be given by a simple proportionality relation of the form $p = w\rho$\footnote{In the literature, $w$ itself is commonly called the `equation of state'.}, being $w$ a constant. In this case, we have

\begin{align}
\dot{\rho} + \alpha \rho \frac{\dot{a}}{a} = 0, \label{preconservation}
\end{align}
where we have defined the proportionality constant $\alpha \equiv 3(1+w)$. Now, it is easy to see that (\ref{preconservation}) simply expresses a conservation law in the form

\begin{align}
\frac{d}{dt}(\rho a^\alpha) = 0 \qquad \Rightarrow \qquad \rho a^\alpha = cte. \label{conservation}
\end{align}

Particularly, for a spatially flat universe ($k\!=\!0$), this will yield a very simple power-law solution to eq \eqref{FrEq1} when $w \neq -1$, $a(t) \propto t^\lambda$, where

\begin{align} 
\lambda = \frac{2}{3(1+w)} \qquad \Leftrightarrow \qquad w = \frac{2 - 3\lambda}{3\lambda}, \label{w(l)}
\end{align}
whereas the $w=-1$ case yields an exponential solution $a(t) \propto e^{Ht}$.

Particularly, for dust and radiation (for which $w=0$ and $w=1/3$, respectively), we find

\begin{subequations}
  \begin{empheq}[left= \empheqlbrace, right={\qquad\quad .}]{align}
  \rho_{dust} &\propto a^{-3}, \qquad \qquad a(t) \propto t^{2/3} \\
  \rho_{rad} &\propto a^{-4}, \qquad \qquad a(t) \propto t^{1/2}
\end{empheq}

In general, one can see that, for any form of matter with positive energy densities $\rho>0$ and nonnegative pressures $p \geq 0$ (or, equivalently, $w\geq 0$ if we already assume the first equality), we have that $\rho$ will always decay at least as fast as $a^{-3}$. For $w=0$, this decay can be thought of as a conserved \emph{total energy} being diluted in a volume that scales as $a^3$ in the expansion. For $w>0$, not only is the energy diluted but the positive pressure also \emph{performs work onto the expansion}, causing a corresponding decrease in the total energy.

Two other important contributions in the Friedmann equations, which can play the role either of an \textit{actual} energy component or an \textit{effective} one, are terms associated with spacial curvature and terms proportional to the metric (the latter, as we have seen in chapters 3 and 4, can arise either as a modification in GR by the insertion of a cosmological constant or as term due to vacuum energy\footnote{
In fact, this property of $\Lambda$, namely, that it can be inserted in either side of Einstein's equation and interpreted either as a geometrical modification of GR or a matter source, was exactly what we used in chapter \ref{renormalization} to take divergent terms in $\braket{T_{ab}}$ and absorb them in the renormalization of gravitational constants $G$ and $\Lambda$ (either incorporating divergencies or finite corrections).});
 they will have equations of state $w=-1/3$ and $w=-1$. Thus, both these species \emph{receive work from the expansion}, and their (effective) energy densities will behave as \vspace{-8pt}

 \begin{empheq}[left= \empheqlbrace, right={\quad .}]{align}
  \rho_k &\propto a^{-2} \\
  \rho_\Lambda &\propto a^0 = cte
 \end{empheq}
\end{subequations} 

Both of these terms then entail (effective) energy densities that tend to become dominant over ordinary matter/energy forms. In fact, it is precisely the latter form that Dark Energy assumes today, being the dominant energy contribution for roughly the last 4 billion years in the history of our universe, and currently corresponding to about 70\% of our total energy density. As of spatial curvature, it seems to be neglegible in any of our cosmological observations, all of which point to a vanishing value of $k$. We will discuss this term in more detail in the next section.

\begin{figure}[H]
\centering
\includegraphics[width=0.6\linewidth]{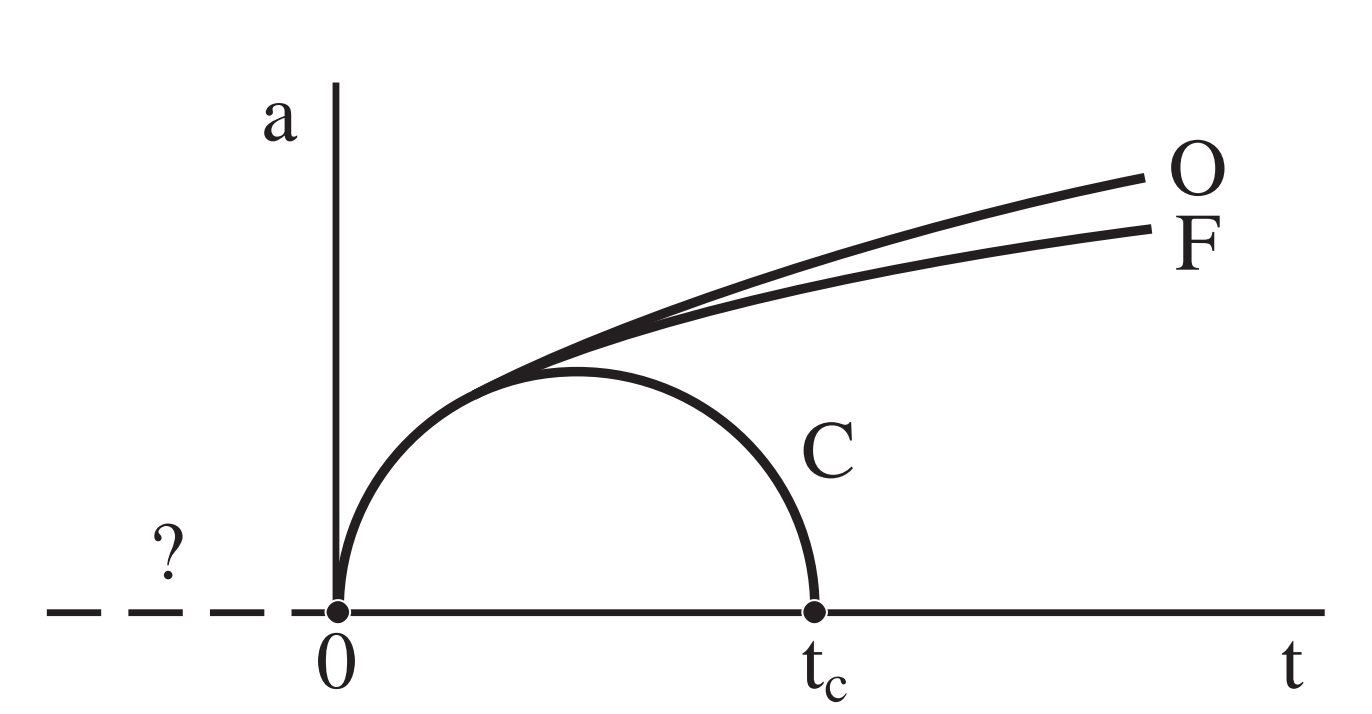}
\caption{Scenarios for FLRW universe filled with ordinary matter ($\Lambda=0$, and likewise for other exotic contributions with $w<-1/3$) for a spherical (\underline{C}losed, $k\!=\!+1$), \underline{F}lat ($k\!=\!0$) or hyperbolic (\underline{O}pen, $k\!=\!-1$) universe. In the spherical case, the universe has a finite spatial volume and recollapses in a finite time $t_c$, whereas in the flat and hyperbolic cases, it has an infinite volume and expands indefinetely. \\ Source: LINDE \cite{andrei}}
\label{standardFLRWuniverse}
\end{figure}

A further observation is that these ordinary conditions of nonnegative energy density and pressures $\rho, p \geq 0$ imply that \emph{a nonempty universe ($\rho>0$) cannot be static}; particularly, if we look at eq. (\ref{FrEq2}), we see that it must always be \emph{decelerating}, as it is expected from the purely attractive character of gravity observed in subcosmological scales (this last conclusion remains unchanged if we add a nonzero spatial curvature term, as one needs $w < -1/3$ to reverse the sign in \eqref{FrEq2}). Under these conditions, one concludes that (i) the universe must be dynamic and if it is expanding ($\dot{a}>0$), it leads to a singularity in a finite time to the past of about $\sim H_0^{-1}$, the well-known \emph{Big Bang} ($H_0$ being the present value of $H \equiv \dot{a}/a$). To the future, it can either continue to expand indefintely or recollapse, depending on the value of $k$. We summarize all three scenarios (with no exotic energy components) in Figure \ref{standardFLRWuniverse}.

The situation can be considerably different when we consider a term in the form of a cosmological constant (regardless of whether it corresponds to an actual energy term or a modification in Einstein equations). Since it has an equation of state $w=-1$, it can actually lead to an \emph{accelerated} expansion (as it has been occurring in our universe for the last 4 billion years). In fact, if one perfectly balances it with an ordinary contribution in a positive curvature scenario, one could actually find a static solution. This was, in fact, Einstein's original motivation to insert $\Lambda$ in his equations. Going from (\ref{EE}) to (\ref{EEwL}), we correspondingly modify the Friedmann equations to:

\begin{subequations}
 \begin{align}
 \Bigl( \frac{\dot{a}}{a}& \Bigl)^{\!\!2} = \frac{\Lambda+\pi\rho}{3} - \frac{k}{a^2}, \label{FrLEq1} \\[2pt]
 \frac{\ddot{a}}{a}& = \frac{\Lambda}{3}-\frac{4\pi}{3}(\rho+3p), \label{FrLEq2}
 \end{align}
\end{subequations}
for which we can obtain a static universe $\dot{a}=0=\ddot{a}$ if $\Lambda>0$ and $k=+1$. It amounts to setting

\begin{align}
\Lambda = 4\pi(\rho+3p), \qquad \text{and} \qquad a = \Bigl[ \frac{3k}{\Lambda+8\pi\rho} \Bigl]^{1/2} = \bigl[ 4\pi(\rho+p) \bigl]^{-1/2}.
\end{align}

This is known as Einstein's Static Universe (ESU). Notice that it is even more symmetric that a generic FLRW universe: it is \emph{spacetime} homogeneous and spatially isotropic (and it is still \emph{locally} symmetric by boosts, but not globally, due its spherical geometry).

To conclude this section, we want to make more explicit the notion of an expanding (or contracting) universe, in terms of the expansion (or contraction) of (instantaneous) scale distances of the isotropic spacelike surfaces associated to the isotropic cosmological frame. Note that, if at a certain time $t$, the distance between two isotropic observers (\textit{e.g.}, 2 galaxies at rest at the cosmological frame) is $R= r a(t)$ (which we can describe by fixed spherical spatial coordinates $(0,0,0)$ and $(r,\theta,\phi)$), then the rate of variation of that distance will be

\begin{align}
\frac{dR}{dt} = r \dot{a} = R \frac{\dot{a}}{a} \equiv R H, \label{hubblelaw}
\end{align}
which is directly proportional to the \emph{intantaneous} geometrical distance $R$, and the proportionality factor is merely the fractional rate of expansion of the universe, $H$, which in cosmology is called the \emph{Hubble Parameter}.

As we have briefly stated earlier, Hubble was the first one to make observations that distant galaxies were in fact moving away from us, with a velocity proportional to their distances. Of course, actual measurements will not correspond to instantaneous geometrical distances and velocities, but actually the ones along our past light cone, as light signals emmited from these galaxies take a finite time to reach us, and they will take different times for different distances, corresponding to different values of $H(t)$. Still, for sufficiently close galaxies $R \ll H_0^{-1}$, the measured velocities will approximately obey a simple proportionality relation:

\begin{align}
\frac{dR}{dt} \simeq H_0 R,
\end{align}
which was indeed found in Hubble's observations in the late 1920's \cite{hubble}, which played an important role in the realization that our universe is in fact expanding.

\subsection{Cosmological Parameters, the $\operatorname{\Lambda CDM}$ model and the Hot Big Bang scenario}

Although the formulation given so far is very useful to describe and calculate predictions for the evolution of our universe, it mostly refers to quantities that are very difficult (or even impossible) to observe directly. In order to draw a closer connection to quantities that are actually observed, which allow us to constrain our cosmological models and test predictions, we start this subsection by constructing a few observationally-oriented cosmological parameters.

A first remark in what concerns cosmological observations is that we do not have causal access to the entirety of spacetime. Obviously, we do not have access to our causal future, nor to spacelike separated regions (particularly, to any instantaneous geometrical distances in $\Sigma_{t_0}$), so that our observations are bounded to probe only the region within our past light cone. 
In fact, since the striking majority of the information we obtain is transported via electromagnetic radiation, our observation space is virtually restricted to a very narrow window \emph{on} our past light cone (since the time scales of any human observations are extremely small compared to those of cosmological phenomena, the `temporal thickness' of the set of past light cones comprising an observation -- or a series of observations -- is neglible for all practical purposes).

As we have mentioned in the previous subsection, radiation propagating in an expanding universe is redshifted. The portion of the redshift effect that is due to cosmic expansion (rather than to any peculiar velocities of the source or the observer relative to the isotropic worldlines) is called the \emph{cosmological redshift}. If we consider a light ray emitted (by an isotropic source) at an event $P_1$, with wavelenght $\lambda_1$, and observed (by an isotropic observer) at an event $P_2$, with wavelength $\lambda_2$, we define this redshift by

\begin{align}
z_{21} \equiv \frac{\lambda_2-\lambda_1}{\lambda_1} = \frac{\omega_1}{\omega_2} - 1. \label{redshift DEF}
\end{align}

\begin{figure}[H]
\centering
\includegraphics[width=0.45\linewidth]{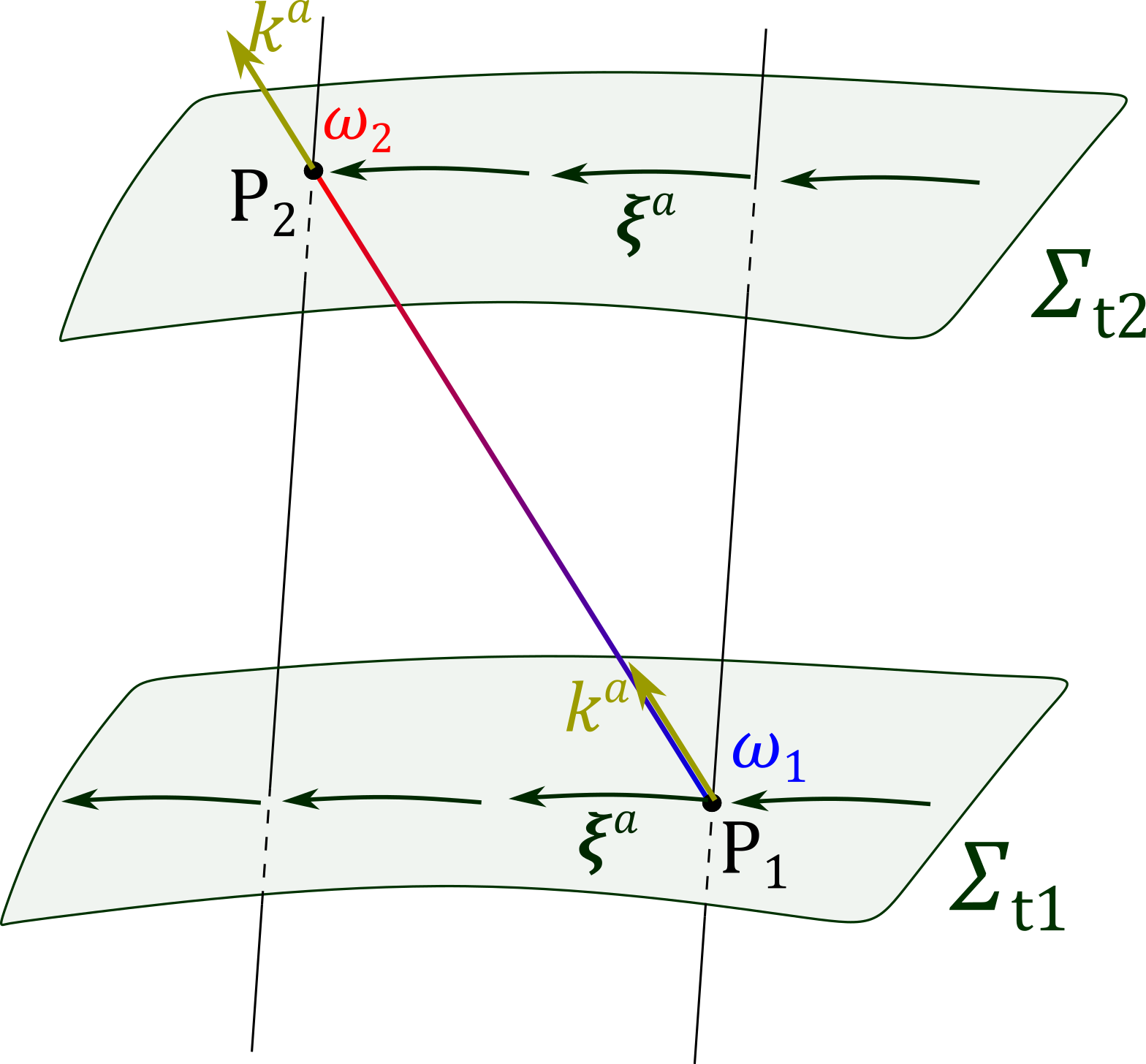}
\caption{ Depiction of a light-ray emmited at an event $P_1$, with a frequency $\omega_1$ and absorbed at an event $P_2$, with a frequency $\omega_2$, as well as the corresponding translation Killing field $\xi^a$ that joins the isotropic worldlines passing through these two events. The cosmological redshift that the light undergoes between those events can be easily computed through the conserved quantity $k_a\xi^a$. \\ Source: By the author. }
\label{Light Ray}
\end{figure}

One particularly convenient way to compute this redshift is by making use of the translation isometries in this spacetime. Let $k^a$ be the null tangent vector to the propagation of the light ray, we pick a translation Killing field $\xi^a$ that is proportional to its projection in the subspaces tangent to $\Sigma_t$, that is (see Figure \ref{Light Ray})

\begin{align}
\xi^a = -h^a_{\;\,b}k^b \propto k^a - (u_bk^b) u^a.
\end{align}

This corresponds to the Killing field that generates the translation that joins the wordlines of the source and the observer. Regardless of the field global normalization, it must scale in time as $a(t)$, as it represents a \emph{spacetime isometry}, and thus it must preserve spacetime distances and angles; in particular since it maps points on each simultaneity surface $\Sigma_t$ to points on \emph{the same} $\Sigma_t$, it must take isotropic worldlines into isotropic worldlines, to preserve their arclength (\textit{i.e.}, their proper-time intervals) between \emph{any} two surfaces $\Sigma_{t_1}$ and $\Sigma_{t_2}$. Thus

\begin{align}
\frac{|\xi_a\xi^a|^{1/2}|_{P_1}}{|\xi_b\xi^b|^{1/2}|_{P_2}} = \frac{a(t_1)}{a(t_2)}. \label{tran a}
\end{align}

Then, making use of the conserved quantity (along the propagation of the light-ray) $k_a\xi^a$ and using that $k^a$ is a null vector, $k_a k^a=0$, so that its dispersion relation is simply $\omega = |\mathbf{k}|$ (\textit{i.e.}, $k^a u_a = - k^a\xi_a/||\mathbf{\xi}||$), we obtain

\begin{align}
\omega_1 &= k_a u_1^a = - k_a \bigl[ \xi^a |\xi_b\xi^b|^{-1/2} \bigl]_{P_1}, \\
\omega_2 &= k_a u_2^a = - k_a \bigl[ \xi^a |\xi_b\xi^b|^{-1/2} \bigl]_{P_2}.
\end{align}

Thus, eq \eqref{tran a} immediately yields the frequency ratio

\begin{align}
\frac{\omega_2}{\omega_1} = \frac{|\xi_a\xi^a|^{1/2}|_{P_1}}{|\xi_b\xi^b|^{1/2}|_{P_2}} = \frac{a(t_1)}{a(t_2)},
\end{align}
and the cosmological redshift \eqref{redshift DEF} between any 2 events is found simply by

\begin{align}
z = \frac{a(t_2)}{a(t_1)}.
\end{align}

This very simple correspondence between redshift and the scale factors allows one to quite directly trace back the history of expansion of the universe by making a large number of measurements from various sources at different distances (provided one has an independent way to measure distances for distant objects\footnote{
 Indeed there are many ingenious way to define and measure distance at various cosmic scales, based on objects of known luminosity or size (which are respectively known as `standard candles' and `standart rules' in the literature), and for which one uses known methods on smaller scales to calibrate measurements on larger ones; this is known as the \emph{cosmic distance ladder} (see \textit{e.g.} \cite{liddle, ryden}). Ahead, we shall define a few operationally useful notions of distance.}).
 Furthermore, if the scale factor $a$ is a monotonic function of time (in our case, a monotonically increasing function of time), $z$ is at a one-to-one corresponce with $t$, making it a directly observable `time parameter' on our past light cone. Being $t_0$ the present time and $a_0\equiv a(t_0)$ our present scale factor, this relation yields

\begin{align}
z(t) = \frac{a_0}{a(t)}, \qquad t \leq t_0,
\end{align}
for which we can find an inverse $t(z)$ by inverting $t$ as a function of $a$ (this is particularly simple in the cases of power-law and exponential expansions).

With this construction, it is observationally more convenient to express other observables and parameters as a function of $z$, rather than $a$. Two important geometrical parameters, which appear directly in the Friedmann equations, are the Hubble parameter $H$ and the deceleration parameter $q$

\begin{align}
H^2 &\equiv \Bigl( \frac{\dot{a}}{a} \Bigl)^2 = \frac{8\pi}{3}\rho - \frac{k}{a^2}, \label{hubble} \\
q &\equiv -\frac{\ddot{a}a}{\dot{a}^2} = -\frac{\ddot{a}/a}{H^2}. \label{deceleration}
\end{align}
(Historically, $q$ was defined with a negative sign precisely due the expectation that the cosmic expansion should necessarily be \textit{decelerated}, so that one would always have $q>0$. However, as Dark Energy actually caused our expansion to be \textit{accelerated}, we presently have $q=q_0<0$.)

Now, in order to extract meaningful information in terms of cosmological redshifts, one must also be able to measure distances independently. For this reason, we give here a few definitions of cosmic distances throughout spacetime, which turn out operationally or observationally useful. The first and most obvious definition of distance we could define are instantaneous geometrical distances. For two worldlines with fixed radial distance coordinates $0$ and $\chi$, respectively, this will be simply given by

\begin{align}
d_G(t) \equiv a(t)\chi.
\end{align}

We note that it is in terms of this distance that we have written the \emph{geometrical} Hubble Law (\eqref{hubblelaw}):

\begin{align}
\frac{d}{dt}d_G(t) = H(t)d_G(t).
\end{align}

Although this particular notion of distance is geometrically intuitive, it does not have any direct observational relevance. A more useful notion of distance could be defined through our past light cone, considering the geometrical distance \emph{at the time $t$ of emission of a light ray}, which is subsequently detected in the present, $t_0$:

\begin{align}
d_{light}(t) = a(t) \int\limits_{t}^{t_0} \frac{dt'}{a(t')} = \frac{1}{1+z}\int\limits_0^z \frac{dz'}{H(z')} \label{dlight}
\end{align}
(where we have switched variables $t \rightarrow z(t)$ along the past light cone).

If one is able to invert equation \eqref{dlight}, and obtain $z(d_{light})$, it is possible to use this relation to write an \emph{observational} Hubble Law, in terms of $d_{light}$ and the Hubble \emph{constant}, $H_0 = H(z=0)$ (employing \eqref{hubble}). Although an exact inverse is not generally analytically possible, it is quite simple to linearize the equation and obtain a first-order approximation for this law:

\begin{align}
z = H_0 d_{light} + \mathcal{O}\bigl( (H_0d_{light})^2 \bigl),
\end{align}
which should be a good approximation well inside the Hubble radius, $d_{light} \ll H_0^{-1}$.

For completeness, we also mention that observationally direct notions of distance can be defined in close relation to the measured light intensity $\mathcal{I}$  of objects of known luminosity $L$, or the measured angular amplitude $\delta \theta$ of objects of known size $\delta l$. Since in Euclidean geometry these quantities relate to the geometrical distance $r$ respectively as

\begin{align}
\mathcal{I} = \frac{L}{4\pi r^2}, \qquad \delta \theta = \frac{\delta l}{r},
\end{align}
we \emph{define} the cosmic distances (in our expanding, potentially curved) associated to these measurements as

\begin{align}
d_L \equiv \sqrt{ \frac{L}{4\pi \mathcal{I}} }, \qquad d_A \equiv \frac{\delta l}{\delta \theta}.
\end{align}

These definitions may seem somewhat awkward in a geometric perspective, but the relevant point is that they can each be \emph{independently} measured in terms of observationally accessible quantities, and then used to compare and test predictions in our cosmological model.

Besides these geometrical parameters, it also proves convenient to define a few parameters for matter, in terms of which we can cast the Friedmann equations in a more observationally convenient form. First, we note that the dimensionless spatial curvature $k$ is not a free parameter, but is rather determined by the energy density. To see this more clearly, we note that if we are in the spatially flat case, $k=0$, then (\ref{hubble}) demands that the energy density assumes a very particular value in respect to the expansion rate, called the \emph{critical density}:

\begin{align}
\rho_c \equiv \Bigl( \frac{3H^2}{8\pi} \Bigl), \label{criticaldens}
\end{align}
in terms of which we define the relative energy density

\begin{align}
\Omega \equiv \rho/\rho_c.
\end{align}

Then, we immediately have that:

\begin{align}
 \begin{cases}
 \Omega < 1 \; \Leftrightarrow \rho<\rho_c  , \quad k=-1 \quad \text{(hyperbolic)} \\
 \Omega = 1 \; \Leftrightarrow \rho=\rho_c  , \quad k=0  \quad \!\!\quad \text{(flat)} \\
 \Omega > 1 \; \Leftrightarrow \rho>\rho_c  , \quad k=+1 \quad \text{(spheric)}
 \end{cases}.
\end{align}

As mentioned in the previous subsection, a useful way to decompose the total energy is to consider several components with partial densities $\rho_j$, such that $\sum_j \rho_j = \rho$, each with a constant equation of state $w_j = p_j/\rho_j$. One may then quite naturally define partial relative densities $\Omega_j$ for each component as:

\begin{align}
\Omega_j \equiv \rho_j/\rho_c, \qquad \quad \sum_j \Omega_j = \Omega,
\end{align}
as well as an \emph{effective partial density} associated with the curvature term

\begin{align}
\Omega_k \equiv -\frac{k}{a^2H^2}
\end{align}
(which we have already seen to correspond to an equation of state $w_j=-1/3$).

One can also define an effective equation of state associated to the \emph{total energy density}:

\begin{align}
w \equiv \frac{p}{\rho} = \frac{\sum_j \rho_jw_j}{\sum_j\rho_j} = \frac{\sum_j \Omega_jw_j}{\Omega},
\end{align}
although, clearly, $w$ will not generally be a constant throughout cosmic evolution even if each $w_j$ is (but it will be approximately constant whenever one single component $j$ dominates over all the others; then $w \simeq w_j$).

With these definitions, one can very conveniently rewrite the Friedmann equations (\ref{FrEq}) as

\begin{subequations} \label{zFrEq}
 \begin{align}
 &\Omega + \Omega_k = 1 ,\label{zFrEq1} \\
 &q = \frac{\Omega}{2}(1+3w). \label{zFrEq2}
 \end{align}
\end{subequations}

Further we can quite simply express the partial densities in terms of $z$ and $\{w_j\}$. Recall that for noninteracting energy components, we had a series of individual conservation laws:

\begin{align}
\rho_j a^{3(1+w_j)} = cte . \label{rho cons}
\end{align}

In terms of the present partial energy densities $\rho_{j0}$ and the redshift $z$, these can be solved simply as

\begin{align}
\rho_j(z) = \rho_{j0}(1+z)^{3(1+w_j)}, \label{rho redsdhift}
\end{align}
from which we immediately obtain the Hubble and deceleration parameters throughout our past light cone:

\begin{align}
H(z) &= H_0 \bigl[ \sum_j\Omega_{j0}(1+z)^{3(1+w_j)} + \Omega_{k0}(1+z)^2 \bigl]^{1/2}, \\
q(z) &= \frac{1}{2}\sum_j \Omega_{j0}(1+3w_j)(1+z)^{3(1+w_j)}.
\end{align}

With these parameters at hand, and armed with increasingly precise and abundant cosmological observations, we are then capable of adjusting and constraining our models, making precise (and independent) determinations of the parameters $H_0$, $\Omega_{k0}$ and $\Omega_{j0}$. These, on their turn allow us to gravitationally infer the matter and energy content of the universe, and both contrast them with other more direct forms of observation (\textit{e.g.}, the mater we \emph{see} electromagnetically) and draw predictions\textbf{/retrodictions} about details of the evolution of the universe and their observational consequences. In what follows, we give a brief account of some of them, outlining the foundations and predictions of the standard ($\operatorname{\Lambda CDM}$) cosmological model\footnote{ For a more detailed account of all those points below, see \cite{liddle, ryden, kolbturner}. }, which is extremely successful in \emph{describing} the great majority of our cosmological observations up to this date\footnote{
One very interesting exception that is arising in the most recent years is the so-called Hubble Constant Tension \cite{eleonora}.}.

Two particularly surprising results are that (i) the observed spatial curvature of our universe is essentially null (within experimental error) $\Omega_{k0} \sim 0$, and (ii) the greater part of energy density today is in the form of a nonobserved, extremely homogeneous energy form (which we call \emph{Dark Energy}) with an equation of state $w\simeq -1$, resembling a cosmological constant $\Lambda$, its relative density being $\Omega_\Lambda \sim 0.7$. Furthermore, of the remaining 30\% -- which are virtually dominated by ``dust'' (nonrelativist, preussureless matter) and concentrate in known astrophysical structures (such as galaxies, clusters and superclusters) -- only about 5\% seem to correspond to baryonic matter, $\Omega_B \sim 0.05$. The remaining 25\% correspond to some unknown species that (like dark energy) does not interact electromagnetically, thus called \emph{(Cold) Dark Matter} (CDM), with $\Omega_{CDM} \sim 0.25$. For this reason, the standard cosmological model is also known as the $\operatorname{\Lambda CDM}$ model. Together, these two dominant (and so far uncomprehended) components form what we call \emph{the dark sector}, comprising around 95\% of all the energy in the observable universe today.

Now, analyzing how the universe was at earlier times, we find immediately from equation \eqref{rho redsdhift} that components with greater values of pressure (of $w_j$) become increasingly more significant as we look further in the past (at increasing values of $z$). Particularly, we see that $\Omega_\Lambda$ decreases in relative importance, coming to a shift where the universe was dominated by cold matter at about $z\sim0.3$. Further, as we go back over 13 billion years, before matter clumped into galaxies and stars and planets could be formed, we come to a point where the universe was dominated by \emph{radiation}, at about $z \sim 3600$; at this time, the temperature of the universe was extremely high (approximately $T \approx 3600 \times 2.7K \approx 10^4K $ \footnote{
 The temperature of radiation scales proportionally to $a^{-1}\propto 1+z$, and the temperature of the cosmic radiation that we observe today is about $2.7K$, as we shall discuss briefly.}).
 Temperatures and densities then become increasingly extreme into the radiation dominated era. If we extrapolate this era all the way back to an initial singularity, as predicted in a (radiation-dominated) FLRW model, we end up with what is called \emph{the Hot Big Bang scenario}\footnote{
 In fact, one does not really require the very singular hyphothesis of a perfectly spatially homogeneous and isotropic universe. It was originally shown by Penrose (and then applied by Hawking in a cosmological context) that, according to General Relativity, singularities should actually form under much more generic conditions; see, for instance, \cite{wald, hawkellis} for singularity theorems. }. Given our knowledge from terrestrial experiments (specially at particle accelerators that reach very high energies), we can more or less safely extrapolate our predictions back to energies of about $kT \sim 10^4 GeV$, which correspond to extremely high redishifts $ z \sim 10^{13}$ and, in the standard Hot Big Bang scenario (for which $a \propto t^{1/2}$ up until a singularity), to very early times: $t \sim 10^{-13}s$ after the Big Bang. 

Our last direct observation window with electromagnetic radiation go back before any galaxies and stars were formed, at a redshift $z \sim 1100$ (roughly $t = 350.000$ years after the Big Bang), in the so-called surface of last scattering: at this time, the universe has undergone a phase transition, becoming so hot that nuclei and electrons cannot be bound together, and form an opaque plasma that constantly scatters photons. It is only after it cools enough for stable atoms to form that the universe became transparent, and this surface of last scattering forms an observable relic of the early universe, the so called Cosmic Microwave Background (CMB). We observe the CMB today as an extremely isotropic radiation from every direction in the background sky, with a distribution that fits extremely well that of a blackbody with a temperature of about $2.73K$ (as it has redshifted for a factor of more than one thousand since it was emmited in a hot plasma). If we correct for a dipole anysotropy (which is attributed to the peciliar velocity of the earth with respect to a cosmic isotropic worldline), we end up with a very homogeneous temperature distribution in it, with very small relative fluctuations $\frac{\delta T}{T} \sim 10^{-5}$. The universe was indeed extremely homogeneous at those early times.

In the next section, then, we begin to investigate the question of why it was so homogeneous (and spatially flat) to start with. Along with the Dark Sector, these are two of the greatest open questions of modern cosmology.

\subsection{Fundamental problems in the $\operatorname{\Lambda CDM}$ model} \label{lcdmP}

The universe that we observe today, of course, is far richer and more complex than a  perfectly homogeneous and isotropic spacetime. Although it is roughly homogeneous and isotropic on very large scales, it becomes richly filled with strutures of various sizes and types as we dwell in smaller ones. For the roughly thirteen billon years that have transcurred since decoupling, matter has been collapsing gravitationally, forming the various structures that we see today, ranging through superclusters, clusters, galaxies, and down to stellar systems and individual celestial bodies like stars and planets. Although the existence of these structures is a commonplace from our perspective as inhabitants of this universe, their formation process turns out to require a very particular adjusting of cosmological parameters, as we can infer in the light of our cosmological models.

On the one hand, they require initial fluctuations in density, which act as seeds for gravitational collapse; these turn out to be given precisely by the tiny fluctuations whose imprint we observe in the CMB today. On the other, they require that the average energy density of the universe is extremely well tuned to the critical density $\Omega \sim 1, \; |\Omega_k| \ll 1$, so that curvature does not quickly dominate over matter, and either recollapses the universe (if $\Omega_k<0$) or makes it expand too fast for structures to form (if $\Omega_k>0$). This is known as the flatness problem in cosmology.

To analyze this problem more, let us consider the spatially spherical case ($k=1, \Omega_k<0$) and estimate the recollapsing time of the universe for a small imbalance in $\Omega_k$ at some given initial time. This leaves us with the question of what would be a reasonable time to impose ``initial conditions'' in the universe. It certainly cannot be at $t=0$, as there should be a singularity there (according to the standard Hot Big Bang model), and, classically, any finite time seems equally arbitrary to impose them. We shall argue in more detail in section \ref{initial conditions} that a relatively natural time to do so should be given by the Planck time $t_p \sim 10^{-43}s$. For the time being, we assume this to be true, and evaluate the evolution of $\Omega_k(t)$ starting from a small imbalance $\Omega_{kp} < 0$ (here, all subscripts $p$ refer to quantities evaluated at $t=t_p$) :

\begin{align}
\Omega_k(t) = 1 -\Omega(t) = \frac{H_p^2a_p^2}{H^2(t)a^2(t)}(1-\Omega_p).
\end{align}

One can see in the Friedmann equations \eqref{FrEq} that the turnpoint between expansion and contraction happens when the matter and curvature terms cancel each other out $\Omega_k = - \Omega$. Since $H$ will vanish at that point, it actually must happen for $\Omega \rightarrow \infty$ and $\Omega_k \rightarrow -\infty$. Before that, during the time that matter is considerably dominant $\Omega \sim 1, |\Omega_k| \ll 1$, considering a radiation-dominated early universe, we have

\begin{align}
H^2(t) \simeq \frac{8\pi}{3} \rho_p \Bigl( \frac{a_p}{a(t)} \Bigl)^4.
\end{align}

Since for an approximately flat, radiation-dominated universe, $a\propto t^{1/2}$, that yields

\begin{align}
\Omega_k(t) = 1 - \Omega(t) \simeq (1-\Omega_p)\frac{t}{t_p} = \Omega_{kp} \frac{t}{t_p}. \label{curvaturek}
\end{align}

If we then estimate ``half the age of the universe'' ($T_U/2$) by a time when $\Omega_k$ and $\Omega$ become comparable, say, at $\Omega_k \sim -1$, and extrapolate (\ref{curvaturek}) up to that point, we obtain

\begin{align}
\Omega_{kp} \sim -\frac{t_p}{T/2}.
\end{align}

Thus, in order that the universe does not recolapse within a very short time $T$, one must tune the energy density at $t_p$ extremely close to critical density. For concreteness, we calculate the upper limits for this imbalance for a few values of $T$:

\begin{align}
 \begin{cases}
  T = 10^{-36}s: |\Omega_{kp}| \simeq \frac{\Omega_p-1}{\Omega_p} \lesssim 5 \times 10^{-8}, \\
  T = 10^{-26}s: |\Omega_{kp}| \simeq \frac{\Omega_p-1}{\Omega_p} \lesssim 5 \times 10^{-18}, \\
  T = 10^{18}s: |\Omega_{kp}| \simeq \frac{\Omega_p-1}{\Omega_p} \lesssim 5 \times 10^{-42}.
 \end{cases}
\end{align}

Particularly, the last case corresponds to the present age of our universe (which is not nearly in a process of recollapse), from which we see that a truly extreme fine-tuning in the curvature would be necessary for the universe to still exist up to this day. On the other hand, if we had a positive initial imbalance in $\Omega_{kp}$, the universe would have expanded drastically faster than it did, not allowing for the formation of any of the structures that we oberve today.

Having noted what seems to be an extreme coincidence regarding the spatial curvature of our universe, we turn our attention to what appears to be another great coincidence: why was the early universe so spatially homogeneous to start with? If we want to avoid the extreme coincidence of merely postulating that it was ``born'' homogeneous (but still with tiny fluctuations that allowed the formation of structures), a reasonable hypothesis would be that it had time to evolve into an equilibrium temperature and density configuration, so that our first observations actually measure this equilibrium profile. However, a problem that emerges in the standard Hot Big Bang scenario is that there are generally \emph{particle horizons}. Those are causal horizons in the past, due to the fact that each isotropic worldline has not (for a radiation-dominated expansion) had time to be in causal contact with all other isotropic worldlines, and thus come to equilibrium with them. This is known as the Horizon Problem. 

To take a closer look at this problem, it is useful to consider a spatially flat FLRW spacetime, and cast its metric in a confomally Minkowskian form:

\begin{align}
ds^2 = a^2(\eta) \bigl[ d\eta^2 - d\mathbf{x}^2 \bigl], \qquad \qquad \eta_0 - \eta = \int^{t_0}_t \frac{dt'}{a(t')}.
\end{align}

Recall that, as conformal transformations preserve the light cones, conformally related spacetimes will have the same causal structure. Thus, this FLRW spacetime will have the same causal structure as (a portion of) Minkowski spacetime. Note then that, in order for all isotropic worldlines to be causally connected at a time $t_0$ (a conformal time $\eta_0$), it is necessary that $\eta$ extends all the way past to $-\infty$ when $t \rightarrow 0$; otherwise, this spacetime will be conformally related to just a portion of Minkowski spacetime for which $\eta > \eta_{sing}$, given by

\begin{align}
\eta_{sing} \equiv \eta_0 - \int_0^{t_0} \frac{dt'}{a(t')}. \label{etasing}
\end{align}
(The value for which we define $\eta_0$ here is arbitrary and irrelevant. The point is whether $\eta_{sing}$ will be a finite time or extend all the way back to $-\infty$.)

\begin{figure}[H]
\centering
\includegraphics[width=0.6\linewidth]{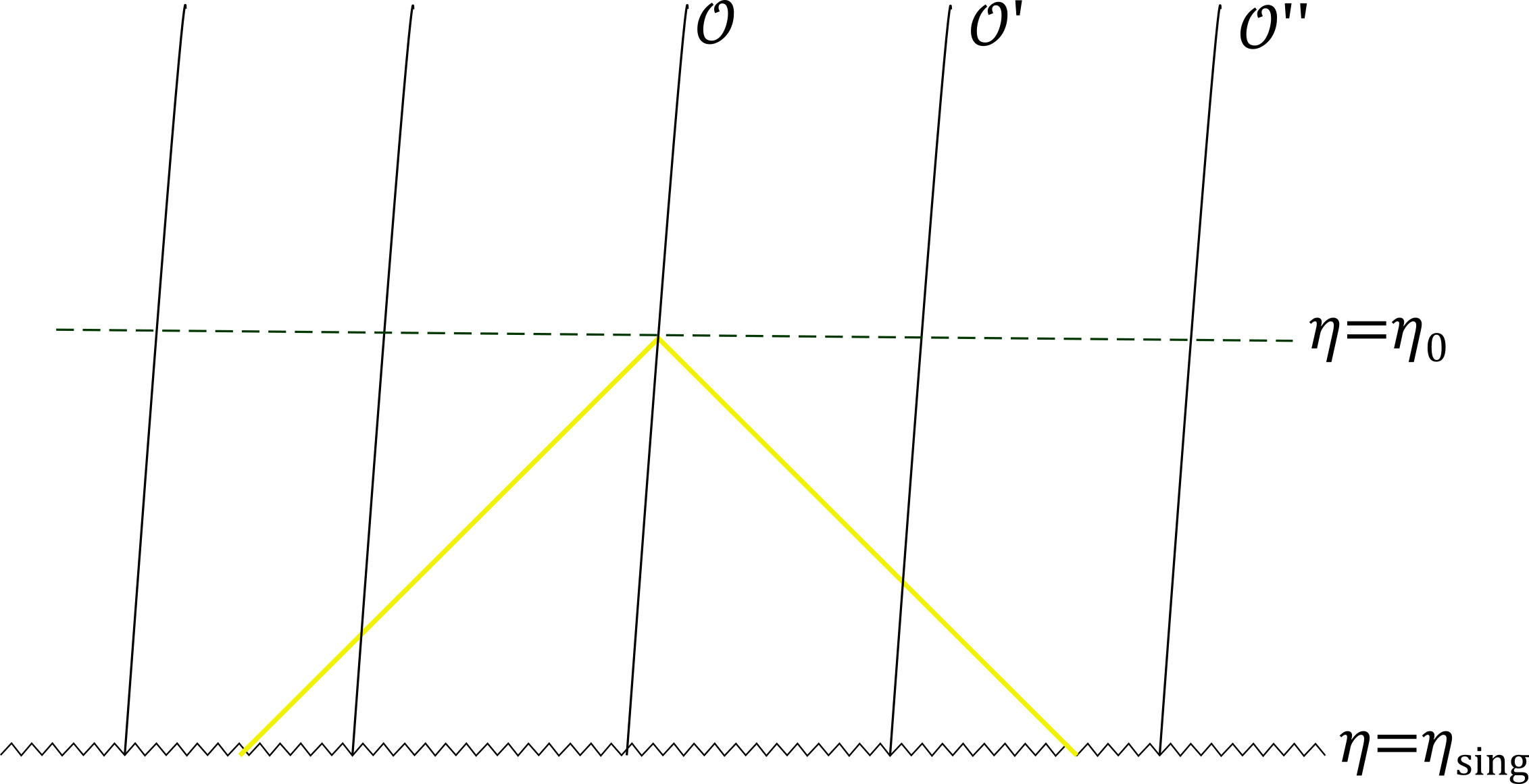}
\caption{Conformally flat FLRW spacetime, represented in conformal coordinates. If $\eta_{sing}$ turns out finite, this space will only be conformally related to a portion of Minkowski spacetime, and distant observers will not have had time to come in causal contact. Tracing the past light cone of an observer $\mathcal{O}$ at the time $\eta_0$, we see that it intersects some of the other worldline (like $\mathcal{O}'$), but not sufficiently distant ones (like $\mathcal{O}''$). \\ Source: By the author. }
\label{Conformal_PICTURE}
\end{figure}

It is apparent from \eqref{etasing} that for a power-law exapansion $a\propto t^\lambda$, with $0< \lambda < 1$, there will be causal horizons. In fact, such horizons will generally occur for any monotonically \emph{decelerated} expansions. In such cases, the Hubble radius $H^{-1}(t)$ of the universe will \emph{increase} with time, and more isotropic worldlines will become causally connected as time passes (that is, increasingly more observers will have had time to interact with one another)\footnote{
In contrast, for \emph{accelerated} expansions, there will be future causal horizons (\textit{i.e.} event horizons) and 2 isotropic observers who were causally connected at early times will later cease to be. This is quite clearly illustrated in the example of an exponentially expanding (de Sitter) space, which will be discussed in section \ref{cinflation}. }.

To analyze the distances across which different points in the far sky were connected at past time $t$, it is useful to evaluate an instanteneous causal radius around an arbitrary isotropic worldline as a function of $t$ (particularly, we will be interested at the time $t_d$ when the CMB was formed). We estimate this causal radius for a dust-dominated universe\footnote{
Note that, indeed, the universe was dust-dominated for most of the time up until the matter-radiation decoupling in the CMB. Furthermore, it makes little difference if we consider $\lambda=2/3$ or $\lambda=1/2$ in this calculation.} ($a\propto t^{2/3}$):

\begin{align}
d_G(t) = c a(t) \int_0^t \frac{dt'}{a(t')} = 3ct
\end{align}
(where we have reincorporated $c \neq 1$ into the formulas to convert to usual distance values).

Thus, for our present universe $t=t_0 \sim H_0^{-1} \sim 10^{10}Yr$, we have

\begin{align}
d_{G}(t_0) \sim 9 \times 10^3 Mpc.
\end{align}

Then, taking the time of matter-radiation decoupling $t_d$ as approximately given by

\begin{align}
\frac{a_0}{a_d} \simeq \biggl( \frac{t_0}{t_d} \biggl)^{\!2/3} \sim 10^3 \qquad \Rightarrow \qquad t_d \sim 3 \times 10^5 Yr,
\end{align}
we may estimate the causal radius at the formation of the CMB:

\begin{align}
d_{G}(t_d) \simeq 3ct_d \sim 0.3Mpc \,.
\end{align}

This represents a geometrical distance on $\Sigma_{t_d}$. We can then calculate the corresponding geometrical distance $D$ at the present time\footnote{This stretctched distance will be observationally meaningful because we are only interested in calculating an angular width from it, and angles are preserved in an isotropic expansion.} $t_0$ (\textit{i.e.}, in $\Sigma_{t_0}$) by streching it by a factor of $\frac{a_0}{a_d}= \left( \frac{t_0}{t_d} \right)^{2/3}$:

\begin{align}
D = \biggl( \frac{t_0}{t_d} \biggl)^{\!2/3} d_{light}(t_d) \sim 300 \operatorname{Mpc} \,.
\end{align}

Thus, the archlenght in the background sky corresponding to 2 causally connected points should be roughly around $D \sim 300 Mpc $. This arc, on its turn, should lie in a sphere of radius $R=d_{light}(t_0) \sim 9 \times 10^3 Mpc $, such that it will correspond to an angle $\theta $ roughly given by

\begin{align}
\theta = \frac{D}{R} \sim \frac{300Mpc}{9000Mpc} = \frac{1}{30} \text{radians} \sim 2^{\circ}.
\end{align}

This means that, according to the standard Hot Big Bang scenario, patches of the CMB separated by more than approximately $2^\circ$ should be causally disconnected! Thus, the fact that the we observe an extremely isotropic CMB would indeed be entirely coincidental, as it could not be attributed to any causal process of thermalization.

For completeness, we also mention the issue of topological defects in grand unification theories (GUTS). Based on the successful unification of the electromagnetic and weak interactions in the so-called electroweak force, at energy scales of about $\sim 1TeV$, it has been proposed that the electroweak and strong forces should become unified at energy scales many orders of magnitude grater, of about $10^{12}TeV$, corresponding to temperatures of around $10^{28}K$. As we have mentioned in the last section, extrapolation of our cosmological model predicts that the universe should indeed have risen to arbitrarily high temperatures, reaching the ones corresponding to grand unification at about $t=10^{-36}s$. The issue when we try to combine these theories with our cosmological model is that, at the point the universe cools enough to undergo a phase transition from a grand unified force to separated strong and electroweak ones, it is expected that topological defects -- particularly, very massive particles (monopoles), with $m_M \sim kT_{GUT} \sim 10^{12}TeV$ -- are produced in a certain abundance; roughly, we can estimate there to be one per causal sphere of radius $d_G(t_{GUT}) = 2t_{GUT}$ (for a radiation-dominated universe, $a \propto t^{1/2}$), such that their numerical density and mass density should have been

\begin{align}
n_{M}(t_{GUT}) \sim (2t_{GUT})^{-3}, \qquad \rho_M \sim \frac{m_M}{(2t_{GUT})^3} \sim 10^{94} TeV/m^3 .
\end{align}

Although astoundingly high, this density should have been relatively insignificant compared to the radiation-dominated critical density at the time. A quick estimate in the standard Hot Big Bang scenario yields

\begin{align}
\rho_{rad}(t_{GUT}) \sim 10^{104} TeV/m^3 \qquad \Leftrightarrow \frac{\rho_M}{\rho_{rad}}\biggl|_{t_{GUT}} \sim 10^{-10}. 
\end{align}

However, since these monopoles would have been extremely heavy, they should behave as noninteracting ``dust'' since very early times, and would become quickly dominating over radiation (the latter was dominant in the very early universe up to about $10^{12}s$). Making use of \eqref{rho cons}, we would estimate for our present universe

\begin{align}
\frac{\Omega_{M}(t_0)}{\Omega_{rad}(t_0)} \simeq 10^{18},
\end{align}
and, since radiation today is about only $10^{-5}$ of the critical density of the universe, we would have

\begin{align}
\Omega_{M0} \equiv \frac{\rho_{M0}}{\rho_{c0}} \sim 10^{13},
\end{align}
in a screaming contradiction with our observations of $\Omega_{total} \sim 1$.

Summarizing, although the $\operatorname{\Lambda CDM}$ model is extremely successful in decribing our observations with very few parameters, it turns out to carry quite significant fundamental issues, either on its own or in conjunction with other physical theories (albeit, in non-tested regimes in the case of monopoles in GUTs). How, then, could we handle all of these issues? Well, it so happens that a single (although considerably long shot) solution to all of them can be found by postulating a very short primordial inflationary period lasting until about $\sim\! 10^{-33}s$ during which the universe would have expanded quasiexponentially by a factor of at least $\sim \!10^{26}$ times. In the next sections, we shall explore the foundations and developments of this quite extreme proposal, both illuminating how it could be realized through field theory in curved spaces, and attempting to draw its connections to observational quantities.

\section{Field Theory and Inflation; spontaneous symmetry breaking} \label{inflationandSSB}

Before we actually dwell in the cosmological developments of inflation, we shall take a moment to describe how a finite inflationary phase could emerge in the joint dynamics of spacetime and matter fields in the first place. We have already seen in the previous chapter that the renormalized vacuum energy of noninteracting quantized fields in curved (de Sitter) spacetimes could give rise to a contribution in the form of cosmological constant -- whose gravitational effects, when taken into account, should ultimately produce an (eternally) exponentially expanding universe.

Then, to be able to grasp how vacuum energy could give rise to a finite-lasting inflationary phase, we are forced to extend our analysis beyond free fields and consider interacting ones. Given the difficulties in fully analyzing quantized interacting fields in curved spacetimes\footnote{ For a further discussion on interacting fields, particular on the self-interacting $\lambda \phi^4$ model, see the final chapter of \cite{birrell} and section 6.7 of \cite{parker}. } and the fact that many relevant features and effects of inflationary cosmology already arise in a classical regime, we will now turn our attention back to classical fields in curved spacetime. Later, we shall consider a perturbative approach to quantize linearized fluctuations of our field.

Throughout the rest of this chapter we shall consider, for concreteness and simplicity, a single self-interacting scalar field $\phi$, minimally coupled to gravity, whose Lagrangian will be generically of the form

\begin{align}
\mathscr{L} = \frac{1}{2}(\nabla^\mu\phi)(\nabla_{\!\mu}\phi) - V(\phi), \label{scalar}
\end{align}
where $V(\phi)$ is a generic potential.

We can immediately derive the Euler-Lagrange equation for this field, which reads

\begin{align}
\Box \phi + V'(\phi) = 0,
\end{align}
where $V' = \frac{dV}{d\phi}$.

Clearly, for the usual harmonic potential $V(\phi)=\frac{m^2}{2}\phi^2$, $m^2\!\geq\!0$ we recover a free scalar field, which can be decomposed in an infinite collection of decoupled harmonic oscillators, with positive and negative frequency modes $u_\mathbf{k}\propto e^{\mp i k_\mu x^\mu} = e^{\mp i \omega t} e^{\pm i \mathbf{k} \!\!\cdot\!\! \mathbf{x}} $, being $\omega$ and $\mathbf{k}$ real.
However, if we make a signal inversion, considering a mass $\mu^2\!=\!-m^2\!<\!0$, we end up with the well-known issues for a theory with a Hamiltonian unbounded from below: there will be no ground state and the dynamics is rendered unstable by the presence of arbitrarily negative energy eigenvalues. Indeed, in a such a case we have the field equations

\begin{align}
\bigl[ \Box - m^2 \bigl] \phi = 0,
\end{align}
for which we still have modes of the kind $u_\mathbf{k} \propto e^{\mp i k_\mu x^\mu}$, but which will have a dispersion relation

\begin{align}
k_\mu k^\mu = (k^0)^2 - \mathbf{k}^2 = -m^2,
\end{align}
such that, for $|\mathbf{k}|<m$, we are forced to consider imaginary frequencies\footnote{In a simmilar spirit, one could also consider imaginary wave vectors $\mathbf{k}$ (which also result in imaginary frequencies). These should be generally forbidden in well-behaved theories so that we do not end up with spatially exponentially divergent modes, but one is forced to consider them if space and time are to be treated in an equal-footing. The asymmetry here arises in the way we split our modes to satisfy an initial condition and boundary problem.} $\omega = \pm k^0 = \pm i\sqrt{m^2-\mathbf{k}^2}$, which lead to (unbounded) exponentially growing modes:

\begin{align}
u_\mathbf{k} \propto e^{\pm \sqrt{m^2-\mathbf{k}^2} \,t } e^{\pm i \mathbf{k} \!\!\cdot\!\! \mathbf{x}}.
\end{align}

Of course, such a plainly pathological model is of little use to us on its own. However, it is useful to transparently reveal the instabilities of a vacuum state surrounding a local maximum, which we will explore in better-behaved model. 
A very interesting nontrivial potential with global minima is the so called $\lambda \phi^4$ model, whose potential is given by

\begin{align}
V(\phi)= V_0 -\frac{m^2}{2}\phi^2 + \frac{\lambda}{4}\phi^4. \label{lf4 potential}
\end{align}

This potential has two symmetrical global minima (corresponding to two stable vacua) at $\phi\!=\!\phi_\pm \!\equiv\!\pm m/\sqrt{\lambda}$ and a local maximum (corresponding to an unstable vacuum) at $\phi\!=\!0$. This relatively simple model turns out to yield a very rich structure: as in the case of a repulsive oscillator (with $\mu^2<0$), 
 it will have exponential instabilities for modes around the local maximum at $\phi=0$; further, if the energy of this field falls below $V_0$, it will be classically confined at one of the potential wells, either around $\phi_+$ or $\phi_-$, which results in a spontaneous symmetry breaking. Particularly, if the field has an energy very close to its absolute minimum, it will have nearly constant values, just making small oscillations around one of its stable vacuum states: $\phi = \phi_\pm + \delta \phi, \, |\delta \phi| \ll |\phi_{\pm}| = m/\sqrt{\lambda}$. This last situation is particularly interesting, because it allows one to break the field in a constant classical contribution plus perturbations, which one can quantize in a linearized regime. Considering, for example, the vacuum at $\phi_0 = \phi_+$, we obtain

\begin{align}
\mathscr{L} =& \frac{1}{2}(\partial_\mu \delta \phi)(\partial^\mu \delta \phi) - \frac{1}{2}(3\lambda \phi_0^2 - m^2) (\delta \phi)^2 - \lambda \phi_0 (\delta \phi)^3 - \frac{\lambda}{4}(\delta \phi)^4 \nonumber \\
 & \; + \phi_0(m^2 - \lambda \phi_0^2) \delta\phi + \biggl( \frac{m^2}{2}\phi_0^2 - \frac{\lambda}{4} \phi_0^4 \biggl).
\end{align}
 
Here, the linear term in $\delta \phi$ will vanish because $\phi_0^2 = \frac{m^2}{\lambda}$, and the constant term can be ignored for quantization purposes (the only nontrivial role that this term may play is gravitational). Then, neglecting terms of cubic order or higher for $\delta \phi$, we can quantize as a field of effective quadratic mass $\mu^2 = 3\lambda\phi_0^2 - m^2 = 2m^2$.

A further aspect, which would have significant cosmological consequences is the possibility that topological defects may arise. At high energies the field can symmetrically explore configurations around both $\phi_+$ and $\phi_-$. As it goes into lower energies however, it is forced to collapse in either one of these regions, spontaneously breaking its symmetry. Over very large, causally disconnected regions, however, one has no reason to expect that the field will uniformly collapse in the same region (either $\phi_+$ or $\phi_-$); more realistically, it should form domains in which $\phi$ has decayed in either value. Then between any two domains, there must be a transitioning region where the field has intermediate values, which will have very high energy densities: such regions are called domain walls. Since this is a simple scalar field, these topological defects between different vacua will be quasi 2-dimensional. In more realistic theories, with different groups of symmetries, the defects may be 1-dimensional (cosmic strings) or 0-dimensional (monopoles).
 
Moreover, if $\phi$ happens to be interacting with other fields in nature, this spontaneously broken classical value may give rise to a mass term for the latter fields\footnote{
 It is beyond the scope of the present work to thoroughly present the exciting topics of symmetry breaking either in wide generality or in its applications to the fields of High Energy Physics (HEP) and Cosmology. Still, the author feels strongly compelled to make a brief discussion in some qualitative aspects of this subject, as they bear an intimate relation with many topics in this text. For a more thorough discussion, the reader is referred to the later chapters of \cite{mandlshawn} for a more pedestrian introduction to this topic in HEP, to \cite{andrei} for a further exposition in a cosmological context (including molopoles and more general topological defects in the universe), or to \cite{strocchi} for a more mathematically rigorous, model-indepent presentation.}. For example, if one considers a massless Dirac field $\psi$ coupled to our scalar field by an interaction term $\mathscr{L}_I = -h \phi \bar{\psi}\psi $ ($h$ being a coupling constant), we have a total Lagrangian 

\begin{align}
\mathscr{L} &= \mathscr{L}_\phi + \mathscr{L}_\psi + \mathscr{L}_I \nonumber \\
 &= \frac{1}{2} (\partial_\mu \phi) (\partial^\mu \phi) + \frac{m^2}{2}\phi^2 - \frac{\lambda}{4}\phi^4 + \bar{\psi}i\gamma^\mu\partial_\mu \psi -h\phi \bar{\psi}\psi
\end{align}
(where, for simplicity, we are considering spacetime to be flat).

Then, again, breaking $\phi$ as $\phi_0 + \delta \phi$ around a minimum at $\phi_0$, we obtain

\begin{align}
\mathscr{L} = \frac{1}{2}(\partial_\mu \delta \phi)(\partial^\mu \delta \phi) + \frac{\mu^2}{2}(\delta \phi)^2 + \bar{\psi}(i\gamma^\mu\partial_\mu - h\phi_0)\psi - h \bar{\psi}\psi \delta \phi + \mathcal{O}\bigl( (\delta\phi)^3 \bigl) + cte,
\end{align}
and, comparing the third term with the massive Dirac Lagrangian, we clearly find that a mass term for the fermions emerges: $m_\psi = h\phi_0 = hm/\sqrt{\lambda}$.

Similarly, in scalar electrodynamics, where one couples the electromagnetic field to a complex scalar field\footnote{
 It is necessary that the field is complex so that Noether Charge will emerge which is associated to gauge transformations, and thus can be identified with electrical charge. } $\chi$, one can build the so-called Abelian Higgs model:

\begin{align}
\mathscr{L} = -\frac{1}{4}F_{\mu\nu}F^{\mu\nu} + D_\mu \chi^* D^\mu \chi + \mu^2\chi^*\chi - \lambda (\chi^*\chi)^2,
\end{align}
being $F_{\mu\nu} = 2\partial_{[\mu}A_{\nu]}$ the Faraday tensor, and $D_\mu$ the Gauge covariant derivative, which acts on charged scalar fields as $D_\mu \chi = (\partial_\mu - ieA_\mu)\chi$ (and $D_\mu \chi^* = (\partial_\mu + ieA_\mu)\chi^*$). This obeys a local gauge symmetry:

\begin{subequations}
 \begin{empheq}[left=\empheqlbrace]{align}
 A_\mu(x) &\rightarrow A'_\mu(x) = A_\mu(x) + \partial_\mu \xi(x), \\
 \chi(x) &\rightarrow \chi'(x) = \chi(x) e^{ie\xi(x)},
 \end{empheq}
\end{subequations}
which one can exploit to make $\chi$ real everywhere. If $\chi(x) = \rho(x)e^{i\theta(x)}$, we can adjust the gauge making $\xi(x) = - \theta(x)$, so that $\chi(x) \rightarrow \chi'(x) = \chi(x)e^{-i\theta(x)} \equiv \frac{1}{\sqrt{2}}\varphi(x)$. In this transformed gauge, the Lagrangian reads:

\begin{align}
\mathscr{L} = -\frac{1}{4}F_{\mu\nu}F^{\mu\nu} + (\partial_\mu \varphi) (\partial^\mu \varphi) + \frac{\mu^2}{2}\varphi^2 - \frac{\lambda}{4} \varphi^4 + \frac{e^2}{2}\varphi^2 A'_\mu A'^\mu .
\end{align}

Then, if we once again split $\varphi = \phi_0 + \delta \varphi$ in a minimum of its potential ($|\varphi_0| = \mu/\sqrt{\lambda}$), we clearly end up with a mass term for the electromagnetic field $A^\mu$:

\begin{align}
m_A^2 = -e^2\varphi_0^2 = -\frac{e^2\mu^2}{\lambda}.
\end{align}

These simple models exemplify (at a merely qualitative level, in our superficial exposition) the types of phenomena that interacting field theories can describe, and are suggestive of the types of phase transition the universe may have undergone during a inflationary period. However, for a description of inflation itself, that is, of a finite period during which the universe has expanded at extremely fast and accelerated rate, we shall focus our attention solely on the scalar field $\phi$, and describe its evolution within a single domain. 

In our $\lambda \phi^4$ model, it is not hard to conceive how one could obtain such a finite inflationary phase: if we adjust the potential constant $V_0$ in \eqref{scalar} such that $V(\phi_\pm) \approx 0$ (and $V(\phi=0) \approx V_0 >0, \; V'(\phi=0)=0 $), it could spend a considerable amount of time near its unstable local maximum, driving a nearly constant relative expansion (\textit{i.e.}, a quasiexponential expansion) and subsequently decay into one of its local minima, giving up energy to ordinary forms of matter and energy; after it decayed, inflation would cease and the expansion would become dominated by other energy forms. In the next section, we shall then concretely consider an inflationary scenario, and derive a more quantitative description of this quasiexponential expansion dynamics driven by a classical scalar field.

\section{Inflationary Cosmology: the chaotic inflation scenario} \label{cinflation}

Historically, there have been a number of proposals for inflationary models and scenarios\footnote{
  A very complete and yet concise account of the history and successive developments of these models and scenarios can be found in section 1.6 of \cite{andrei} (and see the various references therein for particular models and developments).},
  all of which attempted to address the same basic issues raised in the last section by proposing some mechanism through which the universe might have undergone an extremely abrupt period of accelerated expansion in its very early history. Despite these basic similarities, specific models vary widely both in their qualitative features and in their quantitative predictions, which may range as much as dozens of orders of magnitude in several quantities (such as duration of inflation and reheating, magnitude of vacuum energy, magnitude and spectrum of fluctuations, types and abundance of topological defects, etc.), many of which are so far very loosely constrainable by observations\footnote{
  Nevertheless, it  is noteworthy how the observational precision has impressively improved in the last two decades, allowing for progressively better constraints. For a contemporary account of the state-of-the-art and perspectives for such measurements, see \textit{e.g.} \cite{liteBIRD}.}.
 The specific scenario that we shall present here, namely, the chaotic inflation scenario, has many of its developments due to Andrei Linde, and it is more thoroughly presented along with other inflationary models on his book \cite{andrei}, which is the main reference for our present exposition. This scenario not only allows us to solve the motivating issues that pushed us towards inflation in the first place, but it also provides a somewhat `natural' background to the discussion of initial conditions and to the grand questions of why our observable universe has the form it has, as well as a reasonable framework to analyze the primordial fluctiations in the CMB.

For clarity of ideas and computability, we shall restrict the present exposition of this scenario to the case of a real scalar field, with Lagrangian \eqref{scalar}; concretely, the reader may often bear in mind the $\lambda \phi^4$ model, although we shall often make estimates considering different power-law potentials, taking the form

\begin{align}
V(\phi) = \frac{\lambda_n \phi^n}{n M_p^{n-4}}, \label{powerlawV}
\end{align}
being $n>0$, $M_p$ the Planck mass and $0< \lambda_n \ll M_p^{n-4}$. This encompasses the harmonic and quartic potentials (where we make the identifications $m^2 = \lambda_2 M_p^2$ and $\lambda=\lambda_4$).

\subsection{Initial conditions} \label{initial conditions}

Let us now turn our attention to the question of initial conditions. In our previous analysis we have discurred about the great coincidence it would have been to have such a homogeneous universe in the time of decoupling had the universe expanded dominated by radiation ($a\propto t^{1/2}$) all the way back to an initial singularity. Furthermore, it remained an open question why its fluctuations were so small, having a typical relative amplitude of $10^{-5}$ (rather than of any other conceivable value). The problem became even more acute when we noticed how particular the values of some parameters must have been (particularly, how spatial curvatured must have been extremely fine-tuned near $0$) so that our universe could evolve and form structures in the way it did (and, among other things allow for the emergence of life).

In a first sight, when we consider a primordial inflationary scenario, it seems that we can do little better than to push the same problem back in time, reaching an instant for which we must specify \textit{some} initial condition (which should in principle not escape the same issues of fine-tuning to allow the realization of \textit{our} observed universe). We shall argue, however, that one may actually obtain an observable universe such as ours from fairly generic initial conditions. More precise, one may obtain \textit{many realizations} of patches of the universe that look like a FLRW spacetime over extremely large scales -- many orders of magnitude larger than our observable universe --, such that it would be reasonable to assume that \textit{somewhere} there would be a patch such as ours. In a loose analogy with biological evolution, the universe would not have had to have developed in very particular way to end up with very special and complex structures -- the mere fact that it could randomly explore a huge (virtually infinite) sample of different configurations in a sufficiently large space would assure that \emph{some} of these configurations could result in a very special patch of spacetime. Of course, since such a hypothesis refers to regions much larger than our observable universe, many of its fundamental consequences can be no more than unobservable conjectures\footnote{
 Albeit, in the author's opinion, they make for a quite appealing picture compared to the fine-tuning alternative.}.
 However, this does not mean that an inflationary scenario based on it will be devoid of observational consequences \emph{in our observable universe}. Our analysis, then, shall be primarily concerned with the latter (although we would like to stress that a line between one and the other is by no means sharp or \textit{a priori} obvious, and that much can be gained in the latter by pursuing and exploring one's ideas with a good degree of open-mindness, beyond what is obviously verifiable).

Then, to figure more precisely what would be reasonable to postulate as initial conditions for our model, we recur to what we \textit{know} to be the conditions of our observable universe in the very early past. We know that at a point earlier than the formation of the CMB the universe was very hot and came to be dominated by relativistic degrees of freedom (particularly, electromagnetic radiation), contracting back to the past at a rate $a\propto t^{1/2}$ for many orders of magnitude of $a$. We can extrapolate this radiation dominated expansion at least all the way back through primordial nucleosynthesis (first minutes), electroweak unification ($\sim 10^{-10}s$) and even somewhat before that (physics at known energy scales allows us to safely extrapolate back to about $\sim 10^{-13}s$). The standard Hot Big Bang Scenario amounts to assuming that such a rate of expansion extrapolates \emph{all the way back to $t=0$}, where there would be a primordial singularity. However, even if we did not have any of the aformentioned issues that motivate us to postulate an early inflationary phase, one fundamental difficulty remains in extrapolating this analysis back into arbitrarily high curvature and energy scales: quantum gravity. When we reach curvatures as high as the Planck scale, a dynamical description of the universe (including geometry and matter) in terms of a classical continuous spacetime (and matter fields propagating in it) is no longer expected to hold.

Since our description of inflation must apply only in times much earlier than those to which we can safely extrapolate with experimentally verified theories, but yet late enough so that it should be prone to a description in terms or classical spacetimes, we try to impose them at `the edge' of a classical descripton, at Planck time $t_p$. A rather simple, and somewhat physically motivated possibility (given the known homogeneity of the early universe) would be start the inflaton field $\phi$ in a constant (homogeneous) value $\phi_0$ corresponding to a vacuum state. However, we can immediately see that such a condition would be even more singular than a FLRW space (almost) perfectly homogeneous, and that it further presents difficulties of consistency at a quantum level (it cannot take quantum fluctuations into account), and of compatibility with the small inhomogeneities in the early universe, which appear in the CMB and are crucial for the formation of structure in our universe.

In fact, regardless of what the initial state and primordial dynamics of the universe were (and particularly, whether that dynamics depended solely or primarily on a scalar field $\phi$), the energy density of the universe cannot be determined with a precision greater than $M_p l_p^{-3}$, in virtue of the uncertainty principle (here, we shall associate inverse length scales with mass scales, and generally make reference just to the Planck mass -- then we say this energy density cannot be determined with a precision greater than $M_p^4$ ). Thus, instead of assuming this extremely specific (and ultimately inadequade) initial condition, we make more generic (and physically reasonable) hypotheses that, for $t \sim M_p^{-1}$,

\begin{align}
(\partial_0 \phi)(\partial^0 \phi), \, (\partial_i \phi)(\partial^i \phi), \,
V(\phi) \lesssim M_p^4,  \label{subplanckianfield}
\end{align}
and similarly for any scalars derived from spacetime curvature:

\begin{align}
R^2,  R_{ab}R^{ab}, R_{abcd}R^{abcd} \lesssim M_p^4, \quad R_a^{\;b}R_b^{\;c}R_c^{\;a}, R_{abcd}R^{ac}R^{bd} \lesssim M_p^6, \quad \text{etc.} \label{R2}
\end{align}

We stress that, in fact, it is precisely `after the instant' in which condition (\ref{R2}) is satisfied that we may even speak about a dynamical description in a classical spacetime, and specify an initial condition for $\phi$. 

One may then argue that, within these (consistency-binding) intervals given by equations \eqref{subplanckianfield} and \eqref{R2}, there is no \textit{a priori} reason to expect that $(\partial_\mu \phi)(\partial^\mu \phi) \ll M_p^4, \; V(\phi) \ll M_p^4$, or $R^2 \ll M_p^4$ (more precisely, given our ignorance about these initial conditions -- and the physics that governs them -- it should be more or less equally plausible to expect \textit{any} values within physically reasonable restrictions), so that it would be more likely to assume that these quantities had initial conditions with values of the order (\textit{i.e., not much smaller than})

\begin{subequations} \label{almostplanckian}
  \begin{align}
  (\partial_0 \phi)(\partial^0 \phi) \sim (\partial_i \phi)(\partial^i \phi) \sim M_p^4,  \label{bigderivatives} \\
  V(\phi) \sim M_p^4, \label{bigpotential} \\
  R^2 \sim M_p^4 \label{bigR2}
  \end{align}
\end{subequations}
(where \eqref{bigR2} should be understood in the same sense as eq. \eqref{R2}, with ``$R^2$'' standing for any quadratic scalars built from curvature, and all analogous equations with appropriate dimensions for different powers of $R$).

For the time being, we shall assume that the initial conditions are indeed given approximately by eqs. \eqref{almostplanckian} and, in the subsequent discussion, we will try to explore and better understand the implications of this assumption.

\subsection{Quasiexponential expansion and \textit{slow-roll} inflation} \label{inflaton eq}

The treatment of the evolution of the universe under these `generic' initial conditions is still an extremely complicated task. However, motivated by the known conditions of our observable universe and by our requirements for inflation, we are hinted to turning our attention to particularly symplyfing circumstances, namely, \emph{to portions of the universe in which the dynamics yield an approximatly exponentially expanding FLRW universe (i.e. a de Sitter space)}.

A very important symplifying feature of de Sitter spaces is that they have an event horizon at a radius $H^{-1}$ surrounding each isotropic observer $\mathcal{O}$, and that all other isotropic worldlines eventually fall outside this horizon, losing causal contact with $\mathcal{O}$. Similarly to black holes, de Sitter spaces obey so-called ``no-hair'' theorems, which ultimately imply that any effects due to matter and energy that fall outside the horizon of a given domain will be exponentially dampened out and no longer affect the dynamics inside this domain; thus, any spacetimes that are locally approximately like a de Sitter space (whose total stress tensor obeys $T_{ab} \approx \Lambda g_{ab}$) for a large enough region will exponentially approach a de Sitter space. For such a behaviour to be realizable, the domain in which an approximately exponential expansion happens must be bigger than $\sim 2H^{-1}$ (\textit{i.e.} the diameter of a Hubble sphere); as we shall see briefly, this will correspond to the domain for which $T_{ab}$ is dominated by the potential term $V(\phi) \approx cte$. Well, in the `instant' that $V(\phi) \sim M_p$, the Hubble radius will actually be as small as it can possibly be (to be described classically): of the order $H^{-1} \sim M_p^{-1}$.

Conversely, it should be indeed necessary that the expansion in such patches is approximately exponential for the horizon (located at a \emph{dynamical} radius $H^{-1}(t)$) to recede at a sufficiently slow rate so that the primordial inhomogeneities can fall out of it and cease to affect the dynamics inside the relevant causal domain. A rough estimate for this required slowness may be obtained by noting that the recession velocity \emph{of an object located at the horizon} should be of the order $\sim H^{-1}H = 1$ (by a simple application of the Hubble law \eqref{hubblelaw} for $R \approx H^{-1}$) whereas the recession velocity \emph{of the horizon itself} is $|\frac{d}{dt}H^{-1}|=\dot{H}H^{-2}$. Then this condition will be satisfied if $\dot{H} \ll H^2$.

We may then conclude that, in order for inflationary regions to emerge near the Planck epoch with initial conditions \eqref{almostplanckian} (and subsequently grow to considerable sizes), it should be enough that they occur in \emph{any} region of the universe with a minimal size liable to a description in terms of classical spacetime: $l \sim M_p^{-1}$.

We point out here that a particular consequence of condition \eqref{bigpotential} (for potentials obeying the condition \eqref{powerlawV}) will be that initial values $\phi_0$ of the field are tipically very large ($\gg\!\! M_p^4$), so that possible variations in a causally relevant scale should be comparatively small. For example, for the power-law potentials \eqref{powerlawV} with $n=2,4$, $V(\phi_0) \sim M_p^4$, we will have that: \vspace{0pt}

\noindent \strut\vspace{14pt}
\begin{minipage}[t]{0.4\linewidth}
 \begin{empheq}[left=\empheqlbrace]{align*}
  V(\phi) &= \frac{m^2\!\!}{2}\,\phi^2, \;\text{with}\; m\ll M_p \\[4pt]
  V(\phi) &= \frac{\lambda}{4}\;\phi^4, \;\text{with}\; \lambda \ll 1
 \end{empheq}
\end{minipage}%
\begin{minipage}[t]{0.05\linewidth}
  \begin{align*}
  \\[-2pt]
  \qquad \scalebox{2}{$\Rightarrow$}
  \end{align*} 
\end{minipage}%
\hspace{-12pt}%
\begin{minipage}[t]{0.45\linewidth}
\vspace{15pt}
\begin{subequations}
 \begin{empheq}[left=\empheqlbrace]{align}
  \phi_0 &\sim \frac{M_p}{m}M_p \gg M_p, \\[4pt]
  \phi_0 &\sim \lambda^{-\frac{1}{4}}M_p \gg M_p,
 \end{empheq}
\end{subequations}
\end{minipage}

\noindent and, according to \eqref{bigderivatives}, the variation of $\phi$ in a region of the size of the event horizon radius $H^{-1}(\phi) \sim M_p^{-1}$ should not exceed the order of

\begin{align}
\Delta \phi \sim (\partial_i \phi) M_p^{-1} \sim M_p^2M_p^{-1} = M_p \ll \phi_0,
\end{align}
so that the $\phi$ should be relatively homogeneous in typical causal domains.

Furthermore, taking in consideration that we are dealing with a scalar field, so that local anisotropies may appear only from terms $\partial_i \phi$ (and, correspondingly, of the curvature terms to which it couples), we should have that each causal domain in very early spacetime should be locally approximately isotropic. Thus, it should be locally well approximated as a FLRW universe.

Let us then look at the dynamical equations of this universe dominated by a scalar field $\phi$. We have the coupled Einstein equations and Euler-Lagrange equations (here, we insert a Planck mass in the Einstein equations to better vizualize the scales in question for our field):

\begin{subequations} \label{anyroll}
 \begin{align}
 H^2 + \frac{k}{a^2} = \frac{8\pi}{3M_p^2} \left( \frac{\dot{\phi}^2}{2} + \frac{(\mathbf{\nabla}\phi)^2}{2} + V(\phi) \right), \label{arEinstein} \\
 \Box \phi = \ddot{\phi} + 3H\dot{\phi} - \frac{1}{a^2}\nabla^2 \phi = -V'(\phi), \label{arLagrange}
 \end{align}
\end{subequations}
where we once again emphasize that $\Box$ refers to the covariant D'Alembertian while $\nabla^2$ refers to the Laplacian associated with the \textit{static} metric $\tilde{h}_{ij} = a^{-2}h_{ij}$, such that

\begin{align}
\Box \phi &\equiv g^{ab}\nabla_{\!a}\nabla_{\!b} \,\phi, \qquad
\nabla^2 \phi \equiv \tilde{h}^{ij} \tilde{\nabla}_{\!i} \tilde{\nabla}_{\!j} \phi, \qquad \tilde{\nabla}_{\!i} \tilde{h}_{jk} = 0 = \nabla_{\!a} g_{bc}. \label{boxnabla}
\end{align}

We then have that, for a sufficiently uniform field varying in a sufficiently slow manner, more precisely:

\begin{subequations} \label{slowrollC}
 \begin{empheq}[left=\empheqlbrace]{align} 
 \dot{\phi}^2, (\mathbf{\nabla} \phi)^2 \ll V(\phi), \label{slowroll1} \\
 \ddot{\phi}\, , \tfrac{1}{a^2}\! \nabla^2\!\phi \ll V'(\phi), \label{slowroll2}
 \end{empheq}
\end{subequations}
the field equations \eqref{anyroll} can be well approximated by

\begin{subequations} \label{slowroll}
 \begin{align}
 H^2 + \cancelto{\ll \!\!H^2}{\frac{k}{a^2}} = \frac{8\pi}{3M_p^2} V(\phi) , \label{srEinstein} \\
 3H\dot{\phi} = -V'(\phi). \label{srLagrange}
 \end{align}
\end{subequations}

It is then not difficult to see that, for an expanding universe ($\dot{a}>0$) with a not too steep potential slope $V'(\phi)$ near $\phi \approx \phi_0$, the system rapidly evolves to a regime of exponential expansion, in which the curvature term in the LHS of \eqref{srEinstein} becomes negligible (in terms of effective relative densities, it means the evolution rapidly makes $\Omega_\phi \rightarrow 1, \, |\Omega_k| \ll 1$). For reasons that will become apparent below, this is called the \emph{slow-roll regime}, and the conditions \eqref{slowrollC} are called \emph{slow-roll coditions}. In this regime, we have that

\begin{align}
\bigl( V'(\phi) \bigl)^{\!2} = 9H^2 \dot{\phi}^2 \simeq \frac{24\pi}{M_p^2} V(\phi) \dot{\phi}^2 \qquad \Rightarrow \qquad
\dot{\phi}^2 \simeq \frac{M_p^2}{24\pi} \frac{\bigl( V'(\phi) \bigl)^{\!2}}{V(\phi)}.
\end{align}

Particularly, for a power-law potential \eqref{powerlawV}:

\begin{align}
\dot{\phi}^2 = \frac{ n^2 M_p^2}{24\pi} \frac{V(\phi)}{\phi^2}, \label{kin_pot}
\end{align}
so that the general restrictions \eqref{almostplanckian} that we had previously imposed in our potential (which entailed $\phi \gg M_p$) will then assevere that, in the slow-roll regime:

\begin{align}
\phi \gg \frac{n}{4\sqrt{3\pi}}M_p \qquad \Leftrightarrow \qquad \tfrac{1}{2}\dot{\phi}^2 \ll V(\phi). \label{lowkinetic}
\end{align}

This condition then reassures the self-consistency of dynamics withing the slow-roll approximation, and (supplemented by $(\mathbf{\nabla}\phi)^2 \ll V(\phi)$) it asseveres that the kinetic energy will be much smaller than potential energy, so that the stress tensor $T_{\mu\nu}$ will indeed be dominated by the potential term in the slow-roll regime:

\begin{align}
T_{\mu\nu} \approx V(\phi) g_{\mu\nu}.
\end{align}

This means that we shall have precisely the desired equation of state for an inflationary dynamic, namely $p \approx -\rho$, producing a quasiexponential expansion for this patch of the universe.

Of course, one could then ask for how long these conditions hold, whether that is long enough to sustain an inflationary phase at all, and, if so, how much inflation happens while this phase lasts. Well, from the whole set of conditions that we have imposed and derived above, it is not hard to see that the rate of expansion $H$ of the universe will be much larger than the fractional variation rate of $\phi$ (and therefore of $V(\phi)$), as well as than the fractional variation rate of $H$ itself, so that we do indeed obtain an approximate de Sitter space:

\begin{align*}
\frac{3H \dot{\phi}}{H^2\phi} = - \frac{3M_p^2}{8\pi\phi}\frac{V'(\phi)}{V(\phi)} \approx -\frac{3n}{8\pi} \frac{M_p^2}{\phi^2} \ll 1
\end{align*}
(where we have once again estimated for a power-law potential with $n\sim \mathcal{O}(1)$ ). We then have that

\begin{align}
\frac{\dot{\phi}}{\phi}H^{-1} = -\frac{n}{8\pi}\left( \frac{M_p}{\phi} \right)^2 \ll 1 \qquad \Leftrightarrow \qquad \frac{\dot{\phi}}{\phi} \ll H.
\end{align}

Then, taking a time derivative of eq. \eqref{slowroll1} (where we already neglect the spatial curvature term), we obtain

\begin{align*}
2H\dot{H} = \frac{8\pi}{3M_p^2}V'(\phi)\dot{\phi} = \frac{8\pi}{3M_p^2}3H\dot{\phi}^2.
\end{align*}
Thus

\begin{align*}
\dot{H} = 3 \frac{8\pi}{3M_p^2} \cancelto{\ll \!V(\phi)}{ \left( \frac{\dot{\phi}^2\!}{2} \right) } \ll 3H^2.
\end{align*}

If we then drop the $3$ factor in this magnitude comparison, we obtain simply

\begin{align}
\dot{H} \ll H^2,
\end{align}
which was precisely the slowness condition for the recession of the horizon that we required to obtain a period of quasiexponential expansion! Then, for any time interval $\Delta t \lesssim H/\dot{H} \; \bigl(\gg\! H^{-1}\bigl)$, we should have

\begin{align}
\qquad\qquad\qquad\qquad a(t) \approx a_0 e^{Ht}, \qquad t \in [t_p,t_p+\Delta t] , \label{aexp}
\end{align}
with

\begin{align}
H(\phi) = \left[ \frac{8\pi V(\phi)}{3M_p^2} \right]^{1/2}. \label{hubble phi}
\end{align}

\begin{figure}[H]
\centering
\includegraphics[width=0.75\linewidth]{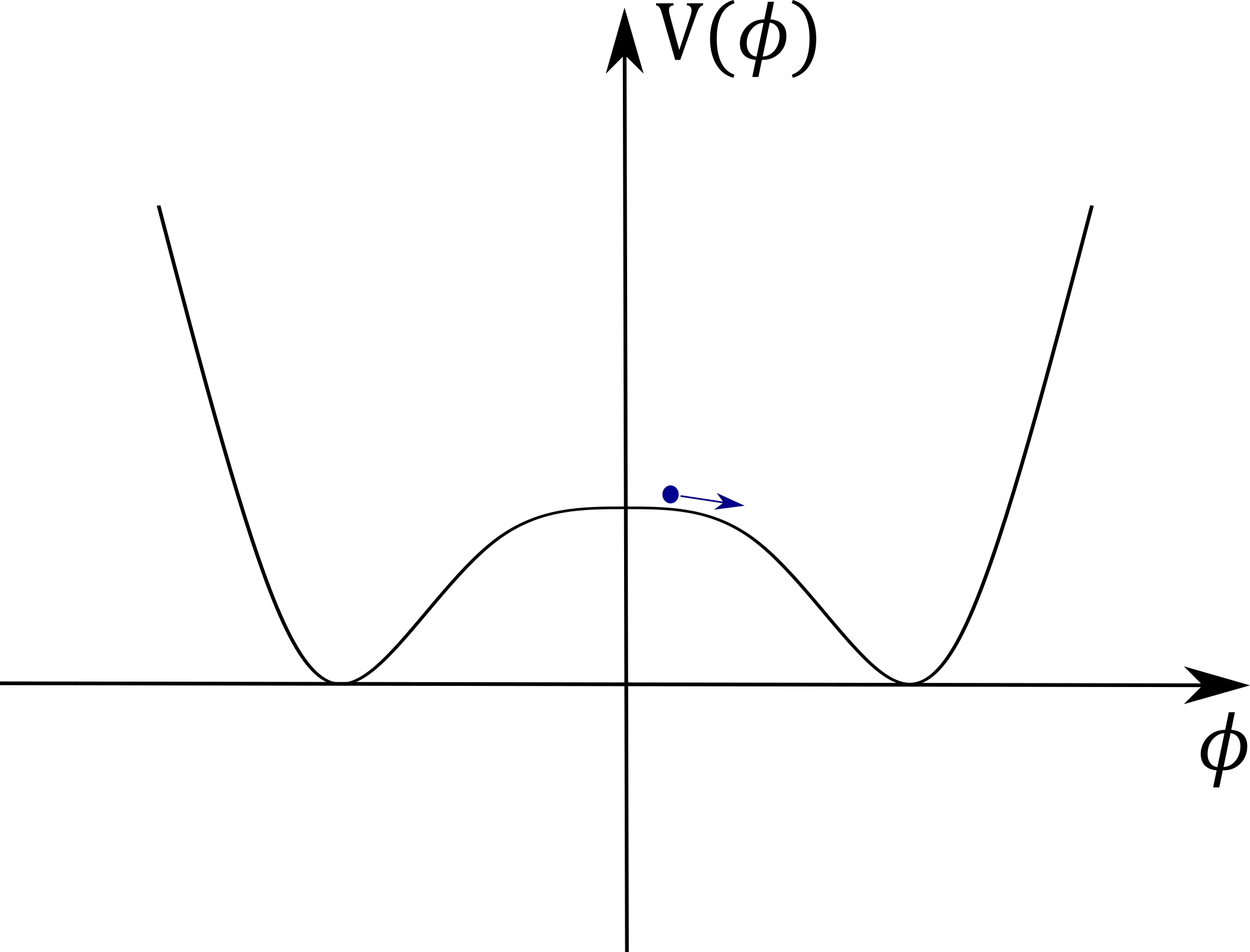}
\caption{Field $\phi$ slowly rolling down its potential well, depicted by a little blue ball which brings out one-dimensional mechanical analog. While the field is near its unstable maximum, it is as though as it is sliding down with terminal velocity in a viscous medium. At later times, when it is considerably far from its maximum, it will perform damped oscillations about its absolute minimum giving up heat to other fields in the universe. \\ Source: By the author. }
\label{slowrolling}
\end{figure}

In the meanwhile, the field $\phi$, governed by equation \eqref{srLagrange}, slowly evolves towards the minimum of its potential; note that this will be just a first order ODE (since we have taken the field to be spatially homogeneous), whose signs are so that $\phi$ will be driven down the potential curve (with speed proportional to its slope). In this regime, our system is entirely analogous to a particle (with position coordinate $\phi$) subject to some viscous friction rolling down a potentiall well with terminal velocity (see Figure \ref{slowrolling}); for this reason we call this expanding regime \emph{slow-roll inflation}.

\subsection{Fluctuations of the inflaton field} \label{sPowerSpectrum}

Having described the \emph{average} behaviour for the evolution of the inflaton $\phi$ (and the corresponding average evolution of the metric), we shall now take a perturbative approach to describe its fluctuations. As it was described in section \ref{inflationandSSB}, our approach will consist of splitting our field in an effectively classical contribution $\phi_0$ -- corresponding to the one that we just described in the previous section, producing an approximately de Sitter background spacetime -- and fluctuations $\delta \phi$, which will propagate on this background, and whose own gravitational effects will be neglected on its dynamics. We write this split as

\begin{align}
\phi(\mathbf{x},t) = \phi_0(t) + \delta \phi(\mathbf{x},t).
\end{align}

Then, assuming the background solution for $\phi_0(t)$ and $a(t)$ given in the last subsection, we can write the linearized equations for $\delta \phi$ from \eqref{arLagrange}:

\begin{align}
\frac{\partial^2(\delta\phi)}{\partial t^2} + 3H\frac{\partial(\delta\phi)}{\partial t} - e^{-2Ht} \nabla^2(\delta \phi) = -V'(\phi_0 + \delta\phi) \simeq -V''(\phi_0)\delta\phi. \label{deltaphi1}
\end{align}

Since in this slow-roll regime $V''(\phi_0)$ will be nearly constant, this approximate equation will be simply the one for a free scalar field, with an effective quadratic mass $\mu^2(\phi_0) \equiv V''(\phi_0)$. Well, this is exactly the equation \eqref{PTdSfe}, that we thoroughly analyzed in section \ref{deSitter}, where $\mu^2$ plays the role of the effective mass $M^2$:

\begin{align}
\bigl[ \partial^2_t + 3H\partial_t - 2e^{-2Ht}\nabla^2 + \mu^2 \bigl] \delta\phi = 0.
\end{align}

Then, these perturbative variations to the field can be quantized by our basic prescription in chapter 3, and we can employ the same field modes as section \ref{deSitter}, which will correspond to the Bunch-Davies vacuum. However, as $\mu^2$ is dictated by the \emph{average field} $\phi_0$, it can have either a positive or a negative sign, depending on the behaviour of its potential; in the slow-regime, it is expected to be slightly negative, as it is decaying near an unstable vacuum (and consistency with the slow-roll conditions imply $\mu^2 \ll H^2$). This will prevent us from straighforwardly transposing the discussion in \ref{deSitter} to analyze the renormalized expectation value of the fluctuatios $\braket{0|(\delta\phi)^2|0}$ (recall that, in our analysis for the renormalization of the power spectrum, we had pathological IR divergences for negative $M^2$).

Then, to be able to extract any meaningful information of this fluctuation spectrum, we shall (i) take advantage of the slow-roll mass consistency condition $\mu^2\ll H^2$ to simplify our field modes in a massless approximation and (ii) argue that, for a space that is not eternally de Sitter, but rather has been inflating for a finite time, it is reasonable to impose an IR cutoff in the spectrum (which shall be implemented at wavelengths many orders of magnitude larger than the hubble radius $H^{-1}$). The first approximation will imply that $\nu\simeq3/2$ in equation \eqref{generalHankel}. This yields a particularly simple form for the field modes, since the Hankel function $H_{3/2}^{(1)}$ can be put in the form

\begin{align}
H_{3/2}^{(1)}(x) = - \sqrt{ \frac{2}{\pi x} }e^{-ix} \left(1 + \frac{i}{x} \right).
\end{align}

Thus we find that the de Sitter adiabatic modes \eqref{de Sitter Adiabatic modes} will be

\begin{align}
h_k(t) = - \frac{iH}{\sqrt{2k^3} } \left( 1 + \frac{ik}{H}e^{-Ht} \right) \exp\Bigl(-\frac{ik}{H}e^{-Ht}\Bigl). \label{de Sitter Adiabatic Masless}
\end{align}

As the universe expands, each of these modes will have its wavelengths exponentially stretched. We then find that, for sufficiently large times, the modes will gradually ``exit the horizon'' (\textit{i.e.}, reach wavelengths greater than the Hubble radius $H^{-1}$), which will be given by each mode by

\begin{align}
ke^{-Ht}<H \Leftrightarrow t > H^{-1}\ln |k/H|.
\end{align}

At this point, we see in equation \eqref{de Sitter Adiabatic Masless} that $h_k$ ceases to oscillate and asymptotically `freezes' in the value

\begin{align}
\tilde{h}_k = -\frac{iH}{\sqrt{2k^3}}
\end{align}
(less of an arbitrary global phase for each mode, which we have not specified). Thus, the dynamical effects of these modes \emph{after inflation} should only manifest when they ``reenter'' the horizon, and their amplitudes at this point should \emph{only depend on their amplitudes at horizon exit}, which will have happened \emph{during} inflation \footnote{Note that these modes \emph{continue to expand their wavelenths after inflation}. However, after the quasiexponential expansion ceases, the Hubble radius $H^{-1}(t)$ will rapidly recede, allowing reentrance.}.

If we then analyze the formal expectation value $\braket{\phi^2(x)}$ in the (Bunch-Davies) de Sitter vacuum, we obtain the following spectral contributions\footnote{ 
The point has been raised by Parker (see \textit{e.g.} \cite{parkerA}) that a physical analysis should consider the \emph{renormalized} value of this spectrum. However, the unsubtracted power spectrum still yields the observed scale-invariant fluctuation spectrum, the main difference being in its associated amplitudes and how they bind the parameters on the inflaton field. Furthermore, the subtracted spectrum is not generally positive-definite, from which difficulties may arise in interpreting it (for a recent discussion in IR divergences and positive definiteness of the spectrum in case of a massless field, see \cite{yang}). It is also worth pointing that we are interested in the IR end of the spectrum, whereas the subtraction procedures such as adiabatic subtraction are in principle designed to correct its behaviour in the UV.}:

\begin{align}
\braket{ 0| \phi^2(x) |0} = \frac{1}{(2\pi)^3} \int d^3 \mathbf{k} \left( \frac{e^{-2Ht}}{2k} + \frac{H^2}{2k^3} \right),
\end{align}
which can be expressed in terms of the physical momentum (as measured by a comoving observer) $\mathbf{p}=\mathbf{k}e^{-Ht}$:

\begin{align}
\braket{ 0| (\delta\phi)^2(x) |0} = \frac{1}{(2\pi)^3} \int \frac{d^3 \mathbf{p}}{p} \left( \frac{1}{2} + \frac{H^2}{2p^2} \right).
\end{align}

In this form, one can immediately recognizes a `Minkowski vacuum' type of contribution in the first term (which is only UV-divergent), whereas the second yields an extra inflationary contribution (which is both IR- and UV-divergent)\footnote{ 
 As we have previously mentioned, one can cover de Sitter spaces with many coordinate systems (includind static coordinates) and build different vacuum modes associated to different de Sitter Symmetries. For an appropriate choice of modes one could interpret this extra IR-divergent term as being due to particles (see chapter 7 of \cite{andrei}), with occupation numbers given by $n(\mathbf{p}) = \frac{H^2}{2p^2}$.}.
 As we mentioned above, we shall be particularly interested in the very long wavelength behaviour $p = ke^{-Ht} \lesssim H$, for which we can neglect the first term. We then argue that, since physically we do not have an eternally inflating de Sitter space extending all the way past to $t \rightarrow - \infty$, but rather a finite inflation phase which cannot be extrapolated past Planck time, it should be reasonable to consider an IR cutoff in the spectrum, restricting our integral to modes with wavelengths within the horizon at $t\sim t_p$ (roughly, with $k = pe^{Ht} > H$). In this case, we obtain the long-wavelength (LW) contribution to the spectrum:

\begin{align}
\braket{(\delta \phi)^2}_{LW} &\approx \frac{H^2}{2(2\pi)^3} \int_{LW} \frac{d^3\mathbf{p}}{p^3} \nonumber \\
 &= \frac{H^2}{4\pi^2} \int\limits_{He^{-Ht}}^H \frac{dp}{p} = \frac{H^2}{4\pi^2} \int\limits_{H}^{He^{Ht}} \frac{dk}{k} \nonumber \\ 
 &=\frac{H^2}{4\pi^2} Ht. \label{LW power spectrum}
\end{align}

Note that this linearly increasing time dependence (which appeared from our cutoff when we restricted the spectrum to modes that were \textit{inside} the horizon before inflation) can be interpreted as reflecting the fact that, as time passes, more modes exit the horizon (each logarithmic interval of $k$ yielding a similar contribution); of course, it should only be considered for time intervals \emph{during} inflation. Then, precisely from these contributions, one could expect to obtain a (logarithmically) scale-invariant power spectrum of fluctuations for long wavelength modes. Particularly, considering the contribution from a limited spectral integral, say, which exited the horizon during 1 e-folding (\textit{i.e.}, during a time interval $\Delta t \sim H^{-1}$), we obtain

\begin{align}
\braket{(\delta \phi)^2}_{\!\Delta k} \approx \frac{H^2}{4\pi^2}. \label{inflaton LW spectrum}
\end{align} 

\subsection{Comments on the end of inflation and its physical imprints} \label{after inflation}

Having derived the conditions for the occurrence of an inflationary period, as well as the approximate dynamical equations well within the inflationary phase, we turn our attention to a few quantities that should be of physical significance \emph{after} inflation, and that possibly yield observational consequences today. They are: (i) the total duration of the inflation, (ii) the expansion factor by which the universe inflates in this period, and (iii) the allocation of the energy of the inflaton field $\phi$ after inflation -- both on average and for its fluctuations -- and how that influences the subsequent dynamics of the observable universe.

These quantities are relevant to determine whether inflation is a viable candidate to solving the problems in the $\operatorname{\Lambda CDM}$ model in the first place (and if it does not create new problems), as well as if it entails any observational consequences other than those that it was designed to fit.

Let us begin by analyzing the duration of the inflationary phase. To do so, we first consider the evolution of the field well within the slow-roll regime, for which $\phi$ evolves by a simple first-order equation \eqref{srLagrange}. Then, by substituting the potential  \eqref{powerlawV}, we end up with a separable equation:

\begin{align}
\dot{\phi} = \sqrt{\frac{n\lambda_n}{24\pi}}M_p^{3-\frac{n}{2}} \phi^{\frac{n}{2}-1},
\end{align}
whose solutions are

\begin{subequations} \label{inflaton(t)}
  \begin{empheq}[left={\empheqlbrace} , right= { \qquad \quad.}]{align}
  \phi(t) &= \phi_0 e^{\sqrt{\frac{\lambda_4}{6\pi}}M_pt}, & n=4 \label{n4} \\
  \phi(t) &= \left[ \phi_0^{\frac{4-n}{2}} - \frac{4-n}{2}\sqrt{\frac{n\lambda_n}{24\pi}}M_p^{3-\frac{n}{2}}t \right]^{\frac{2}{4-n}}, & n \neq 4 \label{nnot4} \\
  &\Rightarrow  \; \phi(t) = \phi_0 - \frac{\sqrt{\lambda_2}M_p^2}{2\sqrt{3\pi}}t, & n = 2 \label{n2}
  \end{empheq}
\end{subequations}

With those, one can make a rough estimate of when inflation ends by analyzing when there is a significant departure from the slow-roll conditions. Particularly, we can analyze when kinetic energy becomes comparable to potential energy: $ V(\phi) \sim \dot{\phi}^2\!/2 $. From equations \eqref{kin_pot} and \eqref{lowkinetic}, see that this condition should be violated when

\begin{align}
\phi \sim \phi_T \equiv \frac{n}{4\sqrt{3\pi}}M_p.
\end{align}

Then, by inverting equations \eqref{inflaton(t)}, we obtain an estimate for the total duration $t_T$ of inflation as a function of the parameters in our potentials. Particularly, for the quadradic \eqref{n2} and quartic \eqref{n4} cases:

\begin{subequations} \label{duration of inflation}
  \begin{empheq}[left=\empheqlbrace]{align}
  t_T &\sim  \frac{2\sqrt{3\pi}}{\sqrt{\lambda_2}M_p^2}\phi_0, & n=2 \label{tn2} \\
  t_T &\sim \sqrt{\frac{6\pi}{\lambda_4}}M_p^{-1} \ln\biggl| \frac{\phi_T}{\phi_0}\biggl|, & n=4 \label{tn4}
  \end{empheq}
\end{subequations}

Furthermore, manipulating equations \eqref{slowroll}, it is not difficult to obtain an exact solution $a\bigl(\phi(t)\bigl)$  for the slow-roll approximation with this potential:

\begin{align*}
\frac{d}{dt} \ln(a) = H &= \frac{8\pi}{3M_p^2} H^{-1} V(\phi) \\
 &= - \frac{8\pi}{3M_p^2} 3 \dot{\phi} \frac{V(\phi)}{V'(\phi)} \\ 
 &= - \frac{8\pi}{M_p^2}  \dot{\phi} \frac{\phi}{n} \\
 &= - \frac{4\pi}{nM_p^2} \frac{d \phi^2\!}{dt} .
\end{align*}

Then, we have simply $d(\ln a) = - \frac{4\pi}{nM_p^2} d(\phi^2) $, which yields

\begin{align}
a(t) = a_0 e^{\frac{4\pi}{nM_p^2}\bigl( \phi_0^2 - \phi^2(t) \bigl)}. \label{agaussian}
\end{align}

Of course, for sufficiently small time intervals (for which $\phi_0^2-\phi^2(t) \approx H(\phi_0)t$), this should be well approximated by \eqref{aexp}. Still, \eqref{agaussian} should give us a more accurate estimate for the total inflation factor $P$ as $\phi$ evolved from $\phi_0 \gg M_p$ to $\phi_T \sim M_p$:

\begin{align}
P \sim e^{ \frac{4\pi}{nM_p^2}\phi_0^2 } \sim e^{ \frac{4\pi}{n} \left( \frac{\lambda_n}{n} \right)^{-2/n} }. \label{inflationfactor}
\end{align}

Once again, we estimate this factor for quadratic and quartic potentials:

\begin{subequations}
  \begin{empheq}[left=\empheqlbrace]{align}
  P_T &\sim e^{4\pi \frac{M_p^2\!}{m^2\!} }, \quad n=2 \\
  P_T &\sim e^{ \frac{\sqrt{2}\pi}{\sqrt{\lambda}} }, \;\;\quad n=4 \label{lf4P}
  \end{empheq}
\end{subequations}

Of course, if we do not have any independent constraints for the potential parameters, little more can be said about the duration of inflation or the inflating factor than the obvious bounds that it should not last up until times where we reach well tested energies, and that it should last long enough to sufficiently dilute spatial curvature and cosmological defects. However, there is a factor that provides us with a more strict estimate of how long inflation should have lasted: the relative density fluctuations in the early universe. As we have seen in the previous section, the average amplitudes of field fluctuations in an exponentially expanding (de Sitter) universe should `freeze out' when their wavelengths stretch up to $H^{-1}$, yielding a spectral contribution to fluctuations \eqref{inflaton LW spectrum}:

\begin{align}
\delta \braket{\phi} \equiv \sqrt{\braket{(\delta\phi)^2}_{\Delta k} } \sim \frac{H(\phi)}{2\pi}. \label{f spec 1}
\end{align}

Then, if we want these field fluctuations to be the ones responsible for the density fluctuations in the CMB, matching an approximately scale-independent spectral amplitude of the order

\begin{align}
\frac{\delta \rho(k)}{\rho} \sim 10^{-5},
\end{align}
we must tune our potential parameters correspondingly. If these fluctuations are indeed produced \emph{during} inflation (due to primordial fluctuations in the inflaton field that later are imprinted in ordinary matter through their interactions) we could make a `handwaving' estimate of their expected amplitudes as follows\footnote{
 For a more thorough derivation, see section 7.5 of \cite{andrei}. Also, for an extensive treatment of fluctuation of various types, for many different field species, we refer the reader to chapter 5 of \cite{weinberg}.} :

\begin{align}
\frac{\delta \rho}{\rho} \sim \frac{\delta V(\phi)}{V(\phi)} = \frac{V'(\phi) \delta \phi}{V(\phi)}.
\end{align}

Then, using equations \eqref{f spec 1} and \eqref{hubble phi}, we obtain the following estimate considering, for simplicity, modes that would have exited the horizon around the end of inflation $\phi \sim \phi_T$:

\begin{align}
\frac{\delta \rho}{\rho}\biggl|_{k \sim H(\phi_T)} &\sim \frac{V'(\phi_T)}{V(\phi_T)} \frac{H(\phi_T)}{2\pi} \nonumber \\
 &= \frac{n}{\phi_T} \frac{4\sqrt{2}}{4\sqrt{3\pi}M_p} \biggl[ \frac{\lambda_n \phi_T^n}{nM_p^{n-4}} \biggl]^\frac{1}{2} \nonumber \\
 &= 4\sqrt{2}n^{\frac{n-1}{2}} \sqrt{\lambda_n}.
\end{align}

For example, estimates in a $\lambda \phi^4$ model (identifying $\lambda=\lambda_4$) yield roughly

\begin{align}
\frac{\delta \rho}{\rho} \sim 10\sqrt{\lambda} \qquad \Rightarrow \qquad \lambda \sim 10^{-12},
\end{align}
which results in a total inflating factor \eqref{lf4P}

\begin{align}
P_T \sim e^{\frac{\sqrt{2}\pi}{\sqrt{\lambda}}} \sim 10^{10^5}, \label{lf4Pl}
\end{align}
and a total duration \eqref{tn4}

\begin{align}
t_T \sim \sqrt{\frac{6\pi}{\lambda}} M_p^{-1} \ln \bigl| \lambda^{-\frac{1}{4}} \bigl| \sim 10^{-35}s. \label{tn4l}
\end{align}

This total time \eqref{tn4l} hints to us that such a model could be adequate to handle the monopole problem (although it is usually required that $t_T$ be one or two orders of magnitude larger)
, whereas \eqref{lf4Pl} show that it would be comfortably enough to resolve the flatness and horizon problems. Indeed, this astounding inflation factor is significantly larger than it would be required to dilute spatial curvature below the fine-tuning alluded in section \ref{lcdmP}, as well as to inflate a region of the Planck size $l_p \sim 10^{-31}m$ up to scales $\delta l \sim 10^{10^5}m $ many orders of magnitude larger than the size of the observable universe $l_U \sim 10^{26}m$, giving observers within the latter enough time to have had causal contact.

Finally, we briefly comment on the subsequent evolution of the universe after the inflationary phase. In principle, if the universe inflated enough to dilute any spatial curvature and monopoles, we also expect ordinary matter and radiation to be brutally diluted. We know, however, that for time scales $t \gtrsim 10^{-13}s$ we had a hot universe essentially dominated by radiation (and, later, with a significant contribution of baryonic matter). To make the two things compatible, we must assume that there will be interactions between the inflaton field and ordinary matter, and that the former will transfer energy to the latter as it decays towards its stable vacuum, in a process that is called \emph{reheating}\footnote{
 As the term \emph{recombination}, that appears in the cosmology literature to refer to the combination of protons and electrons at the time of matter-radiation decoupling and of the formation of the CMB, this term is somewhat misleading. It is suggestive that the universe was also hot before inflation (as recombination suggests that protons and electrons were combined before decoupling), which would be a hyphothesis with no support on observations.} ;
 of course, we need some form of interaction to imprint the primordial fluctuations $\delta\phi$ from the inflaton field in the primordial plasma whose last scattering surface we observe in the CMB today\footnote{For more details on the processes of reheating, see \textit{e.g.} section 7.9 of \cite{andrei} or section 4.2 of \cite{weinberg}.}.

 Note that, after $\phi$ decays from its unstable vacuum, it should subsequently oscillate around its absolute minimum (see Figure \ref{slowrolling}). During these oscillations, interactions with other matter fields should cause it to emit particles, and evolve towards a thermodynamical equilibrium with them. Then, a corresponding upper bound for the reheating temperature can be estimated for an effective number of relativistic degrees of freedom $N^*(T)$ as \cite{kolbturner, andrei}

\begin{align}
\frac{\pi^2}{30} N^*(T_R)(kT_R^4) \sim V(\phi \sim \phi_T).
\end{align}

In this case, if we take for instance $N^* \sim 10^2$ in our quartic model, this would entail

\begin{align}
kT_R \sim \lambda^{1/4}M_p \sim 10^{15}GeV,
\end{align}
which is still around the scale of GUT phase transitions. However, the temperature of reheating generally turns out to be orders of manitude lower to that of thermal equlibrium, due to the inneficiency of reheating if we require the interactions of $\phi$ to other fields in nature to be sufficiently weak \cite{andrei}. In this case, one can transition from an inflationary phase to a usual hot universe described in our standart model, liberated from the fundamental issues of its extrapolation to arbitrarily early times.

%% file: chap6.tex
\chapter{Conclusion}

\label{conclusion}

Vacuum is complex. If there is a single sentence that captures the message of this dissertation, we believe this should be it. As surprising as it may seem to our intuition this concept turns out to spawn an incredibly rich structure, which may be intimately related to some of the most profound questions that we have about our own universe.

When we switch our fundamental perspective from the notion of particles to that of fields, such that the former comes to be conceived just as an emergent manifestation of the latter, we find that a corresponding notion of vacuum as ``a state devoid of particles'' cannot make unambiguous sense. Particularly, in quantum theory, different observers can measure very different particle contents in the same field state, depending on their state of motion (even when they are at the same spacetime region). Moreover, even a fixed inertial observer probing the field in a fixed state may observe very different particle contents at different times, giving rise to the phenomenon of particle creation. Nonetheless, all observers converge on their notions of particle occupations at arbitrarily high-energies, corresponding to increasingly localized short wavelength, which allows for a meaningful, although approximate and nonunique, extension of the concept of vacuum in curved spacetimes, which plays a key role in the renormalization of localized quantities, such as field amplitudes and energy densities.

The vacuum energy density, in particular, is found to play significant roles in many contexts. Even classically it may behave in a nontrivial manner, allowing for the description of a finite inflationary phase for the universe. At a quantum level, however, it reveals an even richer scope of possibilities. As long as one can systematically eliminate the divergencies that appear in our description of quantum field theory, quite surprising physical predictions emerge. Even in a description in flat space, where gravity plays no role whatsoever, it was possible to predict and experimentally verify that the electromagnetic vacuum will present negative pressures and induce an attractive force between two conducting plates. In curved spaces, it is found to give rise to a cosmological-constant kind of term, with constant positive energy and negative pressure of the same magnitude, which could conceivably account for the puzzling cosmic component that we now call Dark Energy. Further still, primordial vacuum fluctuations seem like a promising candidate to explain the tiny fluctuations that we observe in the far background sky, and that were ultimately responsible for the formation of the many structures that we observe in the universe today, including the sun, the earth, and the all life that emerged on it.

Of course, there is still much research to be done in the subject before we can extrapolate from appealing theoretical pictures to making strong claims about the workings of the real world. Particularly, on the observational side, increasingly precise and varied measurements of relics from the early universe should allow for significant, and possibly quite surprising, improvements of our understanding of it. Nonetheless, we hope that the present work may serve as a comprehensible introduction to the theoretical window of such a fascinating subject.

%% file: Apendices.tex
\begin{apendicesenv}

\partapendices 

\chapter{Distributions} \label{distributions}

The subject of distributions is one hard to ignore in physics, and yet it is seldom given a proper treatment in the exposition of topics for which it is relevant (ranging from point-charges in electrostatics, going through bras and kets in ordinary quantum mechanics, and up to its ubiquitous presence in field theory). The present exposition of the topic, far from exhaustive or fully rigorous, aims at laying a few basic definitions and providing a clear picture for its applications in the scope of this dissertation. For a longer but straightforward and physically-oriented exposition, we recommend \cite{friedlander}; or, for a more general and rigorous covering of the topic, see \cite{yvone}.

\textbf{$\bullet$  Distributions as linear functionals.}

The subject of distributions is one of linear algebra. When one is handling an ordinary finite-dimensional vector space $\mathbb{V}$, one fundamental concept is that of linear operators acting on $\mathbb{V}$ to produce scalars. Those operators, in their turn, form another linear space, the \emph{dual space} $\mathbb{V}^*$. $\mathbb{V}^*$ can be easily shown to be isomorphic to $\mathbb{V}$, although there is no natural identification between them.\footnote{On the other hand, the dual of $\mathbb{V}^*$, $\mathbb{V}^{**}$, can be naturally identified with $\mathbb{V}$: given $V \in \mathbb{V}^{**}$ one associates it with the unique vector $v \in \mathbb{V}$ that satisfies $V(\sigma)=\sigma(v), \, \forall \sigma \in \mathbb{V}^*$.}

In this case, let $n=\dim(\mathbb{V})=\dim(\mathbb{V}^*)$ and consider an arbitrary pair of a vector $v\in\mathbb{V}$ and a dual vector $\sigma\in\mathbb{V}^*$. Given a basis $\{\mathbb{e}_i\}$ of $\mathbb{V}$ and its dual basis $\{\mathbb{e}^*_i\}$ in $\mathbb{V}^*$ ($\mathbb{e}^*_i (\mathbb{e}_j) = \delta_{ij}$), we can write $\sigma(v)$ in terms of their components:

\begin{align}
\sigma(v) = \sum_{i=1}^n \sigma_i v^i. \label{discretedual}
\end{align}

However, in the case of infinite-dimensional vector spaces, such as many function spaces, it is no longer generally true that $\mathbb{V}$ and $\mathbb{V}^*$ are isomorphic. In fact $\mathbb{V}^*$ is generally bigger than $\mathbb{V}$: that means one can generally associate to each vector $v \in \mathbb{V}$ a dual vector $u \in \mathbb{V}^*$, but the converse is not always true. Specifically, for function spaces (usually, for ``well-behaved'' functions defined on some open set of a sufficiently smooth manifold $\mathcal{O} \subset \mathcal{M}$ ), these dual vectors are linear \textit{functionals}, and they are called \textit{distributions}.

We shall generally denote the function space under consideration $\mathcal{F}$, whose domain $\mathcal{O}$ is assumed to have a measure $\mu$\footnote{
 For a metric manifold $(\mathcal{M}, g_{ab})$, this just amounts to familiar integrals in $\mathbb{R}^d$: $\int_\mathcal{M} d\mu_g(x)f(x) = \int_{\mathbb{R}^d} d^dx |g(x)|^{\frac{1}{2}} f(x) $.},
 and the space of distributions that act on it $\mathcal{D}$ ($=\mathcal{F}^*$); then, the action of a distribution on a function will generally take the form of a Lebesgue integral. A case of particular interest is when $\mathcal{F}$ is a space of smooth functions on open intervals $U \subset \mathbb{R}^d$, $\mathcal{C}^\infty(U)$ (the case of smooth manifolds can be mapped in (a countable sum of) this one\footnote{See appendix B of \cite{wald} for details of integration on manifolds.});0
 for a brute-force guarantee that one will not have to worry with boundary terms, one often resorts to the subset of these functions with compact support $\mathcal{C}^\infty_0(U)$. Throughout this Appendix, to avoid being cumbersome with technical remarks, we shall always assume our functions to be sufficiently well-behaved so that our assertions make sense (for example, that one may take derivatives to a desired order, that certain integrals converge and that there will be no contributions from boundary terms).

Analogously to (\ref{discretedual}), when we have a distribution $\mathcal{\sigma}\in \mathcal{F}^*$ \emph{that can be identified with a locally integrable function} $\tilde{\sigma}\in\mathcal{F}$, we may write its action on any function $f\in\mathcal{F}$ as

\begin{align}
\sigma[f] = \int d\mu(x) \tilde{\sigma}(x)f(x). \label{continuousdual}
\end{align}

However, that is not always the case. Consider the linear functional $\delta_x$, $x\in\mathcal{O}$, whose action upon a function $f\in\mathcal{F}$ produces the value of $f$ in $x$, that is: $\delta_x[f] \equiv f(x)$. Clearly, there is no locally integrable function $\tilde{\delta}_x$ that satisfies this property in the continuum\footnote{For a slightly longer discussion of this point, see chapter 1 of \cite{friedlander}}. However, we would still like to represent this functional in that form:

\begin{align}
\delta_x[f] = f(x) = \int_{\mathcal{O}} d\mu(y) f(y) \delta(x,y) = \int_U d^dy f(y) \delta(x-y) = \int_U d^dy f(y-x) \delta(y), \label{diracdelta}
\end{align}
so that one often speaks of the ``Dirac delta fuction'' $\delta(x)$, as though one was actually integrating a function in \eqref{diracdelta} -- or, extrapolating further, one speaks of the delta function \textit{outside an integration sign}, where it makes much less formal sense. Our aim here is not to be pedantic about definitions; on the contrary, \emph{we are interested in potentializing their practical use without incurring in any operational pitfalls}. It is usually harmless, and often quite useful, to make informal manipulations with distributions as though they were functions (as it is to ``pass $dx$ multiplying'' in a differential equation, or use ``the wave-function of a particle of momentum $k$'': $\psi_k(x)=(2\pi)^{-\frac{1}{2}}e^{ikx}$). However, there are situations where such approaches break and produce implausible or paradoxical results, and that often happens because one failed to appreciate that he/she is not dealing with a function, but rather with a distribution, and that more care to a particular operation was needed (just like one may be led to the absurd conclusion that $1=2$ by carelessly performing algebraic manipulations involving a division by $0$). A few sensible/troublesome manipulations with distributions include trying to evaluate objects like $\braket{x|x}$, $\int \delta^2(x) f(x)dx$, which would be perfectly natural if one was operating ordinary functions or vectors.

In the context of ordinary nonrelativistic quantum mechanics, one finds that plane waves, the eigenfunctions of the momentum operator, are not square-integrable in $\mathbb{R}^n$, and, as such, they cannot properly represent state vectors in the Hilbert space $\mathcal{H}=\mathcal{E}\subset \mathcal{L}^2(\mathbb{R}^n)$ (we take a subset $\mathcal{E}$, not the entire $\mathcal{L}^2$, to ensure we are restricted to sufficiently well-behaved functions). The case is even worse for the eigenvectors of the position operator\footnote{Note that these are not vectors belonging to $\mathcal{H}$, but rather to a larger space: $\mathcal{H}^{**}$ (the dual of the dual of $\mathcal{H}$).}, which are not even functions to start with.
Nevertheless, all \textit{actual} wave-functions $\psi\in \mathcal{E}$ have well-defined Fourier transforms, as well as (obviously) well-defined values at each point. Thus, although position and momentum eigenstates $\ket{x}: X\ket{x} = x\ket{x}$ and $\ket{p}: P\ket{p} = p\ket{p}$ do not make rigorous sense as state vectors belonging to the Hilbert space $\mathcal{H}$, it is perfectly sensible to evaluate them as distributions: $\forall \ket{\psi} \in \mathcal{H}$,

\begin{align}
\psi(x) &= \braket{x|\psi}, \quad \braket{ x| X | \psi} = x\psi(x), \quad \ket{\psi} = \int dx \ket{x}\braket{x|\psi} = \int dx \ket{x}\psi(x), \\
\bar{\psi}(p) &= \braket{p|\psi}, \quad \braket{ p| P | \psi} = p\bar{\psi}(p), \quad \ket{\psi} = \int dp \ket{p}\braket{p|\psi} = \int dp \ket{p}\bar{\psi}(p).
\end{align}

We also have the famous orthornormality relations, which can be rigorously stated in the distributional sense: $\braket{x|x'} = \delta(x-x')$, $\braket{p|p'} = \delta(p-p')$, $\braket{x|p}= (2\pi)^{-1/2} e^{\pm ipx}$, and which allow one to evaluate any scalar products $\braket{\phi|\psi}$ or operator transformations $\braket{\phi|A|\psi}$ in terms of the $\{ \ket{x} \}$ or $\{ \ket{p} \}$ bases.

\textbf{$\bullet$ Field operators as distributions and applications to 2-point functions:}

Similarly, in the context of QFT, it occurs that the fundamental observables (namely, field configurations or canonical momenta configurations) do not have proper eigenstates in their respective Hilbert Spaces. Nonetheless, one often just writes

\begin{align}
\ket{\phi'}:& \; \phi(x) \ket{\phi'} = \phi'(x) \ket{\phi'}, \quad \forall x\in \Sigma, \\
\ket{\pi'}:& \; \pi(x) \ket{\pi'} = \pi'(x) \ket{\pi'}, \quad \forall x\in \Sigma,
\end{align}
where we used primes $'$ to distinguish between the field (momentum) operator $\phi(x)$ ($\pi(x)$) -- evaluated at any event $x$ belonging to a certain Cauchy surface\footnote{Note that we cannot have a \textit{single} eingenstate of the field operator $\phi(x)$ \textit{in all spacetime}, just like we cannot have a wave-function perfectly localized \textit{at all times}: this would amount to having a sharp (classical) trajectory for $\phi$. Determining it (\textit{or} its momentum) at an entire simultaneity surface $\Sigma$ corresponds to obtaining its maximal information in a quantum description.} $\Sigma$ -- and its eigenvalue $\phi'(x)$ ($\pi'(x)$).

To be more precise, the quantized field observable $\phi(x)$ is actually an \emph{operator-valued distribution} (such that $\braket{\Psi_1|\phi(x)|\Psi_2}$ is an ordinary (number-valued) distribution, $\forall \ket{\Psi_1}, \ket{\Psi_2} \in \mathcal{H}$). (Further considerations an implications of that point can be found in chapter 3 of \cite{fulling})

There are some basic operations that one may perform with distributions. Sum and mulplication by scalars are the most elementary ones, stemming directly from the vector space structure of $\mathcal{D}$. Further uselful operations that one may perform with distributions are \emph{tensor products}, \emph{convolutions} and \emph{taking derivatives}. All of them are well-defined and quite intuitive to handle operationally by recurring to their function correspondents; we refer the reader to \cite{friedlander} for a more detailed definition and examples of each operation (see, respectively, chapters 2, 4 and 5).

Therefore, we have that bilinear objects such as $\phi(x)\phi(x')$, or any combination of them that involves derivatives (such as the terms that appear in the two-point stress tensor $T_{\mu\nu}(x,x')$), are generally well-defined as distributions. They will often show a singular behaviour when one attempts to evaluate them as $x\rightarrow x'$, whereas they are usually regular for $x\neq x'$. Such a behaviour is not at all surprising when we think of the paradigmatic example of Dirac deltas:
If we define a test function $F \in \mathcal{F} \otimes \mathcal{F}$, it is perfectly sensible to evaluate the `double delta' distribution $\Delta(x,x') \equiv \delta_x \otimes \delta_{x'} \in \mathcal{D} \otimes \mathcal{D}$:

\begin{align}
\Delta(x,x')[F] &\equiv \iint d\mu(y) d\mu(y') F(y,y') \delta(x,y) \delta(x',y') \\
 &= F(x,x').
\end{align}

It is also perfectly sensible to evaluate the convolution of two Dirac deltas, $(\delta * \delta)_x$, in a test function $f\in \mathcal{F}$:

\begin{align}
(\delta * \delta)_x[f] &\equiv \int dy f(x-y) (\delta*\delta)(y) \nonumber \\
 &\equiv \iint dy dz f(x-y) \delta(y-z) \delta(z) \nonumber \\
 &= \int dy f(x-y) \delta(y) \nonumber \\
 &= f(x).
\end{align}

So one immediately finds that $(\delta*\delta)_x = \delta_x$ (of course, this last equality is not true for \emph{any} distribution $\sigma$, \textit{i.e.} $\sigma_x \not\equiv (\sigma*\sigma)_x$). Now, \emph{the product of 2 distributions is not generally defined}. Although for any two functions, $g$, and $u$, one could evaluate their action in a test function $f$ like

\begin{align}
(gu)[f] = \int dx \, g(x) u(x) f(x),
\end{align}
the same is \emph{not true} for two distributions $\sigma$ and $\theta$, for which it generally makes no sense to evaluate

\begin{align}
(\sigma \theta)[f] = \int \sigma(x) \theta(x) f(x),
\end{align}
obvious exceptions being the case where one of these distributions can be identified with a function (or when $\operatorname{supp}(\sigma)\cap\operatorname{supp}(\theta) = \emptyset$, for which we could trivially define $\sigma\theta = 0$). Particularly, there is no direct way to make sense of an object like $\delta^2(x)$.

 Then, it should not be surprising that the attempt to directly evaluate expected values such as $\braket{ \Psi | \phi^2(x) | \Psi}$ does not make direct sense, and generally yields divergent results. Much more appaling is the fact that, for a quite large variety of field theories, these divergencies can actually be systematically handled and subtracted to yield meaningful finite physical results (although these procedures often require very sophisticated techniques and cumbersome calculations, and they are not generally free from ambiguities).

\textbf{$\bullet$ Discontinuities and singularities; principal value of distributions:}

In field theory, it is also not unusual that one must handle distributions involving integrals that go directly through singularities in their integrands. In such cases, there is a variety of ways through which one may obtain a meaningful value of the integration, giving rise to ambiguities to such singular distributions. Among the many ways to \textit{define} the action of singular distributions, one conventional one is their so-called \textit{principal value}. Ultimately, evaluating the principal value is one convenient way to cancel out infinities and obtain meaningful finite results; here, we shall give just a superficial glimpse in the subject, applying it to simple distributions that appear in this dissertation (again, we refer the reader to chapter 2 of \cite{friedlander} for a more rigurous and thorough exposition of the subject)

A case of particular interest to us will be calculating the principal value of integrals around order 1 poles. Thus, for a start, let us consider a functional $\sigma \in \mathcal{D}(\mathcal{C}^\infty_0)$ defined through the function $1/x$, which has a singularity at $x=0$. How, then, should we interpret its action on a test function $f \in \mathcal{C}_0^\infty$, $\sigma[f]$? Well, since $\frac{1}{x}$ can be written as $\frac{d}{dx}\ln|x|$ for $x\neq0$, \emph{one particular} way to evaluate it is:

\begin{align}
\sigma[f] &= \int_{-\infty}^{\infty} \!\!dx \frac{f(x)}{x} = - \int_{-\infty}^{\infty} \!\!dx f'(x)\ln|x| \nonumber \\[6pt]
 &\equiv - \lim_{\epsilon \rightarrow 0^+} \biggl[ \int_{-\infty}^{-\epsilon} \!\!dx f'(x)\ln(-x) + \int_\epsilon^{\infty} \!\!dx f'(x)\ln(x)  \biggl] \nonumber \\[6pt]
 &=  \lim_{\epsilon \rightarrow 0^+} \biggl[ \cancelto{\hspace{-65pt} \mathcal{O}(\epsilon\ln\epsilon) \rightarrow 0}{\bigl( f(\epsilon) - f(-\epsilon) \bigl)\ln(\epsilon)} + \int_{-\infty}^{-\epsilon} \!\!dx \frac{f(x)}{x} + \int_\epsilon^{\infty} \!\!dx \frac{f(x)}{x}  \biggl]  \nonumber \\[6pt]
 &=  \lim_{\epsilon \rightarrow 0^+} \biggl[ \int_{-\infty}^{-\epsilon} \!\!dx \frac{f(x)}{x} + \int_\epsilon^{\infty} \!\!dx \frac{f(x)}{x}  \biggl] \nonumber \\[6pt]
 &\equiv \mathcal{P} \left( \int_{-\infty}^{\infty} dx \frac{f(x)}{x} \right),
\end{align}
which is how we \emph{define} the principal value of $\frac{1}{x}$, $\mathcal{P}\bigl( \frac{1}{x} \bigl)$.

Note that (i) the second equality is a particular (arbitrary) way to take the limit in the domain around $x=0$ (the results could be different if we approached $0$ at different rates from the left and the right) and (ii) since the logarithm is also singular at $x=0$ (although it is the derivative of a piecewise continuous function) this integration by parts is also not free of ambiguities. For instance, we could have defined:

\begin{empheq}[left=\empheqlbrace , right={ ,}]{align*}
 \frac{1}{x} &= \frac{d}{dx}\ln(-x), \; &\hspace{-100pt} x<0 \\[4pt]
 \frac{1}{x} &= \frac{d}{dx}\ln(x) + a, &\hspace{-100pt} x>0
\end{empheq}
so that the same procedure would yield

\begin{align}
\sigma[f] &= \int_{-\infty}^{\infty} \!\!dx \frac{f(x)}{x} = - \int_{-\infty}^{0} \!\!dx f'(x)\ln(-x) + \int_0^\infty dx f'(x) \bigl( \ln x + a \bigl) \nonumber \\[6pt]
 &\equiv - \lim_{\epsilon \rightarrow 0^+} \biggl[ \int_{-\infty}^{-\epsilon} \!\!dx f'(x)\ln(-x) + \int_\epsilon^{\infty} \!\!dx f'(x)\ln(x) + a\int_\epsilon^{\infty} \!\!dx f'(x)  \biggl] \nonumber \\[6pt]
 &=  \lim_{\epsilon \rightarrow 0^+} \biggl[ \int_{-\infty}^{-\epsilon} \!\!dx \frac{f(x)}{x} + \int_\epsilon^{\infty} \!\!dx \frac{f(x)}{x}  \biggl] + af(0) \nonumber \\[6pt]
 &\equiv \mathcal{P} \left( \int_{-\infty}^{\infty} dx \frac{f(x)}{x} \right) +af(0).
\end{align}

This gives us the distribution $\sigma(x) \equiv \mathcal{P}(\frac{1}{x})+ a\delta(x)$. Since taking the value $a=0$ is just an arbitrary choice as any other value, so will be the result of the distribution that we try to associate to $1/x$, in regards to its singular region. The point here is that the behaviour of the function in its nonsigular region does not uniquely determine a distribution, and additional information regarding its singular region may be required to define it. For a order 1 pole, the particular way of defining its principal value is adding up the divergent contributions aroud 0 symmetrically, so that they cancel out.

Another interesting application of considering a $1/x$ distribution as the derivative of $ln(x)$ emerges when we consider functions in the complex plane. Let $z$ be a complex number, $z=x+iy=|z|e^{i\arg(z)}$, so that its logarithm is defined as

\begin{align}
\ln(z) = \ln|z| + i\arg(z).
\end{align}

Then, if we take the limit $y\rightarrow 0^+$, we have for all finite $x$ that

\begin{align}
\lim_{\;y\rightarrow 0^+} \ln(z) = \ln|x| + i\pi(1-\Theta(x)) \qquad \Rightarrow \qquad  \frac{d}{dx} \Bigl( \lim_{\;y\rightarrow 0^+} \ln(z) \Bigl) = \frac{1}{x} - i\pi\delta(x).
\end{align}

Similarly, for negative values of $y$:

\begin{align}
\lim_{\;y\rightarrow 0^-} \ln(z) = \ln|x| - i\pi(1-\Theta(x)) \qquad \Rightarrow \qquad  \frac{d}{dx} \Bigl( \lim_{\;y\rightarrow 0^-} \ln(z) \Bigl) = \frac{1}{x} + i\pi\delta(x).
\end{align}

We can then use the complex identity $z^{-1}=\frac{d}{dz}\ln(z)$ to obtain the following distribution in the reals:

\begin{align}
\sigma(x) = \lim_{\epsilon \rightarrow 0^+}\frac{1}{x\pm i\epsilon} \equiv \mathcal{P}\left(\frac{1}{x}\right) \mp i \pi \delta(x). \label{1/iepsilon}
\end{align}

With this identity, one can analyze (particular values of) integrals along the real axis, by displacing their poles infinitesimally in the complex plane (either above or below the axis, depending on the application at hand).

\chapter{Some geometrical derivations} \label{geometry}

Most texts in QFTCS already assume the reader to be familiar, to a fair extent, with both QFT in Minkowski spacetime and General Relativity. In this work, while we do provide a full introductory chapter to QFT, we shall not present a thorough and comprehensive introduction to GR (for that, we refer the reader to the excellent textbook of R. Wald \cite{wald}, where this author personally learned the subject; alternatively, see \cite{hawkellis}). Still, the need was felt to provide an appendix discussing some fundamentals and covering more specific geometric derivations. It should serve both to lay the basic definitions and notations, and to explicitly develop some useful tools for our discussion in the main text, avoiding gaps in our derivations. Besides defining the fundamental geometrical objects used in the formulation of the theory, such as curvature and covariant derivatives, the topics in this appendix include an introduction to the computations of variations in respect to the metric, as well as a few useful geometrical structures, like Lie derivatives, Killing fields, and conformal transformations.

\section{Fundamental building blocks of GR}

The theory of General Relavity, whose original formulation was culminated in Einstein's work, succeeded to incorporate two very simple founding physical principles\footnote{Namely, the local invariance of the speed of light and the equivalence principle.}
 in a geometrical formalism for spacetime. In this formulation, spacetime came to be conceived as a curved, pseudo-Riemannian manifold, whose dynamics are governed by the matter propagating in it.
 Of course, one can trace the roots of this theory back to the simpler geometrical formulation of special relativity, built in Minkowski spacetime $(\mathbb{R}^4, \eta_{ab})$, from which one can find a generalization in curved spacetimes $(\mathcal{M},g_{ab})$, suitable to general relativity.

However, unlike Minkowski spacetime (or even prerelativistic Galilean spacetime), whose affine structure allows for fairly simple geometrical constructions and manipulations with little more than linear algebra and calculus tools, in curved spaces one requires a quite more sophisticated paraphernalia from differential geometry to carry various relevant calculations.

To start with, one can no longer use a single vector space structure to define vectors $v^a$, dual vectors (also called \textit{covectors}) $\omega_a$ and general tensors $T^{abc...}_{\;def...}$ \emph{in the entire spacetime}. Instead, one must build \emph{tangent spaces to each event $x\in\mathcal{M}$}, $\mathbb{V}_x$, and work with tangent vectors $v^a(x)$, dual vectors (also called \emph{cotangent vectors}) $\omega_a(x)$ and general tensors $T^{abc...}_{\;def...}(x)$, with no natural identification between $\mathbb{V}_x$ and $\mathbb{V}_{x'}$ for two distinct events $x\neq x'$.\footnote{Here, we are using the \emph{abstract index notation} (see List of Symbols). Throughout this work, we often switch between concrete (greek) indices and abstract ones (latin, from \textit{a} to \textit{h}), using the former more often in the context of quantum field theory and the latter in `purely geometrical' contexts, maintaining similarities with the literature.}
 This, on its turn, prevents one from having a \emph{unique} geometrical notion of derivatives for tensor fields, which must be specified through further physical postulates. First, one usually requires that it acts symmetrically on scalars, that is:
 
\begin{align}
\nabla_{\!a}\nabla_{\!b} \phi - \nabla_{\!b}\nabla_{\!a} \phi = 0. \label{torsionfree}
\end{align}

Since generally one could have that

\begin{align}
\nabla_{\!a} \nabla_{\!b} \phi - \nabla_{\!b} \nabla_{\!a} \phi = T_{ab}\phi,
\end{align}
being $T_{ab}=-T_{ba}$ the so-called the \emph{torsion tensor}, this is referred to as the \textit{null torsion} condition (or, equivalently, one says that $(\mathcal{M},g_{ab})$ is a \textit{torsion-free space}).
 
Generally, two derivative operators $\nabla_{\!a}$ and $\tilde{\nabla}_{\!a}$ may differ in the following way:

\begin{align}
\tilde{\nabla_{\!b}} v^a = \nabla_{\!b} v^a + C^a_{\;bc} v^c, \label{connection}
\end{align}
where $C^a_{\;bc}$ is called a \textit{connection}. For torsion free spaces, it will be symmetrical in the lower indices: $C^a_{\;bc} = C^a_{\;cb}$.

Thus, the fundamental objects in GR are the metric field $g_{ab}$ and a preferred derivative operator $\nabla_a$ (or, equivalently, a preferred connection). In the standard formulation of GR, motivated by the equivalence principle, one requires the physical derivative operator -- the so-called covariant derivative $\nabla_{\!a}$ -- to be that with respect to which local variations of the metric vanish:

\begin{align}
\nabla_{\!c} \,g_{ab} = 0. \label{metricity}
\end{align}

That is not to say the metric is spacetime homogeneous, but rather that a fundamental notion of a locally nonvarying quantity is defined in closed proximity with it\footnote{In section \ref{liekillcon}, when we define Lie derivatives and construct the notion of continuous isometry groups, we shall ascribe a clearer meaning to the notions of variations of the metric along spacetime from a purely geometrical perspective. }.

Although the formulation so far has been carried in a coordinate-free way, one must often adopt a coordinate system to carry calculations. In practice, one must be able to compute covariant derivatives in terms of the metric components $g_{\mu\nu}$ and ordinary coordinate derivatives $\partial_\mu$. These can be calculated as a particular instance of \eqref{connection}, using the Christoffel Symbols $\Gamma^a_{\;bc}$:

\begin{align}
\nabla_{\!\mu} v^\nu = \partial_\mu v^\nu + \Gamma^\nu_{\mu \alpha} v^\alpha, \label{cconnection}
\end{align}
where $\Gamma^\alpha_{\;\mu\nu}$ can be calculated as a function of metric components as

\begin{align}
\Gamma^\alpha_{\;\mu\nu} = \tfrac{1}{2}g^{\alpha\beta} \bigl[ \partial_{\mu}g_{\nu\beta} + \partial_{\nu}g_{\mu\beta} - \partial_{\beta}g_{\mu\nu} \bigl].
\end{align}

For practical computations, it is very useful to write a contraction of these symbols in termos of \emph{the determinant} of the matrix of metric components, $g$, which reads

\begin{align}
\Gamma^\alpha_{\;\,\alpha\mu} = \tfrac{1}{2} g^{\alpha\beta} \partial_\mu g_{\alpha \beta} = \tfrac{1}{2} g^{-1} \partial_\mu g = \partial_\mu \ln|g|^{\frac{1}{2}} = |g|^{-\frac{1}{2}} \partial_\mu |g|^{\frac{1}{2}}.
\end{align}

The most useful application of this formula for us will be in computing the D'Alembertian for a scalar field. Note that, for a scalar field, while first derivatives will be simply given by partial derivatives, second derivatives will involve Christoffel symbols. We can then write

\begin{align}
\Box \phi \equiv g^{\mu\nu} \nabla_{\!\mu} \nabla_{\!\nu} \phi &= \nabla_{\!\mu} \bigl( g^{\mu\nu} \partial_\nu \phi \bigl) \nonumber \\
  &= \partial_\mu \bigl( g^{\mu\nu} \partial_\nu \phi \bigl) + \Gamma^{\mu}_{\;\, \mu \alpha} g^{\alpha \nu} \partial_\nu \phi \nonumber \\ 
  &= \partial_\mu \bigl( g^{\mu\nu} \partial_\nu \phi \bigl) + g^{\alpha \nu} \partial_\nu \phi \bigl( |g|^{-\frac{1}{2}} \partial_\alpha |g|^{\frac{1}{2}} \bigl) \nonumber \\
  &= |g|^{-\frac{1}{2}}\partial_\mu \bigl( |g|^{\frac{1}{2}} g^{\mu\nu} \partial_\nu \phi \bigl) \label{Dalembertianphi}
\end{align}

Now, we would like to construct an intrinsic notion of curvature for our spacetime, defined uniquely by the metric $g_{ab}$. Ultimately, this curvature will refer to \emph{derivatives} of the metric (up to second order). This may sound very strange at this point, since we postulated the covariant derivative of the metric to be identically null in \eqref{metricity}. It happens that this equation precisely defines how the physical notion of derivatives depend on the metric. Thus, we define the notion of curvature indirectly, through covariant derivatives. To start with, we define the Riemann curvature tensor\footnote{Conventions may somewhat vary in the literature, due both to the choice of metric signature and conventions on different indices.} by

\begin{align}
\nabla_{\!a} \nabla_{\!b} \,\omega_c - \nabla_{\!b} \nabla_{\!a} \,\omega_c = -R_{abc}^{\;\;\;\;\;d} \omega_d, \label{riemman}
\end{align}
so that, indirectly, one can uniquely associate a curvature tensor to a given metric, by equations (\ref{metricity}) and (\ref{riemman}). 

One can also cumpute the curvature components in terms of Christoffel Symbols (that is,  in terms of partial derivatives of the metric components) as

\begin{align}
R_{\mu\nu\alpha}^{\;\;\;\;\;\;\beta} = 2\partial_{[\mu} \,\Gamma^\beta_{\;\,\nu]\alpha} + 2\Gamma^\lambda_{\;\,\alpha[\nu} \,\Gamma^\beta_{\;\,\mu]\lambda}.
\end{align}

It is also worth listing here a few of the symmetries and identities obeyed by the Riemann tensor. They can each be worked with some algebraic effort, and we shall state them without proofs. First we have three independent antissymmetry properties:

\begin{subequations}
\begin{align}
1. &\; R_{abcd} = - R_{bacd}, \\
2. &\; R_{abcd} = - R_{abdc}, \\
3. &\; R_{[abc]d} = 0
\end{align}
(where $2$ makes explicit use that $\nabla_{\!a}$ is the covariant derivative in $(\mathcal{M},g_{ab})$). Together, they imply the following symmetry:

\begin{align}
R_{abcd} = R_{cdab}. \label{Riemman pairs}
\end{align}

Finally, we state the Bianchi identity:

\begin{align}
\nabla_{\![a}R_{bc]de} = 0.
\end{align}

\end{subequations}

Contractions of the Riemann tensor play a key role in GR. Because of all its antissymetry properties, there is only one independent rank $(0,2)$ contraction, which will be the Ricci tensor. We then define the Ricci curvature $R_{ab}$ and the curvature scalar as $R$ as

\begin{align}
R_{ac} &\equiv R_{abc}^{\;\;\;\;\;b}, \\
R &\equiv g^{ac}R_{ac}.
\end{align}

Note that in virtue of eq \eqref{Riemman pairs}, $R_{ab} = R_{ba}$. An important combination of $R_{ab}$ and $R$ is the so-called \emph{Einstein tensor}, defined as

\begin{align}
G_{ab} = R_{ab} - \tfrac{1}{2}R g_{ab}.
\end{align}

It is this tensor that will appear at the left side of Einstein equations. Finally, we note that the Bianchi identity will imply that $G_{ab}$ is covariantly conserved:

\begin{align}
\nabla_{\!a} G^{ab} = 0.
\end{align}

\section{Variations with respect to the metric}

As we have seen thus far, the construction of geometrical quantities from the metric is often very indirect, being determined by the particular way that it fixates the covariant derivative in our spacetime. Thus, it is useful to compile a few results for the systematic computation of variations of these quantities as we vary the metric. Our main interest in computing these variations will be to derive functional derivatives of curvature tensors in the context of field theory. However, another very important and immediate application of such results lies in obtaining perturbative solutions for Einsteins' equations near some known solution, so we make our derivations directly in this latter context. Once we have the desired results at hand, we directly interpret them in terms of infinitesimal variations.

We start by considering a dynamic equation for a generic field variable $g$ (concretely, in our context of interest, this will be Einstein's Equations for the metric), which can be put in the form:

\begin{align}
\mathcal{E}(g) = 0, \label{exact}
\end{align}
where $\mathcal{E}$ is some local differential functional of $g$.

Let us supose now that we know an exact solution $\prescript{0}{}{g}$ of (\ref{exact}), and that we are considering a situation in which the deviation from a certain solution of interest, $g$, with respect to $\prescript{0}{}{g}$ is small. In fact, to express this assertion in a mathematically meaningful way, we assume the existence of a (1-parameter) family of exact solutions of (\ref{exact}), $g(\lambda)$, \textit{i.e.}

\begin{align}
\mathcal{E}\bigl( g(\lambda) \bigl) = 0, \qquad \forall \lambda \in I \label{1-par-family}
\end{align}
(where the domain $I \subset \mathbb{R}$ contains an open interval around $0$), such that:

\begin{empheq}[left= {\empheqlbrace \;}, right={\quad .}]{align}
  (i)& \, g(\lambda) \text{ is a differentiable function of } \lambda \\
  (ii)& \, g(0) = \prescript{0}{}{g}
\end{empheq}

One may then think of $\lambda$ as a parameter that quantifies the deviation of some (unknown) exact solution $g(\lambda)$ to our known solution $\prescript{0}{}{g}$. This way, for arbitrarily small values of $\lambda$, we obtain solutions that will be arbitrarily close to $\prescript{0}{}{g}$. However, since (\ref{exact}) may be too difficult \textbf{(or impossible)} to solve exactly, we may then obtain an approximate solution by noting that

\begin{align}
\frac{d}{d\lambda} \Bigl[ \mathcal{E}\bigl( g(\lambda) \bigl) \Bigl] = 0 \label{der1-par-family}
\end{align}
(since (\ref{1-par-family}) is valid for all $\lambda \in I$). Particularly, this equality must hold for $\lambda=0$. If we then define a perturbation in our field as $\delta g \equiv \lambda \gamma$, that is:

\begin{align}
\gamma \equiv \frac{dg(\lambda)}{d\lambda}\biggl{|}_{\lambda=0},
\end{align}
then, by Leibniz's rule, it is easy to see that (\ref{der1-par-family}) will give us a linear equation for $\gamma$, in the form

\begin{align}
\mathscr{L}(\gamma) = 0 ,\label{linearized}
\end{align}
where $\mathscr{L}$ is a linear local differentiable operator. One then calls (\ref{linearized}) the `linearization of (\ref{exact}) around $\prescript{0}{}{g}$'.

Then, our case of interest will be when $g$ represents the metric field $g_{ab}$ and $\mathcal{E}(g_{ab}) = G_{ab}$, such that \eqref{exact} will represent the exact Einstein Equations in the vacuum\footnote{ 
 For simplicity, our presentation will be focused on the vacuum case \emph{for the unperturbed equations}, but it is straightforward to generalize it to account for matter sources. }
 (which can actually be written simply as $R_{ab}=0$). In this case, let us explicitly derive the \emph{linearized} equations (\ref{linearized}) for the perturbation in the metric.

To achieve that, we must calculate the Ricci tensor $R_{ab}(\lambda)$ associated with the metric $g_{ab}(\lambda)$ in a useful expression. More specifically, we want an expression for it in terms of the background metric $\prescript{0}{}{g}_{ab}$ and with explicit algebraic functions of $\lambda$, so that we may clearly take derivatives with respect to it. The challenge in doing that lies in the fact that the curvature is only indirectly defined in respect to the metric, through covariant derivatives (\ref{riemman}), which are bound to obey eq (\ref{metricity}). We then begin by noting that the covariant derivatives $\prescript{\lambda}{}{\nabla}_{\!a}$ and $\prescript{0}{}{\nabla}_{\!a}$ ($\prescript{\lambda}{}{\nabla}_{\!a}g_{bc}(\lambda) = 0 = \prescript{0}{}{\nabla}_{\!a}\prescript{0}{}{g}_{bc}$) differ when acting on cotangent vectors by a connection in the form:

\begin{align}
\prescript{\lambda}{}{\nabla}_{\!a} \omega_b = \prescript{0}{}{\nabla}_{\!a} \omega_b - C^c_{\;ab}(\lambda)\omega_c .
\end{align}

If we then write $C^c_{\;ab}(\lambda)$ in terms of $\prescript{0}{}{\nabla}_{\!a}$ and $g_{ab}(\lambda)$, we obtain

\begin{align}
C^c_{\;ab}(\lambda) = \tfrac{1}{2}g^{cd} \bigl[ \prescript{0}{}{\nabla}_{\!a}g_{bd}(\lambda) + \prescript{0}{}{\nabla}_{\!b}g_{ad}(\lambda) - \prescript{0}{}{\nabla}_{\!d}g_{ab}(\lambda) \bigl].
\end{align}

Then, with a little algebraic effort, we can compute the Riemann curvature tensor:

\begin{align}
R_{abc}^{\;\;\;\;\;d}(\lambda) = \prescript{0}{}{R}_{abc}^{\;\;\;\;\;d} + 2 \prescript{0}{}{\nabla}_{\![a} C^d_{\;b]c} - 2 C^e_{\;c[a} C^d_{\;b]e},
\end{align}
from which we obtain the Ricci:

\begin{align}
R_{ac}(\lambda) = \prescript{0}{}{R}_{ac} + 2 \prescript{0}{}{\nabla}_{\![a} C^b_{\;b]c} - 2 C^e_{\;c[a} C^b_{\;b]e}.
\end{align}

Then, differentiating this entire expression with respect to $\lambda$ and evaluating it at $\lambda=0$, we obtain

\begin{align}
\dot{R}_{ac} =  2 \prescript{0}{}{\nabla}_{\![a} \dot{C}^b_{\;b]c}, \label{dot R}
\end{align}
where we are using a dot to denote a derivative with respect to $\lambda$ at $\lambda=0$. We note that the term quadratic in $C^c_{\;\,ab}$ will not yield any contribution, since $C^c_{\;\,ab}(\lambda\!=\!0)=0$.

Thus, denoting the metric derivatives by $\gamma$, $\gamma_{ab} \equiv \dot{g}_{ab}$, and by noting that $\prescript{0}{}{\nabla}_{\!a}\prescript{0}{}{g}_{bc}=0$, we can easily compute $\dot{C}^c_{\;ab}$, yielding

\begin{align}
\dot{C}^c_{\;ab} = \frac{1}{2}\prescript{0}{}{g}^{cd} \bigl[ \prescript{0}{}{\nabla}_{\!a}\gamma_{bd} + \prescript{0}{}{\nabla}_{\!b}\gamma_{ad} - \prescript{0}{}{\nabla}_{\!d}\gamma_{ab} \bigl].
\end{align}

Then, subtistituting on \eqref{dot R}, we obtain

\begin{align}
\dot{R}_{ac} = \frac{1}{2} \prescript{0}{}{g}^{bd} \bigl[ \prescript{0}{}{\nabla}_{\!a}\prescript{0}{}{\nabla}_{\!c}\gamma_{bd} + \prescript{0}{}{\nabla}_{\!b}\prescript{0}{}{\nabla}_{\!d}\gamma_{ac} - 2\prescript{0}{}{\nabla}_{\!b}\prescript{0}{}{\nabla}_{\!(c}\gamma_{a)d} \bigl].
\end{align}

At this point, we simplify our notation, dropping the prescript 0 for background quantities and using the background metric to raise and lower indices. In this notation, we obtain:

\begin{align}
\dot{R}_{ac} = \tfrac{1}{2} \nabla_{\!a}\nabla_{\!c}\gamma +\tfrac{1}{2} \nabla^b\nabla_{\!b}\gamma_{ac} - \nabla^b\nabla_{\!(c}\gamma_{a)b},
\end{align}
where we have defined $\gamma \equiv g^{ab}\gamma_{ab}$.

Now, if we multiply this entire equation by an infinitesimal variation $\lambda$, we obtain the form for infinitesimal variations of the curvature, in the familiar notation of chapter \ref{QFTCS}:

\begin{align}
\delta{R}_{ac} = \tfrac{1}{2} g^{bd}\nabla_{\!a}\nabla_{\!c}\delta g_{bd} +\tfrac{1}{2} \nabla^b\nabla_{\!b}\delta g_{ac} - \nabla^b\nabla_{\!(c}\delta g_{a)b} .
\end{align}

As a final remark, we note that variations of the inverse metric $g^{ab}$ are \emph{not} simply given by raising the indexes of $\delta g_{ab}$ with the unperturbed metric. We can calculate these by using that $g^{ab}g_{bc} = \delta^a_{\;c}$, and thus

\begin{align}
0 = \delta( g^{ab}g_{bc} ) = g_{bc} \delta g^{ab} + g^{ab} \delta g_{bc} \qquad \Rightarrow \qquad \delta g_{ab} = - g_{ad} g_{bc} \delta g^{dc}.
\end{align}

\section{Lie derivatives, Killing fields, and conformal transformations} \label{liekillcon}

We have already seen that a crucial operational toolbox to handle curved spaces (in the form of smooth manifolds $(\mathcal{M},g_{ab})$) is their differential structure. We have throughout been using tangent spaces at each point to define tensor fields and compute many local quantities in our theory. Correspondingly, when one must handle \emph{extensive} geometrical quantities and operations (such as finite arclengths or displacements, the parallel transport of vectors and tensors, etc.) one must develop a suitable integral structure to extend differential structures throughout spacetime.

In doing so, a central geometrical structure are \emph{vector fields}, which we can used to build integral orbits (curves) and meaningfully transport local quantities throughout spacetime. Let us consider a differentiable vector field $\xi^a$ defined on the spaces tangent to $\mathcal{M}$ at each event; we can use this field to find integral curves $C \subset \mathcal{M}$ (such that $\xi^a$ will be tangent to them at each event), and we can define a 1-parameter family of diffeomorphisms $\phi: \mathbb{R} \times \mathcal{M} \longrightarrow \mathcal{M}$ which act translating all points along these curves by a variable amount. More precisely, for any parameters $t,s \in \mathbb{R}$, we will have diffeomorphisms obeying

\begin{align}
\phi_{t+s} = \phi_t \circ \phi_s, \qquad \Rightarrow \qquad \phi_0 = \mathbb{1}_\mathcal{M}.
\end{align}

These diffeomorphisms will induce natural maps between tangent vectors (or, more generally, tensors) at point $p$ and tangent vectors (tensors) at points $\phi_t(p)$ along its orbits. These maps are called \textit{pushforwards} (as they ``push tensors forward'' from $p$ to $\phi_t(p)$) and they are denoted $\phi_t^*$:

\begin{align}
T^{a_1...a_l}_{b_1...b_m} \in \mathcal{T}_p(l,m) \;\longrightarrow\; \phi^*_t T^{a_1...a_l}_{b_1...b_m} \in \mathcal{T}_{\phi_t(p)}(l,m) .
\end{align}

With these maps, one can define a particular notion of a derivative to a tensor field along the integral orbits of $\xi^a$, the so-called Lie Derivative. Since we cannot in general subtract tensors defined at different points (\textit{i.e.}, at different tangent spaces), a way to evaluate their difference at the points $p$ and $\phi_t(p)$ is to ``pull the latter back to $p$'' by the induced map $\phi^*_{-t}$. The Lie derivative is thus defined as

\begin{align}
\pounds_\xi (T^{a_1...a_l}_{b_1...b_m})_p = \lim_{t \rightarrow 0} \biggl[ \frac{ (\phi^*_{-t}T^{a_1...a_l}_{b_1...b_m})_p - (T^{a_1...a_l}_{b_1...b_m})_p}{t} \biggl]. \label{lie def}
\end{align}

On the first sight, it may seem that this would just yield a directional derivative along $\xi^a$, $\xi^c\nabla_{\!c}T^{a_1...a_l}_{b_1...b_n}$. Note, however, that we have not imposed any restrictions in the magnitudes or orientations of $\xi^a$ throughout $\mathcal{M}$ (except that they should vary smoothly), such that fixed parameter displacement $t$ may produce displacements of varied magnitudes and directions throughout spacetime\footnote{
 This is particularly clear when we consider, for instance, a rotation: the magnitudes of the displacement for a fixed angular variation $\theta$ will produce larger displacements at larger radii, and go towards different directions for each position.}.
 Thus, a Lie derivative will generally carry information about the variations of $\xi^a$ as well, which will manifest in the form of terms proportional to its covariant derivative.
 
We shall not derive here how to obtain an expression for the Lie derivative in terms of purely geometrical operations (see appendix C of \cite{wald} for a complete and pedagogical derivation), but we quote here the result, which we shall use throughout this thesis:
 
\begin{align}
\pounds_\xi T^{a_1...a_l}_{b_1...b_m} = \xi^c \nabla_{\!c} T^{a_1...a_l}_{b_1...b_m} - \sum_{i=1}^l T^{a_1...c...a_l}_{b_1...b_m} \nabla_{\!c} \xi^{a_i} + \sum_{j=1}^m T^{a_1...a_l}_{b_1...c...b_m} \nabla_{\!b_j} \xi^c. \label{lie cov}
\end{align}

A particular important class of diffeomorphisms in $(\mathcal{M},g_{ab})$ are isometries, that is, transformations that leave the metric field $g_{ab}$ invariant. A vector field that generates a 1-parameter family of isometries $\phi_t^*$, $\phi^*_t g_{ab} = g_{ab}$ is called a \emph{Killing Field}. We immediately see from equation \eqref{lie def} that the Lie derivative of the metric along any Killing field vanishes. Then, eq \eqref{lie cov} yields

\begin{align}
\pounds_\xi g_{ab} = \nabla_{\!a} \xi_b + \nabla_{\!b} \xi_a = 2\nabla_{\!(a} \xi_{b)} = 0, \label{killing equation}
\end{align}
since $\nabla_c g_{ab} = 0$.

In fact, in terms of the Lie derivative in respect to a Killing field, we may then define a precise notion of what it means for the metric to remain constant or vary throughout spacetime, since its covariant derivative is trivially null. We can say that Minkowski spacetime, for instance, has a constant metric \emph{everywhere}, since it is maximally symmetric, and one can get from any event $p$ into any distinct event $q$ by following the integral orbits of a Killing field (\textit{i.e.} by following an isometry). In fact, the same is true for any homogeneous spacetimes, such as Einstein's Static Universe, or de Sitter spaces.

Equation \eqref{killing equation} is called the Killing equation, and, by solving it, one can find the generators of various isometries in one spacetime, if it has any. Another very important family of vector fields in a spacetime are the so-called \emph{conformal} Killing fields. They obey a relation similar to \eqref{killing equation}, called the conformal Killing equation:

\begin{align}
\pounds_\xi g_{ab} = 2\nabla_{\!(a} \xi_{b)} = \lambda g_{ab}. \label{conformal killing equation}
\end{align}

That is, the metric can only vary along a conformal Killing field parallel to itself. Thus, the associated integral transformations, called \emph{conformal transformations}\footnote{In the literature, they are sometimes referred to as \textit{conformal isometries}, but we avoid this term since it may be misleading.} will at most stretch or contract the metric, but will always preserve angles. We note, however, that not all conformal transformations (and not all isometries) must belong to a continuous group generated by a (conformal) Killing field. Generally, one can write a conformal transformation $\psi$ as a spacetime diffeomorphism, whose induced map in the metric ($\psi^*g_{ab}$) will act in the form

\begin{align}
g_{ab}(x) \longrightarrow \tilde{g}_{ab}(x) = \Omega^2(x)g_{ab}(x), \label{conformal transformation}
\end{align}
where $\Omega^2(x)>0$ is a positive function of spacetime.

Conformal transformations are particularly useful in GR because they allow one to distort spacetime distances while preserving all angles. Particularly, it is obvious from \eqref{conformal transformation} that a vector will be spacelike, timelike or null with respect to $\tilde{g}_{ab}$ if, and only if, it is respectively spacelike, timelike or null with respect to $g_{ab}$. This means that any two conformally related spacetimes will have the same causal structure, even though they may have widely different geometries.

For the aforementioned reasons, it can be often more convenient to carry a geometrical or dynamical analysis originally defined in one spacetime in another, conformally related one. For such, it is useful to write geometric tensors from one spacetime in terms of the other. If $\nabla_a$ and $\tilde{\nabla}_a$ are the covariant derivatives related to $g_{ab}$ and $\tilde{g}_{ab}$, respectively, we may express their difference through a connection $C^c_{\;ab}$, defined by

\begin{align}
\tilde{\nabla}_{\!a} \omega_b = \nabla_{\!a}\omega_b - C^c_{\;ab}\omega_c,
\end{align}
such that it can be written as

\begin{align}
C^c_{\;ab} &= \tfrac{1}{2}\tilde{g}^{cd} \bigl[ {\nabla}_{\!a}\tilde{g}_{bd} + {\nabla}_{\!b}\tilde{g}_{ad} - {\nabla}_{\!d}\tilde{g}_{ab} \bigl] \nonumber \\
 &= \Omega^{-1} \bigl[ 2 \delta^c_{(a}\nabla_{\!b)} \Omega - g_{ab} g^{cd} \nabla_d \Omega \bigl].
\end{align}

From this connection, one can derive with some algebraic effort the values of the tensor curvatures $\tilde{R}_{abc}^{\;\;\;\;d}$, $\tilde{R}_{ab}$ and $\tilde{R}$ in terms of their conformally related counterparts and $\Omega$. Particularly, we will be interested in the Ricci curvature and the curvature scale, which read

\begin{align}
  \tilde{R}^b_{\;a} &= \; \Omega^{-2} \bigl\{ R^b_{\;a} - (n-2)g^{bc}\nabla_{\!c}\nabla_{\!a}(\ln\Omega) - \delta^b_{\;a} g^{cd}\nabla_{\!c}\nabla_{\!d}(\ln\Omega) \nonumber \\ 
&  \quad +(n-2)g^{bc}(\nabla_{\!c}\ln\Omega)(\nabla_{\!a}\ln\Omega) - (n-2)\delta^b_{\;a} g^{cd}(\nabla_{\!c}\ln\Omega)(\nabla_{\!d}\ln\Omega) \bigl\}, \\[4pt]
  \tilde{R} \;\; &= \;\Omega^{-2} \bigl\{ R - 2(n-1)g^{cd}\nabla_{\!c}\nabla_{\!d}\ln\Omega - (n-1)(n-2)g^{cd}(\nabla_{\!c}\ln\Omega)(\nabla_{\!d}\ln\Omega) \bigl\}.
\end{align}

Finally, we would like to analyze how our field equations transform upon a conformal transformation. Since we are often interested in working in simpler, conformally related spacetimes than that of direct interest to our problem, it would be particularly convenient to know if there is some conformal scaling that we may perform in our field, $\phi \rightarrow \tilde{\phi} = \Omega^s \phi$ (where $s$ is called a conformal weight), so that the \emph{form} of the field equations remains invariant. More precisely, if our field equations are defined by spacetime differential operator $L_x$, we would like to find an invariance in the form

\begin{align}
L_x \phi(x) = 0 \quad \Leftrightarrow \quad \tilde{L}_x \tilde{\phi}(x) = 0,
\end{align} 
where $\tilde{L}_x$ is the conformally transformed operator.

A particularly simple case is that of massless scalar field for which $L = \Box = g^{ab} \nabla_{\!a} \nabla_{\!b}$. We see that a conformally transformed equation would read

\begin{align}
0 = \tilde{g}^{ab} \tilde{\nabla}_{\!a} \tilde{\nabla}_{\!b} \tilde{\phi} &= \Omega^{-2} g^{ab} \bigl[ \nabla_{\!a}\nabla_{\!b}(\Omega^s \phi) - C^c_{\;ab}\nabla_{\!c}(\Omega^s \phi) \bigl] \nonumber \\
 &= \Omega^{s-2} g^{ab}\nabla_{\!a}\nabla_{\!b}\phi + (2s+n-2)\Omega^{s-3}g^{ab}\nabla_{\!a} \Omega \nabla_{\!b} \phi  \\ & \quad  + s \Omega^{s-3} \phi g^{ab}\nabla_{\!a}\nabla_{\!b}\Omega \nonumber  +  s(n+s-3)\Omega^{s-4}\phi g^{ab} \nabla_{\!a}\Omega \nabla_{\!b} \Omega.
\end{align}

For this equation to be made equivalent with $g^{ab} \nabla_{\!a} \nabla_{\!b} \phi = 0$, we must require that all terms except the first identically vanish. Well, we can immediately see that, for $n \neq 2$, there will be no choice of $s$ that allows for such cancelling. However, a very convenient covariant way of modifying this field equation (which furthermore recovers the same field equation $\Box \phi = 0$ in flat spaces) is by adding a coupling with scalar curvature, making

\begin{align}
L = g^{ab}\nabla_{\!a} \nabla_{\!b} + \xi R, \qquad \xi = cte.
\end{align}

In this case, one can verify that there is indeed a special value of $\xi$ (which will depend on $n$) that will produce a convenient cancellation of terms in the modified equation, namely:

\begin{align}
\xi(n) = \frac{n-2}{4(n-1)}. \label{conformal coupling}
\end{align}

With this value, we can put our equations in a conformally invariant form by choosing the appropriate conformal weight $s= \frac{2-n}{2}$:

\begin{align}
\Bigl[ \Box + \frac{n-2}{4(n-1)}R(x) \Bigl] \phi(x) \rightarrow \Bigl[ \tilde{\Box} + \frac{n-2}{4(n-1)}\tilde{R}(x) \Bigl] \tilde{\phi}(x) = \Omega^{s-2}(x) \Bigl[ \Box + \frac{n-2}{4(n-1)}R(x) \Bigl] \phi(x).
\end{align}

Particularly, for $n=4$, we have $\xi(4)=1/6$ and $s=-1$.

Finally, we note that we cannot maintain conformal invariance if we add a mass term to our equation (unlike the other two terms it will scale as $\Omega^s$, not $\Omega^{s-2}$). One can interpret this fact by noting that $m$ will introduce a natural (inverse) length scale to the theory; since we keep this scale unchanged when we operate a conformal transformation (which distorts distances), this will necessarily provoke a nontrivial distortion in relative scales in our theory.

\end{apendicesenv}